%% file: boost2013_report.tex
\documentclass[pdftex,twocolumn,epjc3]{svjour3}          

\RequirePackage[T1]{fontenc}

\smartqed  

\RequirePackage{graphicx}
\RequirePackage{mathptmx}      
\RequirePackage{flushend}
\RequirePackage[numbers,sort&compress]{natbib}
\RequirePackage[colorlinks,citecolor=blue,urlcolor=blue,linkcolor=blue]{hyperref}

\RequirePackage{amsmath}
\RequirePackage{amssymb}
\RequirePackage{xspace}
\RequirePackage{subfigure}
\RequirePackage{mathrsfs}
\RequirePackage{lineno}
\RequirePackage{authblk}

\journalname{Eur. Phys. J. C}

\def\TeV{\ifmmode {\mathrm{\ Te\kern -0.1em V}}\else
                   \textrm{Te\kern -0.1em V}\xspace\fi}%
\def\GeV{\ifmmode {\mathrm{\ Ge\kern -0.1em V}}\else
                   \textrm{Ge\kern -0.1em V}\xspace\fi}%
\def\MeV{\ifmmode {\mathrm{\ Me\kern -0.1em V}}\else
                   \textrm{Me\kern -0.1em V}\xspace\fi}%
\def\keV{\ifmmode {\mathrm{\ ke\kern -0.1em V}}\else
                   \textrm{ke\kern -0.1em V}\xspace\fi}%
\def\eV{\ifmmode  {\mathrm{\ e\kern -0.1em V}}\else
                   \textrm{e\kern -0.1em V}\xspace\fi}%
\let\tev=\TeV

\newcommand{\pt}{\ensuremath{p_T}\xspace}

\newcommand{\pT}{\ensuremath{p_{\mathrm{T}}}\xspace}

\newcommand{\ptd}{\ensuremath{p_TD}\xspace}

\newcommand{\mtrim}{\ensuremath{m_{\text{trim}}}\xspace}
\newcommand{\mprun}{\ensuremath{m_{\text{prun}}}\xspace}
\newcommand{\mmdt}{\ensuremath{m_{\text{mmdt}}}\xspace}
\newcommand{\msd}{\ensuremath{m_{\mathrm{sd}}}\xspace}
\newcommand{\wmass}{\ensuremath{m_{W}}}
\newcommand{\topmass}{\ensuremath{m_{t}}}
\newcommand{\tautwoone}{\ensuremath{\tau_{21}^{\beta=1}}\xspace}
\newcommand{\tauthreetwo}{\ensuremath{\tau_{32}^{\beta=1}}\xspace}

\newcommand{\kT}{\ensuremath{k_T}\xspace}
\newcommand{\antikt}{{\rm anti-}\kT}

\title{Towards an Understanding of the Correlations in
Jet Substructure}

\subtitle{Report of BOOST2013, hosted by the University of Arizona, 12$^{th}$-16$^{th}$ of August 2013.}
\titlerunning{Boosted objects at the LHC}
\authorrunning{BOOST2013 participants} 

\author[1]{\mbox{D. Adams}}
\author[2]{\mbox{A. Arce}} 
\author[3]{\mbox{L. Asquith}} 
\author[4]{\mbox{M. Backovic}} 
\author[5]{\mbox{T. Barillari}} 
\author[6]{\mbox{P. Berta}} 
\author[7]{\mbox{D. Bertolini}} 
\author[8]{\mbox{A. Buckley}} 
\author[9]{\mbox{J. Butterworth}}
\author[10]{\mbox{R.~C. Camacho Toro}} 
\author[11]{\mbox{J. Caudron}} 
\author[12]{\mbox{Y.-T. Chien}} 
\author[13]{\mbox{J. Cogan}}
\author[9]{\mbox{B. Cooper}} 
\author[14]{\mbox{D. Curtin}} 
\author[15]{\mbox{C. Debenedetti}} 
\author[16]{\mbox{J. Dolen}}
\author[17]{\mbox{M. Eklund}}
\author[11]{\mbox{S. El Hedri}} 
\author[18]{\mbox{S.~D. Ellis}}  
\author[17]{\mbox{T. Embry}} 
\author[19]{\mbox{D. Ferencek}} 
\author[8]{\mbox{J. Ferrando}} 
\author[20]{\mbox{S. Fleischmann}} 
\author[21]{\mbox{M. Freytsis}}  
\author[22]{\mbox{M. Giulini}} 
\author[23]{\mbox{Z. Han}} 
\author[24]{\mbox{D. Hare}} 
\author[25]{\mbox{P. Harris}}
\author[26]{\mbox{A. Hinzmann}} 
\author[27]{\mbox{R. Hoing}} 
\author[12]{\mbox{A. Hornig}} 
\author[28]{\mbox{M. Jankowiak}} 
\author[17]{\mbox{K. Johns}} 
\author[29]{\mbox{G. Kasieczka}} 
\author[27]{\mbox{R. Kogler}} 
\author[17]{\mbox{W. Lampl}}
\author[30]{\mbox{A.~J. Larkoski}} 
\author[12]{\mbox{C. Lee}} 
\author[17]{\mbox{R. Leone}} 
\author[17]{\mbox{P. Loch}} 
\author[21]{\mbox{D. Lopez Mateos}}
\author[31]{\mbox{H. K. Lou}}
\author[32]{\mbox{M. Low}} 
\author[33]{\mbox{P. Maksimovic}}
\author[27]{\mbox{I. Marchesini}}
\author[30]{\mbox{S. Marzani}} 
\author[11]{\mbox{L. Masetti}}
\author[34]{\mbox{R. McCarthy}}
\author[5]{\mbox{S. Menke}} 
\author[32]{\mbox{D.~W. Miller}} 
\author[24]{\mbox{K. Mishra}}
\author[13]{\mbox{B. Nachman}} 
\author[13]{\mbox{P. Nef}}
\author[17]{\mbox{F.~T. O'Grady}}
\author[35]{\mbox{A. Ovcharova}} 
\author[10]{\mbox{A. Picazio}} 
\author[8]{\mbox{C. Pollard}} 
\author[25]{\mbox{B. Potter-Landua}}
\author[25]{\mbox{C. Potter}} 
\author[16]{\mbox{S. Rappoccio}} 
\author[36]{\mbox{J. Rojo}} 
\author[17]{\mbox{J. Rutherfoord}}
\author[25,37]{\mbox{G.~P. Salam}} 
\author[38]{\mbox{R.~M. Schabinger}}
\author[13]{\mbox{A. Schwartzman}} 
\author[21]{\mbox{M.~D. Schwartz}} 
\author[39]{\mbox{B. Shuve}} 
\author[40]{\mbox{P. Sinervo}}
\author[23]{\mbox{D. Soper}}
\author[22]{\mbox{D.~E. Sosa Corral}} 
\author[41]{\mbox{M. Spannowsky}} 
\author[13]{\mbox{E.. Strauss}}
\author[13]{\mbox{M. Swiatlowski}}
\author[30]{\mbox{J. Thaler}} 
\author[25]{\mbox{C. Thomas}} 
\author[42]{\mbox{E. Thompson}} 
\author[24]{\mbox{N.~V. Tran}} 
\author[36]{\mbox{J. Tseng}} 
\author[27]{\mbox{E. Usai}} 
\author[43]{\mbox{L. Valery}}  
\author[17]{\mbox{J. Veatch}} 
\author[44]{\mbox{M. Vos}} 
\author[45]{\mbox{W. Waalewijn}} 
\author[46]{\mbox{J. Wacker}} 
\author[25]{\mbox{C. Young}}

\affil[1]{Brookhaven National Laboratory, Upton, NY 11973, USA}
\affil[2]{Duke University, Durham, NC 27708, USA}
\affil[3]{University of Sussex, Brighton, BN1 9RH, UK}
\affil[4]{CP3, Universite catholique du Louvain, B-1348 Louvain-la-Neuve, Belgium}
\affil[5]{Max-Planck-Institute fuer Physik, 80805 Muenchen, Germany}
\affil[6]{Charles University in Prague, FMP, V Holesovickach 2, Prague, Czech Republic}
\affil[7]{University of California, Berkeley, CA 94720, USA}
\affil[8]{University of Glasgow, Glasgow, G12 8QQ, UK}
\affil[9]{University College London, WC1E 6BT, UK}
\affil[10]{University of Geneva, CH-1211 Geneva 4, Switzerland}
\affil[11]{Universitaet Mainz, DE 55099, Germany}
\affil[12]{Los Alamos National Laboratory, Los Alamos, NM 87545, USA}
\affil[13]{SLAC National Accelerator Laboratory, Menlo Park, CA 94025, USA}
\affil[14]{University of Maryland, College Park, MD 20742, USA}
\affil[15]{University of California, Santa Cruz, CA 95064, USA}
\affil[16]{University at Buffalo, Buffalo, NY 14260, USA}
\affil[17]{University of Arizona, Tucson, AZ 85719, USA}
\affil[18]{University of Washington, Seattle, WA 98195, USA}
\affil[19]{Rutgers University, Piscataway, NJ 08854, USA}
\affil[20]{Bergische Universitaet Wuppertal, Wuppertal, D-42097, Germany}
\affil[21]{Harvard University, Cambridge, MA 02138, USA}
\affil[22]{Universitaet Heidelberg, DE-69117, Germany}
\affil[23]{University of Oregon, Eugene, OR 97403, USA}
\affil[24]{Fermi National Accelerator Laboratory, Batavia, IL 60510, USA}
\affil[25]{CERN, CH-1211 Geneva 23, Switzerland}
\affil[26]{Universitaet Zuerich, 8006 Zuerich, Switzerland}
\affil[27]{Universitaet Hamburg, DE-22761, Germany}
\affil[28]{New York University, New York, NY 10003, USA}
\affil[29]{ETH Zuerich, 8092 Zuerich, Switzerland}
\affil[30]{Massachusetts Institute of Technology, Cambridge, MA 02139, USA}
\affil[31]{Princeton University, Princeton, NJ 08544, USA}
\affil[32]{University of Chicago, IL 60637, USA}
\affil[33]{Johns Hopkins University, Baltimore, MD 21218, USA}
\affil[34]{YITP, Stony Brook University, Stony Brook, NY 11794-3840, USA}
\affil[35]{Berkeley National Laboratory, University of California, Berkeley, CA 94720, USA}
\affil[36]{University of Oxford, Oxford, OX1 3NP, UK}
\affil[37]{LPTHE, UPMC Univ.~Paris 6 and CNRS UMR 7589, Paris, France}
\affil[38]{Universidad Autonoma de Madrid, 28049 Madrid, Spain}
\affil[39]{Perimeter Institute for Theoretical Physics, Waterloo, Ontario N2L 2Y5, Canada}
\affil[40]{University of Toronto, Toronto, Ontario M5S 1A7, Canada}
\affil[41]{IPPP, University of Durham, Durham, DH1 3LE, UK}
\affil[42]{Columbia University, New York, NY 10027, USA}
\affil[43]{LPC Clermont-Ferrand, 63177 Aubiere Cedex, France}
\affil[44]{Instituto de F\'isica Corpuscular, IFIC/CSIC-UVEG, E-46071 Valencia, Spain}
\affil[45]{University of Amsterdam, 1012 WX Amsterdam, Netherlands}
\affil[46]{Stanford Institute for Theoretical Physics, Stanford, CA 94305, USA}

\date{Received: date / Accepted: date}

\begin{document}
\renewcommand\Affilfont{\textnormal\itshape\it\small}
\maketitle

\begin{abstract}
Over the past decade, a large number of jet substructure observables have been
proposed in the literature, and explored at the LHC experiments. Such observables
attempt to utilize the internal structure of jets in order to
distinguish those initiated by quarks, gluons, or by boosted heavy objects, such as top
quarks and $W$ bosons. 
This report, originating from and motivated by the BOOST2013 workshop, presents original particle-level
studies that aim to improve our understanding of the relationships
between jet substructure observables, their complementarity, and their
dependence on the underlying jet properties,
particularly the jet radius and jet transverse momentum. This is explored in the
context of quark/gluon discrimination, boosted $W$ boson tagging and
boosted top quark tagging. 

\end{abstract}

\section{Introduction}
\label{sec:intro}
\input{introduction}

\section{Monte Carlo Samples}
\label{sec:samples}
\input{samples}

\section{Jet Algorithms and Substructure Observables}
\label{sec:algssubstructure}
\input{substructure}


\section{Multivariate Analysis Techniques}
\label{sec:multivariate}
\input{multivariate}

\section{Quark-Gluon Discrimination}
\label{sec:qgtagging}
\input{qgtagging}

\section{Boosted $W$-Tagging}
\label{sec:wtagging}
\input{wtagging}

\section{Top Tagging}
\label{sec:toptagging}
\input{toptagging}

\section{Summary \& Conclusions}
\label{sec:conclusions}
\input{conclusions}

\bibliographystyle{JHEP}       
\bibliography{boost2013_report}   
\clearpage

\end{document}

%% file: introduction.tex
The center-of-mass energies at the Large Hadron Collider are large compared to the heaviest of known particles, even after accounting for parton density functions. With the start of the second phase of operation in 2015, the center-of-mass energy will further increase from 7~\tev{} in 2010-2011 and 8~\tev{} in 2012 to 13~\tev{}. Thus, even the heaviest states in the Standard Model (and potentially previously unknown particles) will often be produced at the LHC with substantial boosts, leading to a collimation of the decay products.  For fully hadronic decays, these heavy particles will not be reconstructed as several jets in the detector, but rather as a single hadronic jet with distinctive internal substructure.  This realization has led to a new era of sophistication in our understanding of both standard Quantum Chromodynamics (QCD) jets, as well as jets containing the decay of a heavy particle, with an array of new jet observables and detection techniques introduced and studied to distinguish the two types of jets.  To allow the efficient propagation of  results from these studies of jet substructure, a series of BOOST Workshops have been held on an annual basis:
SLAC\\ (2009)~\cite{Boost:2009xxold},
Oxford University (2010)~\cite{Boost:2010xxold},
Princeton University (2011)~\cite{Boost:2011xxold},
IFIC Valencia (2012)~\cite{Boost:2012xxold}, 
University of Arizona (2013)~\cite{Boost:2013xxold},
and, most recently, University College London (2014)~\cite{Boost:2014xxold}.
Following each of these meetings, working groups have generated reports
highlighting the most interesting new results, and often including original particle-level studies. Previous BOOST reports can be found at \cite{Abdesselam:2010pt,Altheimer:2012mn,Altheimer:2013yza}.

This report from BOOST 2013 thus views the study and implementation of jet substructure techniques as a fairly mature field, and focuses on the question of the correlations between the plethora of observables that have been developed and employed, and their dependence on the underlying jet parameters, especially the jet radius $R$ and jet transverse momentum (\pt). In new analyses developed for the report, we investigate the separation of a quark signal from a gluon background ($q/g$ tagging), a $W$ signal from a gluon background ($W$-tagging) and a top signal from a mixed quark/gluon QCD background (top-tagging). In the case of top-tagging, we also investigate the performance of dedicated top-tagging algorithms, the HepTopTagger \cite{Plehn:2010st} and the Johns Hopkins Tagger \cite{Kaplan:2008ie}. We study the degree to which the discriminatory information provided by the observables and taggers overlaps by examining the extent to which the signal-background separation performance increases when two or more variables/taggers are combined in a multivariate analysis. Where possible, we provide a discussion of the physics behind the structure of the correlations and the \pt and $R$ scaling that we observe.

We present the performance of observables in idealized simulations without pile-up and detector resolution effects;
the relationship between substructure observables, their correlations, and how these depend on the jet radius $R$ and jet \pt should not be too sensitive to such effects. Conducting  studies using idealized simulations allows us to more clearly elucidate the underlying physics behind the observed performance, and also provides benchmarks for the development of techniques to mitigate pile-up and detector effects. A full study of the performance of pile-up and detector mitigation strategies is beyond the scope of the current report, and will be the focus of upcoming studies.

The report is organized as follows:~in Sections~\ref{sec:samples}-\ref{sec:multivariate}, we describe the methods used in carrying out our analysis, with a description of the Monte Carlo event sample generation in Section~\ref{sec:samples}, the jet algorithms, observables and taggers investigated in our report in Section~\ref{sec:algssubstructure}, and an overview of the multivariate techniques used to combine multiple observables into single discriminants in Section~\ref{sec:multivariate}. Our results follow in Sections~\ref{sec:qgtagging}-\ref{sec:toptagging}, with $q/g$-tagging studies in Section~\ref{sec:qgtagging}, $W$-tagging studies in Section~\ref{sec:wtagging}, and top-tagging studies in Section~\ref{sec:toptagging}. Finally we offer some summary of the studies and general conclusions in Section~\ref{sec:conclusions}.\\

\emph{The principal organizers of and contributors to the analyses presented in this report are:~B.~Cooper, S.~D.~Ellis, M.~Freytsis, A.~Hornig, A.~Larkoski, D.~Lopez Mateos, B.~Shuve, and N.~V.~Tran.}

%% file: samples.tex
Below, we describe the Monte Carlo samples used in the $q$/$g$ tagging,
$W$-tagging, and top-tagging sections of this report. Note
that no pile-up (additional proton-proton
interactions beyond the hard scatter) are included in any samples, and there is no
attempt to emulate the degradation in angular and \pt resolution that
would result when reconstructing the jets inside a real detector; such effects are deferred to 
future study.

\subsection{Quark/gluon and $W$-tagging}

Samples were generated at $\sqrt{s} = 8\TeV$ for QCD dijets, and for $W^+W^-$
pairs produced in the decay of a scalar resonance. The $W$ bosons are
decayed hadronically. The QCD events were split into subsamples of $gg$ and
$q\bar{q}$ events, allowing for tests of discrimination of hadronic $W$ bosons,
quarks, and gluons.

Individual $gg$ and $q\bar{q}$ samples were produced at leading order (LO)
using \textsc{MadGraph5} \cite{Alwall:2011uj}, while $W^+W^-$ samples were generated
using the \textsc{JHU Generator} \cite{Gao:2010qx,Bolognesi:2012mm,Anderson:2013afp}.
Both were generated using \textsc{CTEQ6L1} PDFs \cite{Pumplin:2002vw}. The samples
were produced in exclusive \pt bins of width 100 \GeV, with the slicing parameter
chosen to be the \pt of any final state parton or $W$ at LO. At the parton level,
the \pt bins investigated in this report were 300-400 \GeV, 500-600 \GeV and
1.0-1.1 \TeV. 
The samples were then showered through \textsc{Pythia8}
(version 8.176)~\cite{Sjostrand:2007gs} using the default tune
4C~\cite{Buckley:2011ms}. For each of the various samples ($W,\,q,\,g$) and \pt bins,
500k events were simulated.

\subsection{Top-tagging} \label{sec:top-samples}
Samples were generated at $\sqrt{s}=14\TeV$. Standard Model dijet and top pair
samples were produced with \textsc{Sherpa} 2.0.0 \cite{Gleisberg:2008ta,Schumann:2007mg,Krauss:2001iv,Gleisberg:2008fv,Hoeche:2009rj,Schonherr:2008av}, with matrix elements of up
to two extra partons matched to the shower. The top samples included only
hadronic decays and  were generated in exclusive \pt bins of width 100 \GeV,
taking as slicing parameter the top quark \pt. The QCD samples were generated
with a lower cut on the leading parton-level jet $\pt$, where parton-level jets
are clustered with the \antikt algorithm and jet radii of
$R= 0.4,\,0.8,\,1.2$. The matching scale is selected to be
$Q_{\rm cut}=40,\,60,\,80 \GeV$ for the $p_{T\,\text{min}}=600, 1000$, and
$1500 \GeV$ bins, respectively. For the top samples, 100k events were generated
in each bin, while 200k QCD events were generated in each bin.

%% file: substructure.tex
In Sections~\ref{sec:jetalgs},~\ref{sec:groomers},~\ref{sec:taggers} and~\ref{sec:substructure}, we describe the various jet algorithms, groomers, taggers and other substructure variables used in these studies. Over the course of our study, we considered a larger set of observables, but for presentation purposes we included only a subset in the final analysis, eliminating redundant observables.

We organize the algorithms into four categories:~clustering algorithms,  grooming algorithms, tagging algorithms, and other substructure variables that incorporate information about the shape of radiation inside the jet. We note that this labelling is somewhat ambiguous:~for example, some of the ``grooming'' algorithms (such as trimming and pruning) as well as $N$-subjettiness can be used in a ``tagging'' capacity. This ambiguity is particularly pronounced in multivariate analyses, such as the ones we present here, since a single variable can act in different roles depending on which other variables it is combined with. Therefore, the following classification is  intended only to give an approximate organization of the variables, rather than as a definitive taxonomy.

Before describing the observables used in our analysis, we give our definition of jet constituents. As a starting point, we can think of the final state of an LHC collision event as being described by a list of ``final state particles''. In the analyses of the simulated
events described below (with no detector simulation), these particles include the sufficiently long lived protons, neutrons, photons, pions, electrons and muons
with no requirements on \pT or rapidity. Neutrinos are excluded from the jet analyses.

\subsection{Jet Clustering Algorithms}
\label{sec:jetalgs}

{\bf Jet clustering:}~Jets were clustered using sequential jet clustering algorithms \cite{Bethke:1988zc} implemented in \textsc{FastJet} 3.0.3. Final state particles $i$, $j$ are assigned a mutual distance $d_{ij}$ and a distance to the beam, $d_{i\mathrm{B}}$. The particle pair with smallest $d_{ij}$ are  recombined and the algorithm repeated until the smallest distance is from a particle $i$ to the beam, $d_{i\mathrm{B}}$, in which case $i$ is set aside and labelled as a jet. The distance metrics are defined as
\begin{eqnarray}
d_{ij} &=& \mathrm{min}(p_{Ti}^{2\gamma},p_{Tj}^{2\gamma})\,\frac{\Delta R_{ij}^2}{R^2},\\
d_{i\mathrm{B}} &=& p_{Ti}^{2\gamma},
\end{eqnarray}
where $\Delta R_{ij}^2=(\Delta \eta_{ij})^2+(\Delta\phi_{ij})^2$, with $\Delta \eta_{ij}$ being the separation in pseudorapidity of particles $i$ and $j$, and $\Delta \phi_{ij}$ being the separation in azimuth. In this analysis, we use the \antikt algorithm ($\gamma=-1$) \cite{Cacciari:2008gp}, the Cambridge/Aachen (C/A) algorithm ($\gamma=0$) \cite{Dokshitzer:1997in,Wobisch:1998wt_IP}, and the \kT algorithm ($\gamma=1$) \cite{Catani:1993hr,Ellis:1993tq}, each of which has varying sensitivity to soft radiation in the definition of the jet.

This process of jet clustering serves to identify jets as (non-overlapping) sub-lists of final state particles within the original event-wide list.  The particles 
on the sub-list corresponding to a specific jet are labeled the ``constituents'' of that jet, and most of the tools described here process this sub-list of 
jet constituents in some specific fashion to determine some property of that jet.  The concept of constituents of a jet can be generalized to a more detector-centric version where the constituents are, for example, tracks and calorimeter cells, or to a perturbative QCD version where the constituents are partons (quarks and gluons).  These different descriptions are not identical, but are closely related.  We will focus on the MC based analysis of simulated events, while drawing 
insight from the perturbative QCD view.  Note also that, when a detector (with a magnetic field) is included in the analysis, there will generally be a minimum \pT
requirement on the constituents so that realistic numbers of constituents will be smaller than, but presumably still proportional to, the numbers found in the analyses 
described here.\\

\noindent {\bf Qjets:}~We also perform non-deterministic jet clustering \cite{Ellis:2012sn,Ellis:2014eya}. Instead of always clustering the particle pair with smallest distance $d_{ij}$, the pair selected for combination is chosen probabilistically according to a measure
\begin{equation}
P_{ij} \propto \,e^{-\alpha \,(d_{ij}-d_{\rm min})/d_{\rm min}},
\end{equation}
where $d_{\rm min}$ is the minimum distance for the usual jet clustering algorithm at a particular step. This leads to a different cluster sequence for the jet each time the Qjet algorithm is used, and consequently different substructure properties. The parameter $\alpha$ is called the rigidity and is used to control how sharply peaked the probability distribution is around the usual, deterministic value. The Qjets method uses statistical analysis of the resulting distributions to extract more information from the jet than can be found in the usual cluster sequence.

\subsection{Jet Grooming Algorithms}
\label{sec:groomers}

 {\bf Pruning:}~Given a jet, re-cluster the constituents using the C/A algorithm. At each step, proceed with the merger as usual unless both
 \begin{equation}
 \frac{\mathrm{min}(p_{Ti},p_{Tj})}{p_{Tij}} < z_{\rm cut}\,\,\,\mathrm{and}\,\,\,\Delta R_{ij} > \frac{2m_j}{p_{Tj}} R_{\rm cut},
 \end{equation}
 in which case the merger is vetoed and the softer branch  discarded. The default parameters used for pruning \cite{Ellis:2009me} in this report are $z_{\rm cut}=0.1$ and $R_{\rm cut}=0.5$, unless otherwise stated. One advantage of pruning is that the thresholds used
 to veto soft, wide-angle radiation scale with the jet kinematics, and so the algorithm is expected to perform comparably over a wide range of momenta.\\

 \noindent {\bf Trimming:}~Given a jet, re-cluster the constituents into subjets of radius $R_{\rm trim}$ with the \kT algorithm. Discard all subjets $i$ with 
 \begin{equation}
 p_{Ti} < f_{\rm cut} \, p_{TJ}.
 \end{equation}
 The default parameters used for trimming \cite{Krohn:2009th} in this report are $R_{\rm trim}=0.2$ and $f_{\rm cut}=0.03$, unless otherwise stated.\\
 
   \noindent {\bf Filtering:}~Given a jet, re-cluster the constituents into subjets of radius $R_{\rm filt}$ with the C/A algorithm. Re-define the jet to consist of only the hardest $N$ subjets, where $N$ is determined by the final state topology and is typically one more than the number of hard prongs in the resonance decay (to include the leading final-state gluon emission) \cite{Butterworth:2008iy}. While we do not independently use filtering, it is an important step of the HEPTopTagger to be defined later.\\
 
 \noindent {\bf Soft drop:}~Given a jet, re-cluster all of the constituents using the C/A algorithm. Iteratively undo the last stage of the C/A clustering from $j$ into subjets $j_1$, $j_2$. If
 \begin{equation}
 \frac{\mathrm{min}(p_{T1},p_{T2})}{p_{T1}+p_{T2}} < z_{\rm cut} \left(\frac{\Delta R_{12}}{R}\right)^\beta,
 \end{equation}
 discard the softer subjet and repeat. Otherwise, take $j$ to be the final soft-drop jet \cite{Larkoski:2014wba}. Soft drop has two input parameters, the angular exponent $\beta$ and the soft-drop scale $z_{\rm cut}$. In these studies we use the default $z_{\rm cut}=0.1$ setting, with $\beta=2$.

\subsection{Jet Tagging Algorithms}
\label{sec:taggers}

\noindent {\bf Modified Mass Drop Tagger:}~Given a jet, re-cluster all of the constituents using the C/A algorithm. Iteratively undo the last stage of the C/A clustering from $j$ into subjets $j_1$, $j_2$ with $m_{j_1}>m_{j_2}$. If either
\begin{equation}
m_{j_1} > \mu \, m_j\,\,\,\mathrm{or}\,\,\, \frac{\mathrm{min}(p_{T1}^2,p_{T2}^2)}{m_j^2}\,\Delta R_{12}^2 < y_{\rm cut},
\end{equation}
then discard the branch with the smaller transverse mass $m_T = \sqrt{m_i^2 + p_{Ti}^2}$, and re-define $j$ as the branch with the larger transverse mass. Otherwise, the jet is tagged. If de-clustering continues until only one branch remains, the jet is considered to have failed the tagging criteria \cite{Dasgupta:2013ihk}. In this study we use by default $\mu = 1.0$ (i.e. implement no mass drop criteria) and $y_{\rm cut} = 0.1$. With respect to the singular parts of the splitting functions, this describes the same algorithm as running soft drop with $\beta = 0$. \\

\noindent {\bf Johns Hopkins Tagger:}~Re-cluster the jet using the C/A algorithm. The jet is iteratively de-clustered, and at each step the softer prong is discarded if its $p_{\rm T}$ is less than $\delta_p\,p_{\mathrm{T\,jet}}$. This continues until both prongs are harder than the $p_{\rm T}$ threshold, both prongs are softer than the $p_{\rm T}$ threshold, or if they are too close ($|\Delta\eta_{ij}|+|\Delta\phi_{ij}|<\delta_R$); the jet is rejected if either of the latter conditions apply. If both are harder than the $p_{\rm T}$ threshold, the same procedure is applied to each: this results in 2, 3, or 4 subjets. If there exist 3 or 4 subjets, then the jet is accepted: the top candidate is the sum of the subjets, and $W$ candidate is the pair of subjets closest to the $W$ mass \cite{Kaplan:2008ie}. The output of the tagger is the mass of the top candidate ($m_t$), the mass of the $W$ candidate ($m_W$), and $\theta_{\rm h}$, a helicity angle defined as the angle, measured in the rest frame of the $W$ candidate, between the top direction and one of the $W$ decay products. The two free input parameters of the John Hopkins tagger in this study are $\delta_p$ and $\delta_R$, defined above, and their values are optimized for different jet kinematics and parameters in Section~\ref{sec:toptagging}.\\

\noindent {\bf HEPTopTagger:}~Re-cluster the jet using the C/A algorithm. The jet is iteratively de-clustered, and at each step the softer prong is discarded if $m_1/m_{12}>\mu$ (there is not a significant mass drop). Otherwise, both prongs are kept. This continues until a prong has a mass $m_i < m$, at which point it is added to the list of subjets. Filter the jet using $R_{\rm filt}=\mathrm{min}(0.3,\Delta R_{ij})$, keeping the five hardest subjets (where $\Delta R_{ij}$ is the distance between the two hardest subjets). Select the three subjets whose invariant mass is closest to $m_t$ \cite{Plehn:2010st}. The top candidate is rejected if there are fewer than three subjets or if the top candidate mass exceeds 500 GeV. The output of the tagger is $m_t$, $m_W$, and $\theta_{\rm h}$ (as defined in the Johns Hopkins Tagger). The two free input parameters of the HEPTopTagger in this study are $m$ and $\mu$, defined above, and their values are optimized for different jet kinematics and parameters in Section~\ref{sec:toptagging}.\\

\noindent {\bf Top-tagging with Pruning or Trimming:}~In the studies presented in Section~\ref{sec:toptagging} we add a $W$ reconstruction step to the pruning and trimming algorithms, to enable a fairer comparison with the dedicated top tagging algorithms described above. Following the method of the BOOST 2011 report \cite{Altheimer:2012mn}, a $W$ candidate is found as follows:~if there are two subjets, the highest-mass subjet is the $W$ candidate (because the $W$ prongs end up clustered in the same subjet), and the $W$ candidate mass, $m_W$, the mass of this subjet; if there are three subjets, the two subjets with the smallest invariant mass comprise the $W$ candidate, and $m_W$ is the invariant mass of this subjet pair. In the case of only one subjet, the top candidate is rejected. The top mass, $m_t$, is the full mass of the groomed jet.\\

\subsection{Other Jet Substructure Observables} \label{sec:substructure}

The jet substructure observables defined in this section are calculated using jet constituents prior to any grooming. This approach has been used in several analyses in the past, for example~\cite{Khachatryan:2014hpa, Aad:2014haa}, whilst others have used the approach of only considering the jet constituents that survive the grooming procedure~\cite{ATL-PHYS-PUB-2014-004}. We take the first approach throughout our analyses, as this approach allows a study of both the hard and soft radiation characteristic of signal vs.~background. However, we do include the effects of initial state radiation and the underlying event, and unsurprisingly these can have a non-negligible effect on variable performance, particularly at large \pt and jet $R$. This suggests that the differences we see between variable performance at large \pt/$R$ will be accentuated in a high pile-up environment, necessitating a dedicated study of pile-up to recover as much as possible the ``ideal'' performance seen here. Such a study is beyond the scope of this paper. \\

\noindent {\bf Qjet mass volatility:}~As described above, Qjet algorithms re-cluster the same jet non-deterministically to obtain a collection of interpretations of the jet. For each jet interpretation, the pruned jet mass is computed with the default pruning parameters. The mass volatility, $\Gamma_{\rm Qjet}$, is defined as \cite{Ellis:2012sn}
\begin{equation}
\Gamma_{\rm Qjet} = \frac{\sqrt{\langle m_J^2 \rangle-\langle m_J\rangle^2}}{\langle m_J\rangle},
\end{equation}
where averages are computed over the Qjet interpretations. We use a rigidity parameter of $\alpha=0.1$ (although other studies suggest a smaller value of $\alpha$ may be optimal \cite{Ellis:2012sn,Ellis:2014eya}), and 25 trees per event for all of the studies presented here.\\

\noindent {\bf $N$-subjettiness:}~$N$-subjettiness \cite{Thaler:2010tr} quantifies how well the radiation in the jet is aligned along $N$ directions. To compute $N$-subjettiness, $\tau_N^{(\beta)}$, one must first identify $N$ axes within the jet. Then,
\begin{equation}
\tau_N^{\beta} = \frac{1}{d_0} \sum_i p_{Ti} \,\mathrm{min}\left( \Delta R_{1i}^\beta,\ldots,\Delta R_{Ni}^\beta\right),
\end{equation}
where distances are between particles $i$ in the jet and the axes,
\begin{equation}
d_0 = \sum_i p_{Ti}\,R^\beta
\end{equation}
and $R$ is the jet clustering radius. The exponent $\beta$ is a free parameter. There is also some choice in how the axes used to compute $N$-subjettiness are determined. The optimal configuration of axes is the one that minimizes
$N$-subjettiness; recently, it was shown that the ``winner-take-all'' (WTA) axes can be easily computed and have superior performance compared to other minimization techniques \cite{Larkoski:2014uqa}. We use both the WTA (Section~\ref{sec:toptagging}) and one-pass \kT optimization axes (Sections~\ref{sec:qgtagging} and~\ref{sec:wtagging}) in our studies.

Often, a  powerful discriminant is  the ratio,
\begin{equation}
\tau_{N,N-1}^{\beta} \equiv \frac{\tau_N^{\beta}}{\tau_{N-1}^{\beta}}.
\end{equation}
While this is not an infrared-collinear (IRC) safe observable, it is calculable \cite{Larkoski:2013paa} and can be made IRC safe with a loose lower cut on $\tau_{N-1}$.\\

\noindent {\bf Energy correlation functions:}~The transverse momentum version of the energy correlation functions are defined as \cite{Larkoski:2013eya}:
\begin{equation}
\mathrm{ECF}(N,\beta) = \sum_{i_1 < i_2<\ldots<i_N \in j} \left(\prod_{a=1}^N p_{T i_a}\right)\left( \prod_{b=1}^{N-1} \prod_{c=b+1}^N \Delta R_{i_b i_c}\right)^\beta,
\end{equation}
where $i$ is a particle inside the jet. It is preferable to work in terms of dimensionless quantities, particularly the energy correlation function double ratio:
\begin{equation}
C_N^{\beta} = \frac{\mathrm{ECF}(N+1,\beta)\,\mathrm{ECF}(N-1,\beta)}{\mathrm{ECF}(N,\beta)^2}.
\end{equation}
This observable measures higher-order radiation from leading-order substructure. Note that $C_2^{\beta=0}$ is identical to the variable \ptd introduced by CMS in~\cite{Chatrchyan:2012sn}.

%% file: multivariate.tex
\noindent

Multivariate techniques are used to combine multiple variables into a
single discriminant in an optimal manner. The extent to which the discrimination power
increases in a multivariable combination indicates to what extent the
discriminatory information in the variables overlaps. There exist alternative
 strategies for studying correlations in discrimination power, such as 
 ``truth matching''~\cite{Larkoski:2014pca}, but these are not explored here.

In all cases, the multivariate technique used to combine variables is
a Boosted Decision Tree (BDT) as implemented in the TMVA
package~\cite{Hocker:2007ht}. An example of the BDT settings used in
these studies, chosen to reduce the effect of overtraining, is given
in~\cite{Hocker:2007ht}. The BDT implementation including gradient
boost is used.
Additionally, the simulated data were split into training and testing samples and comparisons of the BDT output were compared to ensure that the BDT performance was not affected by overtraining.  

%% file: qgtagging.tex
In this section, we examine the differences between quark- and
gluon-initiated jets in terms of substructure variables. At a fundamental level, the primary difference between quark-  and gluon-initiated jets is the color charge of the initiating parton, typically expressed in terms of the
ratio of the corresponding Casimir factors $C_F/C_A = 4/9$.  
Since the quark has the smaller color charge, it radiates less than a
corresponding
gluon and the naive expectation is that the resulting quark jet will contain
fewer constituents than the corresponding gluon jet. The differing color
structure of the two types of jet will also be realized in the detailed
behavior of their radiation patterns.  
We determine the extent to which the substructure observables capturing these differences are correlated, 
providing some theoretical understanding of these
variables and their performance. 
The motivation for these studies
arises not only from the
desire to ``tag'' a jet as originating from a quark or gluon, but also
to improve our  understanding of the quark and gluon components of the
QCD backgrounds relative to boosted resonances.  While recent studies
have suggested that quark/gluon tagging efficiencies depend highly on
the Monte Carlo generator used \cite{Aad:2014gea,Gallicchio:2012ez}, we are more interested in
understanding the scaling performance with $\pt$ and $R$, and the
correlations between observables, which are expected to be treated
consistently within a single shower scheme.

Other examples of recent analytic studies of the correlations between jet observables
relevant to quark jet versus gluon jet discrimination can be found in
\cite{Larkoski:2013paa, Larkoski:2014tva, Larkoski:2014pca, Procura:2014cba}.

\subsection{Methodology and Observable Classes}

These studies use the $qq$ and $gg$ MC samples described  in Section~\ref{sec:samples}. 
The showered events were clustered with \textsc{FastJet}
3.03 using
the \antikt~algorithm with jet radii of $R = 0.4,\, 0.8,\, 1.2$. In
both signal (quark) and background (gluon) samples, an upper and lower cut on
the leading jet $\pt$ is applied after showering/clustering, to ensure
similar $\pt$ spectra for signal and background in each \pt bin. The bins
in leading jet \pt that are considered are 300-400 GeV, 500-600 GeV,
1.0-1.1 TeV, for the 300-400 GeV, 500-600 GeV,
1.0-1.1 TeV parton \pt slices respectively. 
Various jet grooming approaches are applied to the jets, as described in Section~\ref{sec:substructure}. 
Only leading and subleading jets in each sample are used. The
following observables are studied in this section:
\begin{itemize}
\item Number of constituents ($n_{\rm constits}$) in the jet.
\item Pruned Qjet mass volatility, $\Gamma_{\rm Qjet}$.
\item 1-point energy correlation functions, $C_1^{\beta}$ with $\beta=0,\,1,\,2$.
\item 1-subjettiness, $\tau_1^{\beta}$ with $\beta=1,\,2$. The $N$-subjettiness axes are computed using one-pass $k_t$ axis optimization.
\item Ungroomed jet mass,  $m$.
\end{itemize}
For simplicity, we hereafter refer to quark-initiated jets (gluon-initiated jets) as quark jets (gluon jets).

We will demonstrate that, in terms of their jet-by-jet correlations and their ability to separate quark jets from gluon jets, 
the above observables fall into five Classes.  The first three observables, $n_{\rm constits}$, 
$\Gamma_{\rm Qjet}$ and $C_1^{\beta=0}$, each constitutes a Class of its own  (Classes I to III) in the sense that they each carry some independent information 
about a jet and, when combined, provide substantially better quark jet and gluon jet separation than any one observable alone.  Of the remaining
observables, $C_1^{\beta=1}$ and $\tau_1^{\beta=1}$ comprise a single class (Class IV) because their distributions are  similar 
for a sample of jets, their jet-by-jet values are highly correlated, and they exhibit very similar power to separate 
quark jets and gluon jets (with very similar dependence on the jet parameters $R$ and $p_T$); this separation power is not improved
when they are combined.  The fifth class (Class V) is composed of $C_1^{\beta=2}$, $\tau_1^{\beta=2}$ and the (ungroomed) jet mass.  Again the 
jet-by-jet correlations are strong (even though the individual observable distributions are somewhat different), the quark versus gluon separation power is very similar
(including the $R$ and $p_T$ dependence), and little is achieved by combining more than one of the Class V observables.  This class structure is
not surprising given that the observables within a class exhibit very similar dependence on the kinematics of the underlying jet constituents.
For example, the members of Class V are constructed from of a sum over pairs of constituents using products of the energy of each member 
of the pair times the angular separation squared for the pair (this is apparent for the ungroomed mass when viewed in terms of a mass-squared with small angular separations).  
By the same argument, the Class IV and Class V observables will be seen to be more similar than any other pair of classes, differing only in the
power ($\beta$) of the dependence on the angular separations, which  produces small but detectable differences.  
We will return to
 a more complete discussion of jet masses in Section~\ref{sec:qg_mass}.

\subsection{Single Variable Discrimination}

\begin{figure*}
\centering
\subfigure[$n_ {\rm constits}$]{\includegraphics[width=0.30\textwidth]{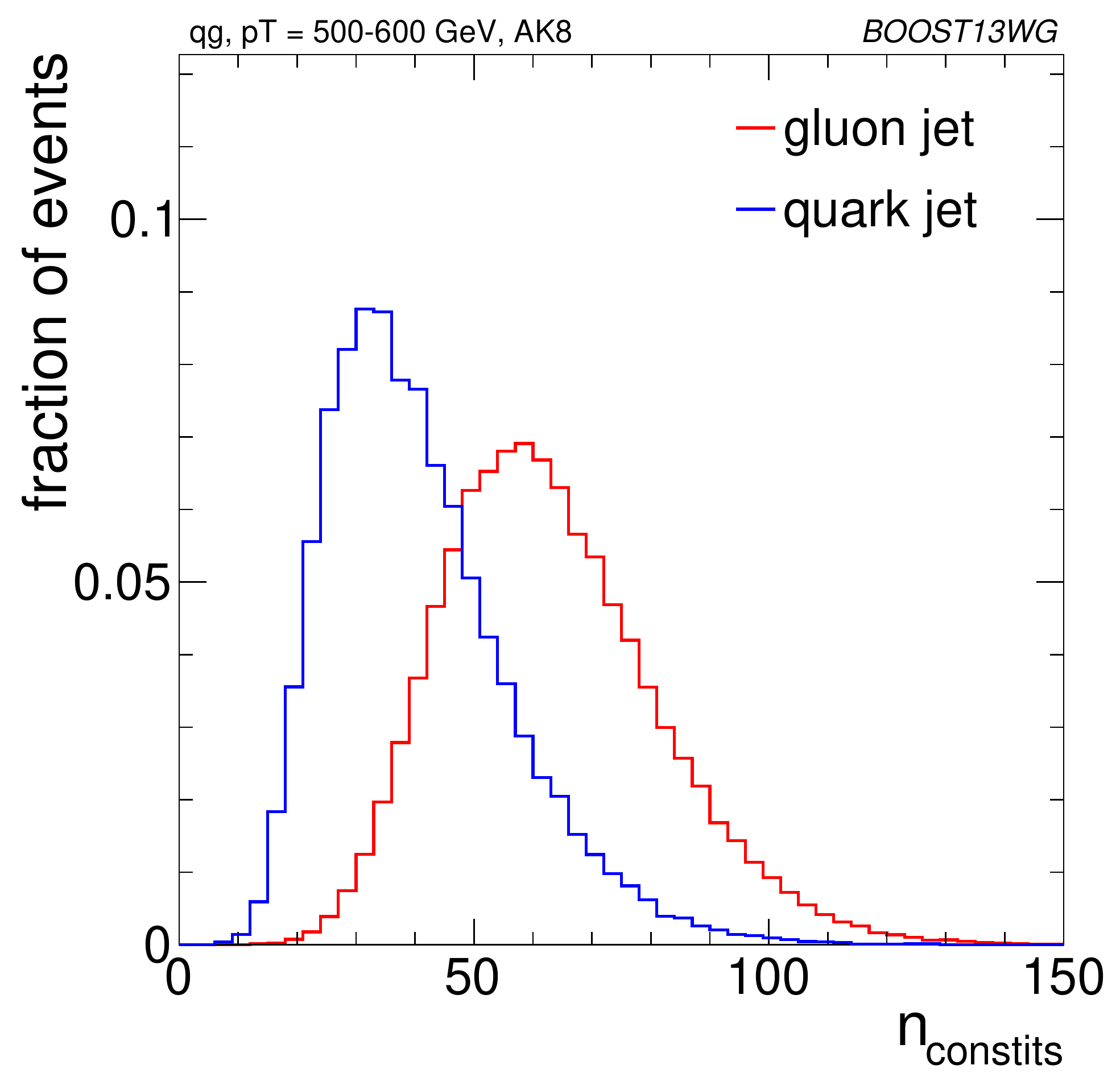}}
\subfigure[$\Gamma_{\rm Qjet}$]{\includegraphics[width=0.30\textwidth]{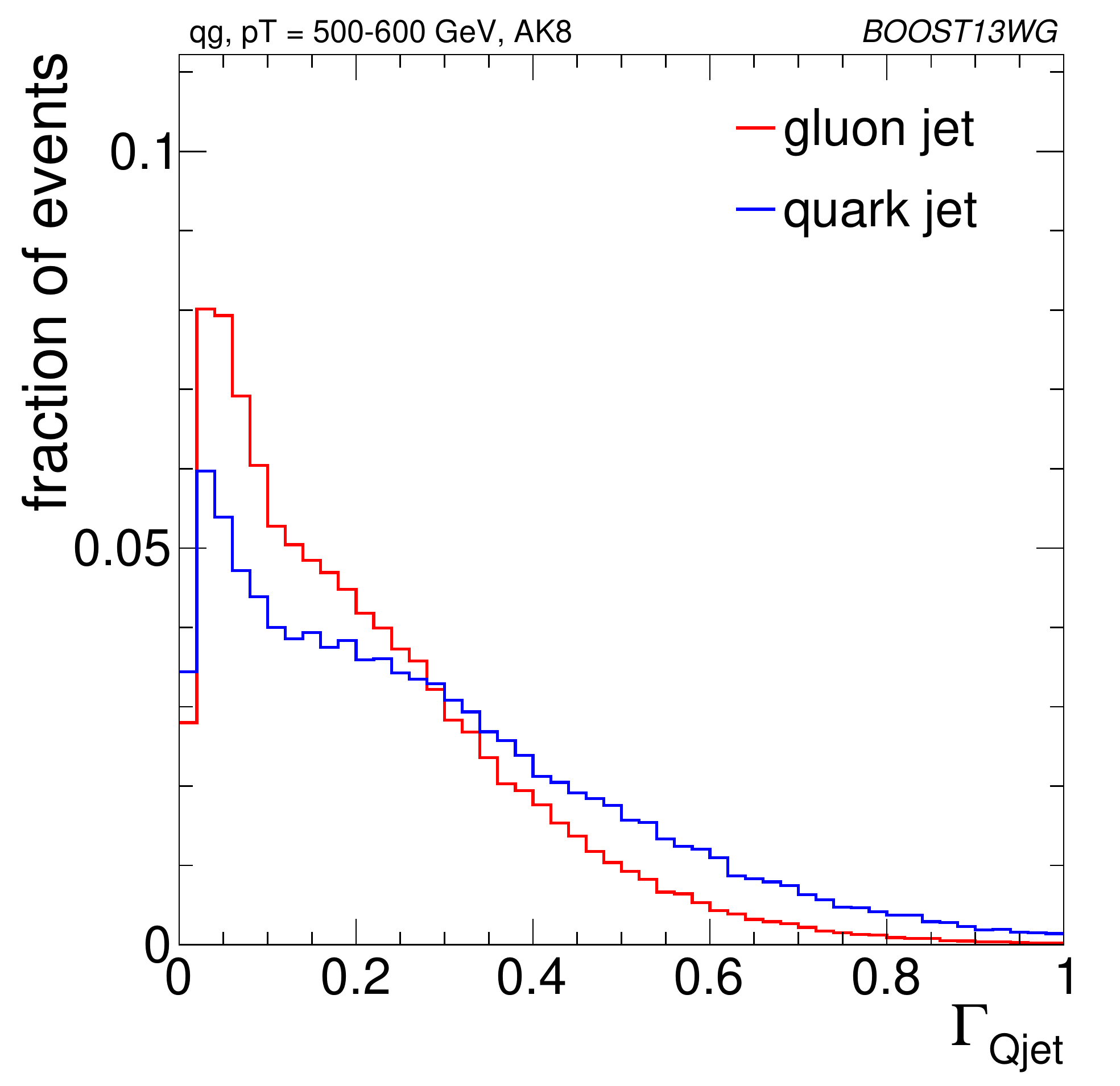}}
\subfigure[$C_1^{\beta=0}$]{\includegraphics[width=0.30\textwidth]{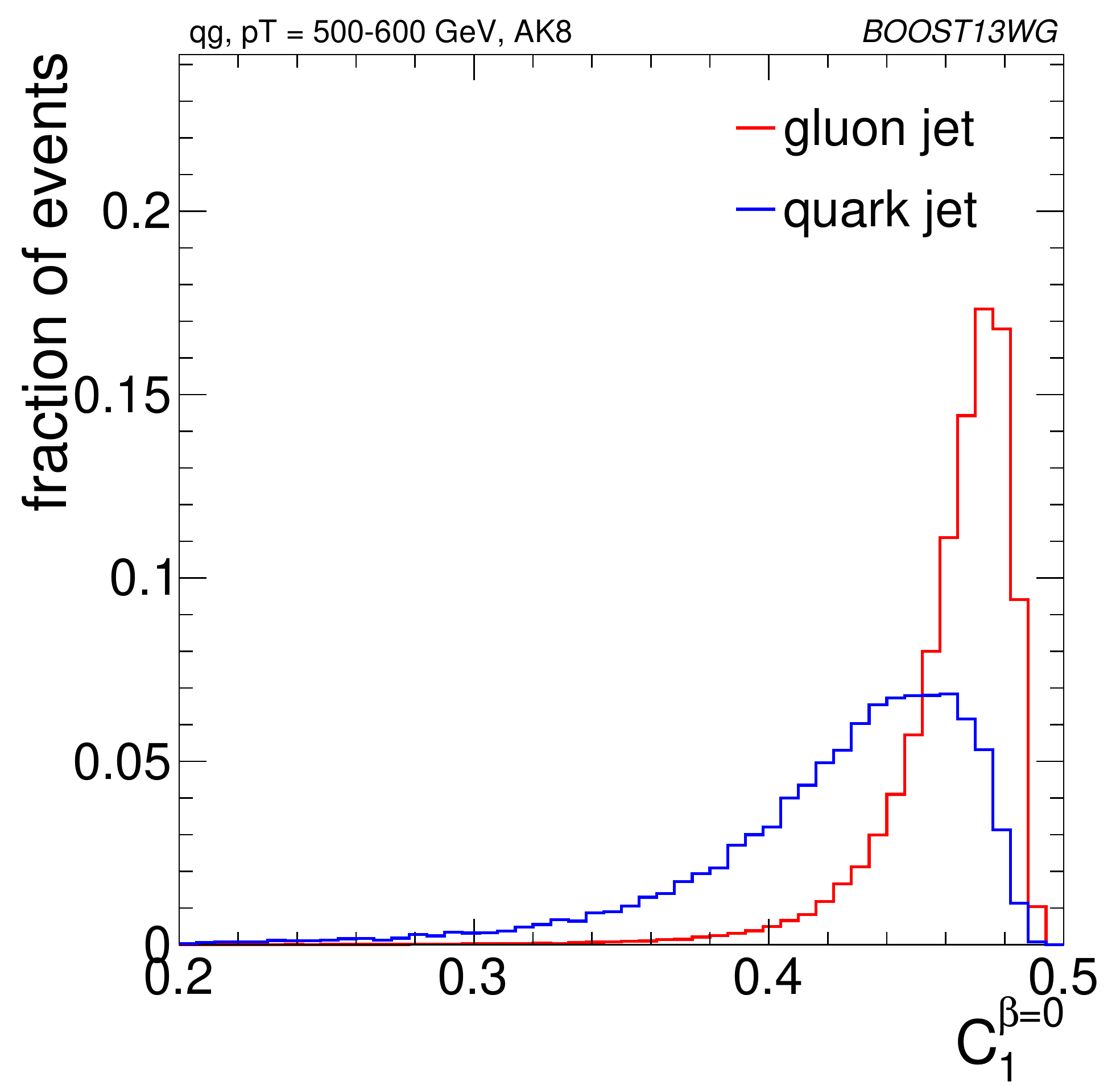}}\\
\subfigure[$C_1^{\beta=1}$]{\includegraphics[width=0.30\textwidth]{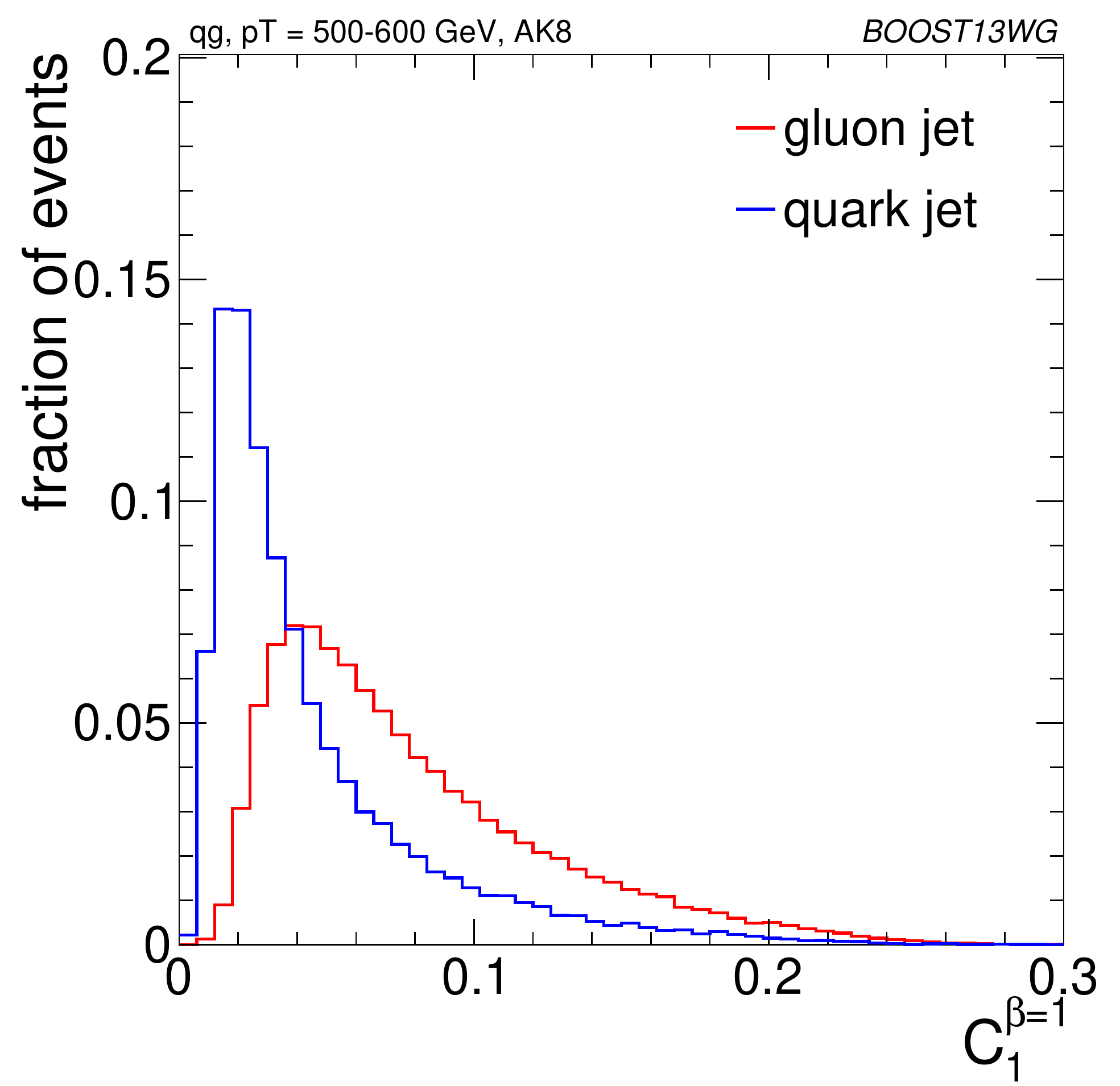}}
\subfigure[$\tau_{1}^{\beta=1}$]{\includegraphics[width=0.30\textwidth]{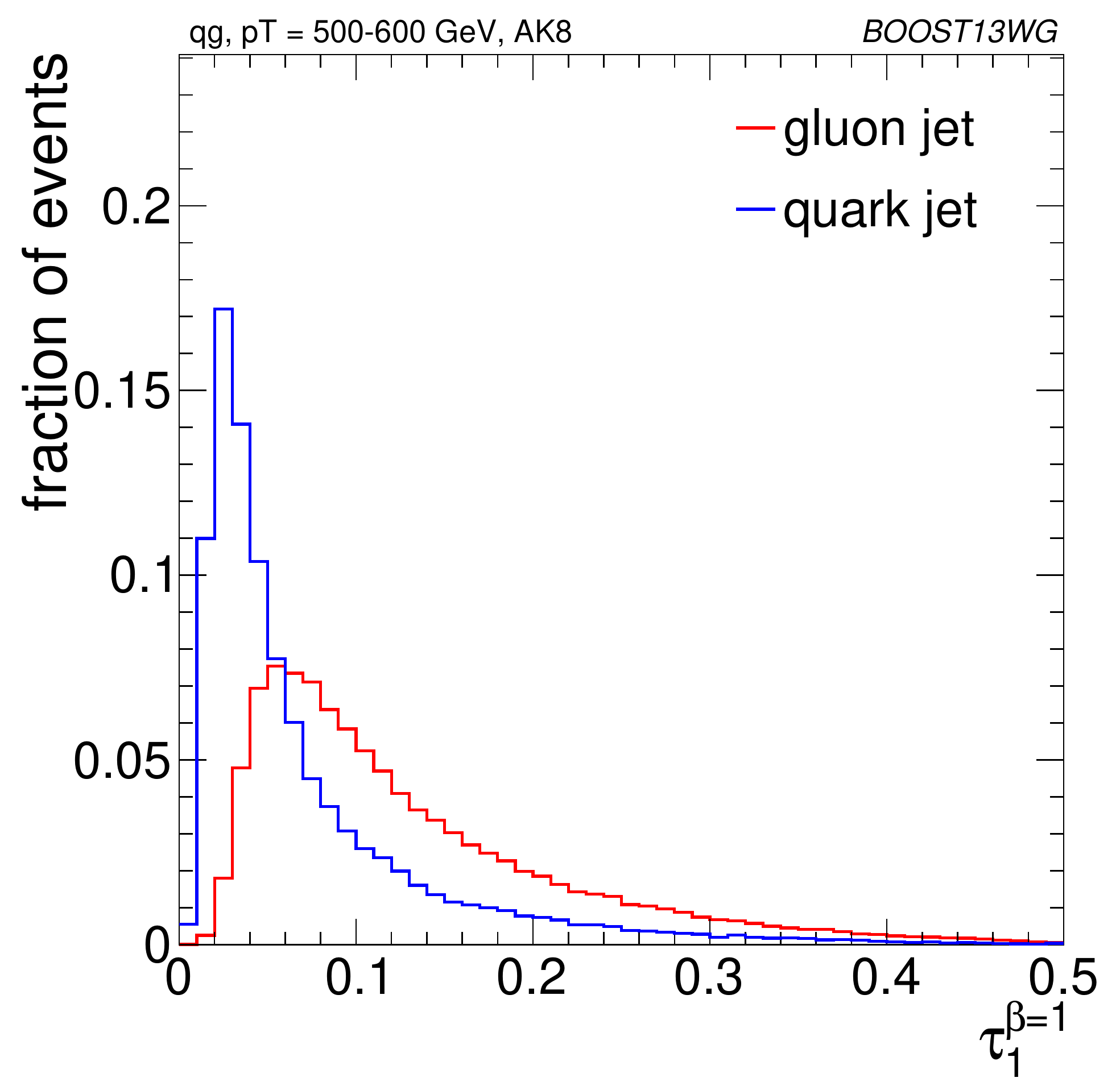}}\\
\subfigure[$C_1^{\beta=2}$]{\includegraphics[width=0.30\textwidth]{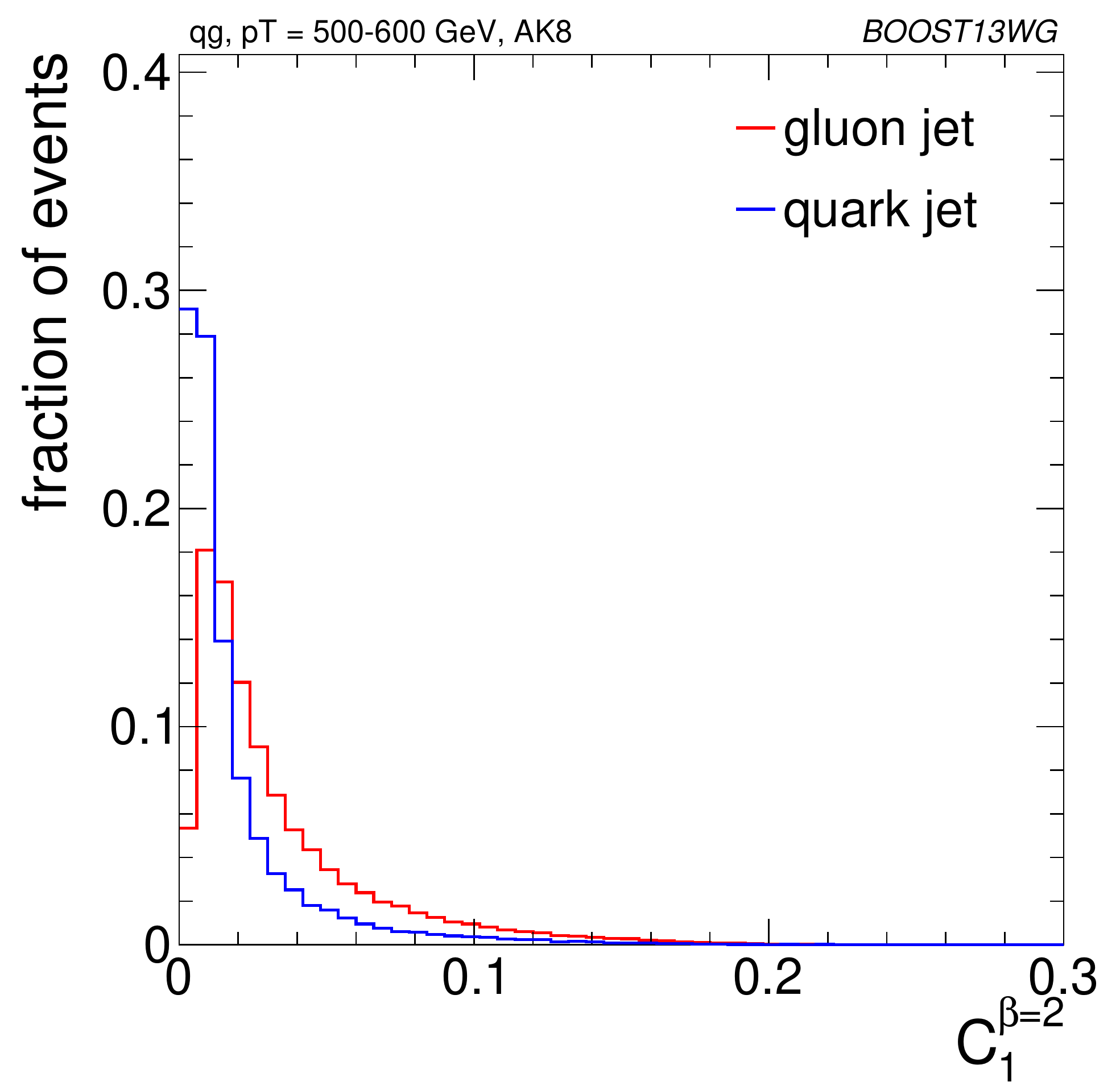}}
\subfigure[$\tau_{1}^{\beta=2}$]{\includegraphics[width=0.30\textwidth]{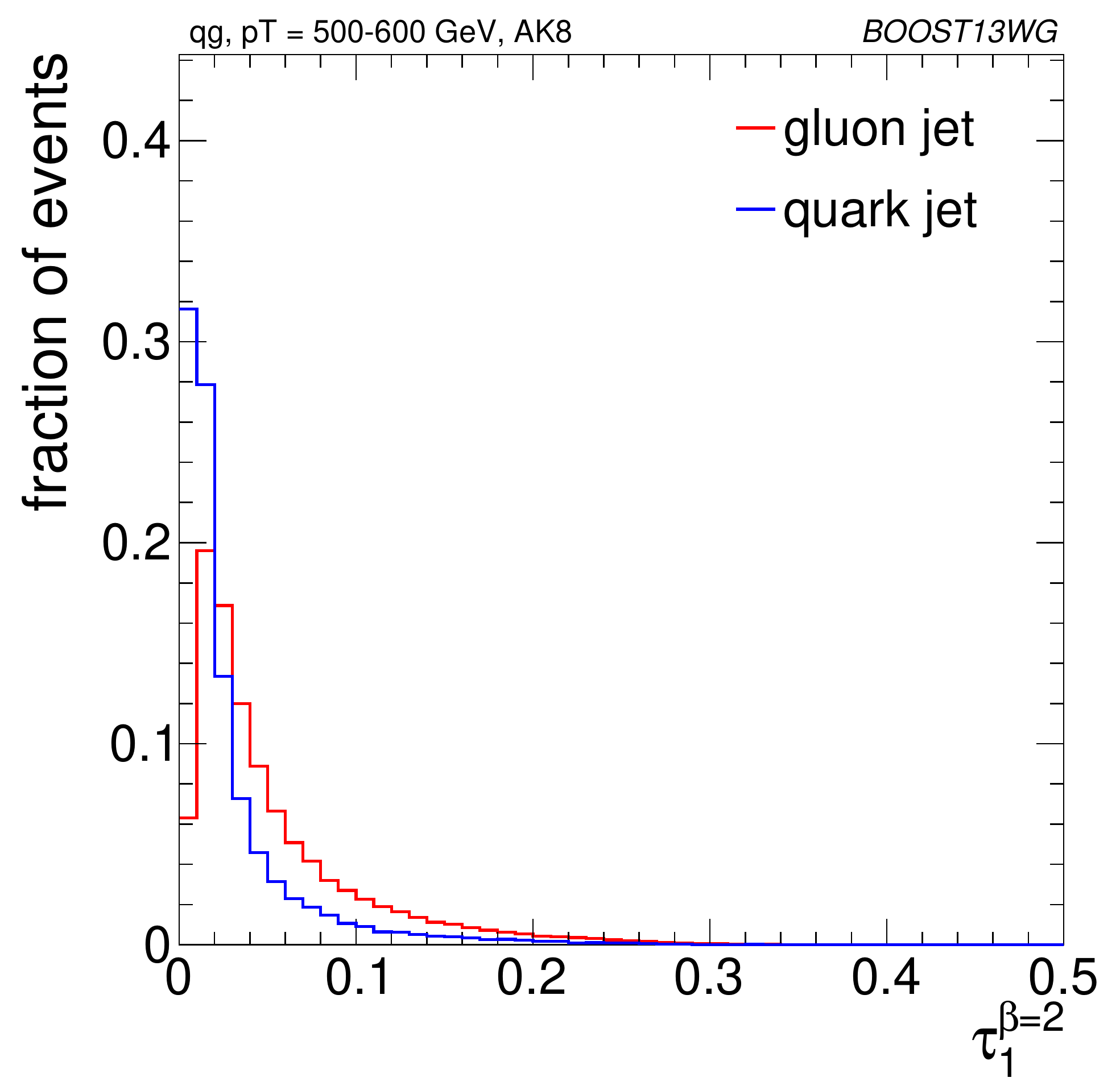}}
\subfigure[Ungroomed mass]{\includegraphics[width=0.30\textwidth]{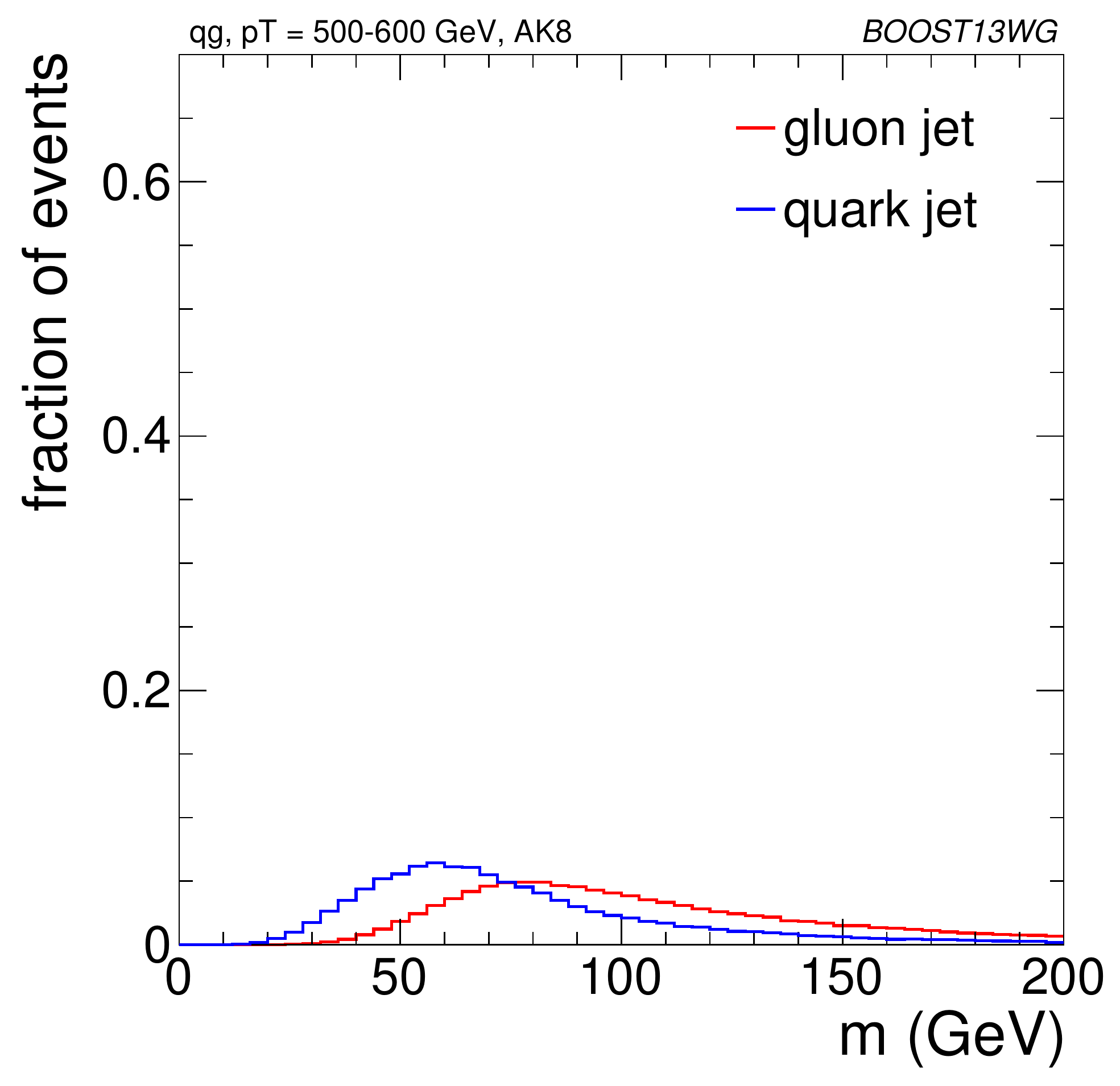}}
\caption{Comparisons of quark and gluon distributions of different substructure variables, organized by Class, for leading jets in the 
$\pt=500-600 \GeV$ bin using the anti-\kT $R=0.8$ algorithm. The first three plots are Classes I-III, with Class IV in the second row, and Class V in the third row.}
\label{fig:qg_pt500_subst_AKt_R08}
\end{figure*}

In Figure~\ref{fig:qg_pt500_subst_AKt_R08} are shown the quark and
gluon distributions of different substructure observables in the
$\pt=500-600\GeV$ bin for $R=0.8$ jets. These distributions illustrate some of the distinctions between the Classes made above. 
The fundamental difference between quarks and gluons, namely their color charge and consequent amount of radiation in the jet, is clearly indicated in Figure~\ref{fig:qg_pt500_subst_AKt_R08}(a), 
suggesting that simply counting constituents  provides good separation between quark and gluon jets.  In fact, among the observables
considered, one can see by eye that $n_{\rm constits}$ should provide the highest separation
power, \textit{i.e.}, the quark and gluon distributions are most distinct, as was originally noted in \cite{Gallicchio:2011xq,Gallicchio:2012ez}. 
Figure~\ref{fig:qg_pt500_subst_AKt_R08} further suggests
that $C_1^{\beta=0}$ should provide the next best separation, followed by $C_1^{\beta=1}$, as was also
found by the CMS and ATLAS Collaborations \cite{CMS:2013kfa,Aad:2014gea}.   

\begin{figure*}
\centering
\includegraphics[width=0.48\textwidth]{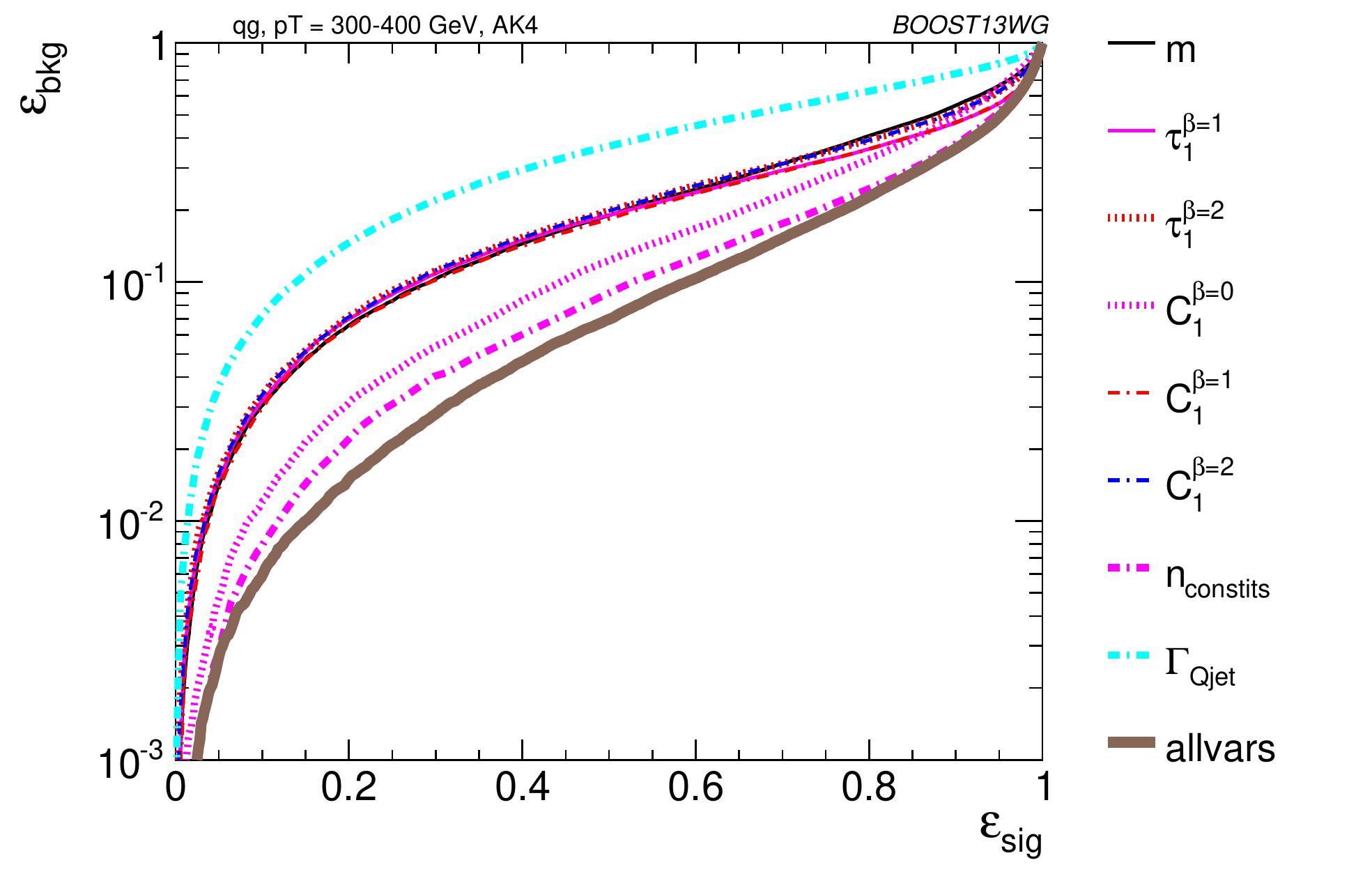}
\includegraphics[width=0.48\textwidth]{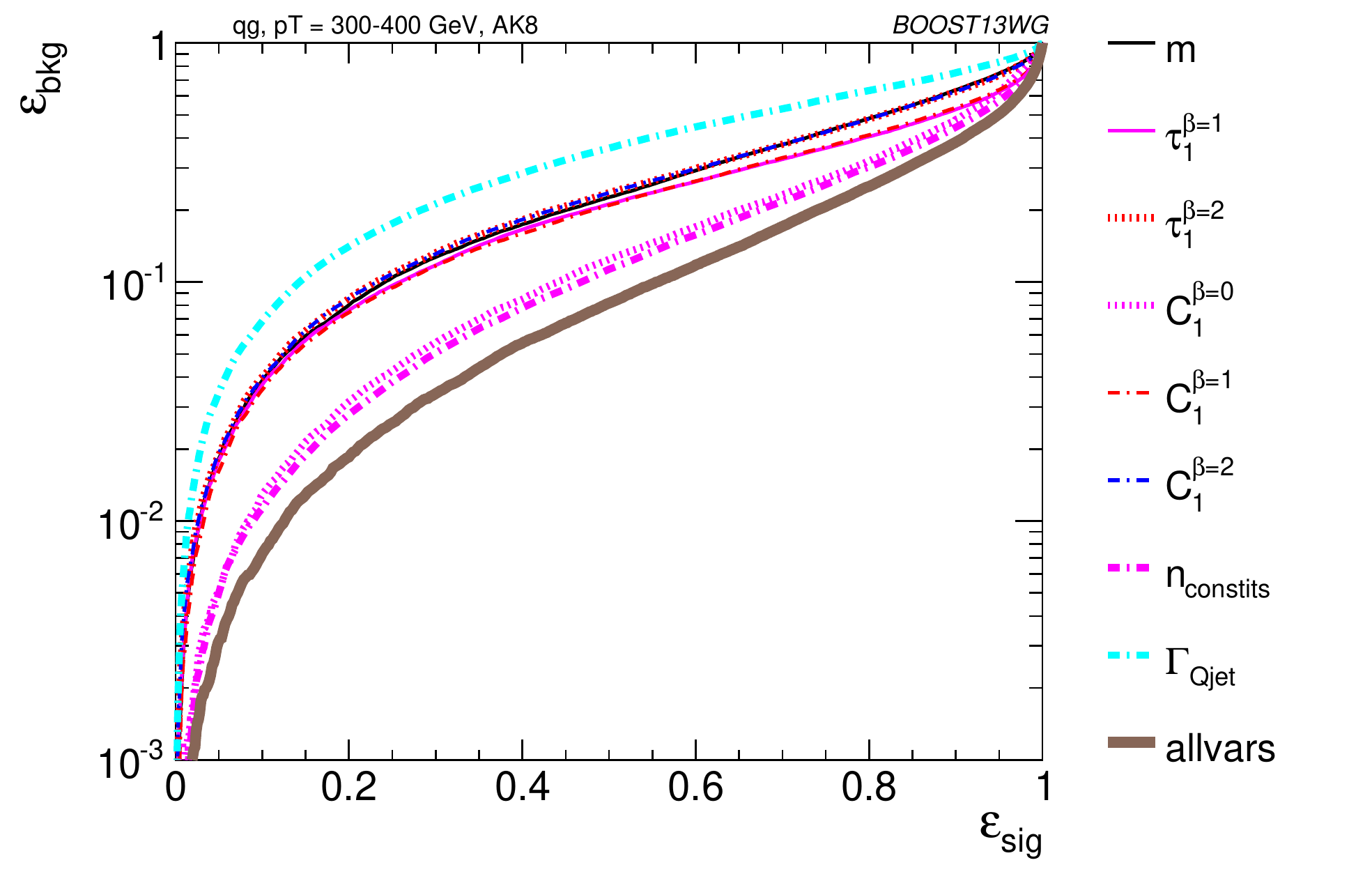}
\includegraphics[width=0.48\textwidth]{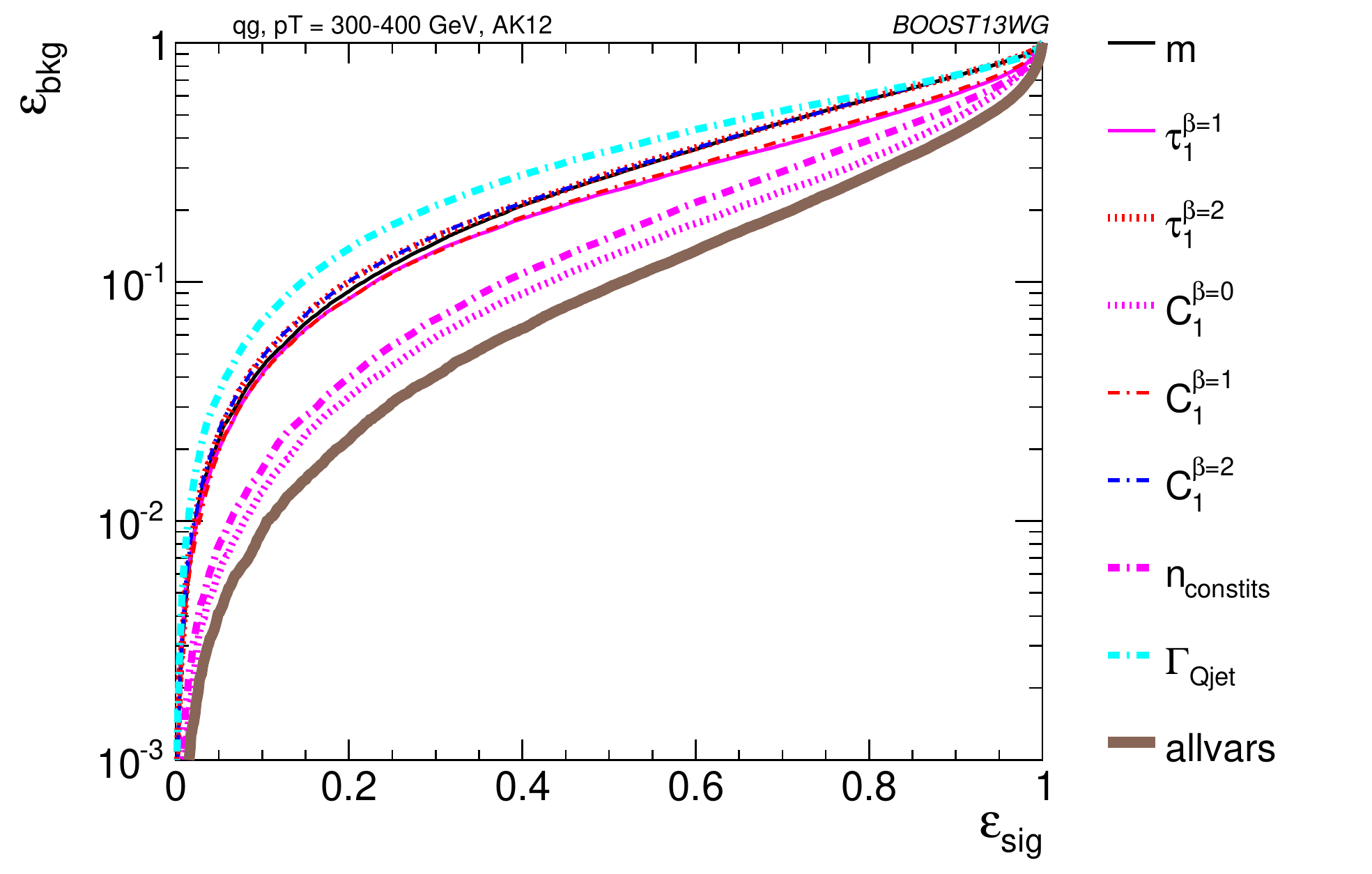}
\caption{The ROC curve for all single variables considered for
  quark-gluon discrimination in the \pt 300-400 \GeV bin using the
  \antikt $R=0.4$ (top-left), 0.8 (top-right) and 1.2 (bottom) algorithm.
}
\label{fig:qg_pt300_single}
\end{figure*}

To more quantitatively study the power of each observable as a
discriminator for quark/gluon tagging, Receiver Operating Characteristic (ROC) curves are built by scanning each distribution
and plotting the background efficiency (to select gluon jets) vs.~the signal efficiency (to select quark jets). 
Figure~\ref{fig:qg_pt300_single} shows these ROC curves for all of the
substructure variables shown in 
Figure~\ref{fig:qg_pt500_subst_AKt_R08} for $R=0.4, 0.8$ and $1.2$ jets (in the $\pt=300$-$400\GeV$
bin). In addition, the ROC curve for a tagger built from a BDT
combination of all the variables (see Section~\ref{sec:multivariate}) is shown.
As suggested earlier, $n_{\rm constits}$ is the best performing variable for all $R$ values, although $C_1^{\beta=0}$ is not far behind, particularly
for $R=0.8$. Most other variables have similar performance, with the main exception of $\Gamma_{\rm Qjet}$, which shows significantly worse
discrimination (this may be due to our choice of
rigidity $\alpha = 0.1$, with other studies suggesting that a smaller value,
such as $\alpha = 0.01$, produces better results \cite{Ellis:2012sn,Ellis:2014eya}). The combination of all variables shows somewhat better discrimination than any individual observable, and
we  give a more detailed discussion in Section~\ref{sec:qg_combi} of the correlations between the observables and their impact on the combined discrimination power.

\begin{figure*}
\centering
\subfigure[$n_ {\rm constits}$]{\includegraphics[width=0.30\textwidth]{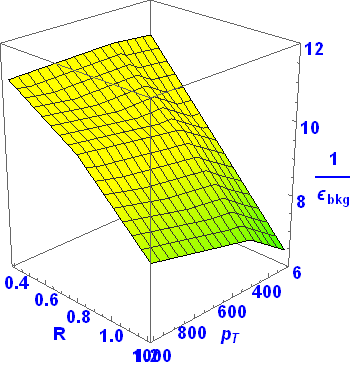}}
\subfigure[$\Gamma_{\rm Qjet}$]{\includegraphics[width=0.30\textwidth]{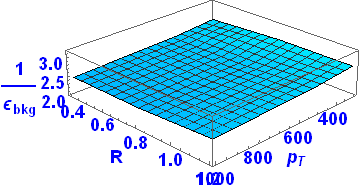}}
\subfigure[$C_1^{\beta=0}$]{\includegraphics[width=0.30\textwidth]{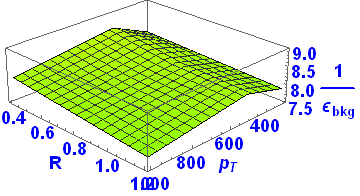}}\\
\subfigure[$C_1^{\beta=1}$]{\includegraphics[width=0.30\textwidth]{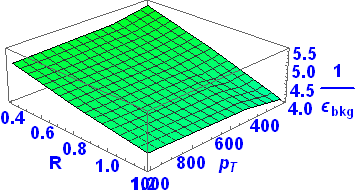}}
\subfigure[$\tau_{1}^{\beta=1}$]{\includegraphics[width=0.30\textwidth]{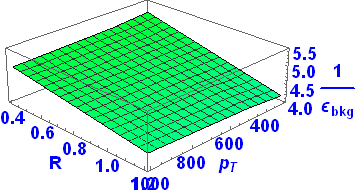}}\\
\subfigure[$C_1^{\beta=2}$]{\includegraphics[width=0.30\textwidth]{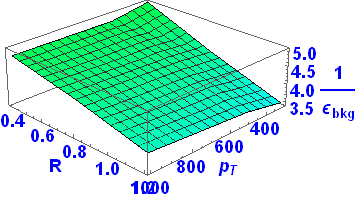}}
\subfigure[$\tau_{1}^{\beta=2}$]{\includegraphics[width=0.30\textwidth]{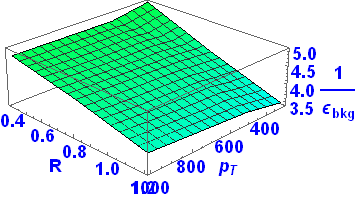}}
\subfigure[Ungroomed mass]{\includegraphics[width=0.30\textwidth]{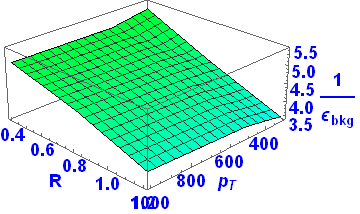}}\\
\caption{Surface plots of $1/\epsilon_\text{bkg}$ for all single variables considered for
  quark-gluon discrimination as functions of $R$ and $\pt$. The first three plots are Classes I-III, with Class IV in the second row, and Class V in the third row. }
\label{fig:qg_surface_single}
\end{figure*}

We now examine how the performance of the substructure observables varies with $\pt$ and $R$.  To present the results in a ``digestible'' fashion
we  focus on the gluon jet ``rejection'' factor, $1/\epsilon_\text{bkg}$, for a quark signal efficiency, $\epsilon_\text{sig}$, of $50\,\%$.
We can use the values of $1/\epsilon_\text{bkg}$ generated for the 9 kinematic points introduced above ($R = 0.4, 0.8, 1.2$ and 
the 100 GeV $\pt$ bins with lower limits
$p_T = 300\, \text{GeV}$, $500\, \text{GeV}$, $1000\,\text{GeV}$) to generate surface plots.  The surface plots in Figure~\ref{fig:qg_surface_single}
indicate both the level of gluon rejection
and the variation with $\pt$ and $R$ for each of the studied single observable. 
The color shading in these plots is defined so that a value of $1/\epsilon_\text{bkg}\simeq 1$ yields the color ``violet'', while 
$1/\epsilon_\text{bkg}\simeq 20 $ yields the color ``red''.   The ``rainbow'' of colors in between vary linearly with $\log_{10} (1/\epsilon_\text{bkg})$. 

 We organize our results by the classes introduced in the previous subsection:

\noindent{\bf Class I:}~The sole constituent of this class is $n_{\rm constits}$. We see in  Figure~\ref{fig:qg_surface_single}(a) that, as expected,  the numerically largest rejection rates occur
for this observable, with the rejection factor ranging from 6 to 11 and 
varying rather dramatically with $R$.  As $R$ increases the jet collects more constituents from the underlying event, which are the same
for quark and gluon jets, and the separation power decreases.  At large $R$, there is some improvement with increasing $\pt$ due to the 
enhanced QCD radiation, which is different for quarks vs.~gluons.  

\noindent{\bf Class II:}~The variable $\Gamma_{\rm Qjet}$ constitutes this class. Figure~\ref{fig:qg_surface_single}(b) confirms the 
limited efficacy of this single
observable (at least for our parameter choices) with a rejection rate only in the range 2.5 to 2.8.  On the other hand, this 
observable probes a very different
property of jet substructure, \textit{i.e.}, the sensitivity to detailed changes in the grooming procedure, and this difference is suggested
by the distinct $R$ and $\pt$ dependence illustrated in  Figure~\ref{fig:qg_surface_single}(b).  The rejection rate increases with increasing $R$
and decreasing $\pt$, since the distinction between quark and gluon jets for this observable arises from the relative importance of the one
``hard'' gluon emission configuration.  The role of this contribution is enhanced for both decreasing $\pt$ and increasing $R$. 
This general variation with $\pT$ and $R$ is the opposite of what is exhibited in all of the other single variable plots in Figure~\ref{fig:qg_surface_single}.

\noindent{\bf Class III:}~The only member of this class is $C_1^{\beta=0}$. Figure~\ref{fig:qg_surface_single}(c) indicates that this observable  can itself provide a rejection rate in the range
7.8 to 8.6 (intermediate between the two previous observables), and again with distinct $R$ and $\pt$ dependence.  In this case the rejection
rate decreases slowly with increasing $R$, which follows from the fact that $\beta = 0$ implies no weighting of $\Delta R$ in the definition of $C_1^{\beta=0}$, greatly reducing 
the angular dependence.
The rejection rate peaks at intermediate $\pt$ values, an effect visually enhanced by the limited number of 
$\pt$ values included.

\noindent{\bf Class IV:}~Figures~\ref{fig:qg_surface_single}(d) and (e)  confirm
the very similar properties of the observables $C_1^{\beta=1}$ and $\tau_1^{\beta=1}$ (as already suggested in
Figures~\ref{fig:qg_pt500_subst_AKt_R08}(d) and (e)). They have
 essentially identical rejection rates (4.1 to 5.4) and identical $R$ and $\pt$ dependence (a slow decrease with increasing $R$ and an even
slower increase with increasing $\pt$).  

\noindent{\bf Class V}:~The observables $C_1^{\beta=2}$, $\tau_1^{\beta=2}$, and $m$ have similar rejection rates in the 
range 3.5 to 5.3, as well as very similar $R$ and $\pt$ dependence (a slow decrease with increasing $R$ and an even
slower increase with increasing $\pt$).  

Arguably, drawing a distinction between the Class IV and Class V observables is a fine point, 
but the color shading does suggest some
distinction from the slightly smaller rejection rate in Class V.  Again the strong similarities between the plots within the second and third rows in 
Figure~\ref{fig:qg_surface_single} speaks to the common properties of the observables within the two classes.

In summary, the overall discriminating power between quark and gluon jets tends to
 decrease with increasing $R$, except for the $\Gamma_{\rm Qjet}$ observable, presumably in large part due to the
 contamination from the underlying event. Since the construction of the $\Gamma_{\rm Qjet}$ observable explicitly
involves pruning away the soft, large angle constituents, it is not surprising that it exhibits different $R$ dependence. 
In general the discriminating power increases slowly and monotonically
with $\pt$ (except for the $\Gamma_{\rm Qjet}$ and $C_1^{\beta=0}$ observables). This is presumably due to the overall
increase in radiation from high $\pt$ objects, which accentuates the differences in the quark and gluon color charges and 
 providing some increase in discrimination.  
In the following section, we study the effect of combining multiple observables.

\begin{figure*}
\centering
\subfigure[$C_1^{\beta=1}+\tau_{1}^{\beta=1}$]{\includegraphics[width=0.30\textwidth]{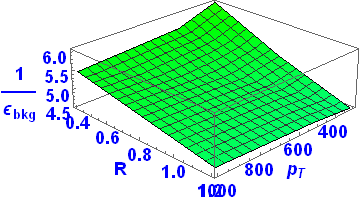}}
\subfigure[$C_1^{\beta=2}+\tau_{1}^{\beta=2}$]{\includegraphics[width=0.30\textwidth]{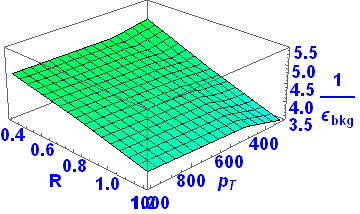}}
\caption{Surface plots of $1/\epsilon_\text{bkg}$ for the indicated pairs of variables from (a) Class IV and (b) Class V considered for
  quark-gluon discrimination as functions of $R$ and $\pt$. }
\label{fig:qg_surface_pairA}
\end{figure*}

\subsection{Combined Performance and Correlations}\label{sec:qg_combi}
Combining multiple observables in a BDT can give further improvement over cuts on a single variable. Since the improvement from combining correlated observables is expected to be inferior to that from combining uncorrelated observables, studying the performance of multivariable combinations gives insight into the correlations between substructure variables and the physical  
features allowing for quark/gluon discrimination. Based on our discussion of the correlated properties
of observables within a single class, we expect little improvement in the rejection rate when combining observables from the same class,
and substantial improvement when combining observables from different classes.  Our classification of observables for quark/gluon tagging  therefore motivates the study of
particular combinations of variables for use in experimental analyses.
 
 To quantitatively study the improvement obtained from multivariate analyses, we build quark/gluon taggers from
every pair-wise combination of variables studied in the previous section; we also compare the pair-wise performance with the all-variables combination.
  To illustrate the results achieved in this way, we use the same  2D 
surface plots as in Figure~\ref{fig:qg_surface_single}.  
Figure~\ref{fig:qg_surface_pairA} shows pair-wise plots for variables in (a) Class IV and (b) Class V, respectively.  Comparing to the corresponding plots
in Figure~\ref{fig:qg_surface_single}, we see that combining $C_1^{\beta=1}+\tau_{1}^{\beta=1}$ provides a 
small ($\sim10\%$) improvement in the rejection rate  with essentially no change in the $R$ and $\pt$ dependence, while combining $C_1^{\beta=2}+\tau_{1}^{\beta=2}$   
yields a rejection rate that is essentially identical to the single observable rejection rate for all $R$ and $\pt$ values (with a similar conclusion if one
of these observables is replaced with the ungroomed jet mass $m$).  
This  confirms the expectation that the
observables within a single class effectively probe the \textit{same} jet properties.

\begin{figure*}
\centering
\subfigure[$n_{\rm constits}+\Gamma_{\rm Qjet}$]{\includegraphics[width=0.28\textwidth]{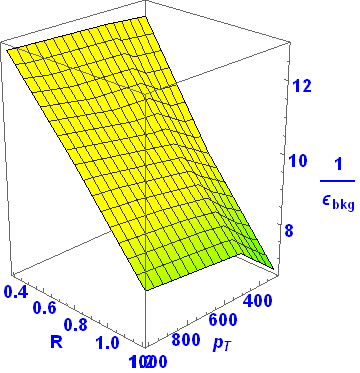}}
\subfigure[$n_{\rm constits}+ C_1^{\beta=1}$]{\includegraphics[width=0.28\textwidth]{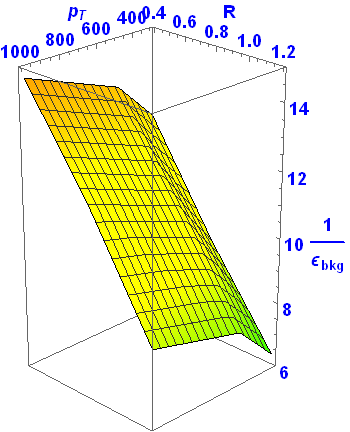}}
\subfigure[$n_{\rm constits}+ C_1^{\beta=2}$]{\includegraphics[width=0.28\textwidth]{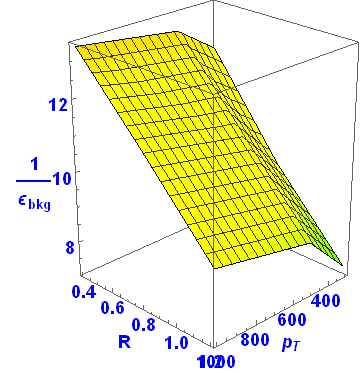}}\\
\subfigure[$n_{\rm constits}+ C_1^{\beta=0}$]{\includegraphics[width=0.28\textwidth]{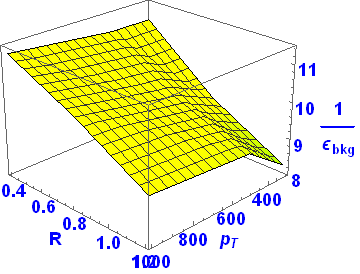}}
\subfigure[$n_{\rm constits}+ \tau_1^{\beta=1}$]{\includegraphics[width=0.28\textwidth]{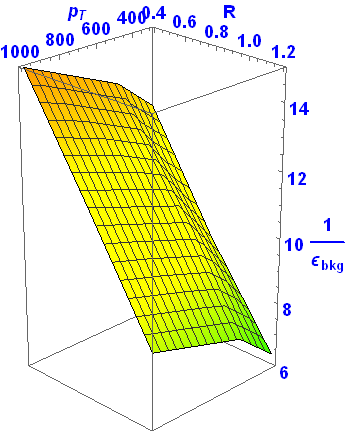}}
\subfigure[$\Gamma_{\rm Qjet}+ C_{1}^{\beta=0}$]{\includegraphics[width=0.28\textwidth]{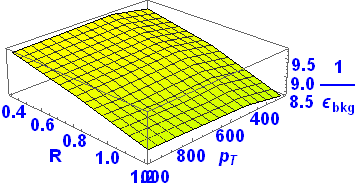}}\\
\subfigure[$\Gamma_{\rm Qjet}+ C_{1}^{\beta=1}$]{\includegraphics[width=0.28\textwidth]{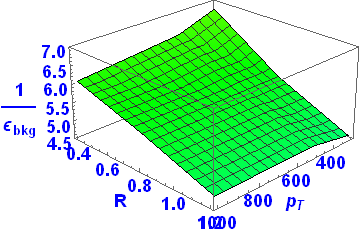}}
\subfigure[$\Gamma_{\rm Qjet}+ C_{1}^{\beta=2}$]{\includegraphics[width=0.28\textwidth]{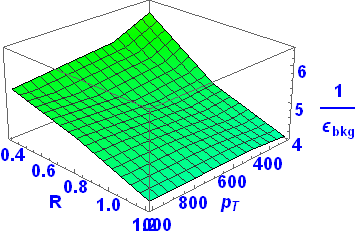}}
\subfigure[$ C_{1}^{\beta=0}+ C_{1}^{\beta=1}$]{\includegraphics[width=0.28\textwidth]{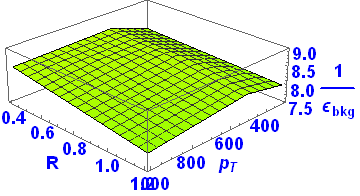}}\\
\subfigure[$ C_{1}^{\beta=0}+ C_{1}^{\beta=2}$]{\includegraphics[width=0.28\textwidth]{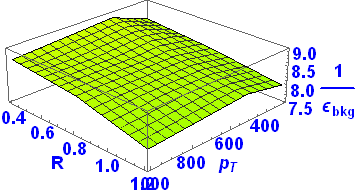}}
\subfigure[$ C_{1}^{\beta=1}+ C_{1}^{\beta=2}$]{\includegraphics[width=0.28\textwidth]{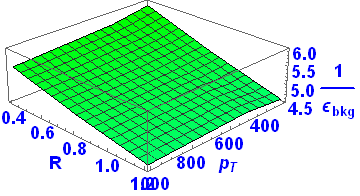}}
\subfigure[All]{\includegraphics[width=0.28\textwidth]{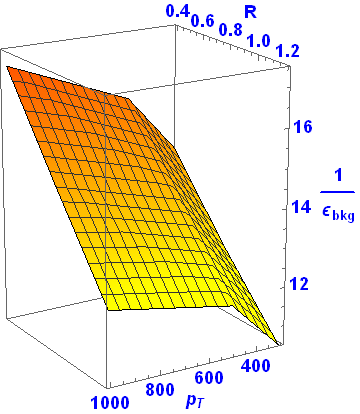}}
\caption{Surface plots of $1/\epsilon_\text{bkg}$ for the indicated pairs of variables from different classes considered for
  quark-gluon discrimination as functions of $R$ and $\pt$. }
\label{fig:qg_surface_pairB}
\end{figure*}

Next, we consider  cross-class pairs of observables  in Figure~\ref{fig:qg_surface_pairB}, where, except in the one case noted below,
we use only a single observable from each class for illustrative purposes.
Since $n_{\rm constits}$ is the best performing single variable, the largest rejection
rates are obtained from combining another observable with $n_ {\rm constits}$ (Figures~\ref{fig:qg_surface_pairB}(a) to (e)).  
In general, the rejection rates are larger for the pair-wise case
than for the single variable case.  In particular, the pair $n_{\rm constits}+ C_1^{\beta=1}$ in Figure~\ref{fig:qg_surface_pairB}(b)
yields rejection rates in the range 6.4 to 14.7 with the largest
values at small $R$ and large $\pt$.  As expected, the pair  $n_{\rm constits}+ \tau_1^{\beta=1}$ in Figure~\ref{fig:qg_surface_pairB}(e)
yields very similar rejection rates (6.4 to 15.0), since $C_1^{\beta=1}$ and $\tau_1^{\beta=1}$ are both in Class IV. 
The other pairings with $n_{\rm constits}$ yield smaller 
rejection rates and smaller dynamic ranges.  The pair $n_ {\rm constits}+ C_1^{\beta=0}$ (Figure~\ref{fig:qg_surface_pairB}(d)) exhibits
the smallest range of rates (8.3 to 11.3), suggesting that the differences between these two observables serve to substantially 
reduce the $R$ and $\pt$ dependence for the pair.
The other pairs shown exhibit
similar behavior.  

The $R$ and $\pt$ dependence of the pair-wise combinations is generally similar to the single observable with the most dependence 
on $R$ and $\pt$.  The smallest $R$ and $\pt$ variation always occurs 
when pairing with $C_1^{\beta=0}$.  Changing any of the observables in these pairs with a different observable in the same class (\textit{e.g.},
$C_1^{\beta=2}$ for $\tau_1^{\beta=2}$) produces very similar results.  

Figure~\ref{fig:qg_surface_pairB}(l) shows the
performance of a BDT combination of all the current observables,
with rejection rates in the range 10.5 to 17.1.  The performance is
very similar to that observed for the pair-wise $n_{\rm constits}+
C_1^{\beta=1}$ and $n_{\rm constits}+ \tau_1^{\beta=1}$ combinations,
but with a somewhat narrower range and slightly larger maximum
values. This suggests that almost all of the available information to
discriminate quark and gluon-initiated jets is captured by $n_{\rm
  constits}$ and $C_1^{\beta=1}$ or $\tau_1^{\beta=1}$ variables; this confirms 
  the finding that near-optimal performance can be obtained with a pair of variables
  from \cite{Gallicchio:2011xq}.

\begin{figure*}
\centering
\includegraphics[width=0.48\textwidth]{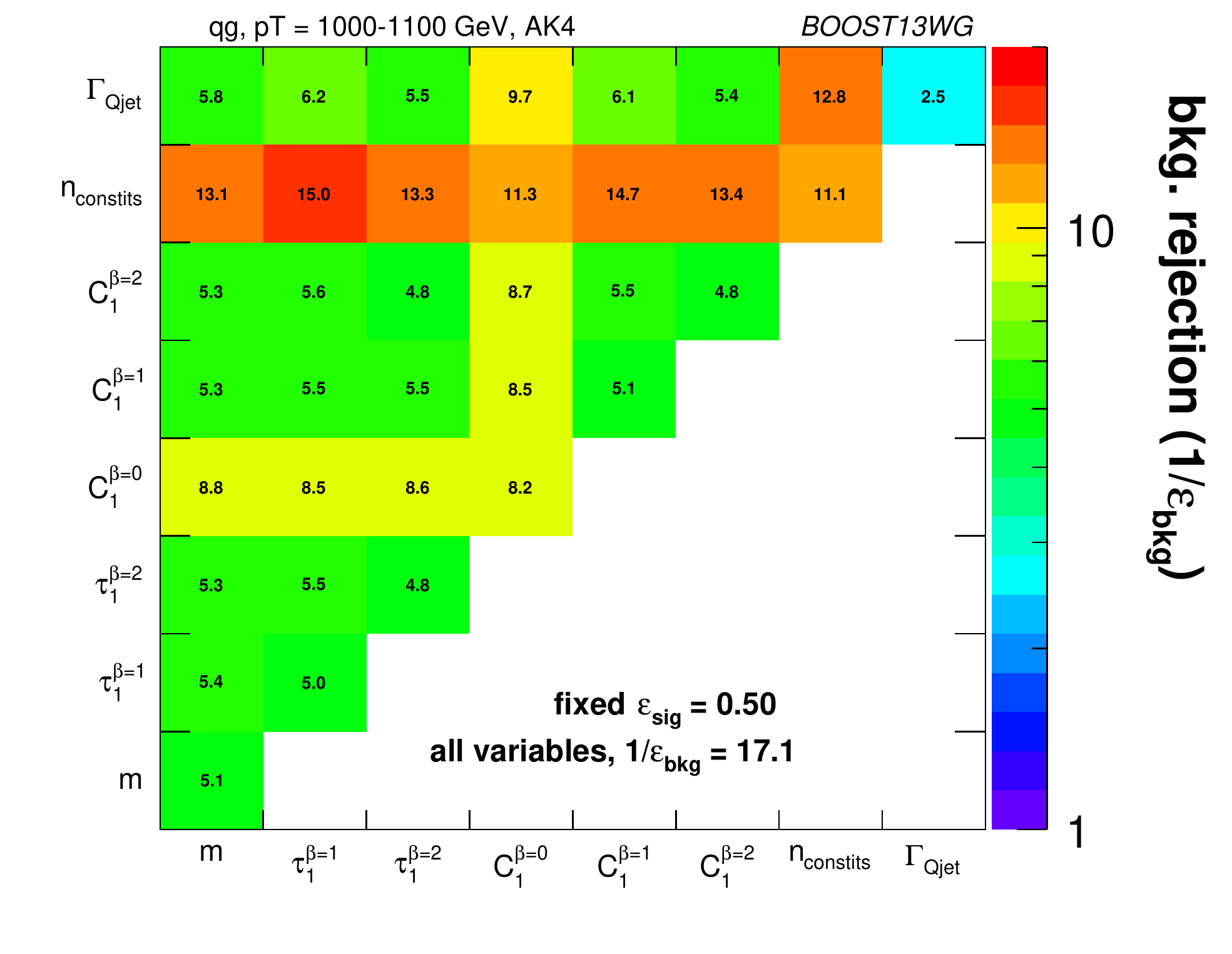}
\includegraphics[width=0.48\textwidth]{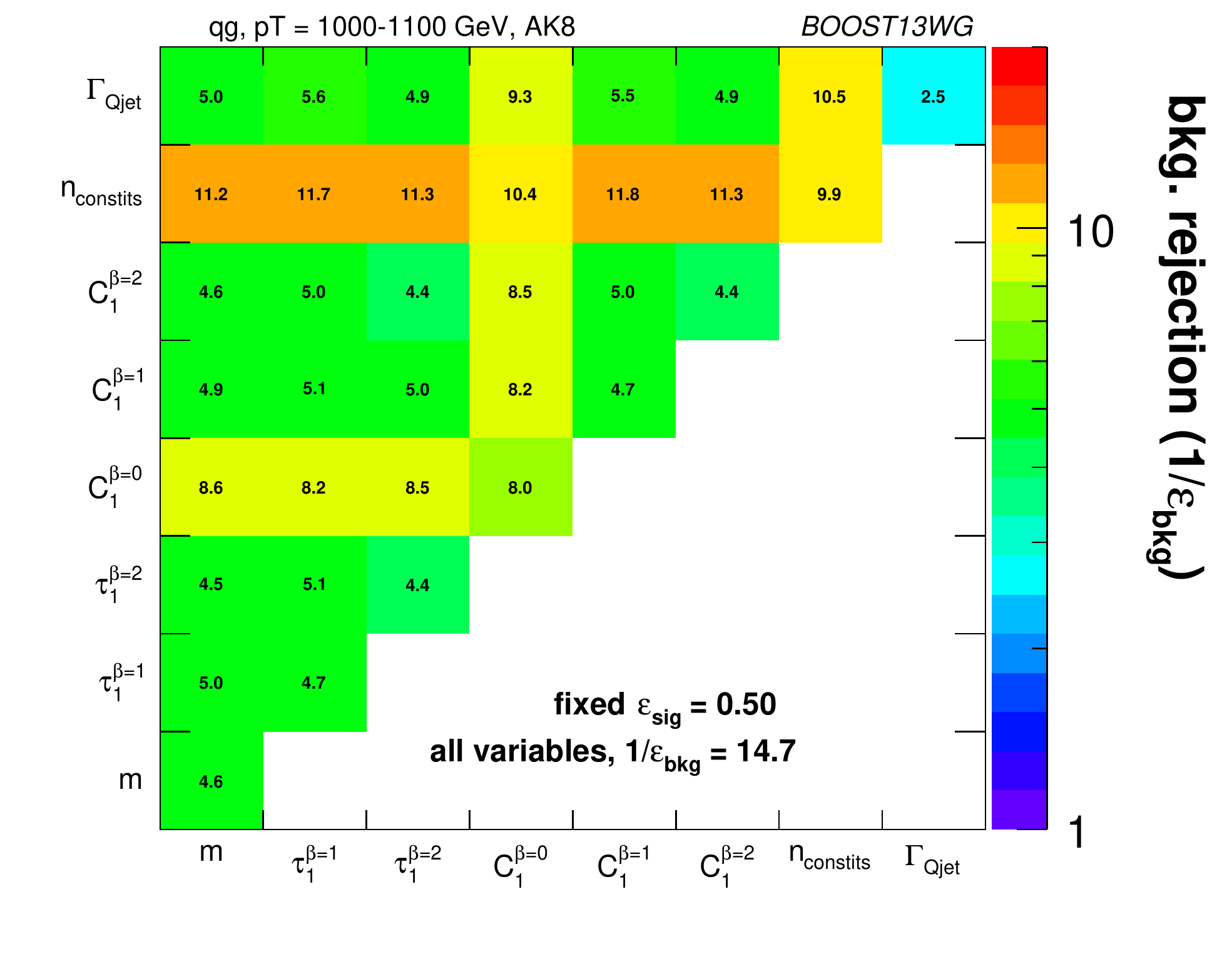}
\includegraphics[width=0.48\textwidth]{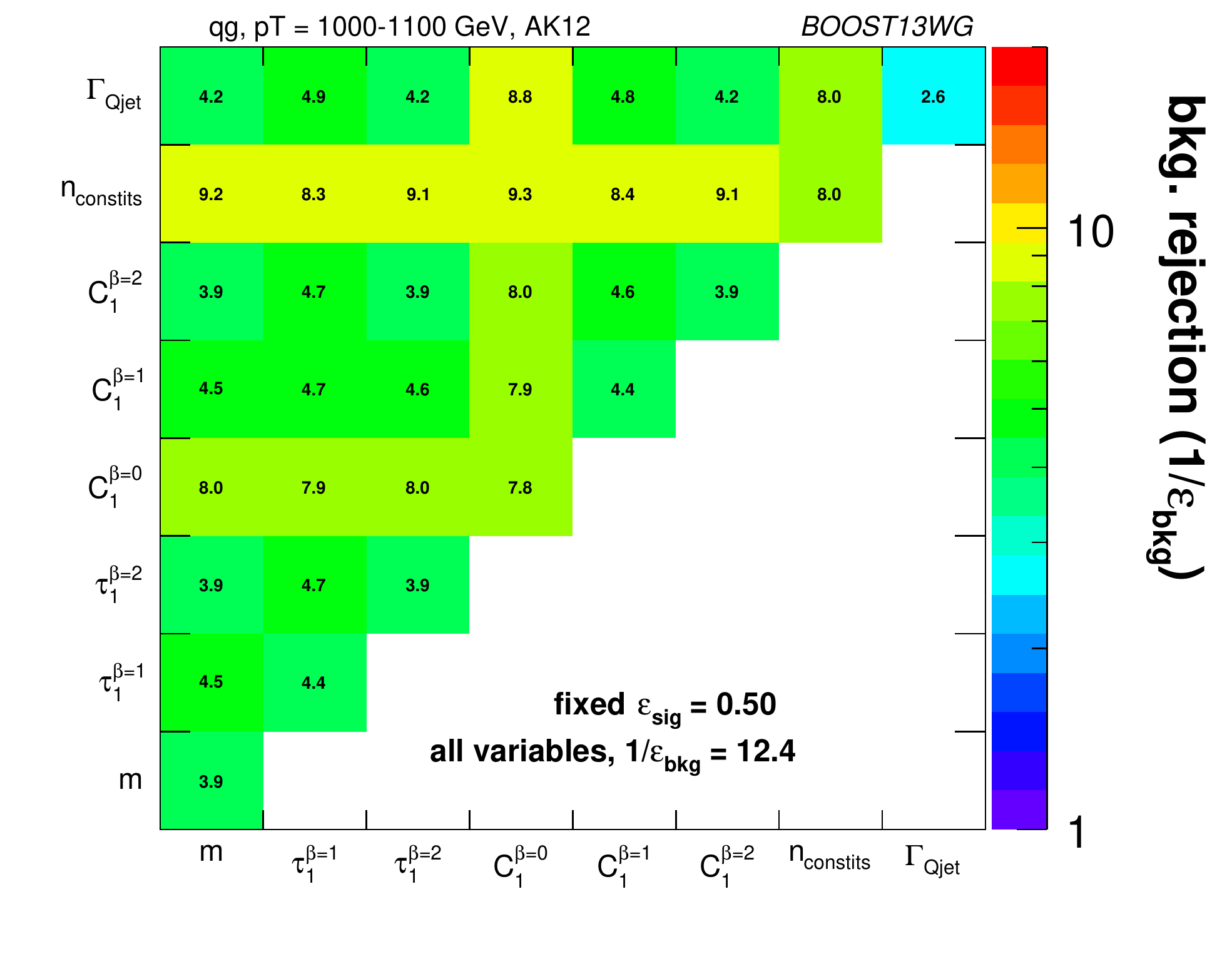}
\caption{Gluon rejection defined as $1/\epsilon_{\rm gluon}$ when using each 2-variable combination 
as a tagger with 50\% acceptance for quark jets. Results are shown for
jets with $\pt=1-1.1 \TeV$ and
for (top left) $R=0.4$; (top right) $R=0.8$; (bottom) $R=1.2$. The rejection obtained with a tagger that uses all variables is also shown
in the plots. }
\label{fig:qg_pt1000_comb}
\end{figure*}

\begin{figure*}
\centering
\includegraphics[width=0.48\textwidth]{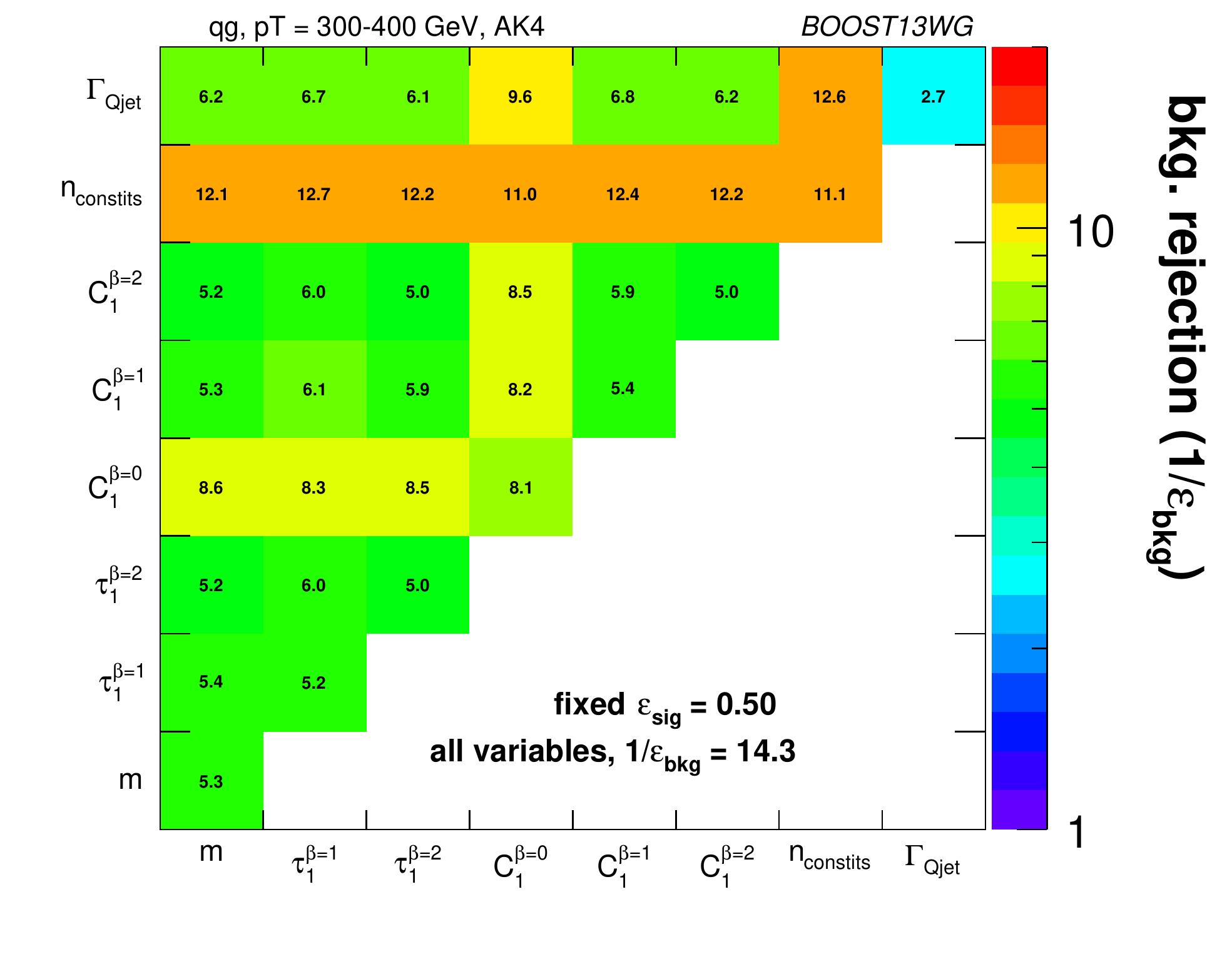}
\includegraphics[width=0.48\textwidth]{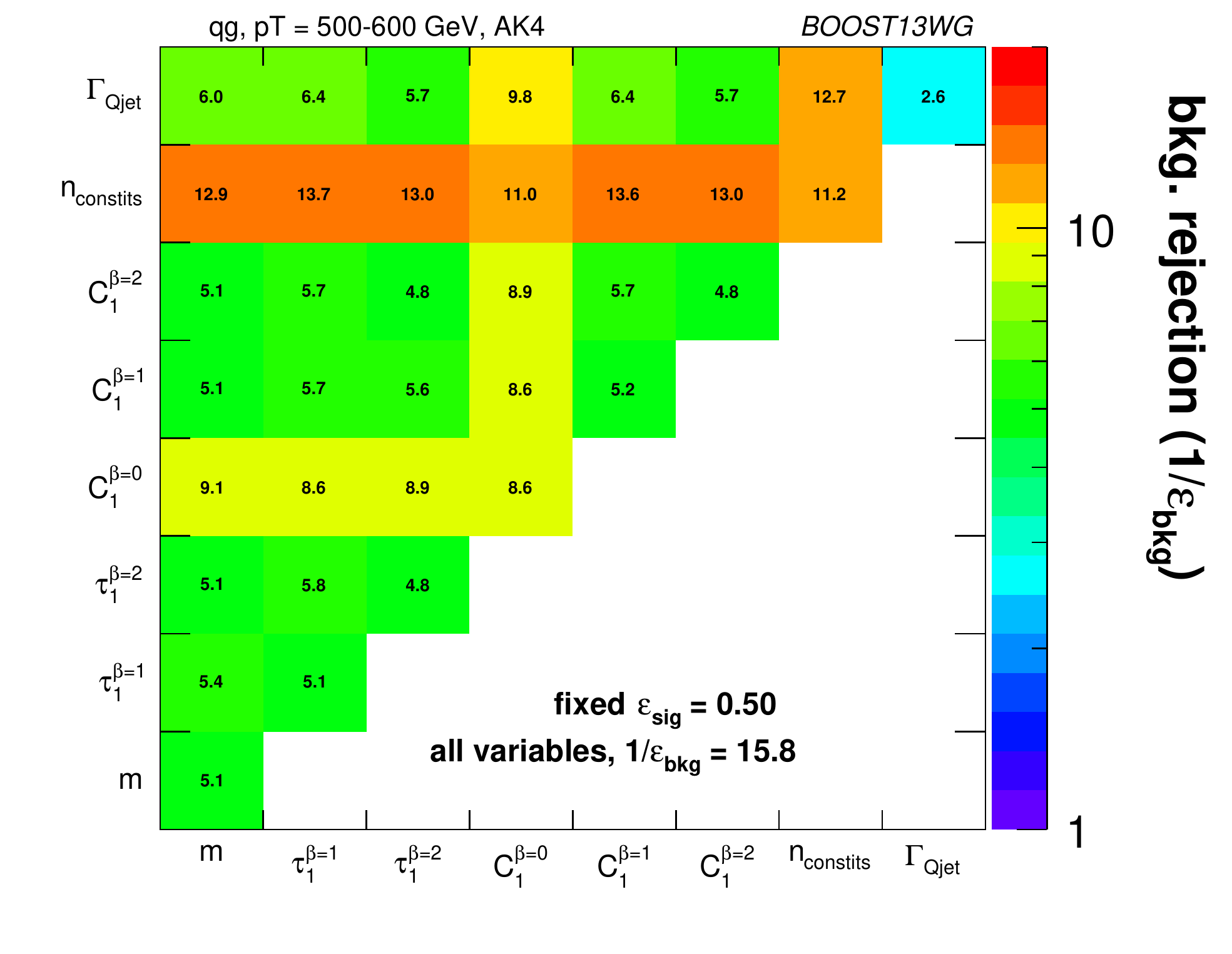}
\includegraphics[width=0.48\textwidth]{./Figures/QGTagging/pT1000/AKtR04/effBkg2D.pdf}
\caption{Gluon rejection defined as $1/\epsilon_{\rm gluon}$ when using each 2-variable combination 
as a tagger with 50\% acceptance for quark jets. Results are shown for R=0.4 jets with (top left) $\pt=300-400 \GeV$, 
(top right) $\pt=500-600 \GeV$ and (bottom) $\pt=1-1.1 \TeV$. The rejection obtained with a tagger that uses all variables is also shown
in the plots. }
\label{fig:qg_akt4_comb}
\end{figure*}

Some features are more easily seen with an alternative presentation of
the data. In Figures~\ref{fig:qg_pt1000_comb} and
\ref{fig:qg_akt4_comb} we fix $R$ and $\pt$ and simultaneously show the single- and pair-wise observables performance in a single matrix.  
The numbers in each cell are the same rejection rate for gluons used earlier,
 $1/\epsilon_\text{bkg}$, with $\epsilon_\text{sig}= 50\,\% $ (quarks).  Figure~\ref{fig:qg_pt1000_comb} shows the results for $\pt=1-1.1 \TeV$ and
$R =0.4,0.8,1.2$, while  Figure~\ref{fig:qg_akt4_comb} is for $R = 0.4$ and the 3 $\pt$ bins.  The single observable rejection rates appear on the 
diagonal, and the pairwise results are off the diagonal.  The largest pair-wise rejection rate, as already suggested by Figure~\ref{fig:qg_surface_pairB}(e),
appears at large $\pt$ and small $R$ for the pair $n_ {\rm constits}+ \tau_1^{\beta=1}$ 
(with very similar results for $n_ {\rm constits}+ C_1^{\beta=1}$). 
The correlations indicated by the shading\footnote{The connection between the value of the rejection rate and the shading color in 
Figures~\ref{fig:qg_pt1000_comb} and \ref{fig:qg_akt4_comb} is the same as that in Figures~\ref{fig:qg_surface_single} to \ref{fig:qg_surface_pairB}.}   
should be largely understood as indicating the organization of the observables into the now-familiar
classes.  The all-observable (BDT) result appears as the number at the lower right in each plot.

\subsection{QCD Jet Masses}\label{sec:qg_mass}

\begin{figure*}
\centering
\subfigure[Quark jets]{\includegraphics[width=0.40\textwidth]{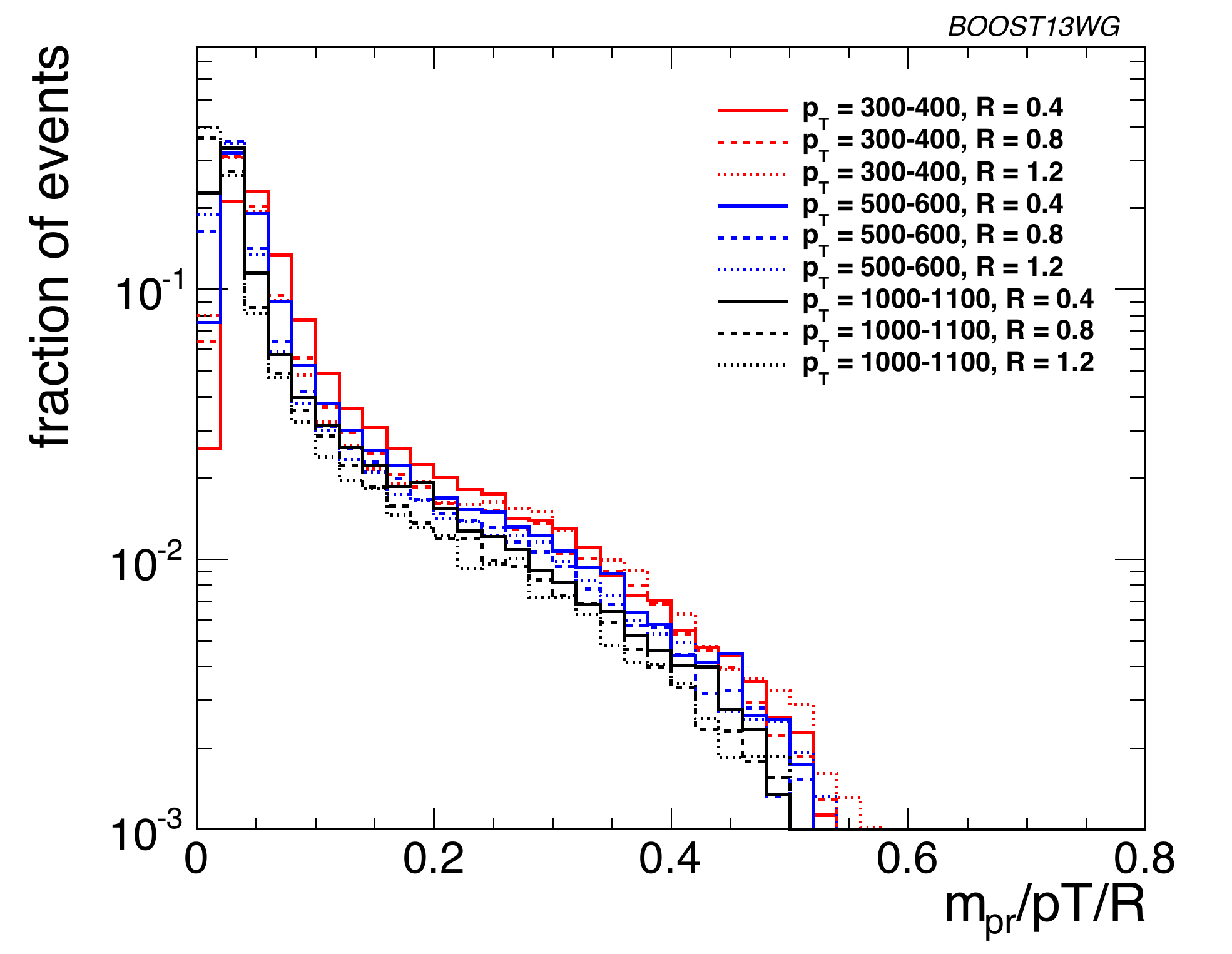}}
\subfigure[Gluon jets]{\includegraphics[width=0.40\textwidth]{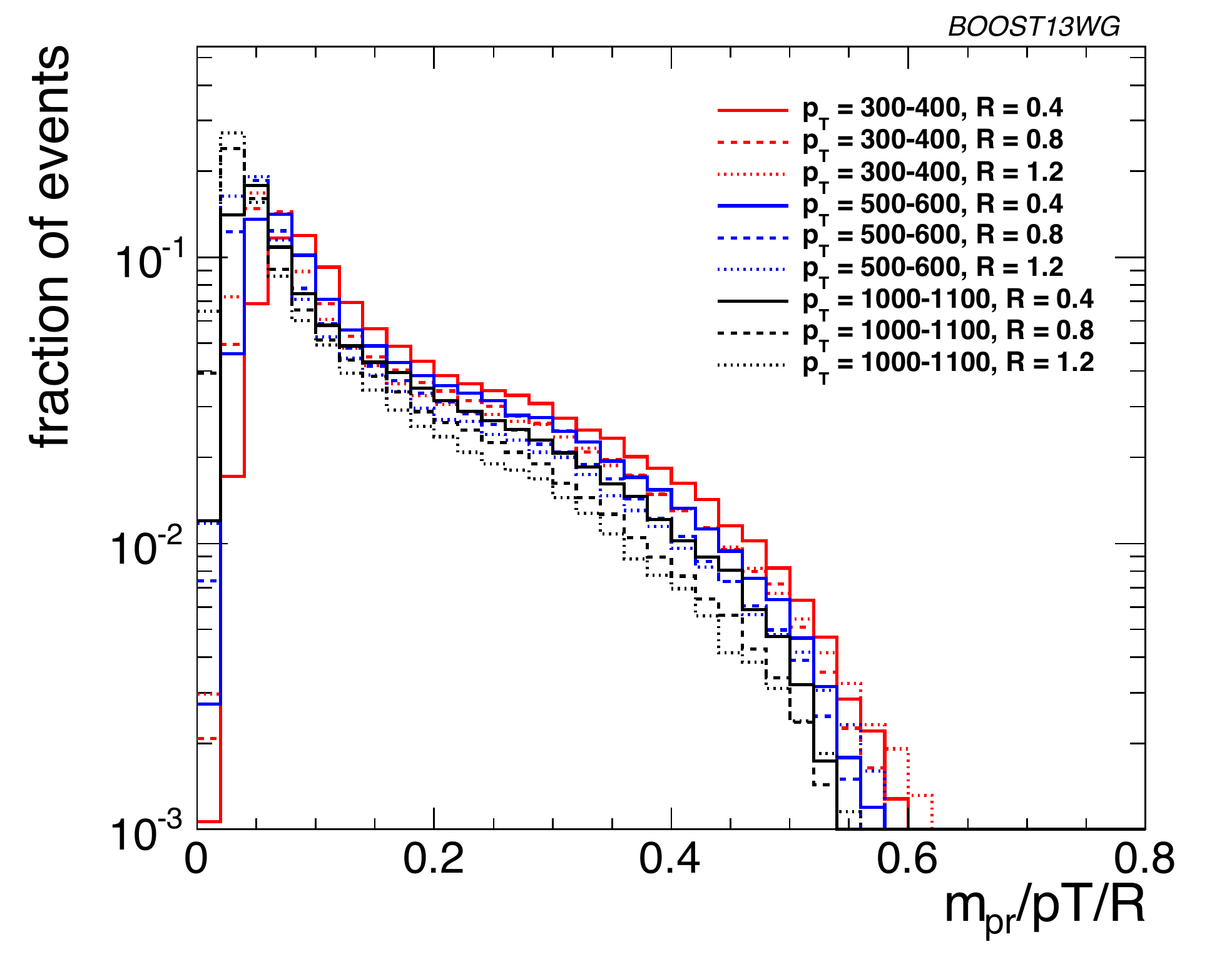}}
\caption{Comparisons of quark and gluon ungroomed mass distributions versus the scaled variable $m/p_T/R$. }
\label{fig:qg_masses_log}
\end{figure*}

To close the discussion of $q/g$-tagging, we provide some insight into the behavior of the masses of QCD jets initiated by both kinds of partons,
 with and without grooming.  Recall that, in practice, an
identified jet is simply a list of constituents, \textit{i.e.}, final
state particles.  To the extent
that the masses of these individual constituents can be neglected (due to the constituents being relativistic), each constituent has a ``well-
defined'' 4-momentum from its energy and direction.  It follows that the 4-momentum of the jet  is simply the sum of the 4-momenta of the constituents and its square is the jet mass squared.
Simply on dimensional grounds,
we know that jet mass must have an overall linear scaling with $\pt$, with the remaining $\pt$ dependence arising predominantly from the running of the coupling,
$\alpha_s(\pt)$.  The $R$ dependence is also crudely linear as the jet mass scales approximately with the largest angular opening between any 2 constituents,
which is set by $R$.

To demonstrate this universal behavior for jet mass, we first note that if we consider the mass distributions for many kinematic points (various values of $R$ and $\pt$),
we observe considerable variation in behaviour. This variation, however, can largely be removed by plotting versus the scaled variable $m/p_T/R$.  The mass distributions for quark and gluon jets versus $m/p_T/R$ for all of our kinematic points 
are shown in Figure~\ref{fig:qg_masses_log}, where
we use a logarithmic scale on the y-axis to clearly exhibit the behavior of these distributions over a large dynamic range.  We observe that the distributions
for the different kinematic points do approximately scale as expected, \textit{i.e.}, the simple arguments above capture most of the variation with $R$ and $\pt$.
We will consider shortly an explanation of the residual non-scaling. A
more rigorous quantitative understanding of jet mass distributions
requires all-orders calculations in QCD, which have been performed for
groomed and ungroomed jet mass spectra at high logarithmic accuracy, both in the context of direct QCD resummation~\cite{Li:2012bw,Dasgupta:2012hg,Dasgupta:2013ihk,Dasgupta:2013via} and Soft Collinear Effective Theory~\cite{Chien:2012ur,Jouttenus:2013hs,Liu:2014oog}. 

\begin{figure*}
\centering
\subfigure[Quark jets]{\includegraphics[width=0.40\textwidth]{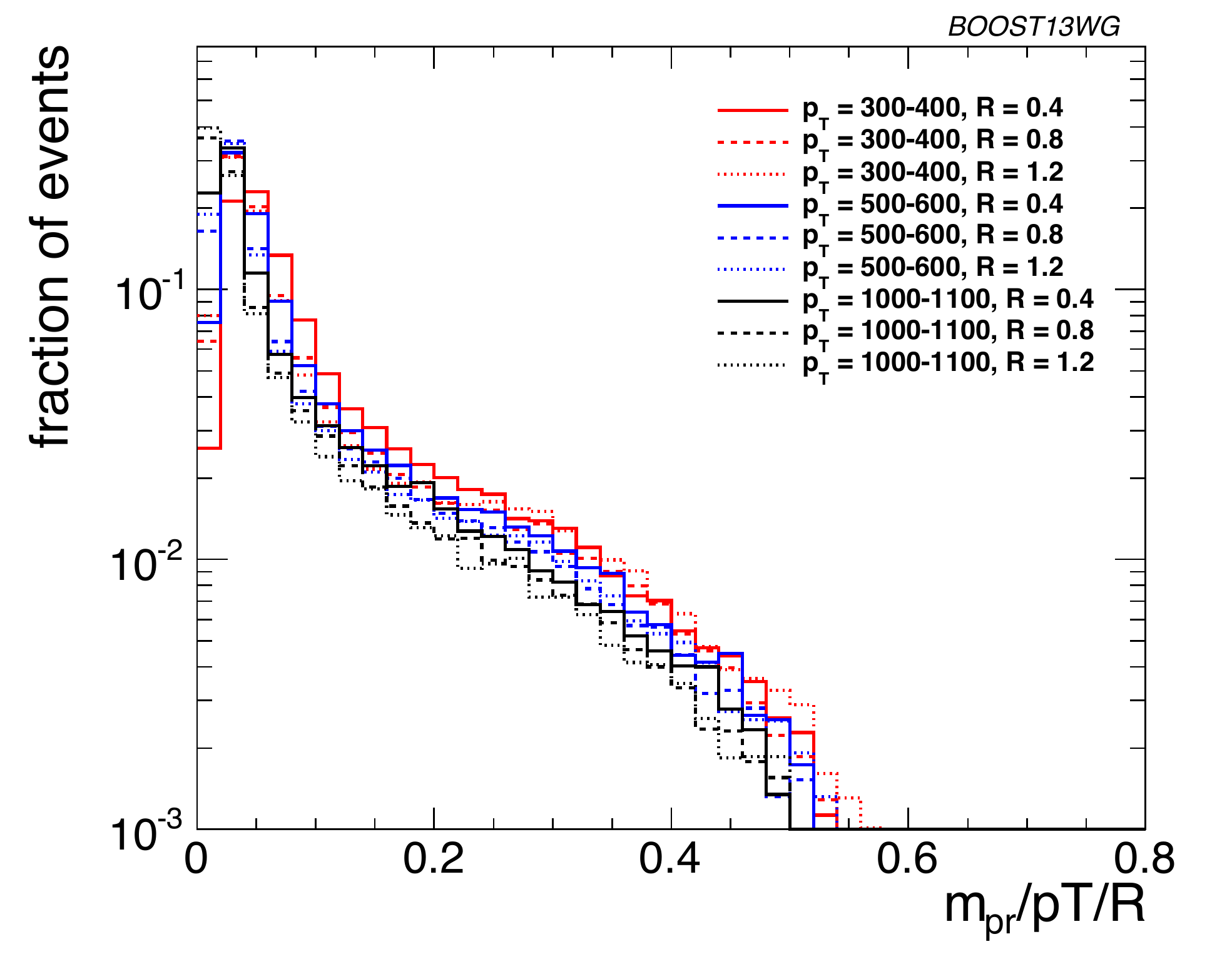}}
\subfigure[Gluon jets]{\includegraphics[width=0.40\textwidth]{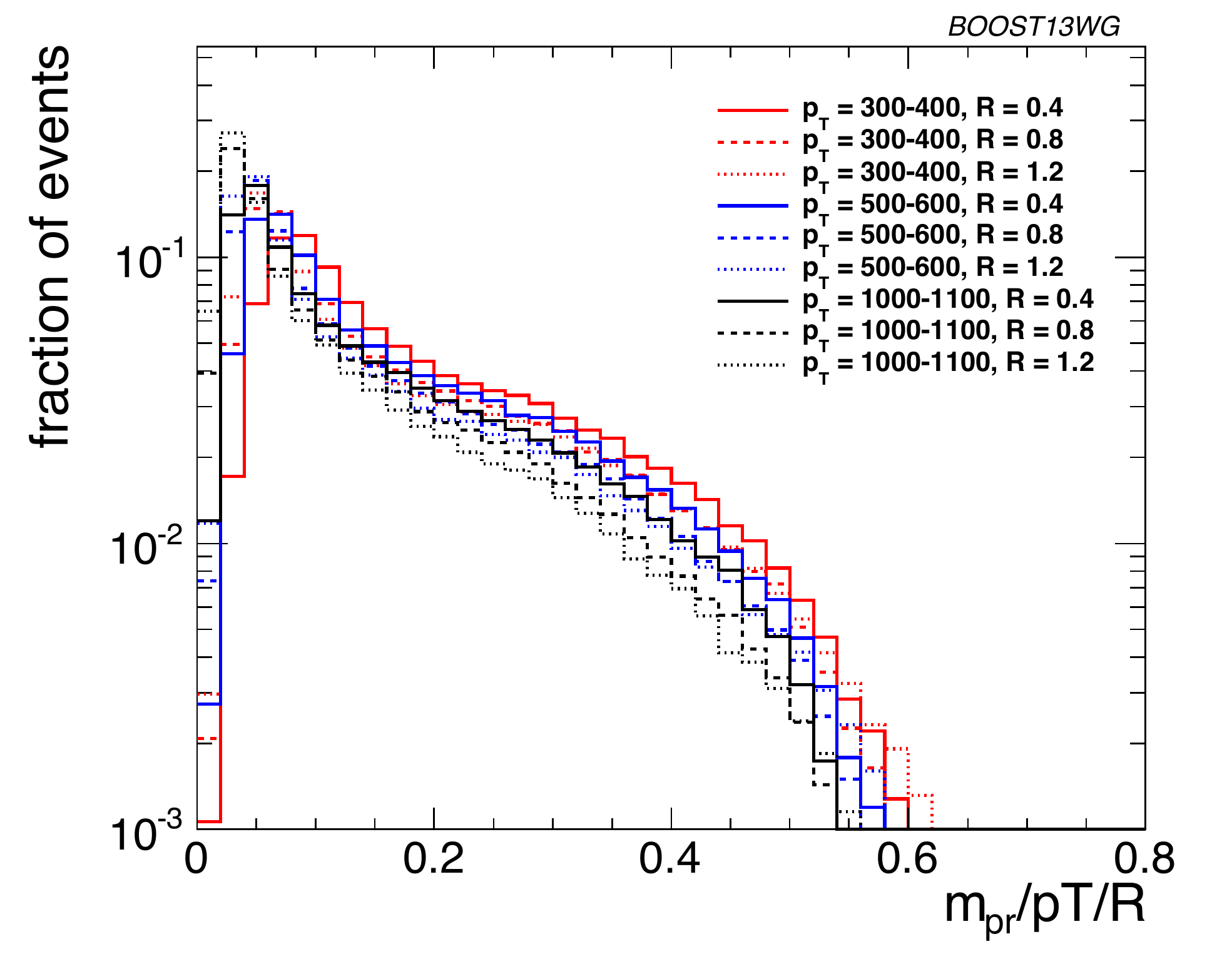}}
\caption{Comparisons of quark and gluon pruned mass distributions versus the scaled variable $m_\text{pr}/p_T/R$. }
\label{fig:qg_prmasses_log}
\end{figure*}

Several features of Figure~\ref{fig:qg_masses_log} can be easily understood.  The distributions all cut off rapidly for $m/p_T/R > 0.5$, which is understood as
the precise limit (maximum mass) for a jet composed of just 2 constituents.  As expected from the soft and collinear singularities in QCD, the mass distribution peaks 
at small mass values.  The actual peak is ``pushed'' away from the origin by the so-called Sudakov form factor.  Summing the corresponding logarithmic structure 
(singular in both $\pt$ and angle) to all orders in perturbation theory yields a distribution that is highly damped  as the mass vanishes.  In words, there is precisely 
\textit{zero} probability that a color parton emits \textit{no} radiation (and the resulting jet has zero mass). Above the Sudakov-suppressed part of phase space, there are two 
structures in the distribution:~the ``shoulder'' and the ``peak''.
 The large mass shoulder ($0.3 < m/p_T/R < 0.5$) is driven largely by the presence of a single large angle, energetic
emission in the underlying QCD shower, \textit{i.e.}, this regime is quite well described by low-order perturbation theory\footnote{The shoulder label will become more clear when examining groomed jet mass distributions.} 
In contrast, we can think of the peak region as corresponding to multiple soft emissions.  This simple, necessarily approximate picture 
provides an understanding of the bulk of the
differences between the quark and gluon jet mass distributions.  Since the probability of the single large angle, energetic emission is proportional to the color charge,
the gluon distribution should be enhanced in this region by a factor of about $C_A/C_F = 9/4$, consistent with what is observed in Figure~\ref{fig:qg_masses_log}.
Similarly the exponent in the Sudakov damping factor for the gluon jet mass distribution  is enhanced by the same factor, 
leading to a peak ``pushed'' further from the origin.  Therefore, compared to a quark jet, the gluon jet mass 
distribution exhibits a larger average jet mass, with a larger relative contribution arising from the perturbative shoulder region and a small mass peak that is further
from the origin.

Together with  the fact that the number of constituents in the jet is also larger (on average) for the gluon jet simply because a gluon will radiate more than a quark, these features explain much of what we observed earlier in terms of the effectiveness of the various observables to separate quark jets from gluons jets. They also give us insight into the difference in the distributions for the observable 
$\Gamma_{\rm Qjet}$. Since the shoulder is dominated by a single large angle, \textit{hard} emission, it is minimally impacted by pruning,
which is designed to remove the large angle, \textit{soft} constituents (as shown in more detail below). Thus, jets in the shoulder exhibit small volatility and they are a larger
component in the gluon jet distribution.  Hence gluon jets, on average, have smaller values of  $\Gamma_{\rm Qjet}$ than quark jets as in 
Figure~\ref{fig:qg_pt500_subst_AKt_R08}(b).  Further, this feature of gluon jets is distinct from the fact that there are more constituents, explaining why
$\Gamma_{\rm Qjet}$ and $n_ {\rm constits}$ supply largely independent information for distinguishing quark and gluon jets. 

To illustrate some of these points in more detail, Figure~\ref{fig:qg_prmasses_log} exhibits the same jet mass distributions  
\textit{after pruning}~\cite{Ellis:2009me,Ellis:2009su}.  Removing the large angle,  soft constituents
moves the peak in both of the distributions from  $m/p_T/R \sim 0.1 - 0.2$ to the region around $m/p_T/R \sim 0.05$.  This explains why pruning works to reduce the
QCD background when looking for a signal in a specific jet mass bin.  The shoulder feature at higher mass is much more apparent after pruning, as is the larger shoulder for
the gluon jets.  A quantitative (all-orders) understanding of groomed mass distributions is also possible. For instance, resummation of the pruned mass distribution was achieved in~\cite{Dasgupta:2013ihk,Dasgupta:2013via}. Figure~\ref{fig:qg_prmasses_log} serves to confirm the physical understanding of the relative behavior of $\Gamma_{\rm Qjet}$ for quark and gluon jets.

Our final topic in this section is the residual $R$ and $\pt$ dependence exhibited in Figures~\ref{fig:qg_masses_log} and \ref{fig:qg_prmasses_log}, which
indicates a deviation from the naive linear scaling that has been removed by using the scaled variable $m/p_T/R$.  
A helpful, intuitively simple, if admittedly imprecise, model of a jet is to separate the constituents of the jet into ``hard'' (with $\pt$'s that are of order the jet 
$\pt$) versus ``soft'' (with $\pt$'s small and fixed compared to the jet $\pt$), and ``large'' angle (with an angular separation from the jet direction of order $R$)
versus ``small'' angle (with an angular separation from the jet direction smaller than and not scaling with $R$) components. 
As described above the Sudakov damping factor excludes constituents that are very soft or very small angle (or both).
In this simple picture 
perturbative large angle, hard constituents appear rarely, but, as described above, they characterize the large mass jets that appear in the
``shoulder'' of the jet mass distribution where the mass scales approximately linearly with the jet $\pt$ and with $R$.  
The hard, small angle constituents are somewhat more numerous
and contribute to a jet mass that does not scale with $R$.  The soft constituents are much more numerous (becoming more numerous
with increasing jet $\pt$) and contribute to a jet mass that scales like
$\sqrt{p_{T,\text{jet}}}$.  The small angle, soft constituents contribute to a jet mass that does not scale with $R$, while the large angle, soft
constituents do contribute to a jet mass that scales like $R$  and grow in number approximately linearly in $R$ (\textit{i.e.}, with the area
of the annulus at the outer edge of the jet).  This simple picture allows at least a qualitative explanation of the behavior observed in 
Figures~\ref{fig:qg_masses_log} and \ref{fig:qg_prmasses_log}.
    
As already suggested, the residual $\pt$ dependence can be understood as arising primarily from the slow decrease of the strong coupling $\alpha_s(\pt)$ as $\pt$ increases.  This leads to a corresponding decrease in the (largely perturbative) shoulder regime for both distributions
at higher $\pt$, \textit{i.e.}, a decrease in the number of hard, large angle constituents.  
At the same time, and for the same reason, the Sudakov damping is less strong with increasing $\pt$ and the peak moves in towards the origin.
While the number of soft constituents increases with increasing jet $\pt$, their contributions to the scaled jet mass distribution shift to smaller
values of $m/\pt$ (decreasing approximately like $1/\sqrt{\pt}$).
Thus the overall impact of increasing $\pt$ for both distributions is a (gradual) shift to smaller values of $m/p_T/R$.  This is just what is observed in 
Figures~\ref{fig:qg_masses_log} and \ref{fig:qg_prmasses_log}, although the numerical size of the effect is reduced in the pruned case.

The residual $R$ dependence is somewhat more complicated.  
The perturbative large angle, hard constituent contribution
largely scales in the variable $m/p_T/R$, which is why we see little residual $R$ dependence in either figure at higher masses ($m/p_T/R > 0.4$).  
The contribution of the small angle constituents (hard and soft) contribute at fixed $m$ and thus shift to the left versus the scaled variable as $R$ increases.  
This presumably explains the small shifts in this direction at small mass observed in both figures.
The large angle, soft constituents contribute to mass values that scale like $R$, and, as noted above, tend to increase in number as $R$ increases 
(\textit{i.e.}, as the area of the jet grows).  Such contributions  yield a scaled jet mass
distribution that shifts to the right with increasing $R$  and presumably explain the behavior at small $\pt$ in Figure~\ref{fig:qg_masses_log}.  Since pruning
largely removes this contribution, we observe no such behavior in Figure~\ref{fig:qg_prmasses_log}. 

 \subsection{Conclusions}\label{sec:qg_concl}

In Section~\ref{sec:qgtagging} we have seen that a variety of jet observables
 provide information about the jet that can be employed  to effectively separate quark-initiated from gluon-initiated jets.  Further,
when used in combination, these observables can provide superior separation. Since the improvement depends on the correlation between observables,
we use the multivariable performance to separate the observables into different classes, with each class containing highly correlated observables.  
We saw that the best performing single observable is simply the number of
constituents in the jet, $n_ {\rm constits}$,  while the largest
further improvement comes from combining with $C_1^{\beta =1}$ (or
$\tau_1^{\beta=1}$). The performance of this combined tagger is
strongly dependent on \pt and $R$, with the best performance being
observed for smaller $R$ and higher \pt. The smallest
$R$ and $\pt$ dependence arises from combining $n_ {\rm constits}$ with $C_1^{\beta =
  0}$. Some of the commonly used observables for $q/g$ tagging are highly correlated
and do not provide extra information when used together. 
We have found that adding further variables to the $n_ {\rm constits}$
+ $C_1^{\beta =1}$  or $n_ {\rm constits}$
+ $\tau_1^{\beta =1}$ BDT combination
results in only a small improvement in performance, suggesting that almost all of the available information to
discriminate quark and gluon-initiated jets is captured by $n_{\rm
  constits}$ and $C_1^{\beta=1}$ (or $\tau_1^{\beta=1}$)
variables. In addition to demonstrating these correlations, we have provided a discussion of the 
physics behind the structure of the correlation.  Using the jet mass as an example, we have given arguments to explicitly explain the differences between jet observables 
initiated by each type of parton.

Finally, we remind the reader that the numerical results were derived for a particular color
configuration ($qq$ and $gg$ events), in a particular implementation of the parton shower and hadronization. Color
connections in more complex event configurations, or different Monte Carlo programs,
may well exhibit somewhat different efficiencies and rejection factors. The value of our results
is that they indicate a subset of variables expected to be rich in information about
the partonic origin of final-state jets. These variables can  be expected to act as valuable
discriminants in searches for new physics, and could also be used to define 
model-independent final-state measurements which would nevertheless be sensitive to the
short-distance physics of quark and gluon production.
 

%% file: wtagging.tex
In this section, we study the discrimination of a boosted, hadronically decaying $W$ boson (signal) against a gluon-initiated jet
background, comparing the performance of various groomed jet
masses and substructure variables.
A range of different distance parameters for the \antikt jet
algorithm are explored, in a range of different leading jet \pt bins.
This allows us to determine the
performance of observables as a function of jet radius and jet boost, and to see
where different approaches may break down. The groomed
mass and substructure variables are then combined in a BDT as described in Section~\ref{sec:multivariate}, and the performance of the resulting BDT discriminant
explored through ROC curves to understand the degree to which
variables are correlated, and how
this changes with jet boost and jet radius. Using BDT combinations of substructure variables to improve W tagging has been studied earlier in 
\cite{Cui:2010km}.

\subsection{Methodology}

These studies use the $WW$ samples as signal and the dijet $gg$ as background, described previously in Section~\ref{sec:samples}. Whilst only gluonic backgrounds
are explored here, the conclusions regarding the dependence of the
performance and correlations on the jet boost and radius are not
expected to be substantially different for quark backgrounds; we will
see that the differences in the substructure properties of quark- and
gluon-initiated jets, explored in the last section, are significantly
smaller than the differences between $W$-initiated and gluon-initiated jets.

As in the $q/g$ tagging studies, the showered events were clustered with \textsc{FastJet}
3.03 using
the \antikt~algorithm with jet radii of $R = 0.4,\, 0.8,\, 1.2$. In
both signal and background samples, an upper and lower cut on
the leading jet $\pt$ is applied after showering/clustering, to ensure
similar $\pt$ spectra for signal and background in each \pt bin. The bins
in leading jet \pt that are considered are 300-400 \GeV, 500-600 \GeV,
1.0-1.1 \TeV, for the 300-400 \GeV, 500-600 \GeV,
1.0-1.1 \TeV parton \pt slices respectively. The jets then have various
grooming algorithms applied and substructure observables
reconstructed as described in
Section~\ref{sec:substructure}. The substructure observables
studied in this section are: 

\begin{itemize}
\item Ungroomed, trimmed (\mtrim), and pruned (\mprun) 
jet  masses.
\item Mass output from the modified mass drop tagger (\mmdt).
\item Soft drop mass with $\beta=2$ (\msd). 
\item 2-point energy correlation function ratio $C_2^{\beta=1}$  (we also studied $\beta=2$ but do not show its results because it showed poor discrimination power).
\item $N$-subjettiness ratio $\tau_2 / \tau_1$ with $\beta=1$ ($\tau_{21}^{\beta=1}$) and with axes computed using one-pass $k_t$ axis optimization (we also studied $\beta=2$ but did not show its results because it showed poor discrimination power).
\item Pruned Qjet mass volatility, $\Gamma_{\rm Qjet}$.
\end{itemize}

\subsection{Single Variable Performance}

In this section we  explore the performance of the various groomed
jet mass and substructure variables in separating signal
from background.
Since we
have not attempted to optimise the grooming parameter settings of
each grooming algorithm, we do not place much emphasis
here on the relative performance of the groomed masses, but instead
concentrate on how their performance changes depending on the
kinematic bin and jet radius considered. 

\begin{figure*}
\centering
\subfigure[Ungroomed mass]{\includegraphics[width=0.30\textwidth]{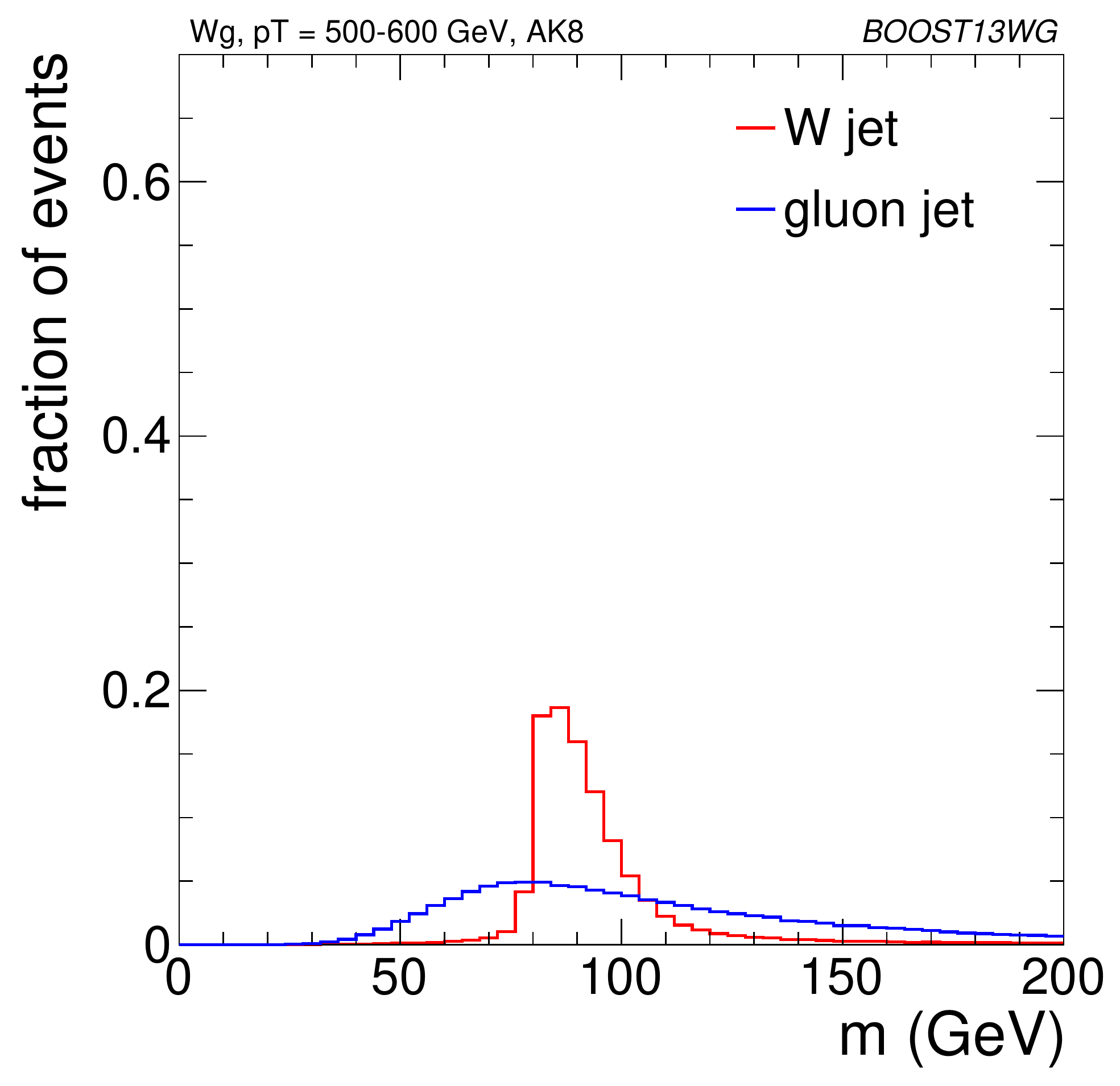}}
\subfigure[Pruned mass]{\includegraphics[width=0.30\textwidth]{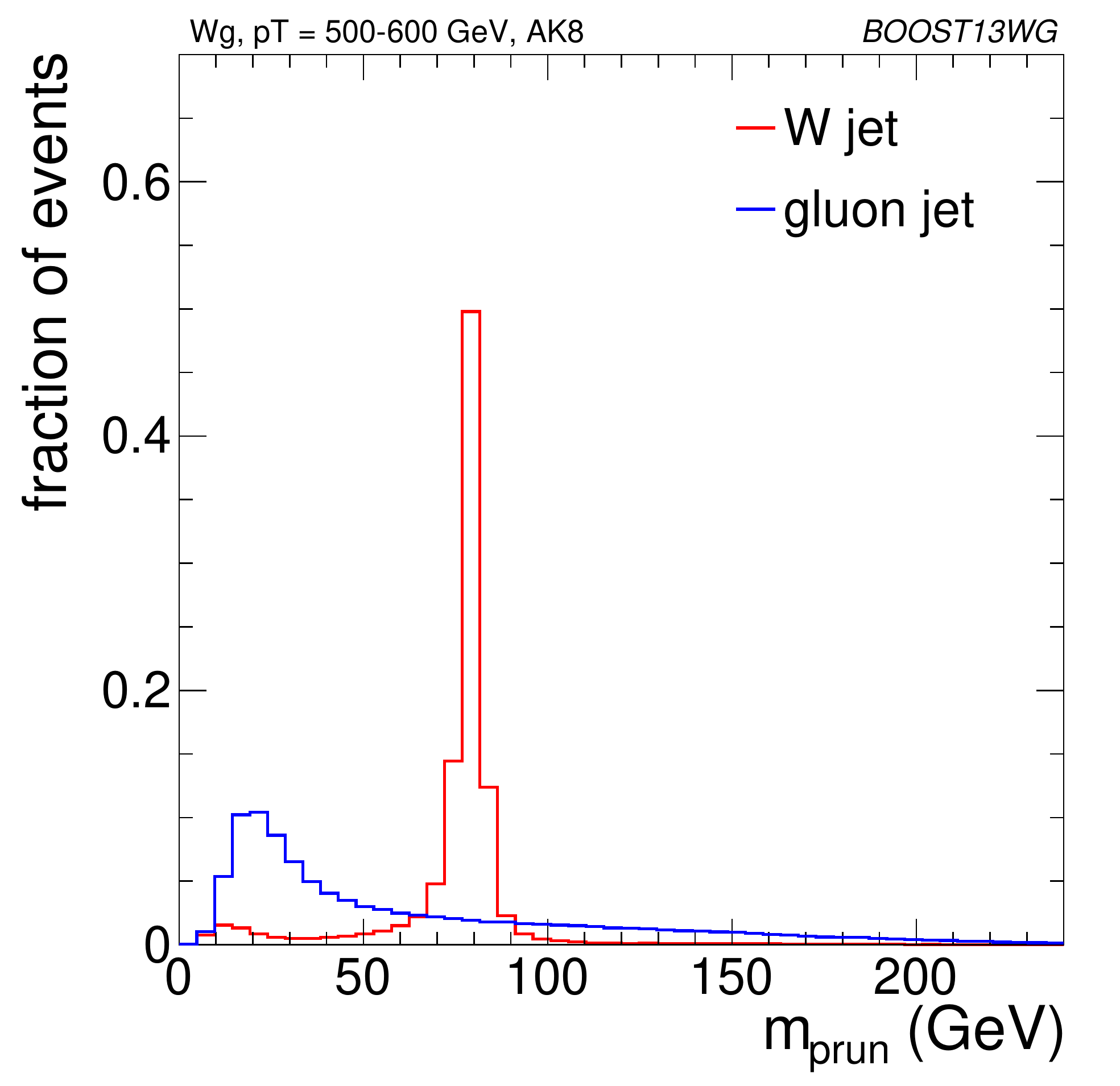}}
\subfigure[Trimmed mass]{\includegraphics[width=0.30\textwidth]{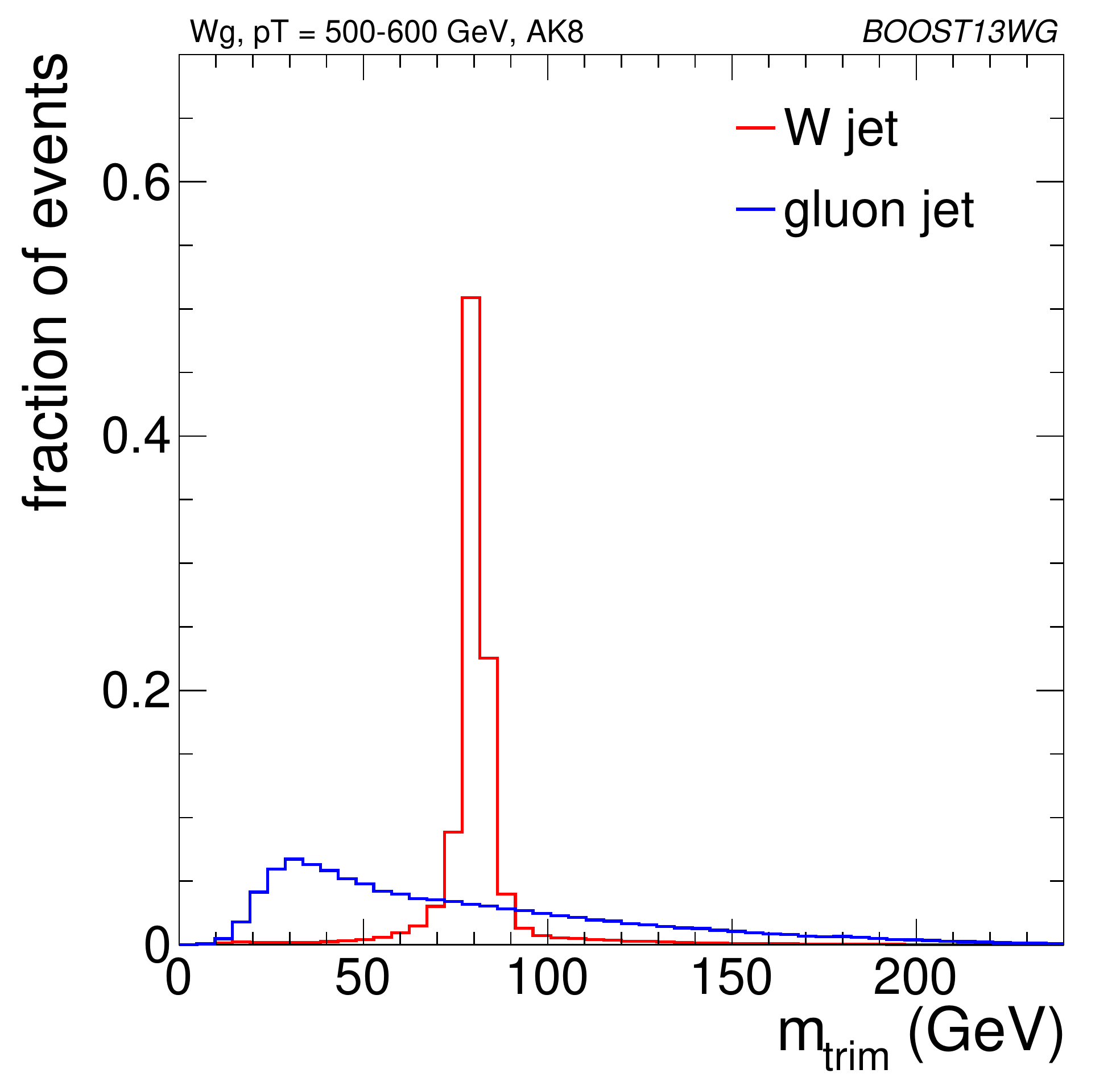}}\\
\subfigure[mMDT mass]{\includegraphics[width=0.30\textwidth]{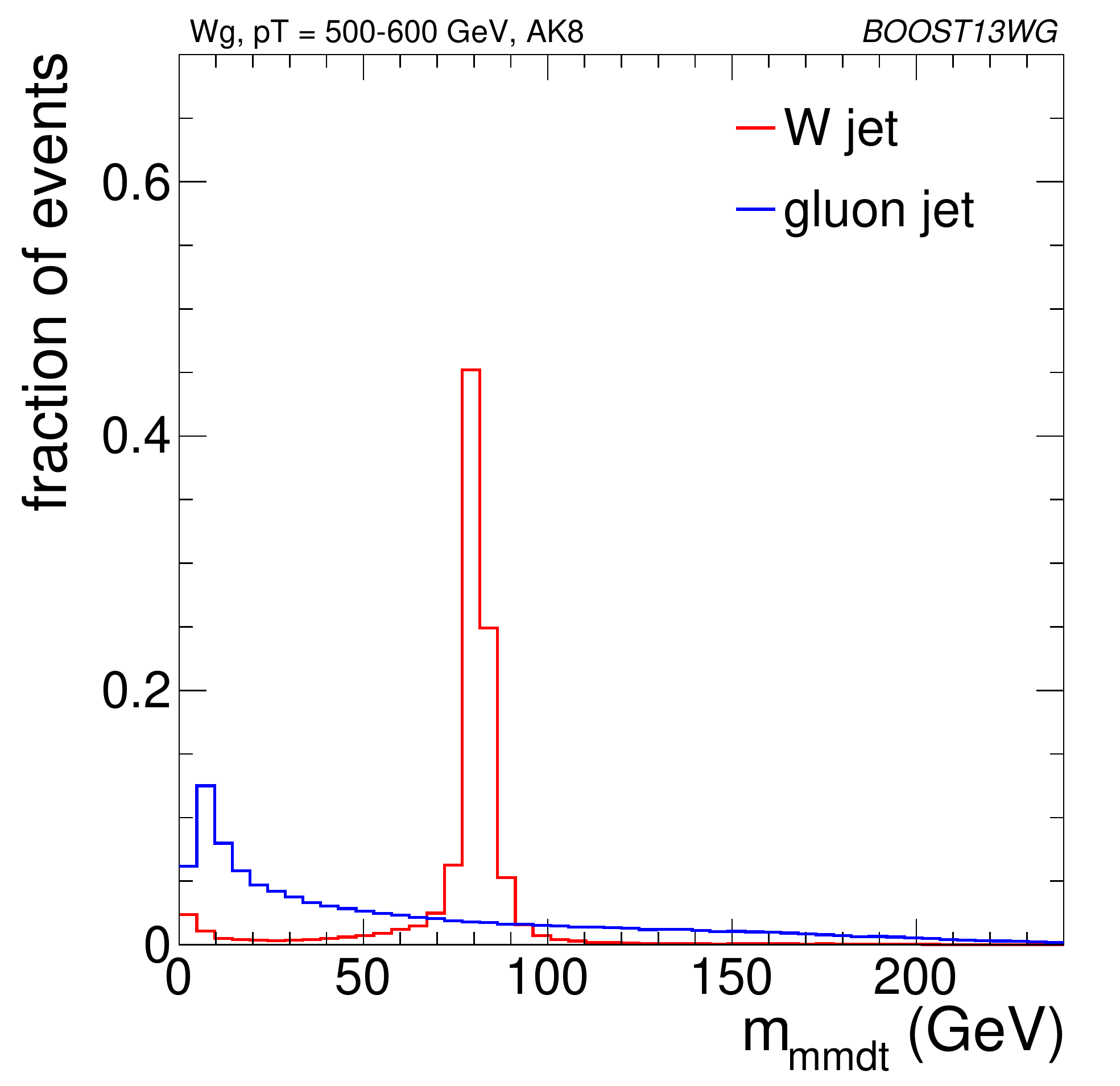}}
\subfigure[Soft-drop $\beta=2$ mass]{\includegraphics[width=0.30\textwidth]{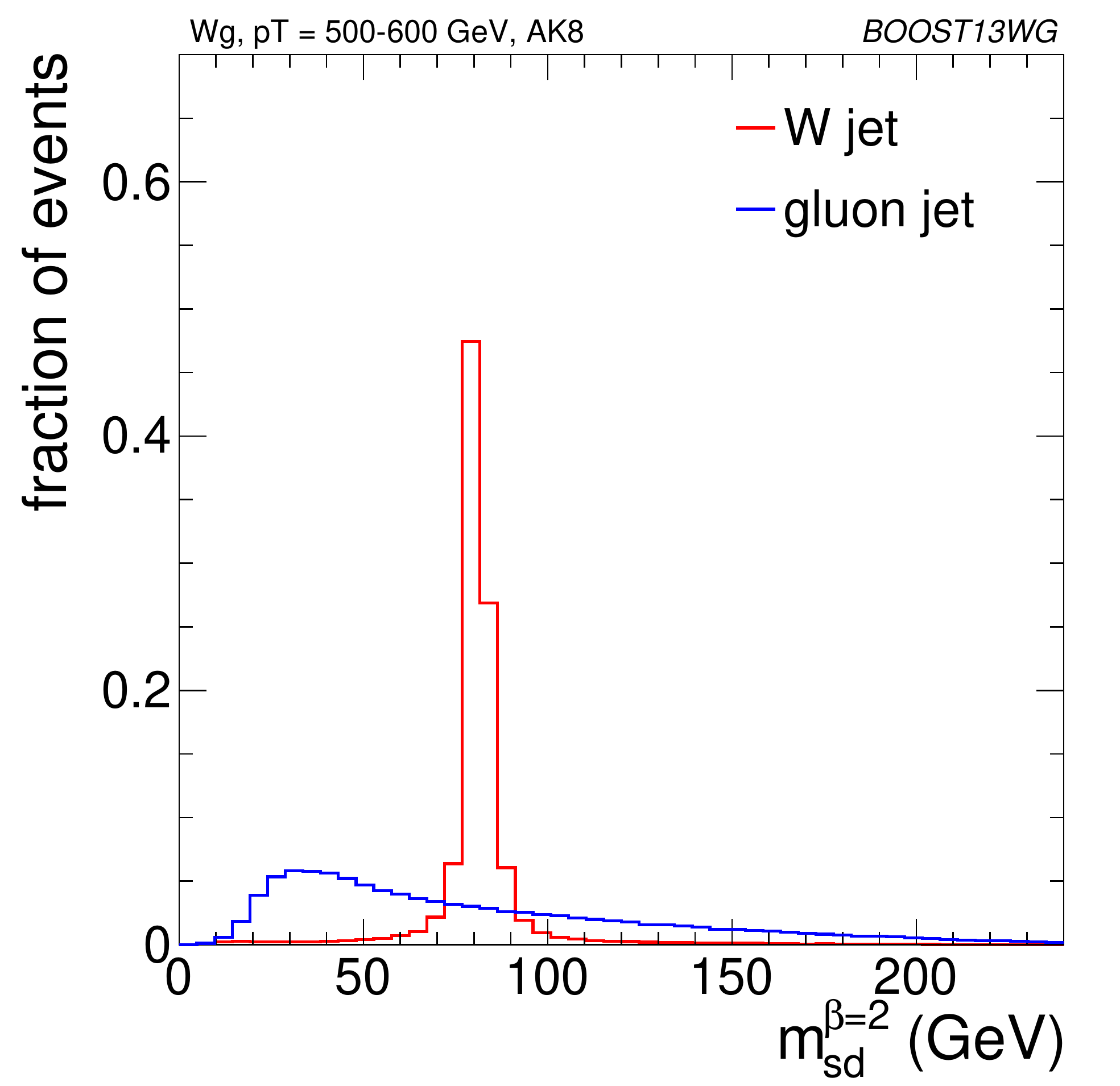}}
\caption{Leading jet mass distributions in the $gg$ background and
  $WW$ signal samples in the \pt = 500-600 \GeV bin using the \antikt $R=0.8$ algorithm.}
\label{fig:pt500_mass_AKt_R08}
\end{figure*}

\begin{figure*}
\centering
\subfigure[$C_2^{\beta=1}$]{\includegraphics[width=0.30\textwidth]{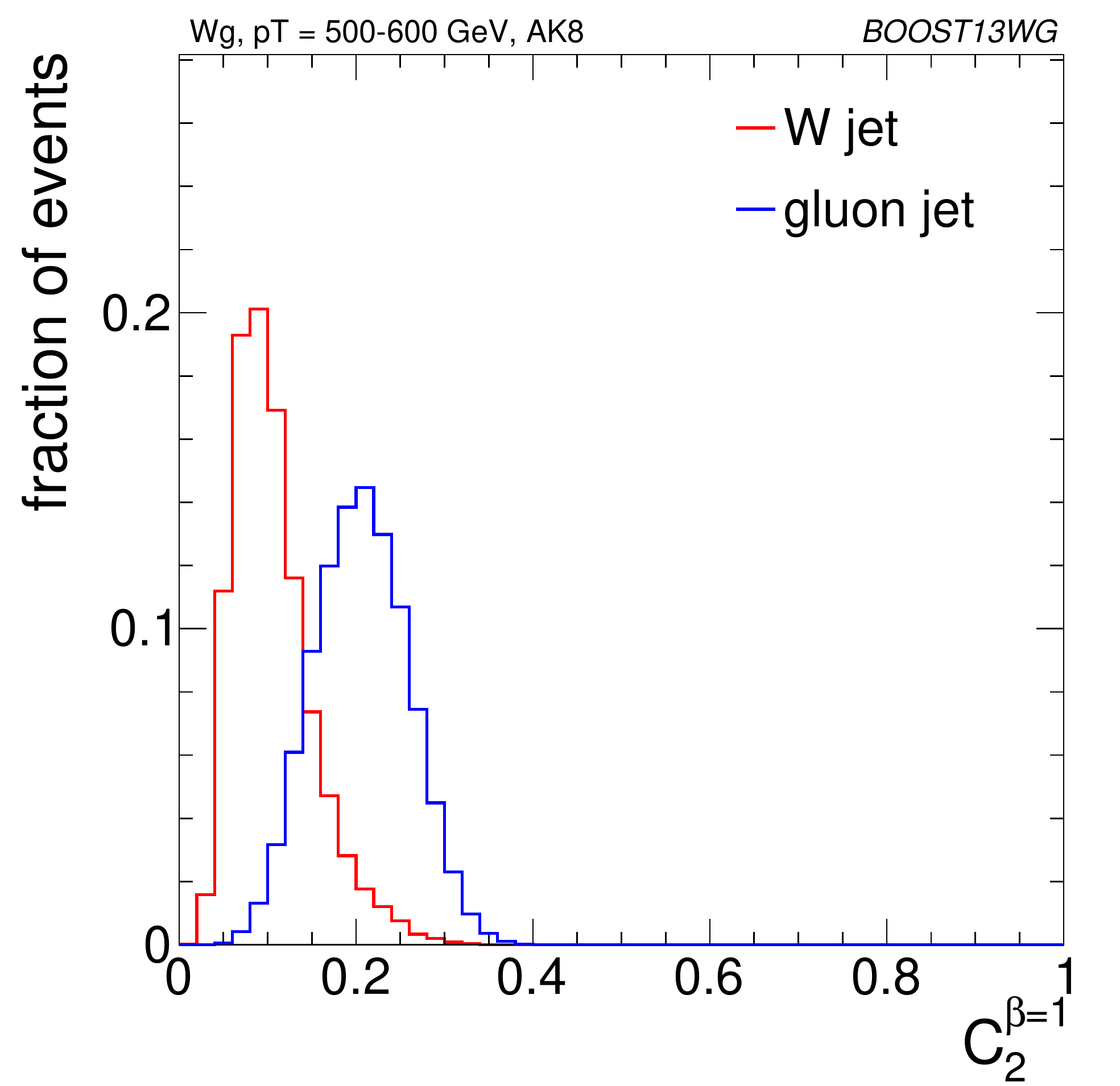}}
\subfigure[$C_2^{\beta=2}$]{\includegraphics[width=0.30\textwidth]{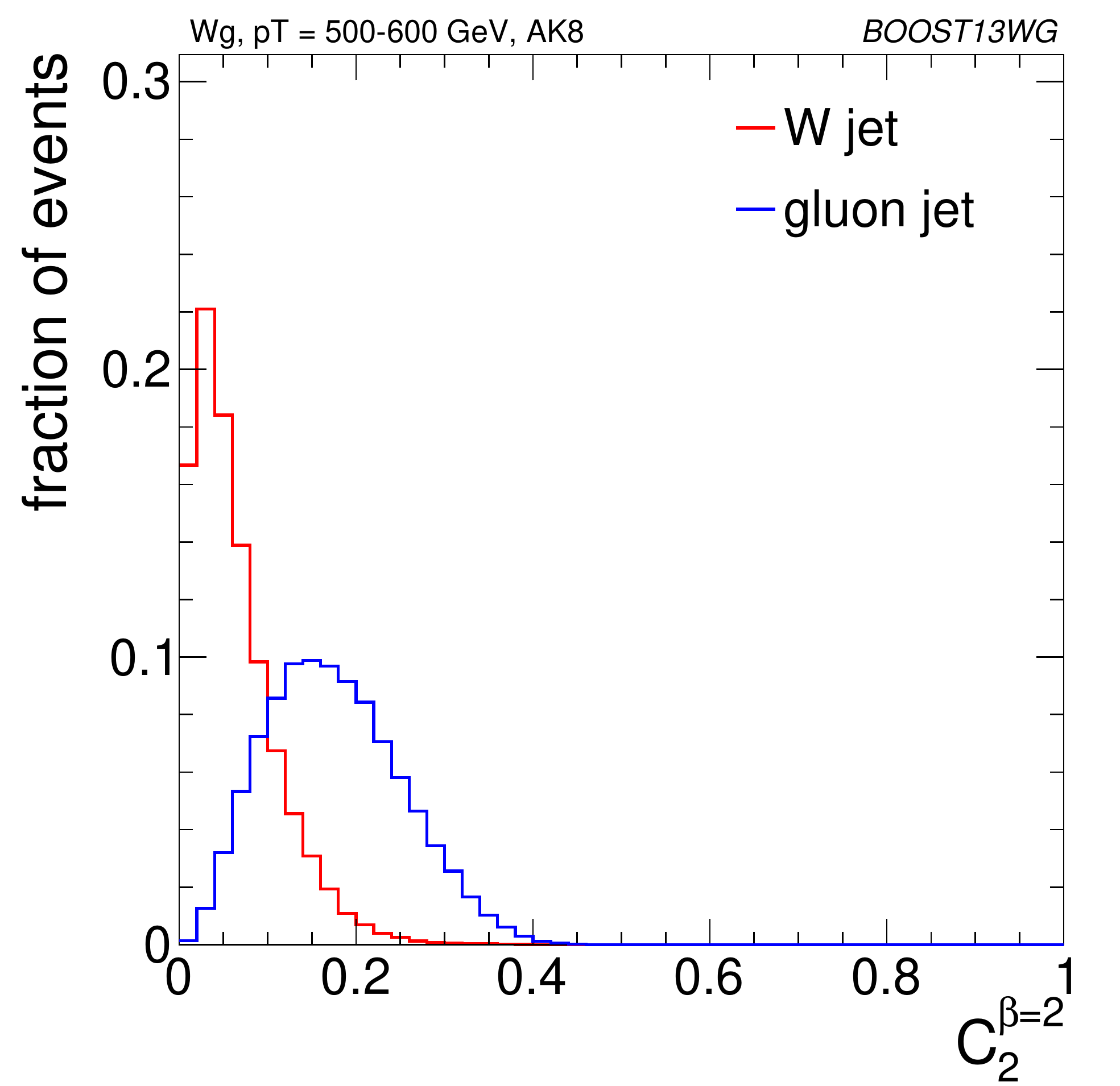}}
\subfigure[$\Gamma_{Qjet}$]{\includegraphics[width=0.30\textwidth]{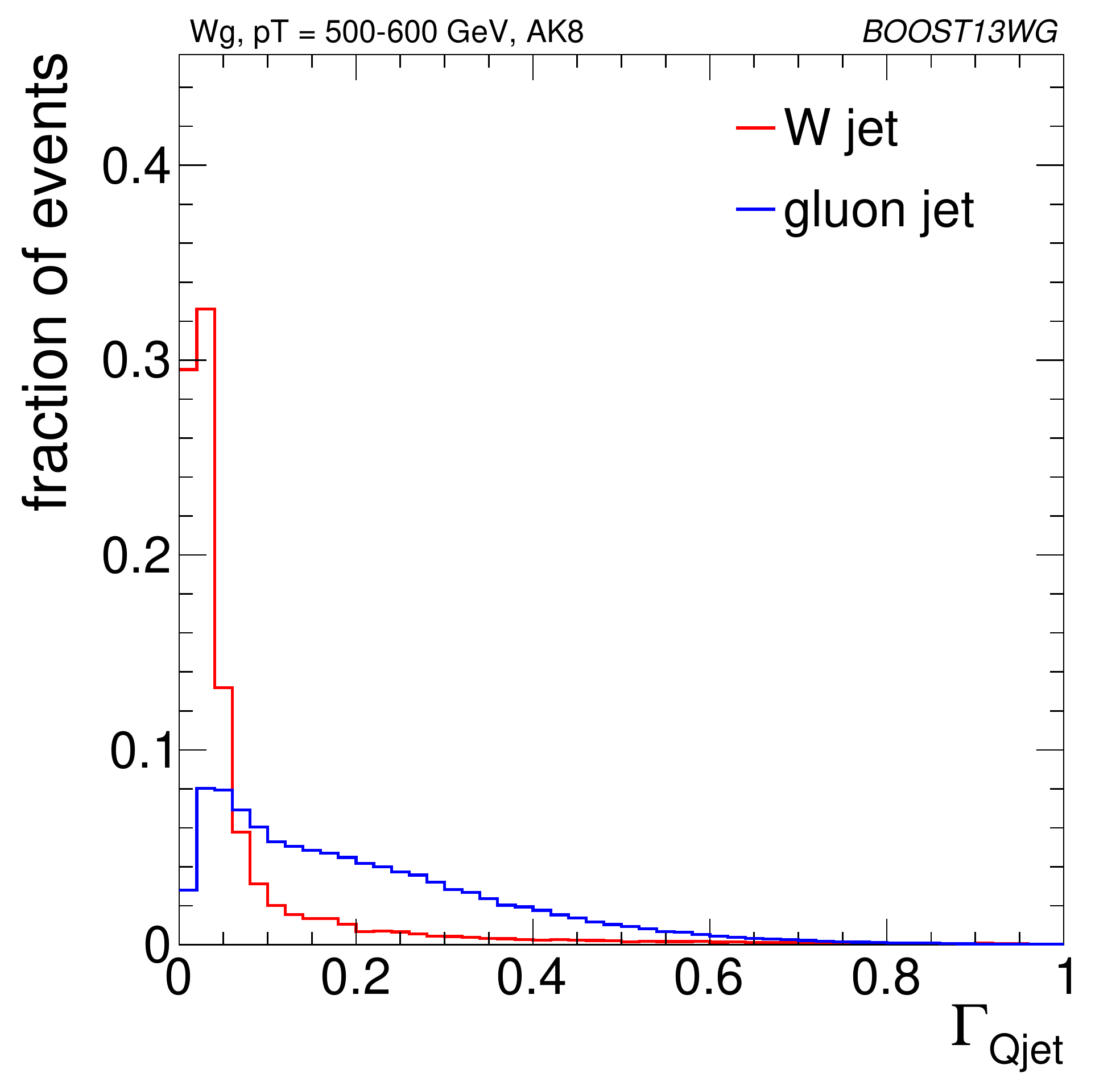}}\\
\subfigure[$\tau_{21}^{\beta=1}$]{\includegraphics[width=0.30\textwidth]{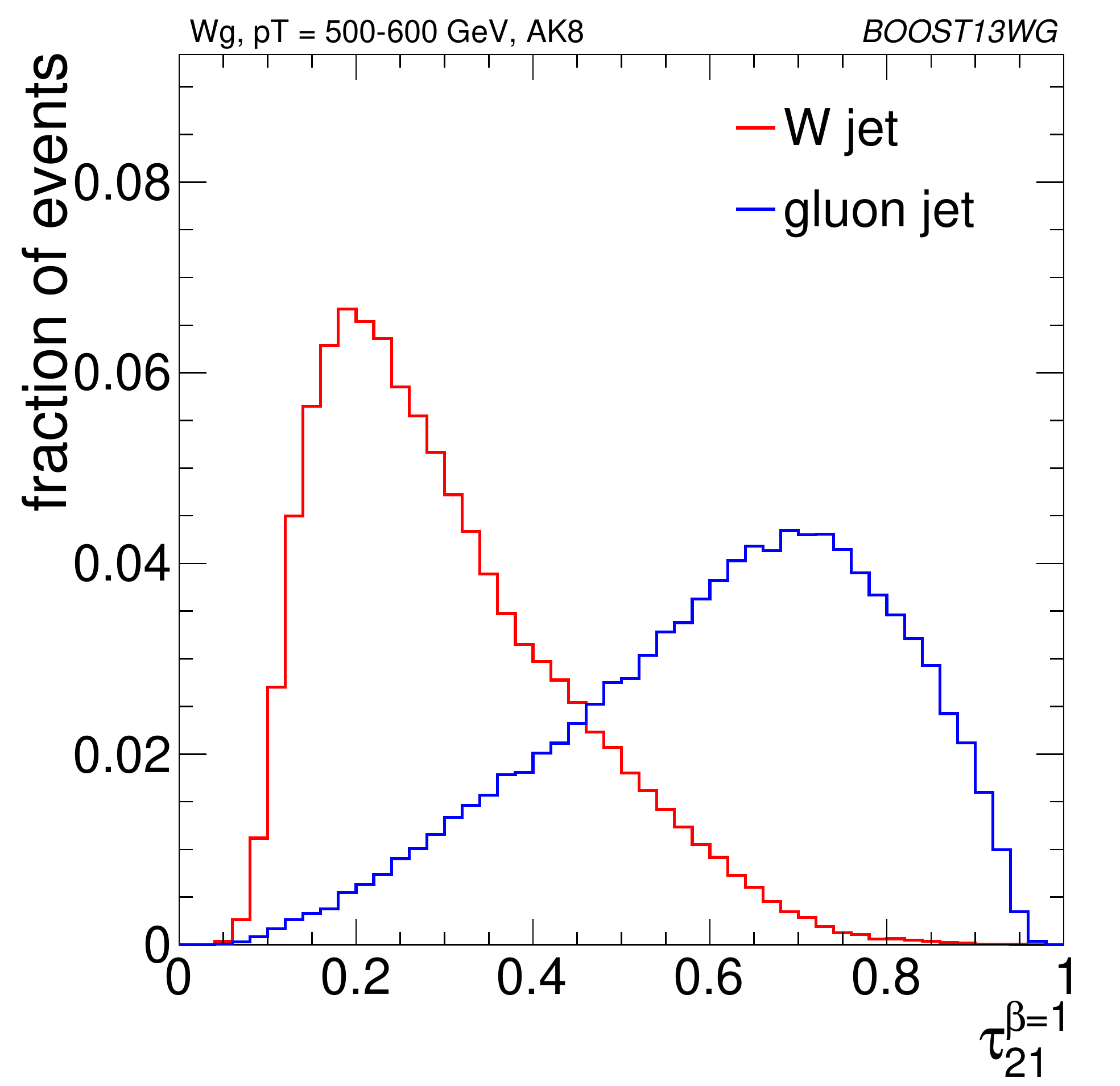}}
\subfigure[$\tau_{21}^{\beta=2}$]{\includegraphics[width=0.30\textwidth]{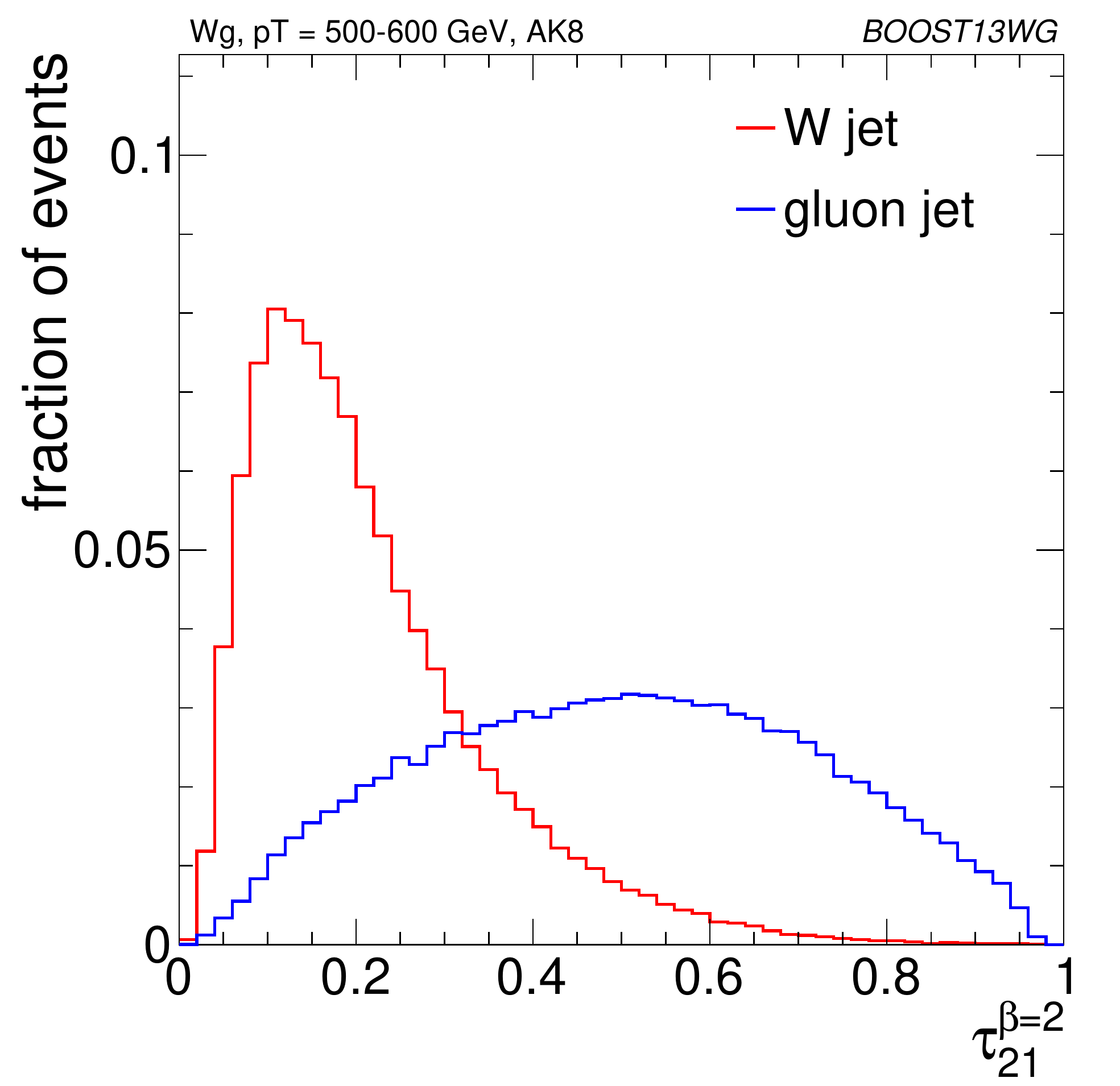}}
\caption{Leading jet substructure variable distributions in the $gg$
  background and $WW$ signal samples in the \pt = 500-600 \GeV bin using the \antikt $R=0.8$ algorithm.}
\label{fig:pt500_subst_AKt_R08}
\end{figure*}

Figure~\ref{fig:pt500_mass_AKt_R08}  compares the signal and
background in terms of the different groomed masses explored for the
\antikt $R=0.8$ algorithm in the \pt = 500-600 \GeV bin. One can clearly see
that, in terms of separating signal and background, the groomed masses
are significantly more performant than the ungroomed \antikt $R=0.8$
mass. Using the same jet radius and \pt bin, Figure~\ref{fig:pt500_subst_AKt_R08}
compares signal and background for the different substructure variables
studied. 

\begin{figure*}
\centering
\subfigure[\antikt $R=0.8$, \pt = 300-400 \GeV bin]{\includegraphics[width=0.48\textwidth]{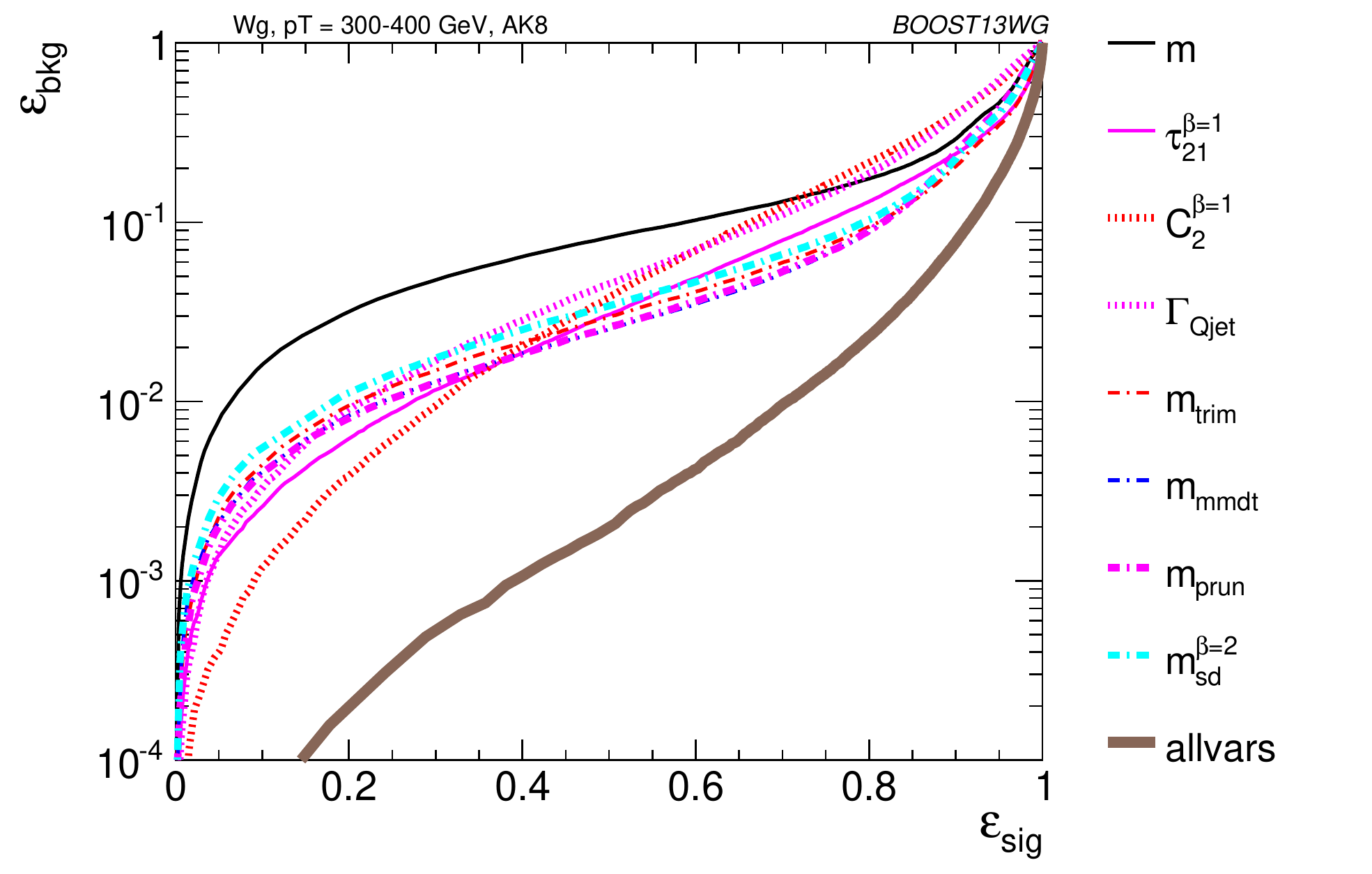}}
\subfigure[\antikt $R=1.2$, \pt = 300-400 \GeV bin]{\includegraphics[width=0.48\textwidth]{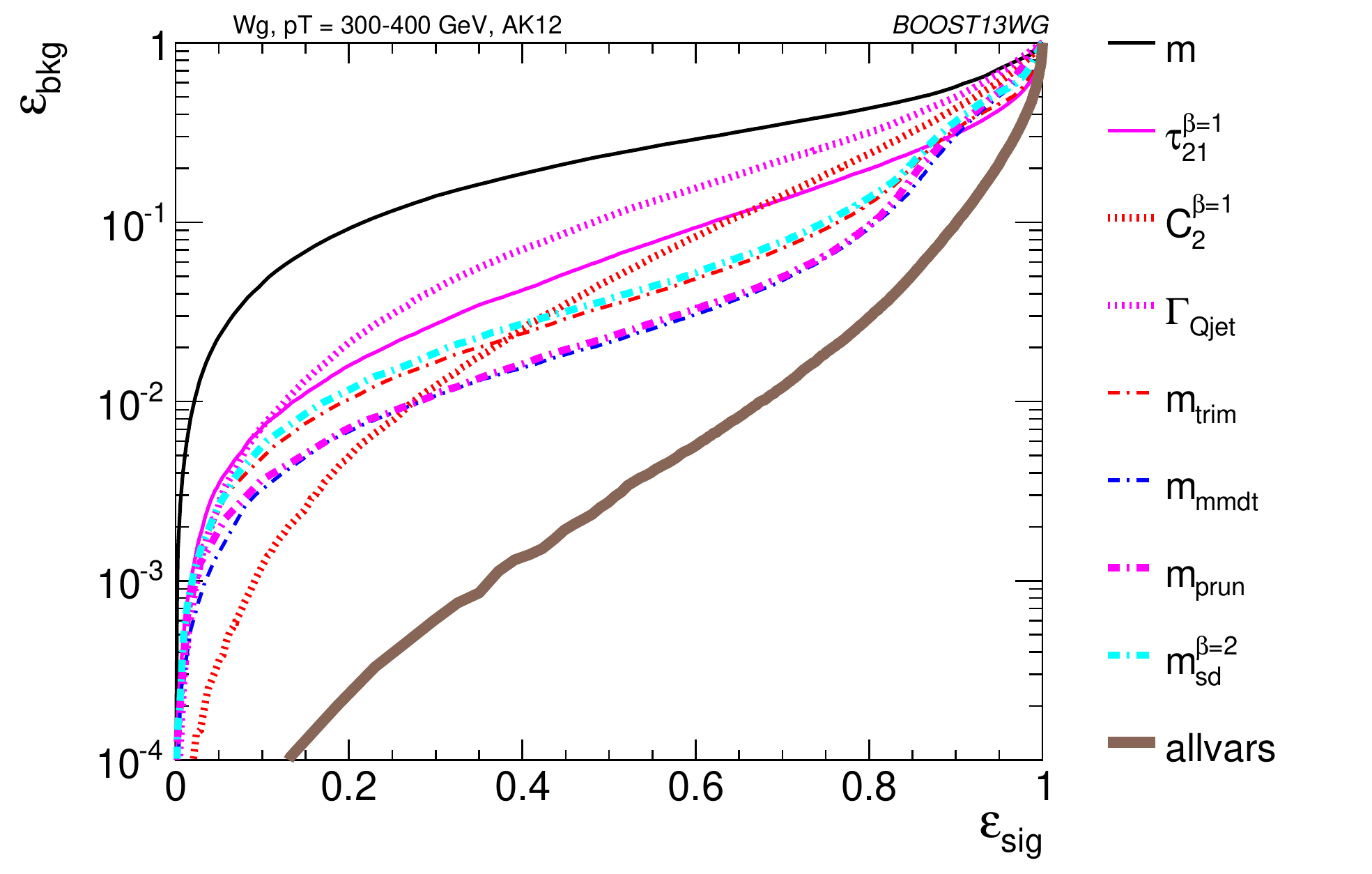}}
\caption{ROC curves for single variables considered for $W$
tagging in the \pt = 300-400 \GeV bin using the \antikt $R=0.8$ algorithm and $R=1.2$ algorithm, along with a BDT combination of all variables (``allvars'').}
\label{fig:pt300_single}
\end{figure*}

\begin{figure*}
\centering
\subfigure[\antikt $R=0.8$, \pt = 500-600 \GeV bin]{\includegraphics[width=0.48\textwidth]{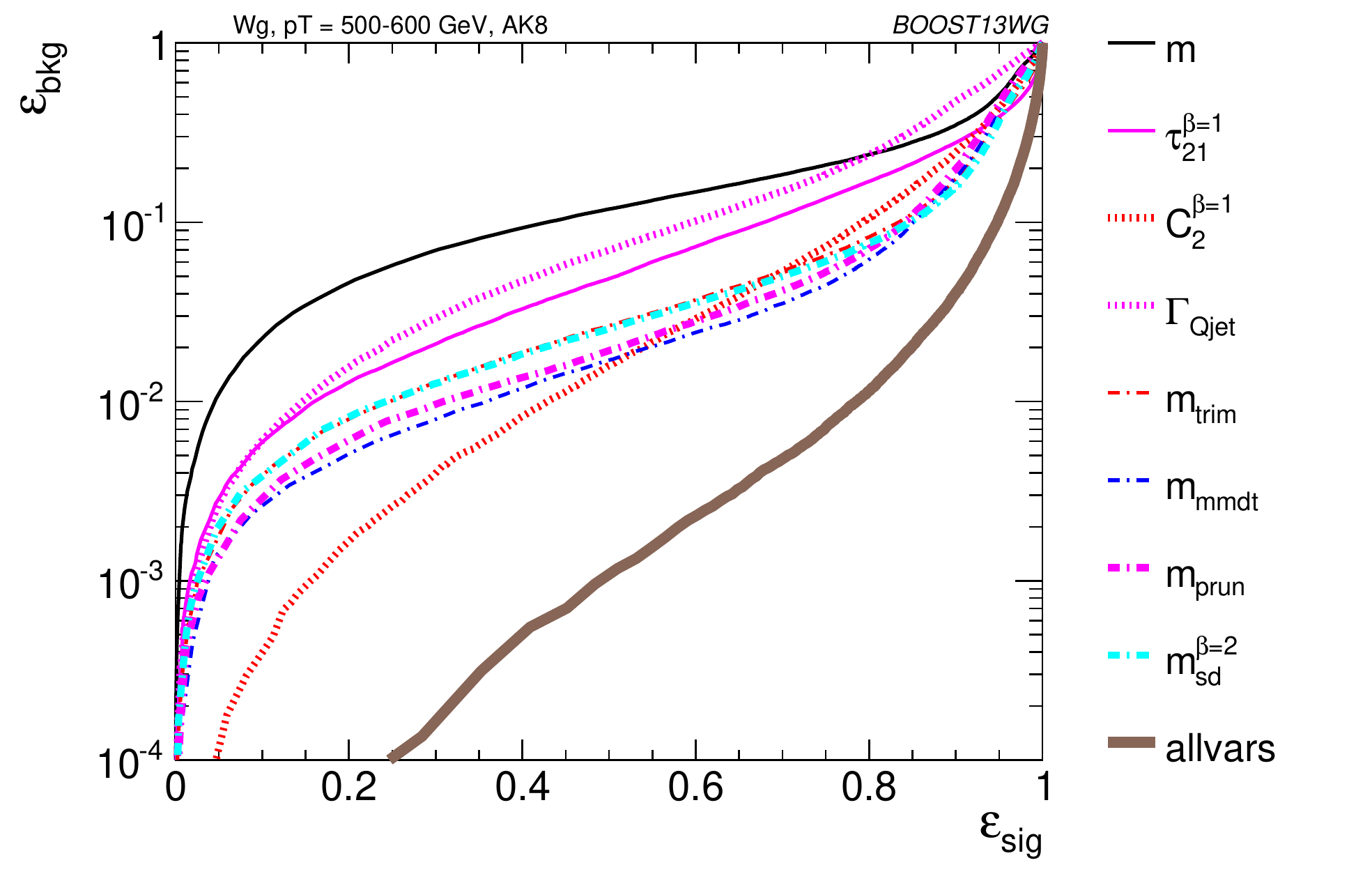}}
\subfigure[\antikt $R=1.2$, \pt = 500-600 \GeV bin]{\includegraphics[width=0.48\textwidth]{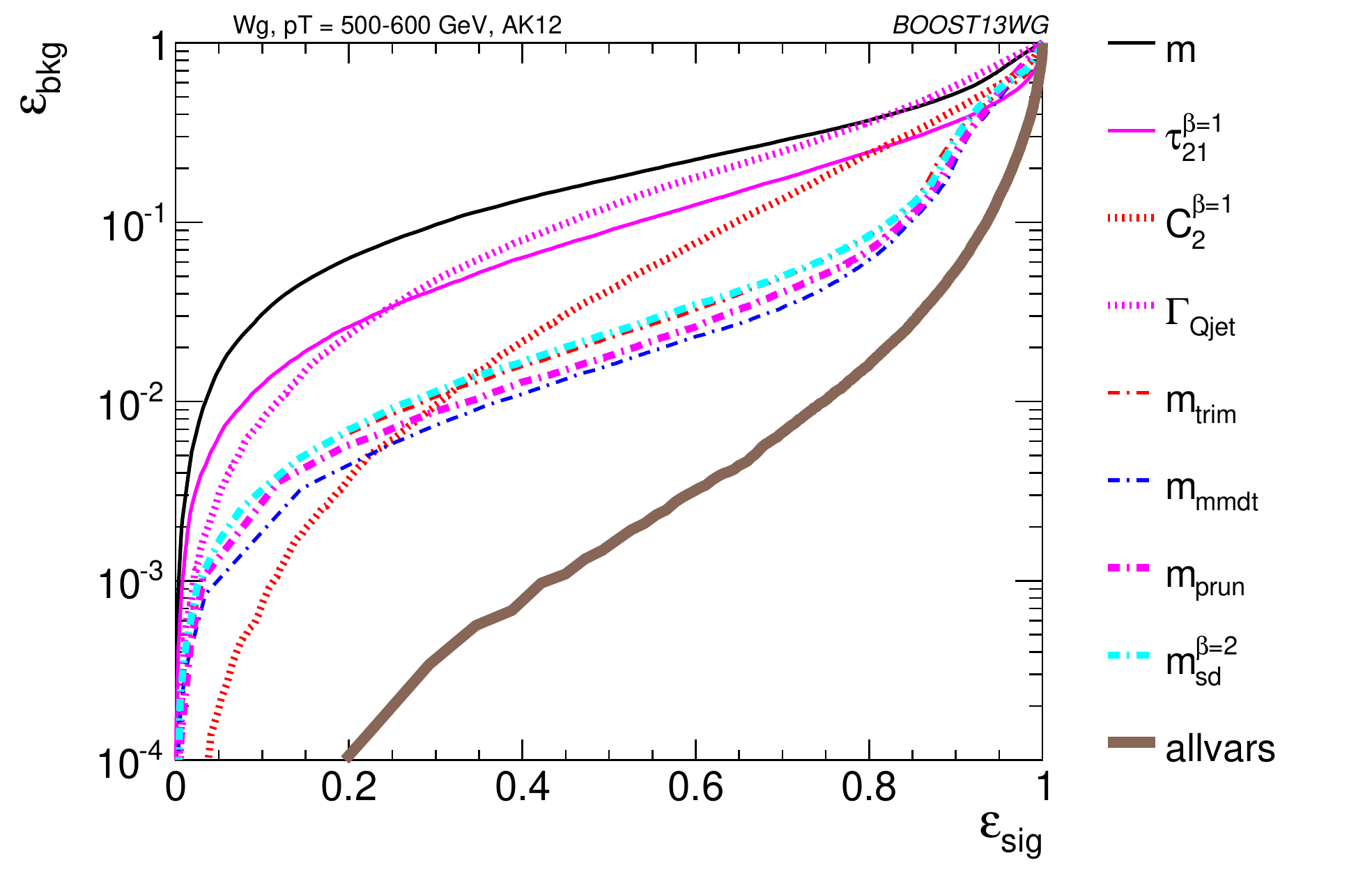}}
\caption{ROC curves for single variables considered for $W$
tagging in the \pt = 500-600 \GeV bin using the \antikt $R=0.8$ algorithm and $R=1.2$ algorithm, along with a BDT combination of all variables (``allvars'')}
\label{fig:pt500_single}
\end{figure*}

\begin{figure*}
\centering
\subfigure[\antikt $R=0.4$, \pt = 1.0-1.1 \TeV bin]{\includegraphics[width=0.48\textwidth]{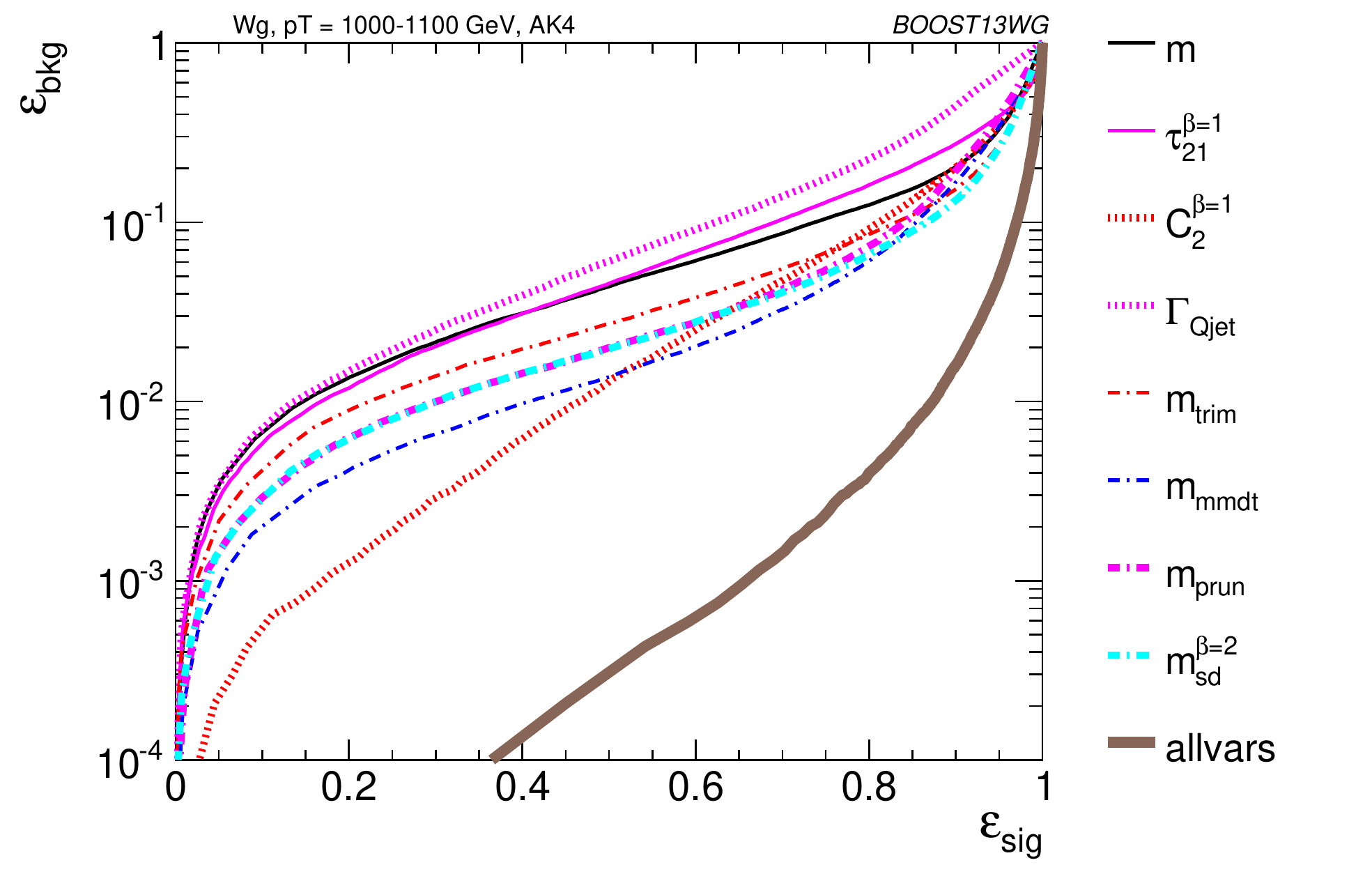}}
\subfigure[\antikt $R=0.8$, \pt = 1.0-1.1 \TeV bin]{\includegraphics[width=0.48\textwidth]{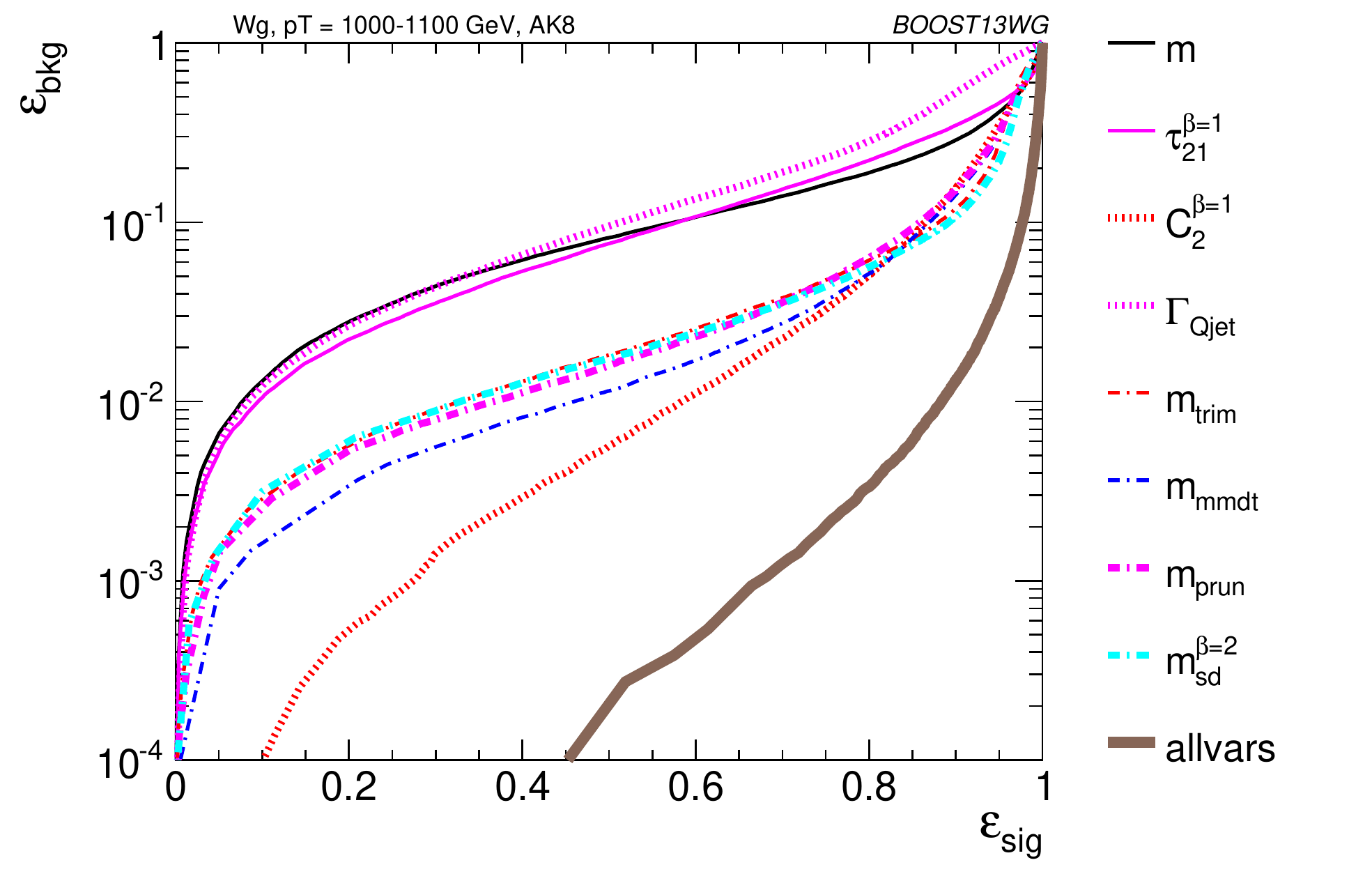}}\\
\subfigure[\antikt $R=1.2$, \pt = 1.0-1.1 \TeV bin]{\includegraphics[width=0.48\textwidth]{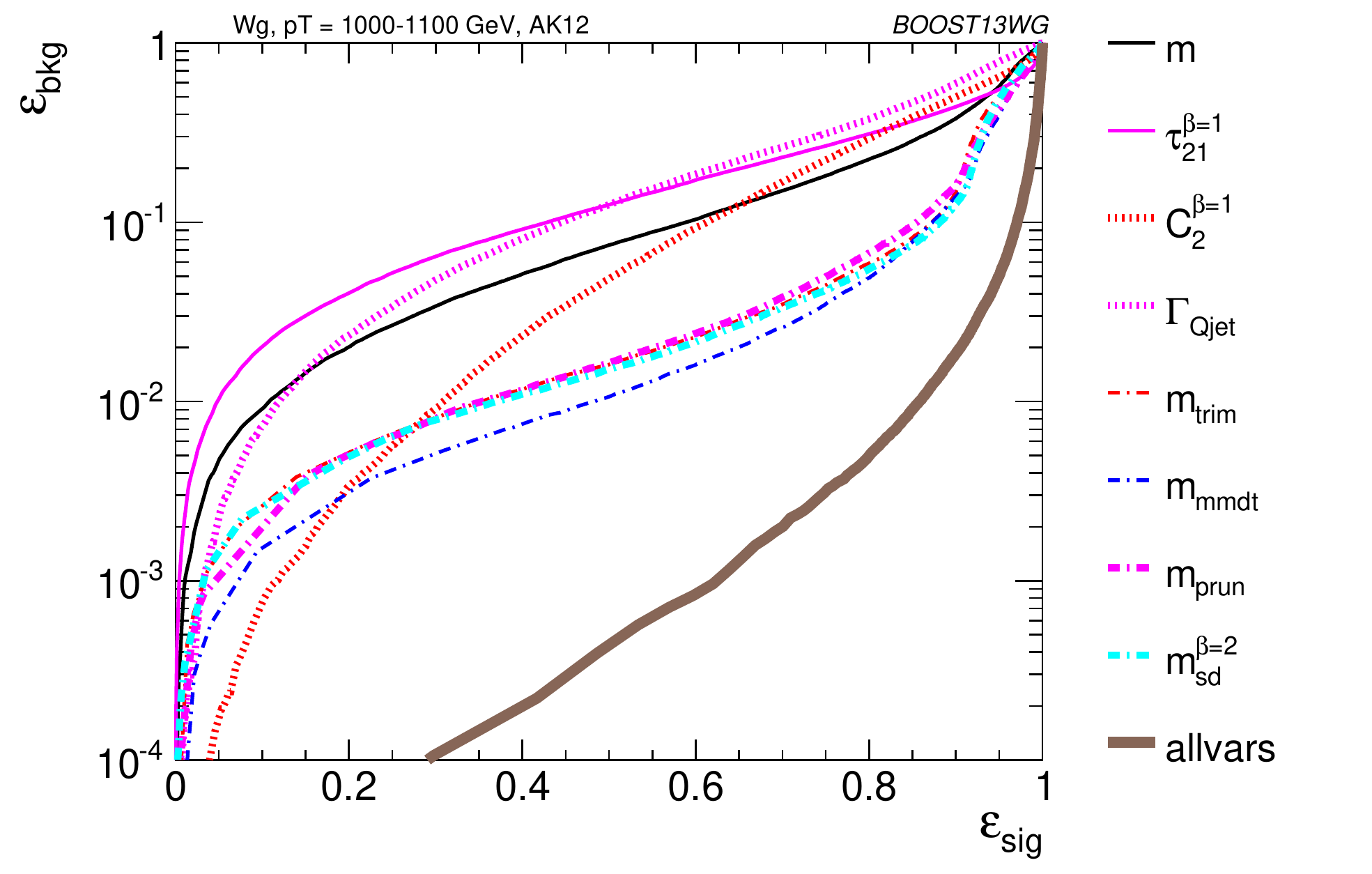}}
\caption{ROC curves for single variables considered for $W$
tagging in the \pt = 1.0-1.1 \TeV bin using the \antikt $R=0.4$ algorithm,
anti-\kT $R=0.8$ algorithm and $R=1.2$ algorithm, along with a BDT combination of all variables (``allvars'')}
\label{fig:pt1000_single}
\end{figure*}

Figures~\ref{fig:pt300_single},~\ref{fig:pt500_single}
and~\ref{fig:pt1000_single} show the single variable ROC curves
for various \pt bins and values of $R$. The single variable performance is also
compared to the ROC curve for a BDT combination of all the variables
(labelled ``allvars''). In all cases, the ``allvars'' option
is significantly more performant than any of the individual single variables
considered, indicating that there is considerable complementarity
between the variables, and this is explored further in Section~\ref{sec:w_combined}.

\begin{figure*}
\centering
\subfigure[\antikt $R=0.8$, \pt = 300-400 \GeV bin]{\includegraphics[width=0.48\textwidth]{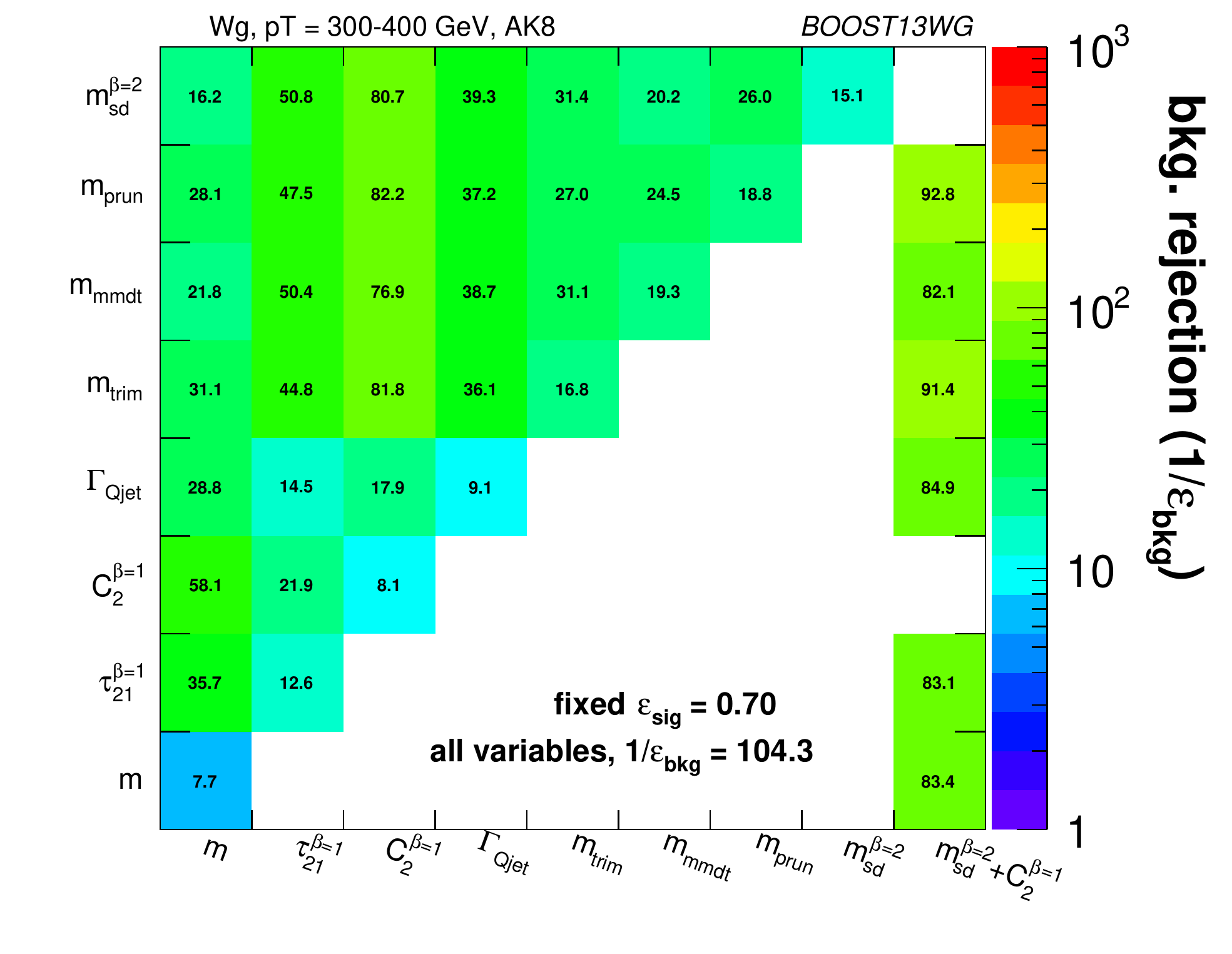}\label{fig:pt300_comb2D_08}}
\subfigure[\antikt $R=1.2$, \pt = 300-400 \GeV bin]{\includegraphics[width=0.48\textwidth]{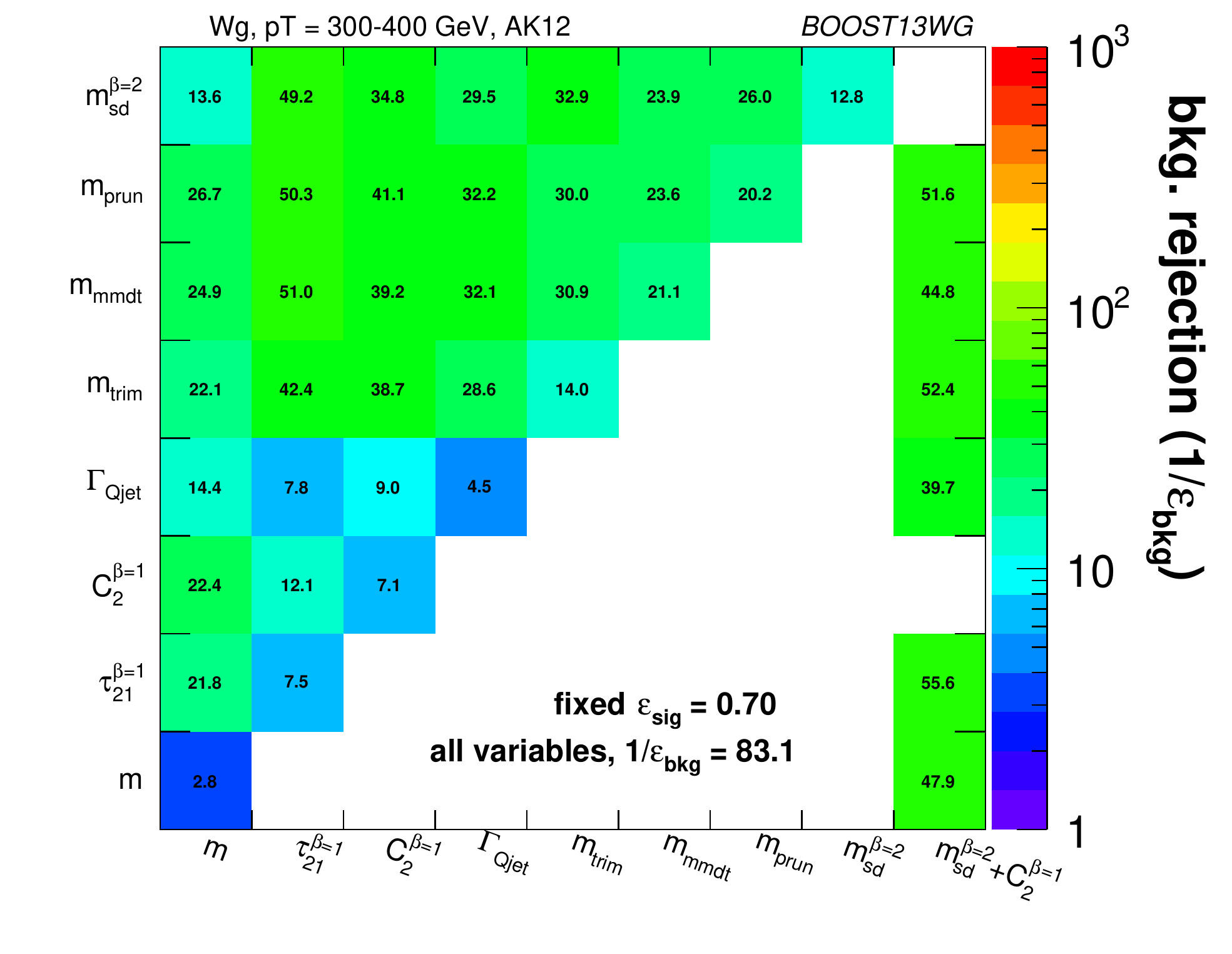}\label{fig:pt300_comb2D_12}}
\caption{
The background rejection
for a fixed signal efficiency (70\%) of each BDT combination of
each pair of variables considered, in the \pt = 300-400 \GeV bin using the anti-\kT $R=0.8$
algorithm and $R=1.2$ algorithm. Also shown is the background
rejection for three-variable combinations involving $m_{sd}^{\beta=2} +
C_2^{\beta=1}$, and for a BDT combination of all of the variables considered.
}
\label{fig:pt300_comb2D}
\end{figure*}

\begin{figure*}
\centering
\subfigure[\antikt $R=0.8$, \pt = 500-600 \GeV bin]{\includegraphics[width=0.48\textwidth]{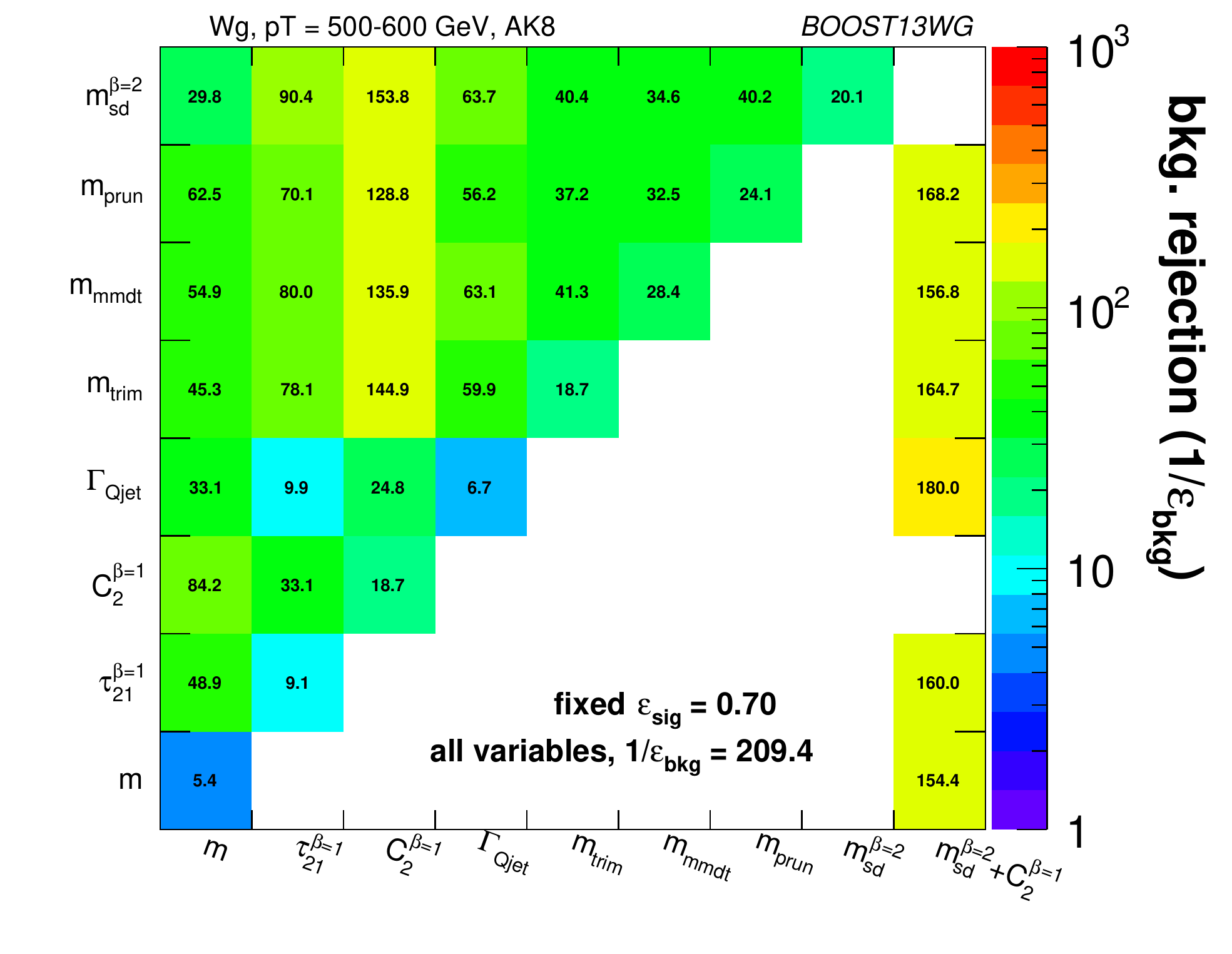}\label{fig:pt500_comb2D_08}}
\subfigure[\antikt $R=1.2$, \pt = 500-600 \GeV bin]{\includegraphics[width=0.48\textwidth]{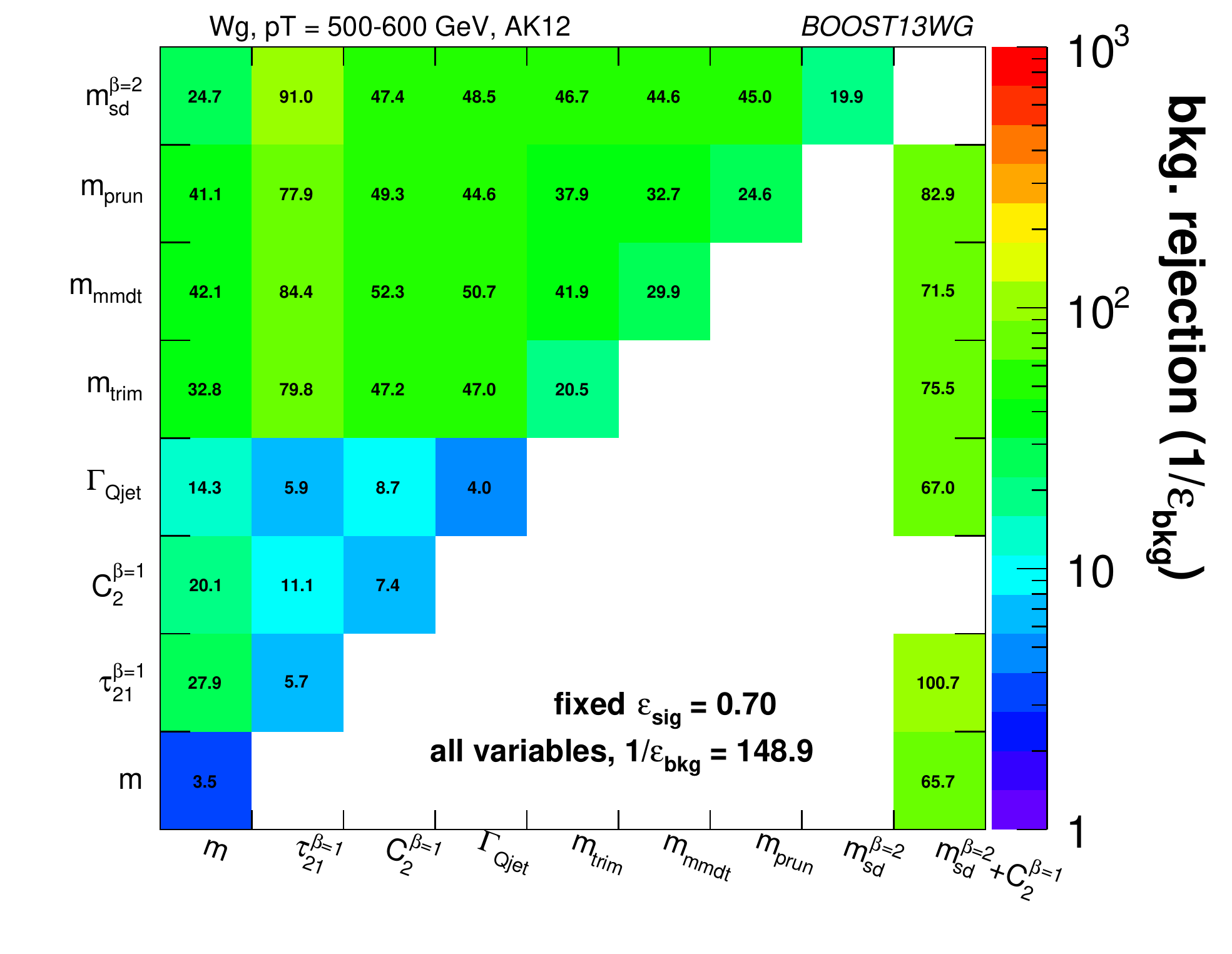}\label{fig:pt500_comb2D_12}}
\caption{The background rejection
for a fixed signal efficiency (70\%) of each BDT combination of
each pair of variables considered, in the \pt = 500-600 \GeV bin using the \antikt $R=0.8$
algorithm and $R=1.2$ algorithm. Also shown is the background
rejection for three-variable combinations involving $m_{sd}^{\beta=2} +
C_2^{\beta=1}$, and for a BDT combination of all of the variables considered.}
\label{fig:pt500_comb2D}
\end{figure*}

\begin{figure*}
\centering
\subfigure[\antikt $R=0.4$, \pt = 1.0-1.1 \TeV bin]{\includegraphics[width=0.48\textwidth]{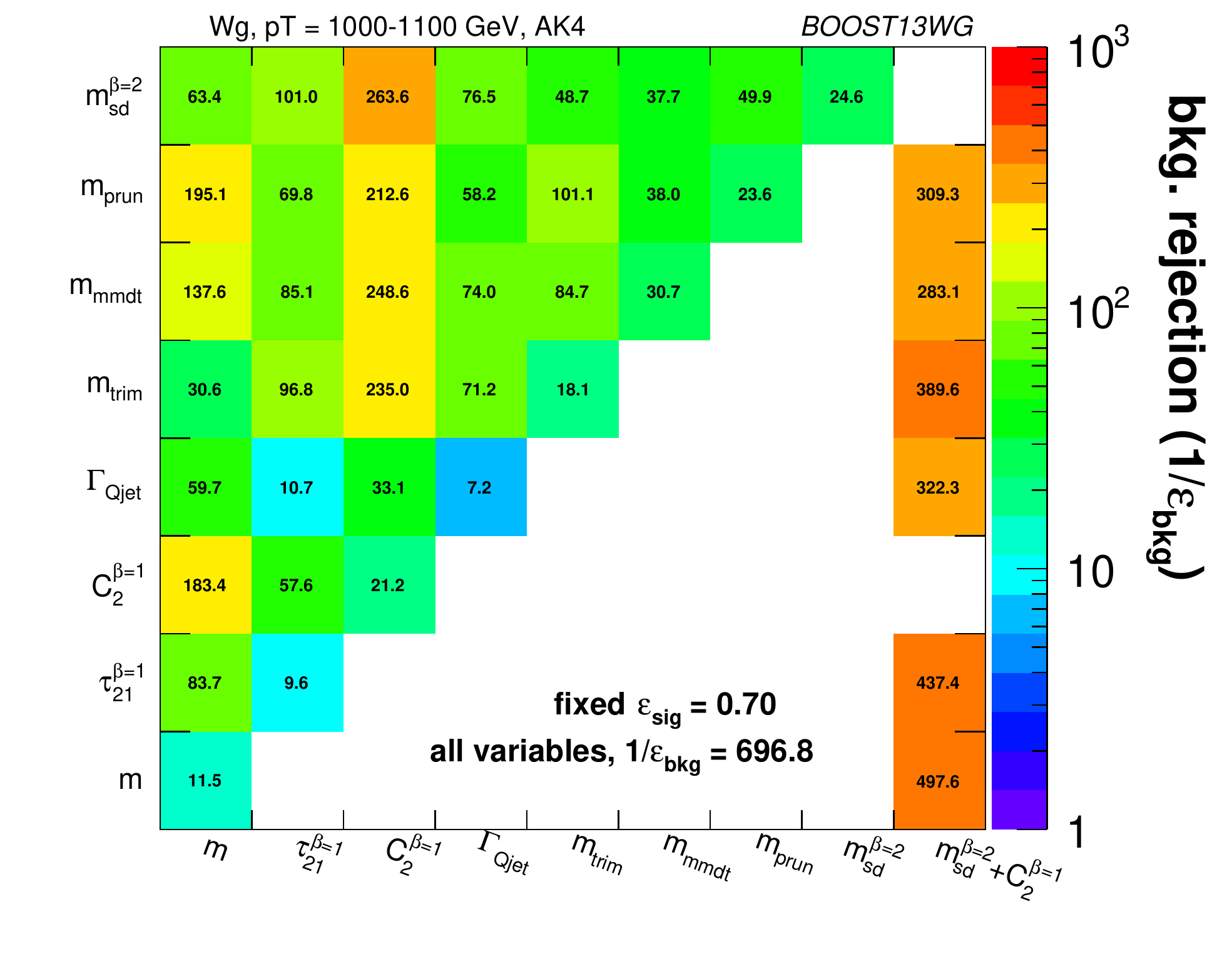}\label{fig:pt1000_comb2D_04}}
\subfigure[\antikt $R=0.8$, \pt = 1.0-1.1 \TeV bin]{\includegraphics[width=0.48\textwidth]{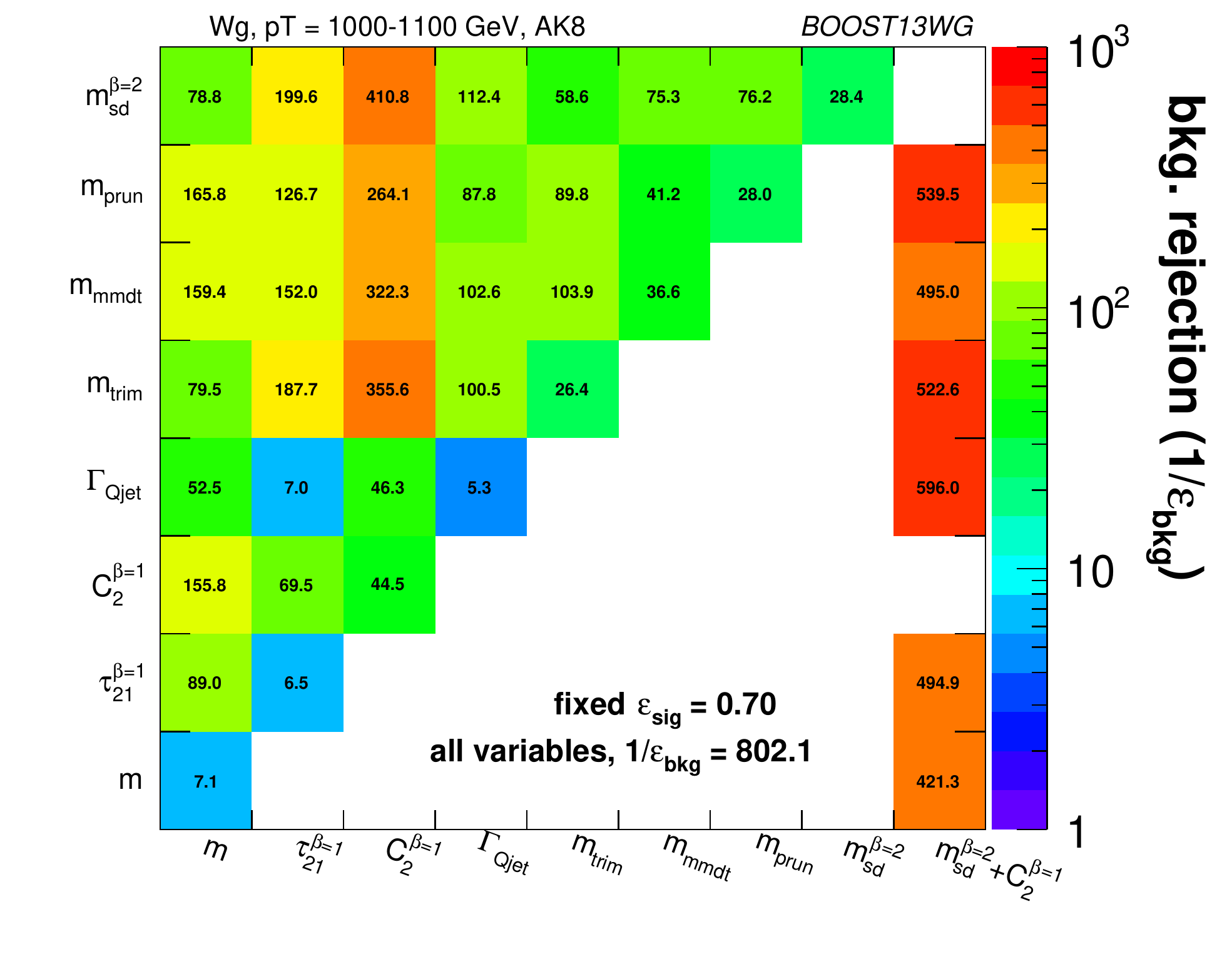}\label{fig:pt1000_comb2D_08}}\\
\subfigure[\antikt $R=1.2$, \pt = 1.0-1.1 \TeV bin]{\includegraphics[width=0.48\textwidth]{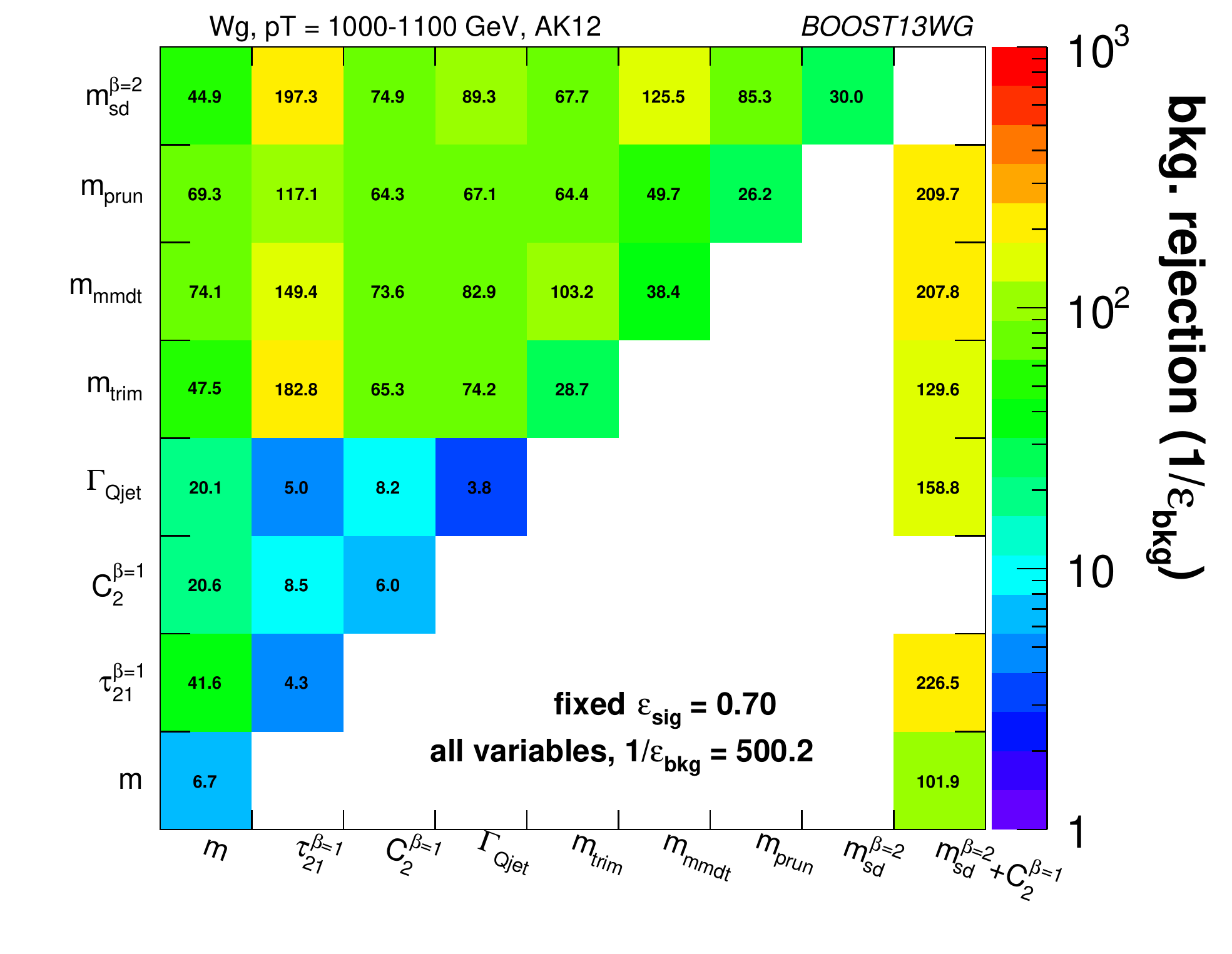}\label{fig:pt1000_comb2D_12}}
\caption{The background rejection
for a fixed signal efficiency (70\%) of each BDT combination of
each pair of variables considered, in the \pt = 1.0-1.1 \TeV bin using
the anti-\kT $R=0.4$, $R=0.8$ and $R=1.2$ algorithm. Also shown is the background
rejection for three-variable combinations involving $m_{sd}^{\beta=2} +
C_2^{\beta=1}$, and for a BDT combination of all of the variables considered.}
\label{fig:pt1000_comb2D}
\end{figure*}

 In
Figures~\ref{fig:pt300_comb2D},~\ref{fig:pt500_comb2D}
and~\ref{fig:pt1000_comb2D} the same information is shown in a format
that more readily allows for a quantitative comparison of performance
for different $R$ and \pt; matrices are presented which give the
background rejection for a signal efficiency of 70\%\footnote{Note
  that we here choose to report the rejection for a higher signal
  efficiency than the 50\% that was used in the $q/g$ tagging studies
  of Section~\ref{sec:qgtagging}, because the
rejection rates in $W$ tagging are considerably higher.} for single variable cuts, as well as two- and three-variable BDT combinations. The results are shown separately for each \pt bin and jet radius
considered.  Most relevant for our immediate discussion, the diagonal entries of these plots show the background
rejections for a single variable BDT using the labelled observable, and can thus be examined to get a
quantitative measure of the individual single variable performance,
and to study how this changes with jet radius and momenta. The off-diagonal
entries give the performance when two variables (shown on the x-axis and on the y-axis, respectively) are combined in
a BDT. The
final column of these plots shows the background rejection
performance for three-variable BDT combinations of $m_{sd}^{\beta=2} +
C_2^{\beta=1} + X$. These results will be discussed later in Section~\ref{sec:Wtagallvars}. 

In general, the most performant single variables are
the groomed masses. However, in certain kinematic bins and for certain
jet radii, $C_2^{\beta=1}$ has a background rejection that is
comparable to or better than the groomed masses. 

\begin{figure*}
\centering
\subfigure[\antikt $R=1.2$, \pt = 300-400 \GeV bin]{\includegraphics[width=0.4\textwidth]{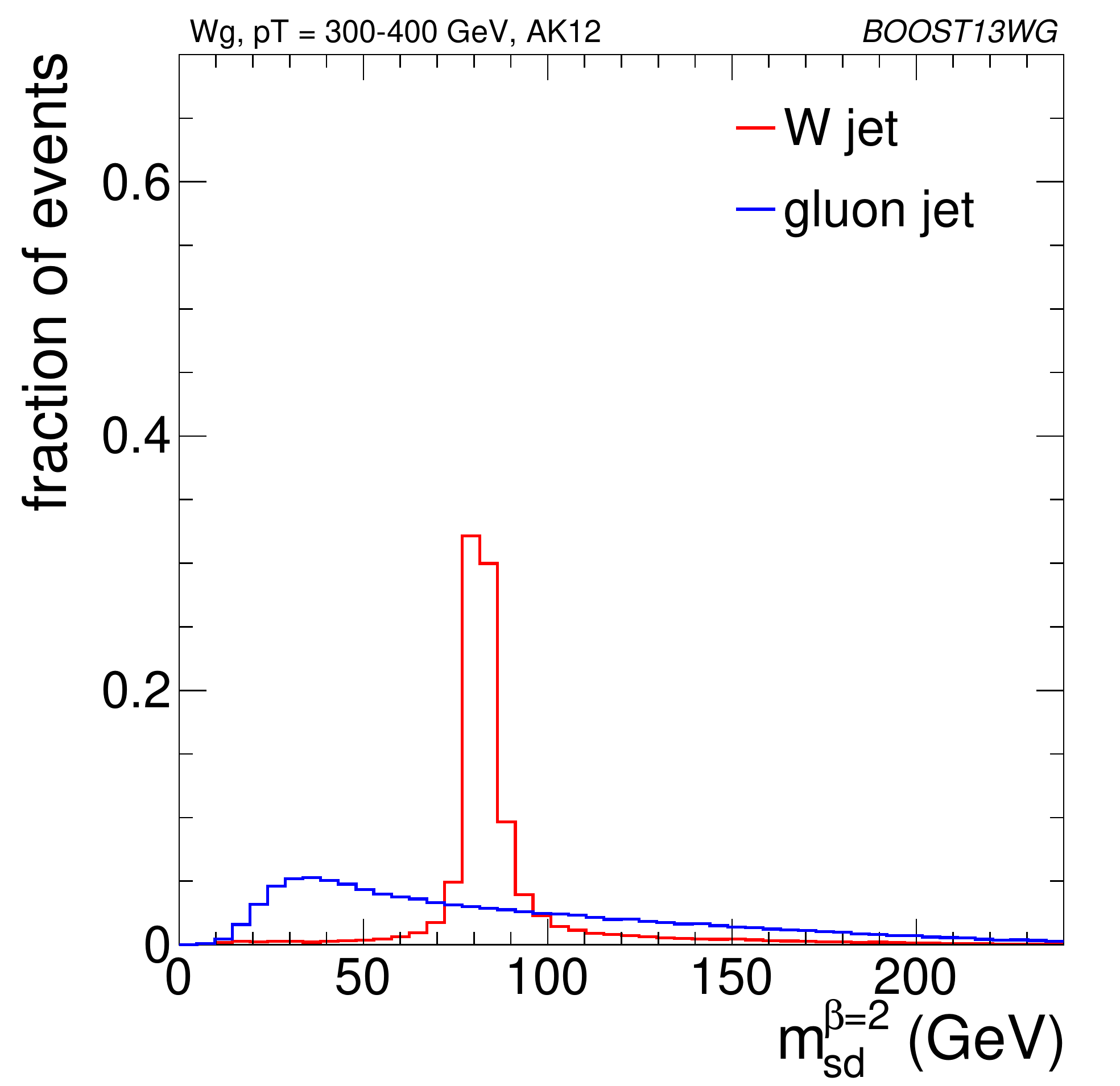}}
\subfigure[\antikt $R=1.2$, \pt = 1.0-1.1 \TeV bin]{\includegraphics[width=0.4\textwidth]{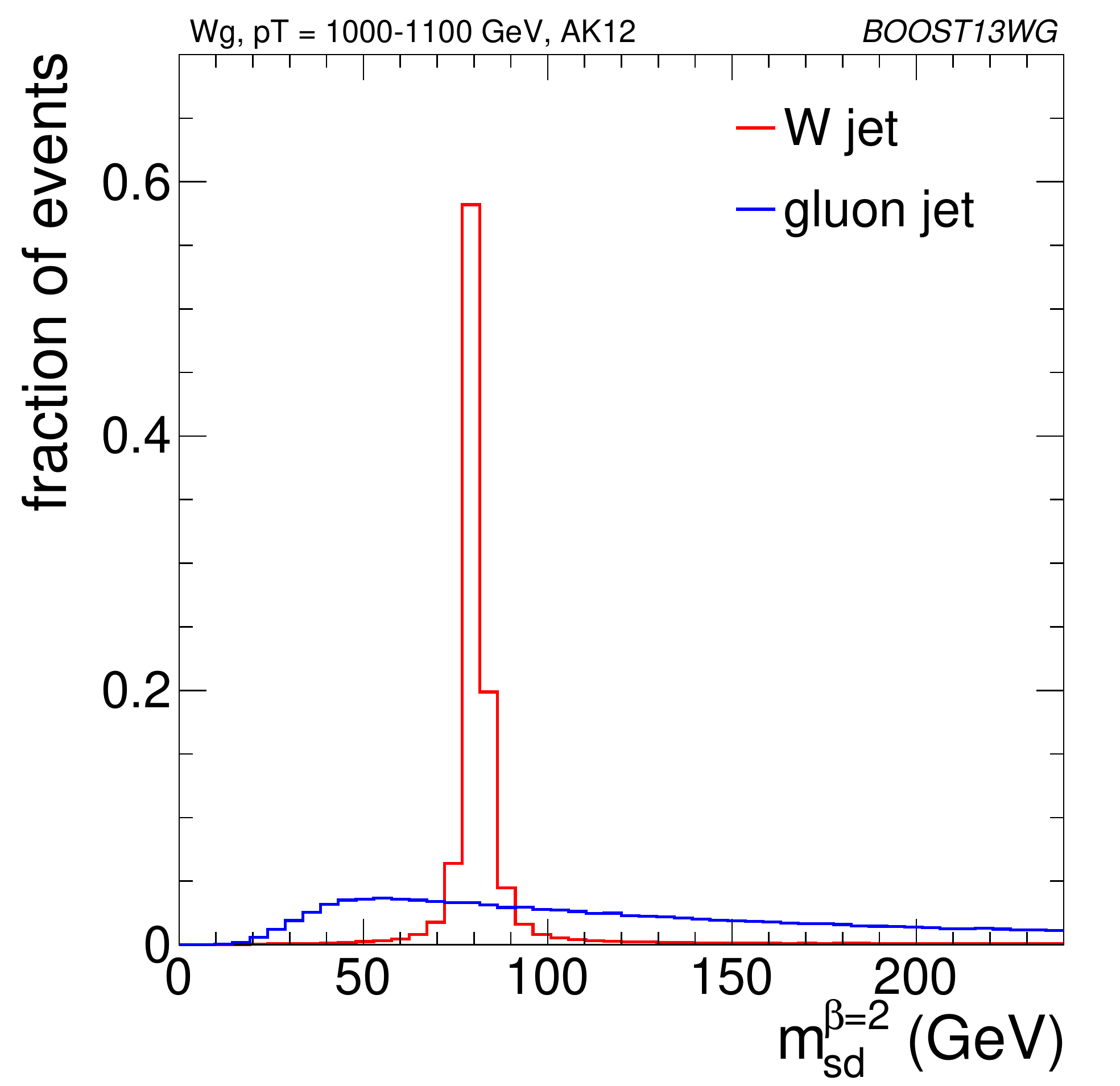}}
\subfigure[\antikt $R=1.2$, \pt = 300-400 \GeV bin]{\includegraphics[width=0.4\textwidth]{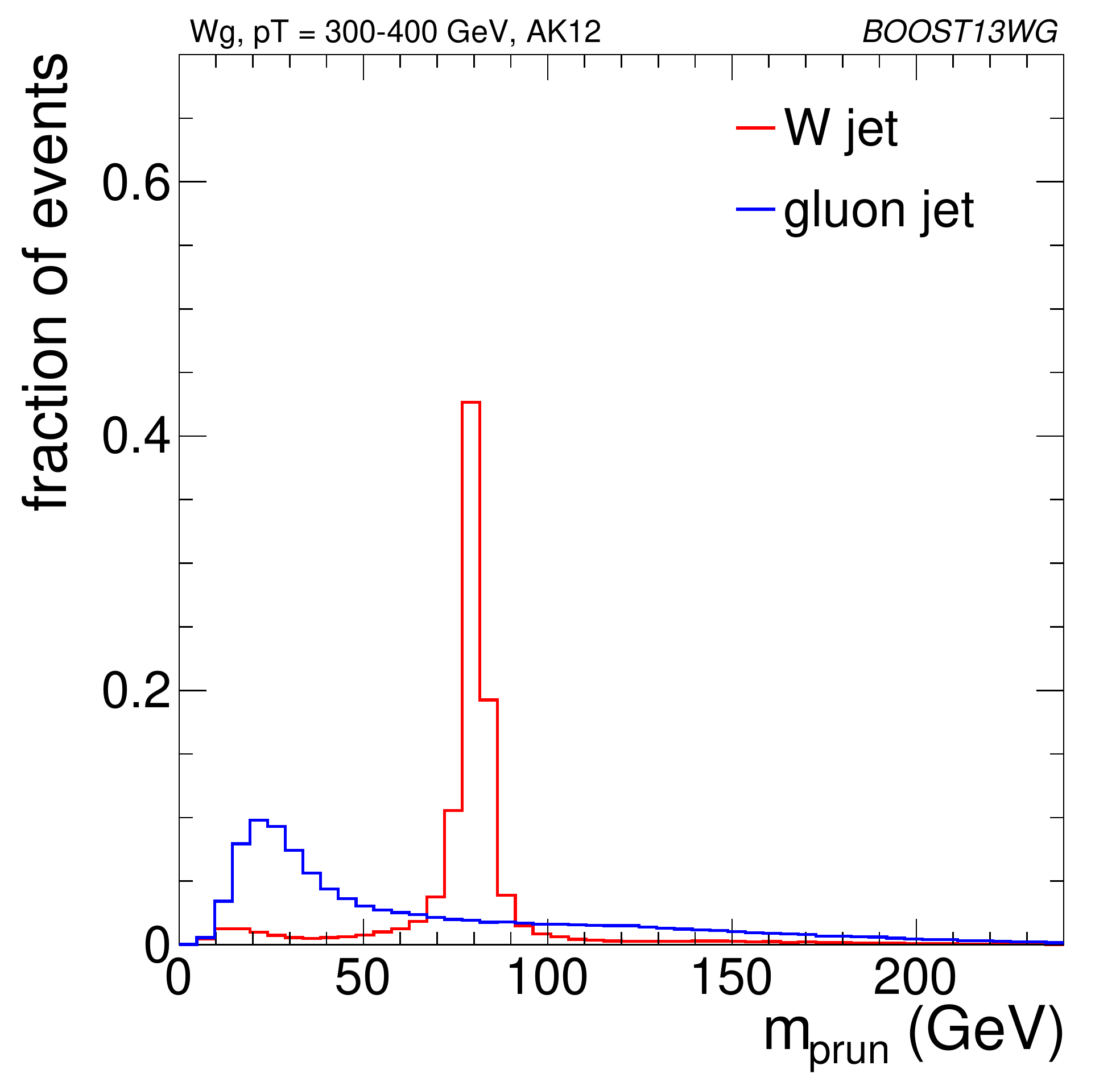}}
\subfigure[\antikt $R=1.2$, \pt = 1.0-1.1 \TeV bin]{\includegraphics[width=0.4\textwidth]{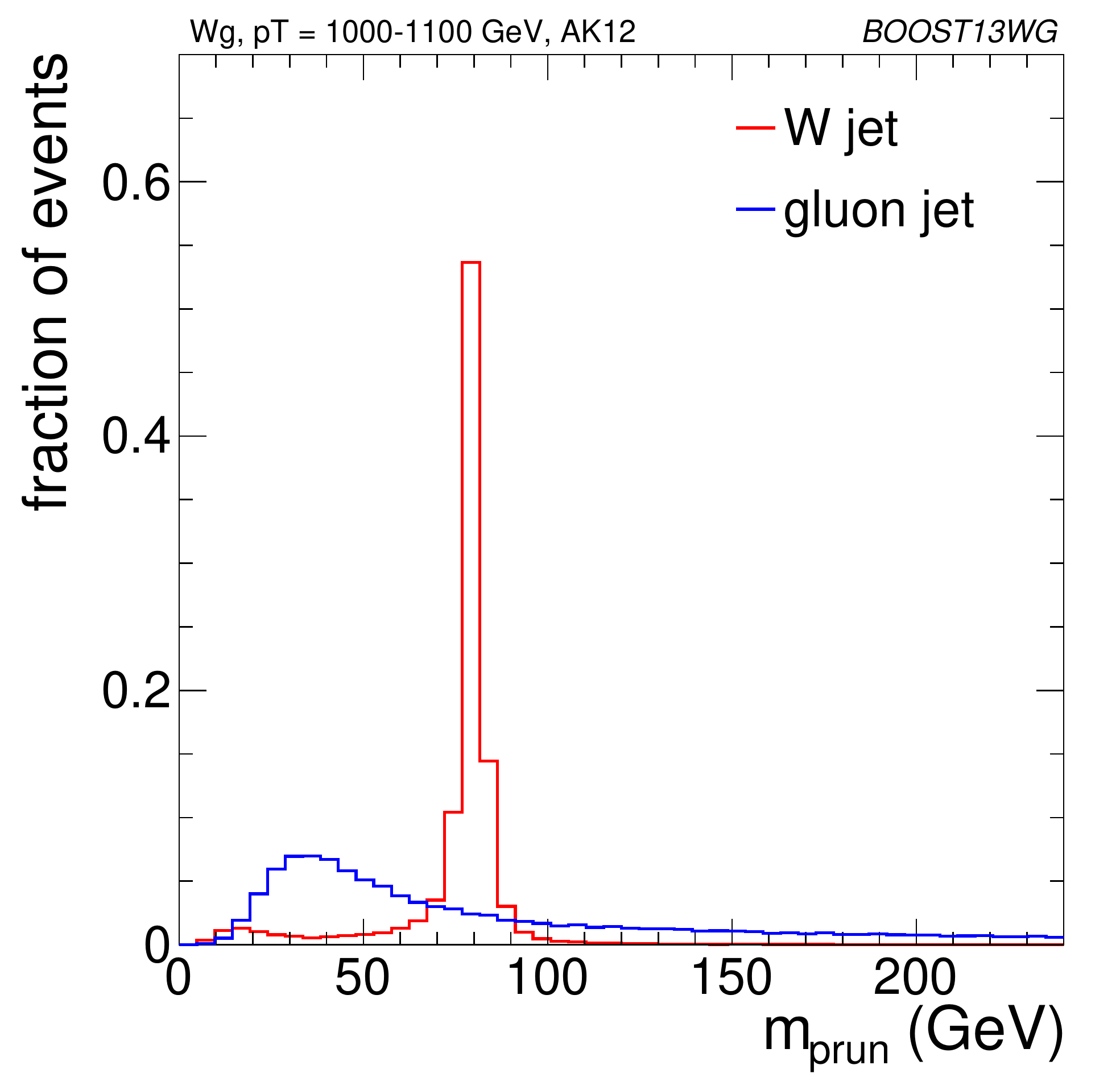}}
\caption{The Soft-drop $\beta=2$ and pruned groomed mass distribution for signal
  and background $R=1.2$ jets in two different \pt bins.}
\label{fig:ptdepend_groomedmass}
\end{figure*}

We first examine the variation of performance with jet \pt. By comparing
Figures~\ref{fig:pt300_comb2D_08},~\ref{fig:pt500_comb2D_08}
and~\ref{fig:pt1000_comb2D_08}, we can see how the background
rejection performance varies with increased momenta whilst keeping the jet
radius fixed to $R=0.8$. Similarly, by comparing Figures~\ref{fig:pt300_comb2D_12},~\ref{fig:pt500_comb2D_12}
and~\ref{fig:pt1000_comb2D_12} we can see how performance evolves with
\pt for $R=1.2$. For both $R=0.8$ and $R=1.2$ the background rejection power of
the groomed masses increases with increasing \pt, with a factor 1.5-2.5 increase in rejection in going from the 300-400 GeV to
1.0-1.1 TeV bins. In Figure~\ref{fig:ptdepend_groomedmass} we show the
\msd and \mprun groomed masses for signal and background in the \pt = 300-400 and \pt = 1.0-1.1 \TeV bins for $R=1.2$ jets. Two effects result in the improved
performance of the groomed mass at high \pt. Firstly, as is evident
from the figure, the resolution of the signal peak after grooming
improves, because the groomer finds it easier to pick out the hard
signal component of the jet against the softer components of the
underlying event when the signal is boosted. Secondly, it follows
from Figure~\ref{fig:qg_prmasses_log} and the discussion in Section~\ref{sec:qg_mass}
that, for increasing \pt, the perturbative shoulder of the gluon
distribution decreases in size, and
thus there is a slight decrease (or at least no increase) of the background contamination in the
signal mass region (m/$p_{T}$/R $\sim$ 0.5). 

\begin{figure*}
\centering
\subfigure[\antikt $R=0.8$, \pt = 300-400 \GeV bin]{\includegraphics[width=0.4\textwidth]{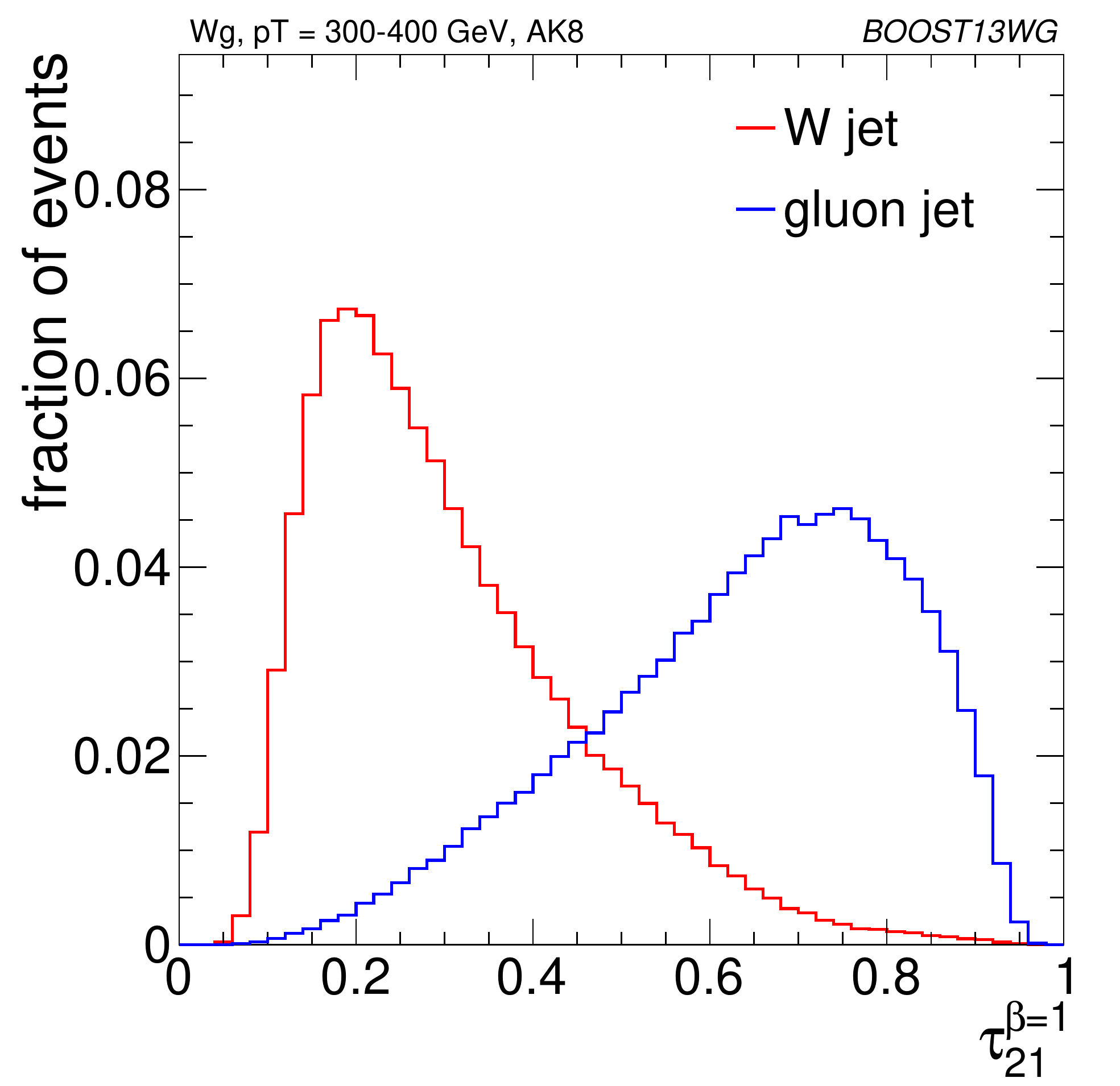}}
\subfigure[\antikt $R=0.8$, \pt = 1.0-1.1 \TeV bin]{\includegraphics[width=0.4\textwidth]{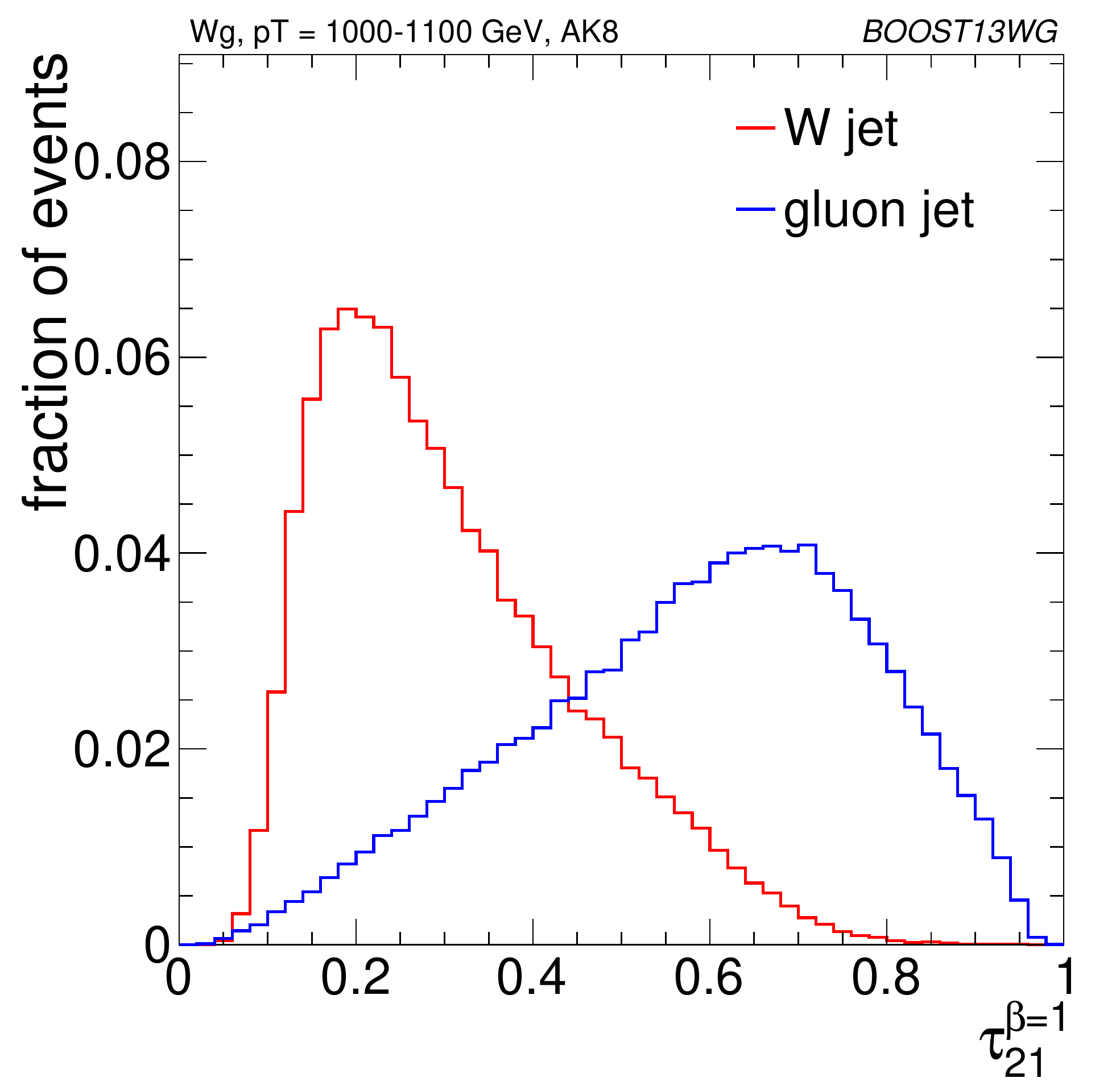}}
\subfigure[\antikt $R=0.8$, \pt = 300-400 \GeV bin]{\includegraphics[width=0.4\textwidth]{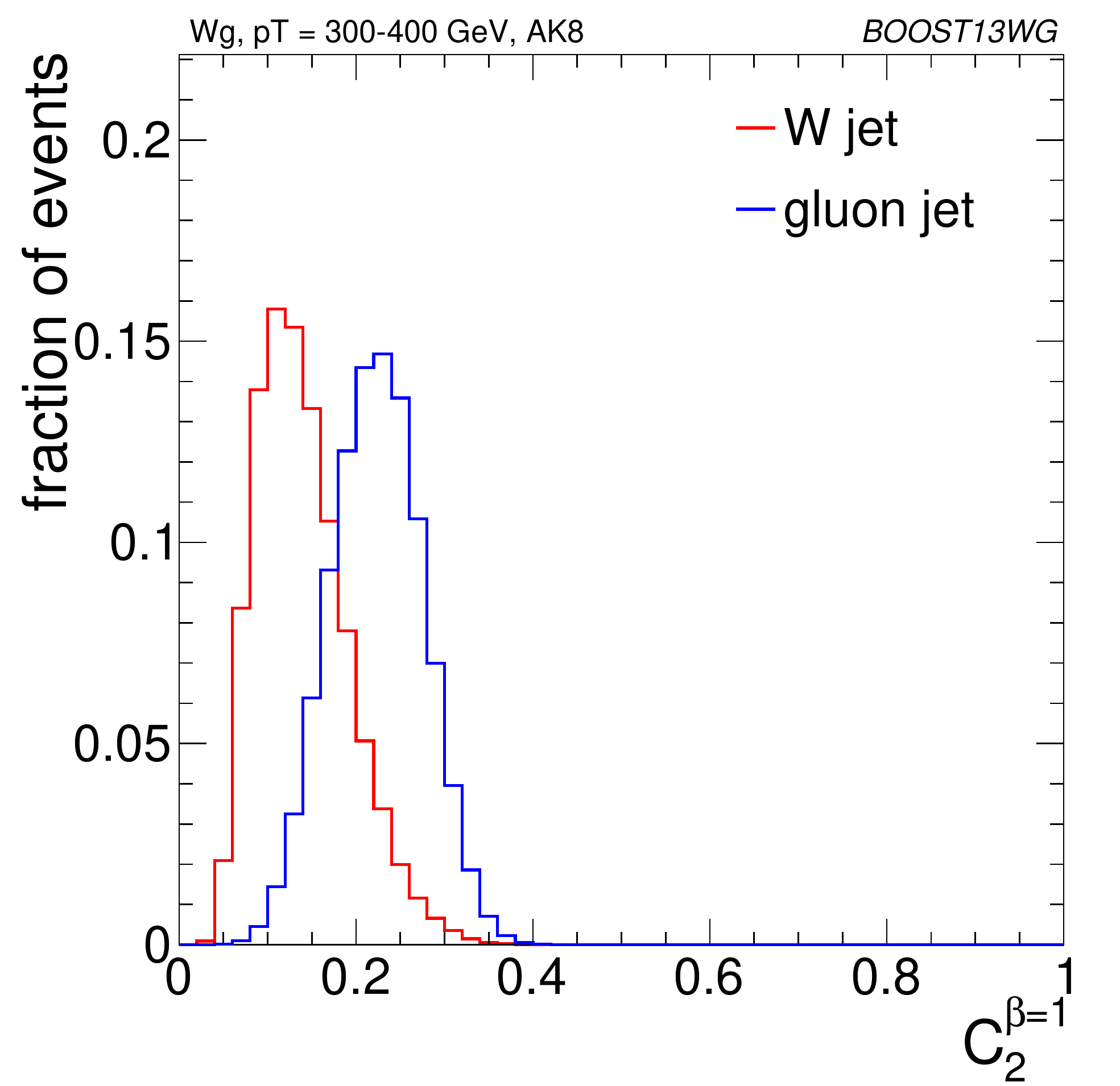}}
\subfigure[\antikt $R=0.8$, \pt = 1.0-1.1 \TeV bin]{\includegraphics[width=0.4\textwidth]{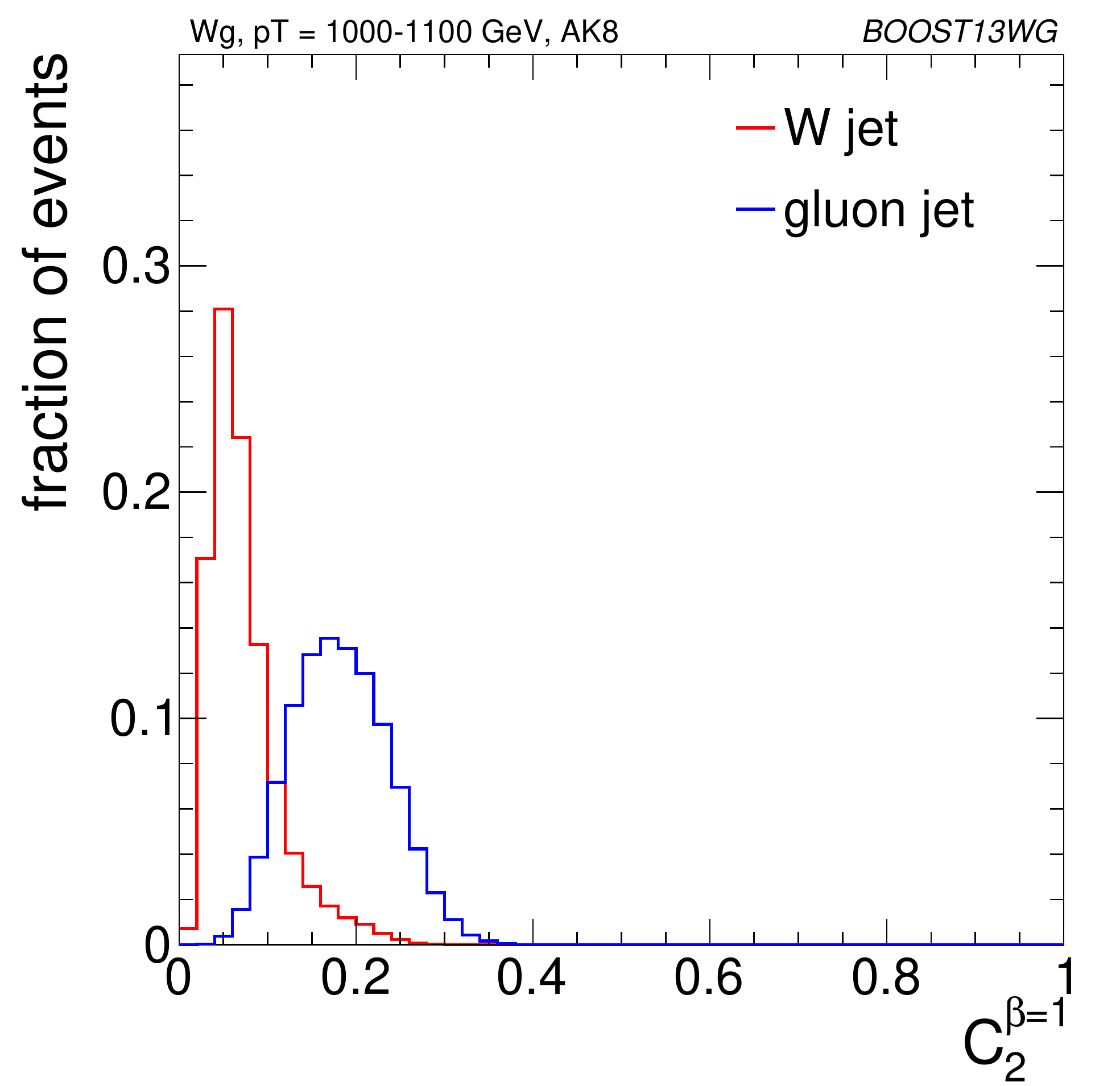}}
\caption{The $\tau_{21}^{\beta=1}$ and $C_2^{\beta=1}$ distributions for signal
  and background $R=0.8$ jets in two different \pt bins.}
\label{fig:ptdepend_substructure}
\end{figure*}

However, one can see from the Figures~\ref{fig:pt300_comb2D_12},~\ref{fig:pt500_comb2D_12}
and~\ref{fig:pt1000_comb2D_12} that the $C_2^{\beta=1}$, $\Gamma_{\rm Qjet}$ and
$\tau_{21}^{\beta=1}$ substructure variables behave somewhat
differently. The background rejection power of the $\Gamma_{\rm Qjet}$ and
$\tau_{21}^{\beta=1}$ variables both decrease with increasing \pt, by
up to a factor two in going from the 300-400 \GeV to
1.0-1.1 \TeV bins. Conversely the rejection power of $C_2^{\beta=1}$
dramatically increases with increasing \pt for $R=0.8$, but does not
improve with \pt for the larger jet radius $R=1.2$. In Figure~\ref{fig:ptdepend_substructure} we show the
$\tau_{21}^{\beta=1}$ and $C_2^{\beta=1}$ distributions for signal and background in the \pt
300-400 \GeV and \pt = 1.0-1.1 \TeV bins for $R=0.8$ jets. For
$\tau_{21}^{\beta=1}$ one can see that, in moving from  lower to 
higher \pt bins, the signal peak remains fairly unchanged, whereas the
background peak shifts to smaller $\tau_{21}^{\beta=1}$ values,
reducing the discriminating power of the variable. This is expected,
since jet substructure methods explicitly relying on the identification of hard
prongs would expect to work best at low \pt, where the prongs would
tend to be more separated. However, $C_2^{\beta=1}$ does not rely on
the explicit identification of subjets, and one can see from
Figure~\ref{fig:ptdepend_substructure} that the discrimination power
visibly increases with increasing \pt. This is in line with the
observation in \cite{Larkoski:2013eya} that $C_2^{\beta=1}$ performs best when $m/\pt$ is small.
The negative correlation between the discrimination power of $\Gamma_{\rm Qjet}$ and increasing
\pt can be understood in similar terms.  As discussed in Section~\ref{sec:qg_mass}, the low volatility
component of a gluon jet, the ``shoulder'', is enhanced as \pt increases leading to a background (QCD)
volatility distribution more peaked at low values.  In contrast the signal (W) jets will include more relatively soft 
radiation as \pt increases leading to a more volatile configuration. 
Thus, as \pt increases, the signal jets will exhibit a somewhat broader volatility distribution, while
the background jets will exhibit a somewhat narrower volatility distribution, \textit{i.e.}, the
distributions become more similar reducing the discriminating power of $\Gamma_{\rm Qjet}$. 

\begin{figure*}
\centering
\subfigure[\antikt $R=0.4$, \pt = 1.0-1.1 \TeV bin]{\includegraphics[width=0.4\textwidth]{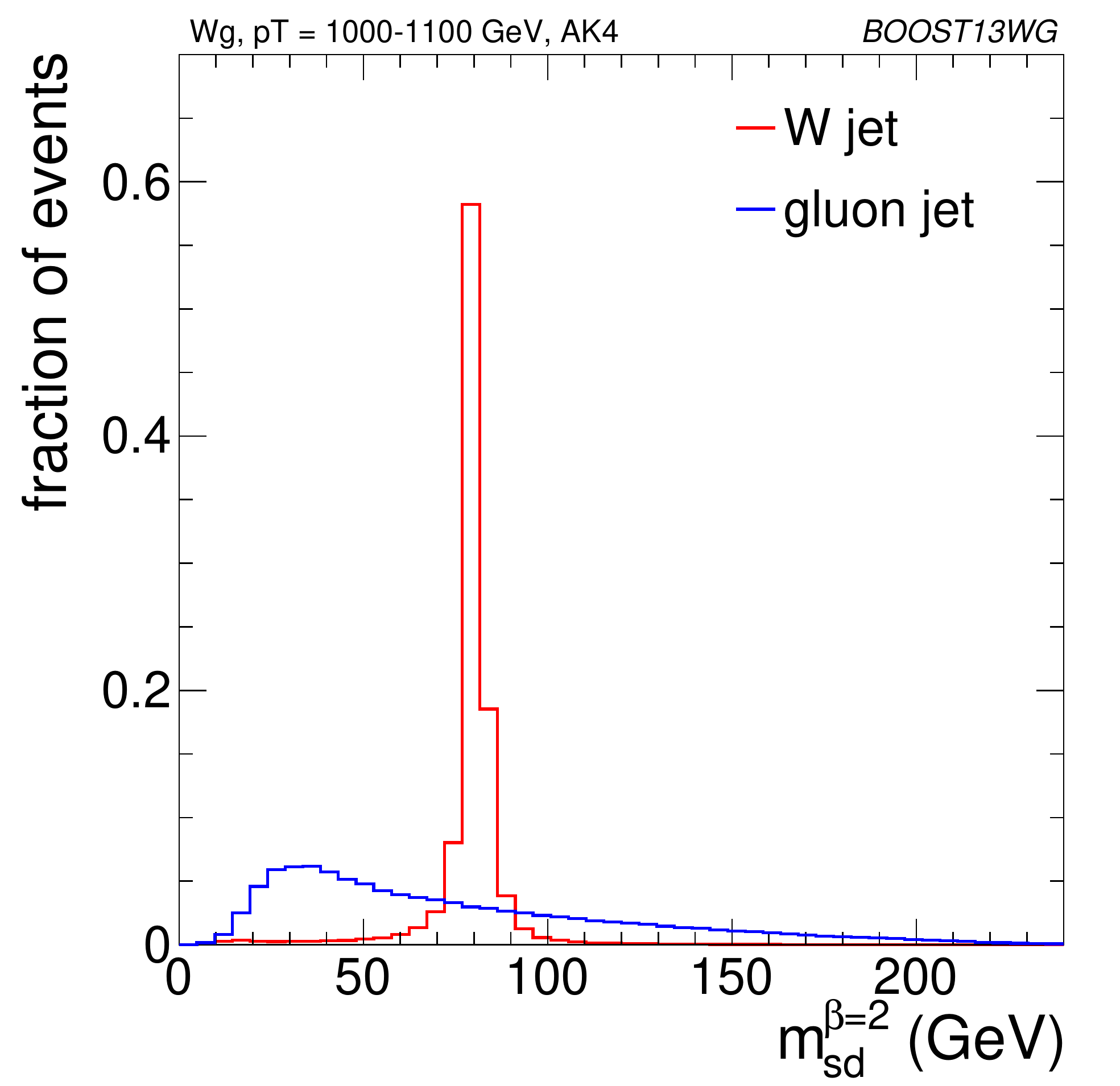}}
\subfigure[\antikt $R=1.2$, \pt = 1.0-1.1 \TeV bin]{\includegraphics[width=0.4\textwidth]{./Figures/WTagging/pT1000/AKtR12/h_mass_sdb2.pdf}}
\subfigure[\antikt $R=0.4$, \pt = 1.0-1.1 \TeV bin]{\includegraphics[width=0.4\textwidth]{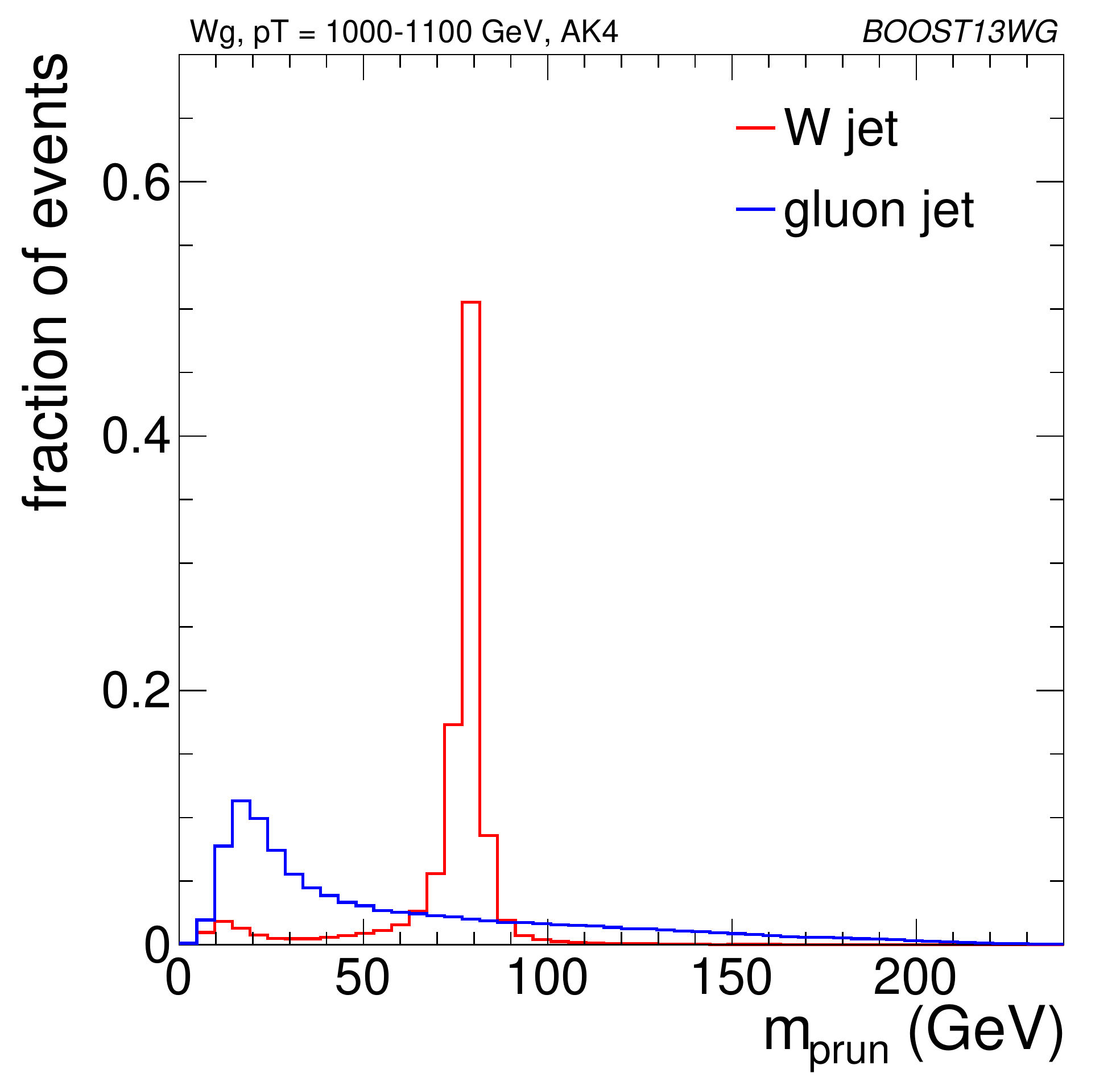}}
\subfigure[\antikt $R=1.2$, \pt = 1.0-1.1 \TeV bin]{\includegraphics[width=0.4\textwidth]{./Figures/WTagging/pT1000/AKtR12/h_mass_prun.pdf}}
\caption{The Soft-drop $\beta=2$ and pruned groomed mass distribution for signal
  and background $R=0.4$ and $R=1.2$ jets in the \pt = 1.0-1.1 \TeV bin.}
\label{fig:Rdepend_groomedmass}
\end{figure*}

We now compare the performance of different jet radius parameters in the same \pt bin by comparing the individual sub-figures
 of Figures~\ref{fig:pt300_comb2D},~\ref{fig:pt500_comb2D}
and~\ref{fig:pt1000_comb2D}. To within
$\sim$~25\%, the background rejection power of the groomed masses remains
constant with respect to the jet
radius. Figure~\ref{fig:Rdepend_groomedmass} shows how the groomed
mass changes for varying jet radius in the \pt = 1.0-1.1 \TeV bin. One
can see that the signal mass peak remains unaffected by the increased
radius, as expected, since grooming removes the soft contamination
which could otherwise increase the mass of the jet as the radius
increased. The gluon background in the signal mass region also remains
largely unaffected, as follows from Figure~\ref{fig:qg_prmasses_log}
and the discussion in Section~\ref{sec:qg_mass},
where it is shown that there is very little dependence of the groomed gluon mass
distribution on $R$ in the signal region ($m/\pt/R \sim$ 0.5). 

\begin{figure*}
\centering
\subfigure[\antikt $R=0.8$, \pt = 1.0-1.1 \TeV bin]{\includegraphics[width=0.4\textwidth]{./Figures/WTagging/pT1000/AKtR08/h_tau21_b1.pdf}}
\subfigure[\antikt $R=1.2$, \pt = 1.0-1.1 \TeV bin]{\includegraphics[width=0.4\textwidth]{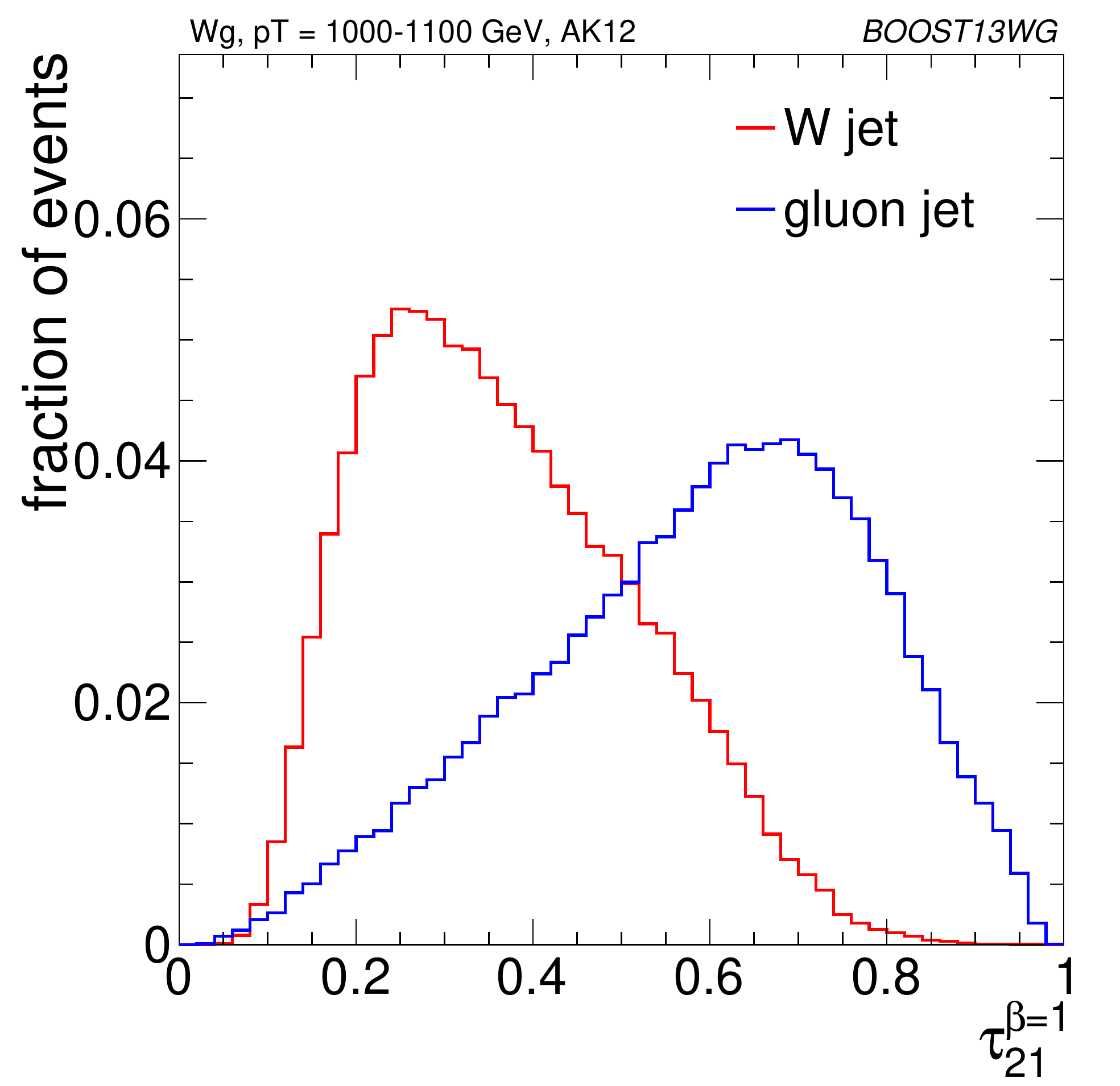}}
\subfigure[\antikt $R=0.8$, \pt = 1.0-1.1 \TeV bin]{\includegraphics[width=0.4\textwidth]{./Figures/WTagging/pT1000/AKtR08/h_c2_b1.pdf}}
\subfigure[\antikt $R=1.2$, \pt = 1.0-1.1 \TeV bin]{\includegraphics[width=0.4\textwidth]{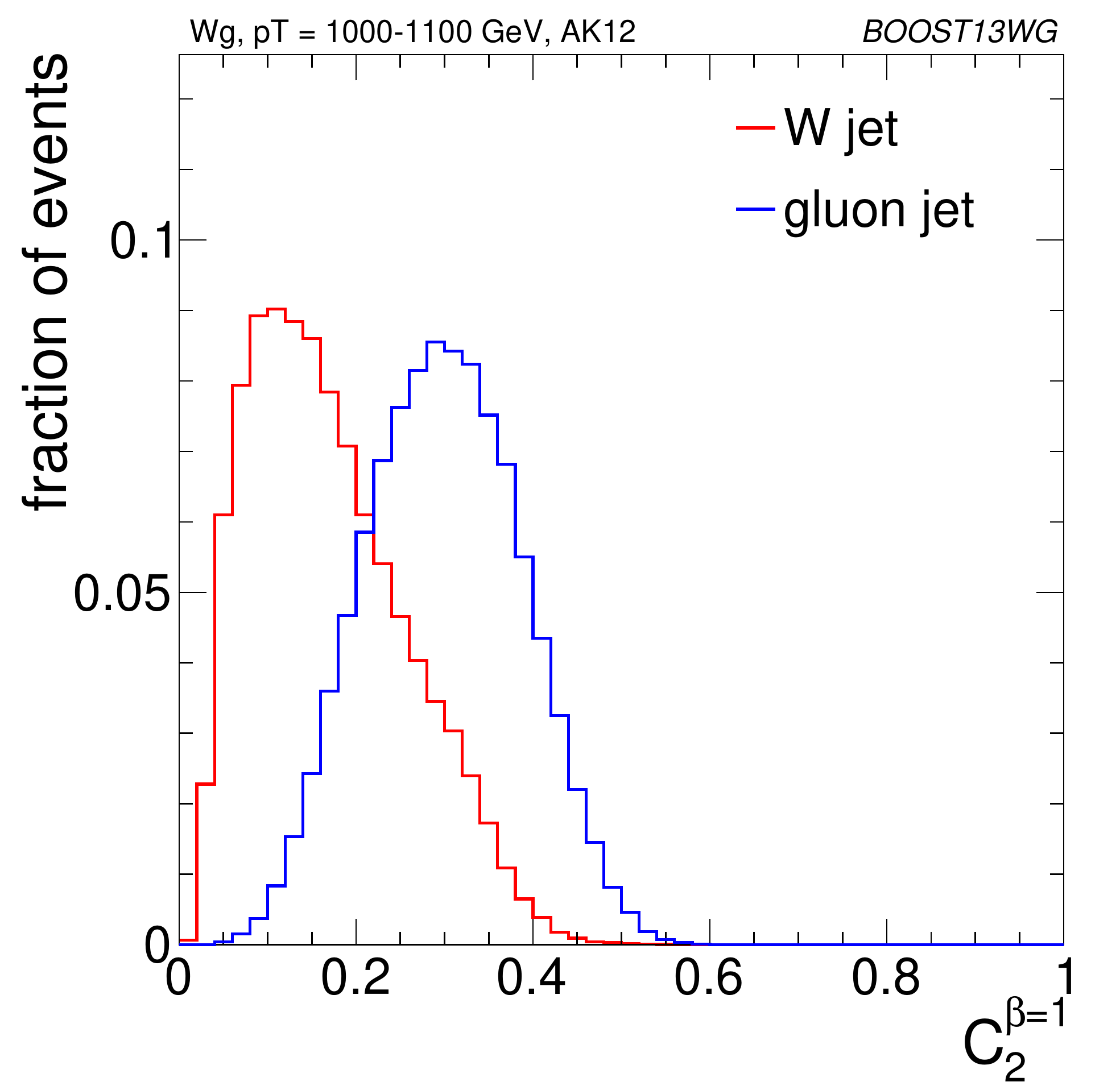}}
\caption{The $\tau_{21}^{\beta=1}$ and $C_2^{\beta=1}$ distributions for signal
  and background $R=0.8$ and $R=1.2$ jets in the \pt = 1.0-1.1 \TeV bin.}
\label{fig:Rdepend_substructure}
\end{figure*}

However, we again see rather
different behaviour versus $R$ for the substructure variables. In all \pt bins
considered, the most performant substructure variable, $C_2^{\beta=1}$,
performs best for an \antikt distance parameter of $R=0.8$. The
performance of this variable is dramatically worse for the larger jet
radius of $R=1.2$ (a factor seven worse background rejection in
the \pt = 1.0-1.1 \TeV bin), and substantially worse for $R=0.4$. For the other
jet substructure variables considered, $\Gamma_{\rm Qjet}$ and
$\tau_{21}^{\beta=1}$, their background rejection
power also reduces for larger jet radius, but not to the same
extent. Figure~\ref{fig:Rdepend_substructure} shows the
$\tau_{21}^{\beta=1}$ and $C_2^{\beta=1}$ distributions for signal and
background in the \pt = 1.0-1.1 \TeV bin for $R=0.8$ and $R=1.2$ jet
radii. For the larger jet radius, the
$C_2^{\beta=1}$  distribution of both signal and background gets wider,
and consequently the discrimination power decreases. For
$\tau_{21}^{\beta=1}$ there is comparatively little change in the distributions
with increasing jet radius. The increased sensitivity of $C_{2}$ to
soft wide angle radiation in comparison to $\tau_{21}$ is a known
feature of this variable~\cite{Larkoski:2013eya}, and a useful feature
in discriminating coloured versus colour singlet jets. However, at
very large jet radii ($R\sim1.2$), this feature becomes
disadvantageous; the jet can pick up a significant amount of initial
state or other uncorrelated radiation, and $C_{2}$ is more sensitive
to this than is $\tau_{21}$.  This uncorrelated radiation has no (or
very little) dependence on whether the jet is $W$- or gluon-initiated, and so sensitivity to this radiation means that the 
discrimination power will decrease. A similar description applies to the 
variable $\Gamma_{\rm Qjet}$, and the story is very similar to that
for $\Gamma_{\rm Qjet}$ with increasing \pt.  At larger $R$ the low volatility ``shoulder'' is enhanced in the
QCD background jet, leading to a narrower volatility distribution.  For the W jet,
the larger $R$ includes more uncorrelated radiation in the jet, leading to a broader
volatility distribution.  So, as with increasing \pt, increasing $R$ results in volatility distributions
for signal and background jets that are more similar and $\Gamma_{\rm Qjet}$ exhibits
reduced discrimination power.

\subsection{Combined Performance}\label{sec:w_combined}

Studying the improvement in performance (or lack thereof) when combining single variables into a multivariate analysis gives insight into the correlations
among jet observables. The off-diagonal entries in Figures~\ref{fig:pt300_comb2D},~\ref{fig:pt500_comb2D}
and~\ref{fig:pt1000_comb2D} can be used to compare the performance
of different BDT two-variable combinations, and see how this varies as
a function of \pt and $R$. By comparing the background rejection
achieved for the two-variable combinations to the background rejection
of the ``all variables'' BDT, one can also understand how 
discrimination can be improved by adding further variables to the
two-variable BDTs.

In general the most powerful two-variable
combinations involve a groomed mass and a non-mass substructure
variable ($C_2^{\beta=1}$, $\Gamma_{\rm Qjet}$ or
$\tau_{21}^{\beta=1}$). Two-variable combinations of the substructure
variables are not as powerful in comparison.  Which particular mass +
substructure variable combination is the most
powerful depends strongly on the \pt and $R$ of the jet, as discussed
in the sections to follow.

There is also modest improvement in
the background rejection when different groomed masses are combined,
 indicating that there is complementary information between the
different groomed masses (first shown in \cite{Soper:2010xk}). In addition, there is an improvement in the
background rejection when the groomed masses are combined with the
ungroomed mass, indicating that grooming removes some useful
discriminatory information from the jet. These observations are
explored further in the section below.

Generally, the $R=0.8$ jets offer the best two-variable
combined performance in all \pt bins explored here. This is despite
the fact that in the highest \pt = 1.0-1.1 \TeV bin the average
separation of the quarks from the $W$ decay is much smaller than 0.8,
and well within 0.4. This conclusion could of course be susceptible to
pile-up, which is not considered in this study. It is in marked
contrast to the $R$ dependence of the $q/g$ tagging performance
shown in Section~\ref{sec:qgtagging}, where a monotonic improvement
in performance with reducing $R$ is observed.

\subsubsection{Mass + Substructure Performance}

As already noted, the largest background rejection at 70\% signal
efficiency are in general achieved using those two-variable BDT combinations
which involve a groomed mass and a non-mass substructure variable. We now investigate
the \pt and $R$ dependence of the performance of these combinations.

For both $R=0.8$ and $R=1.2$ jets, the rejection power of these two-variable
combinations increases substantially with increasing \pt, at least
within the \pt range considered here.

\begin{figure*}
\centering
\subfigure[\pt = 300-400 \GeV]{\includegraphics[width=0.78\textwidth]{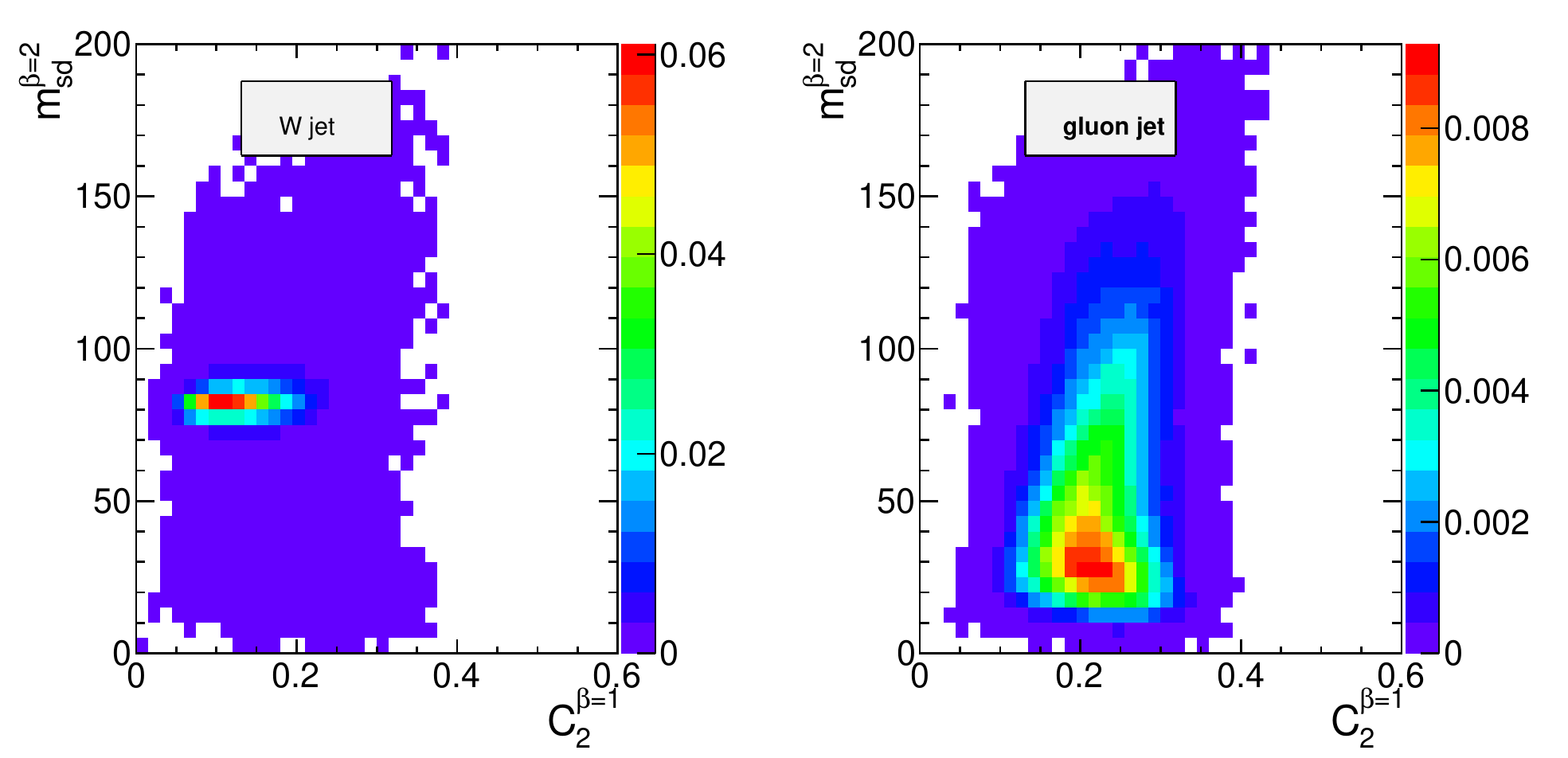}\label{fig:pt300_2d_c2b1_AKt_R08}}
\subfigure[\pt = 500-600 \GeV]{\includegraphics[width=0.78\textwidth]{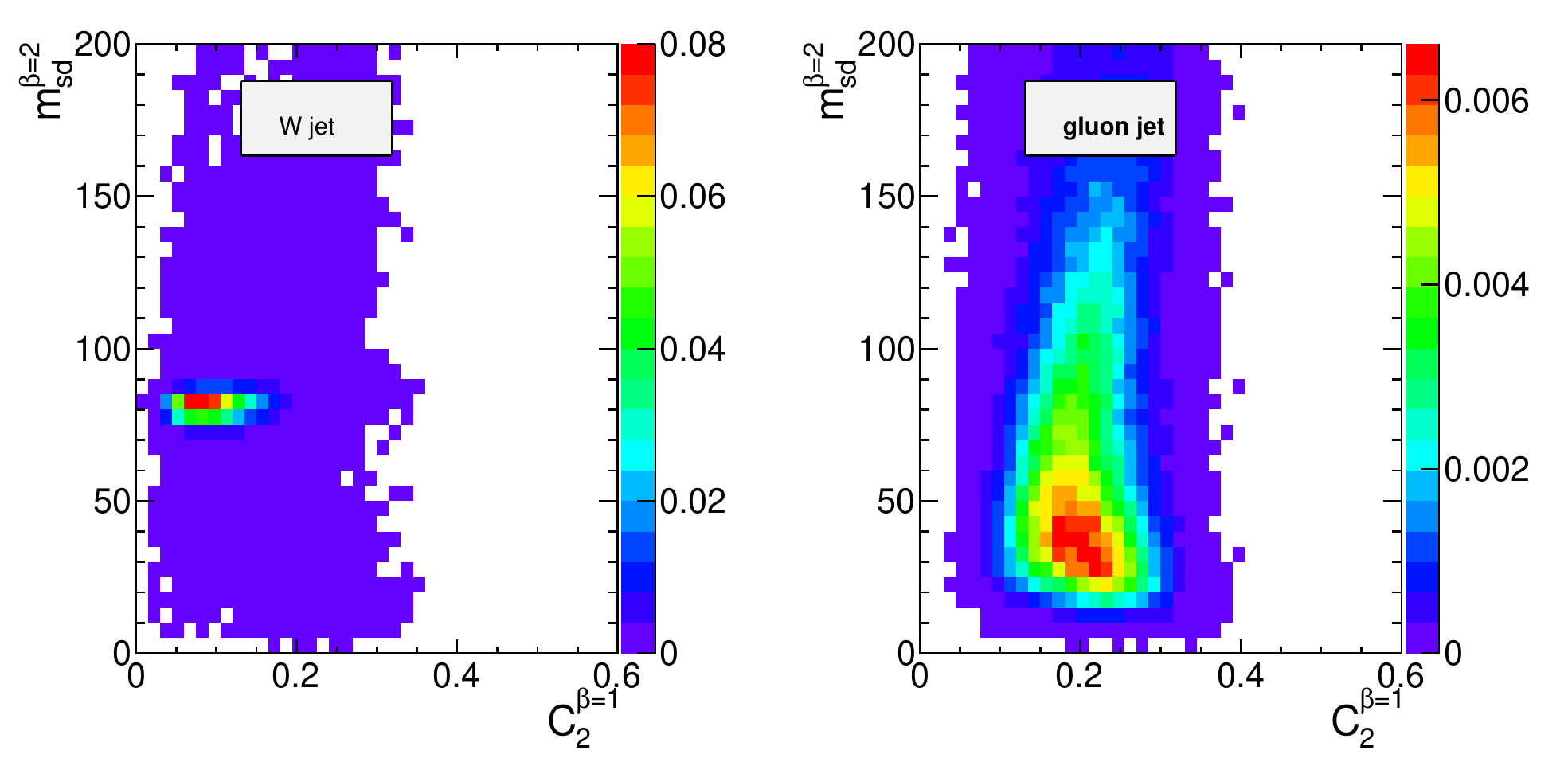}\label{fig:pt500_2d_c2b1_AKt_R08}}
\subfigure[\pt = 1.0-1.1 \TeV]{\includegraphics[width=0.78\textwidth]{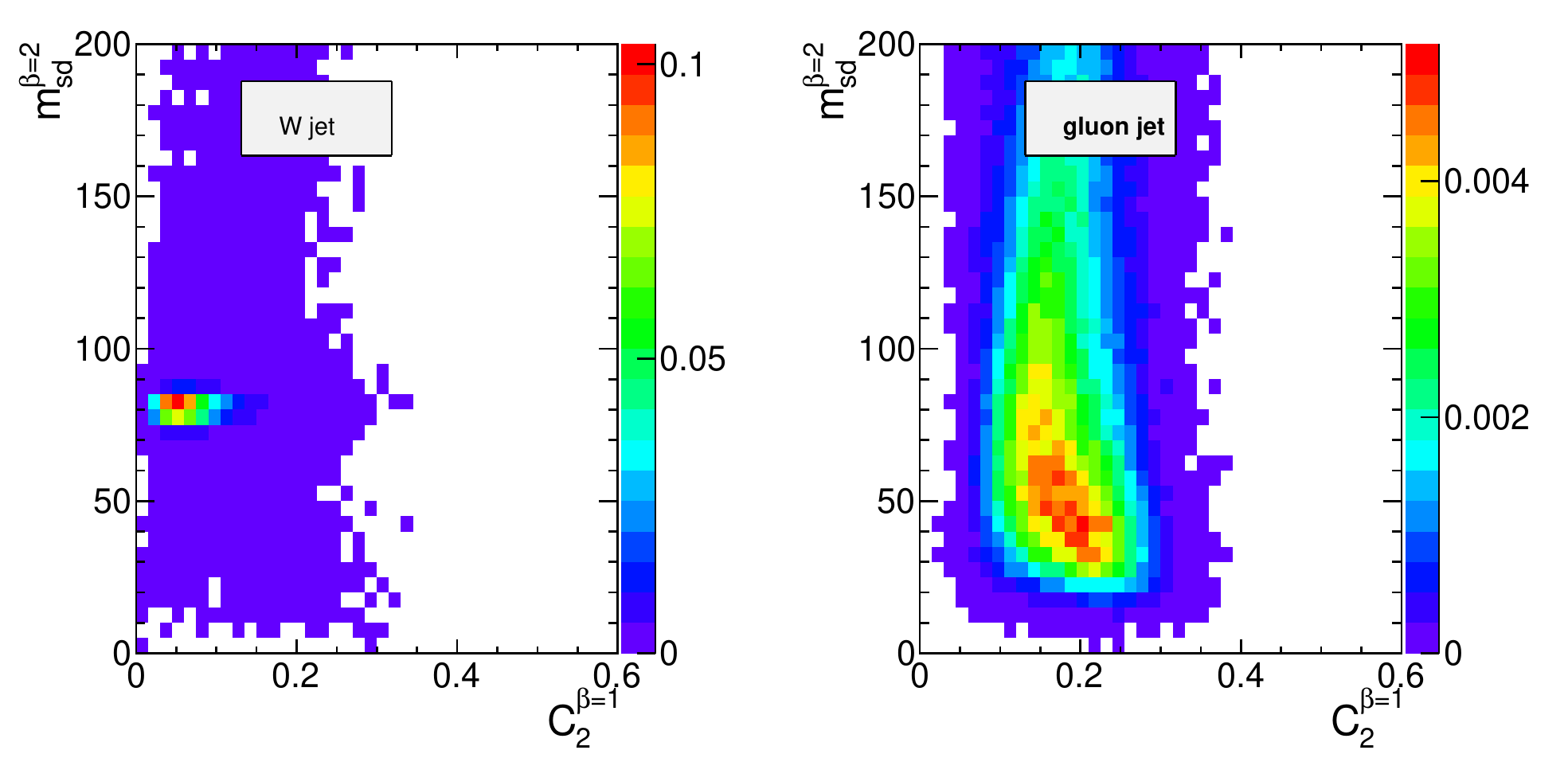}\label{fig:pt1000_2d_c2b1_AKt_R08}}
\caption{2-D histograms of $m_{sd}^{\beta=2}$ versus $C_2^{\beta=1}$
  distributions for $R=0.8$ jets in the various \pt bins considered, shown separately for signal and background.}
\label{fig:2d_c2b1_AKt_R08}
\end{figure*}

For a jet radius of $R=0.8$, across the full \pt range considered, the
groomed mass + substructure variable combinations with the
largest background rejection are those which
involve $C_2^{\beta=1}$. For example, in combination with
\msd, this produces a five-, eight- and fifteen-fold
increase in background rejection compared to using the groomed mass
alone. In Figure~\ref{fig:2d_c2b1_AKt_R08} are shown 2-D histograms of
\msd versus $C_2^{\beta=1}$ for $R=0.8$ jets in the various \pt bins
considered, for both signal and background. The relatively low degree of
correlation between \msd versus $C_2^{\beta=1}$ that
leads to these large improvements in background rejection can be
seen. What little correlation exists is rather
non-linear in nature, changing from a negative to a positive
correlation as a function of the groomed mass, something which helps
to improve the background rejection in the region of the $W$ mass peak.

\begin{figure*}
\centering
\subfigure[$R=0.4$]{\includegraphics[width=0.78\textwidth]{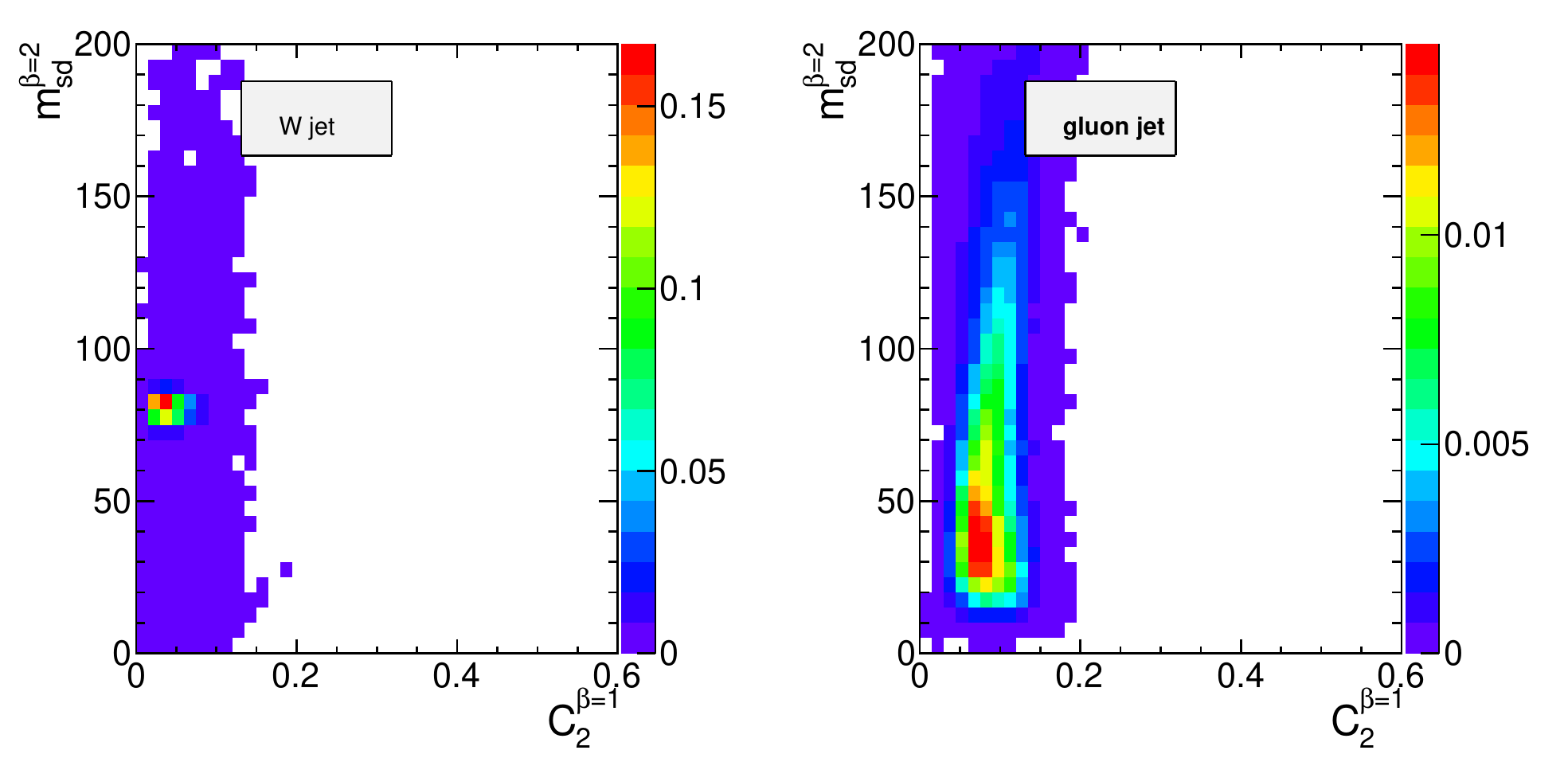}\label{fig:pt1000_2d_c2b1_AKt_R04}}
\subfigure[$R=0.8$]{\includegraphics[width=0.78\textwidth]{./Figures/WTagging/pT1000/AKtR08/h2d_jc2_b1_j_mass_sdb2_WW_onSame.pdf}\label{fig:pt1000_2d_c2b1_AKt_R08_b}}
\subfigure[$R=1.2$]{\includegraphics[width=0.78\textwidth]{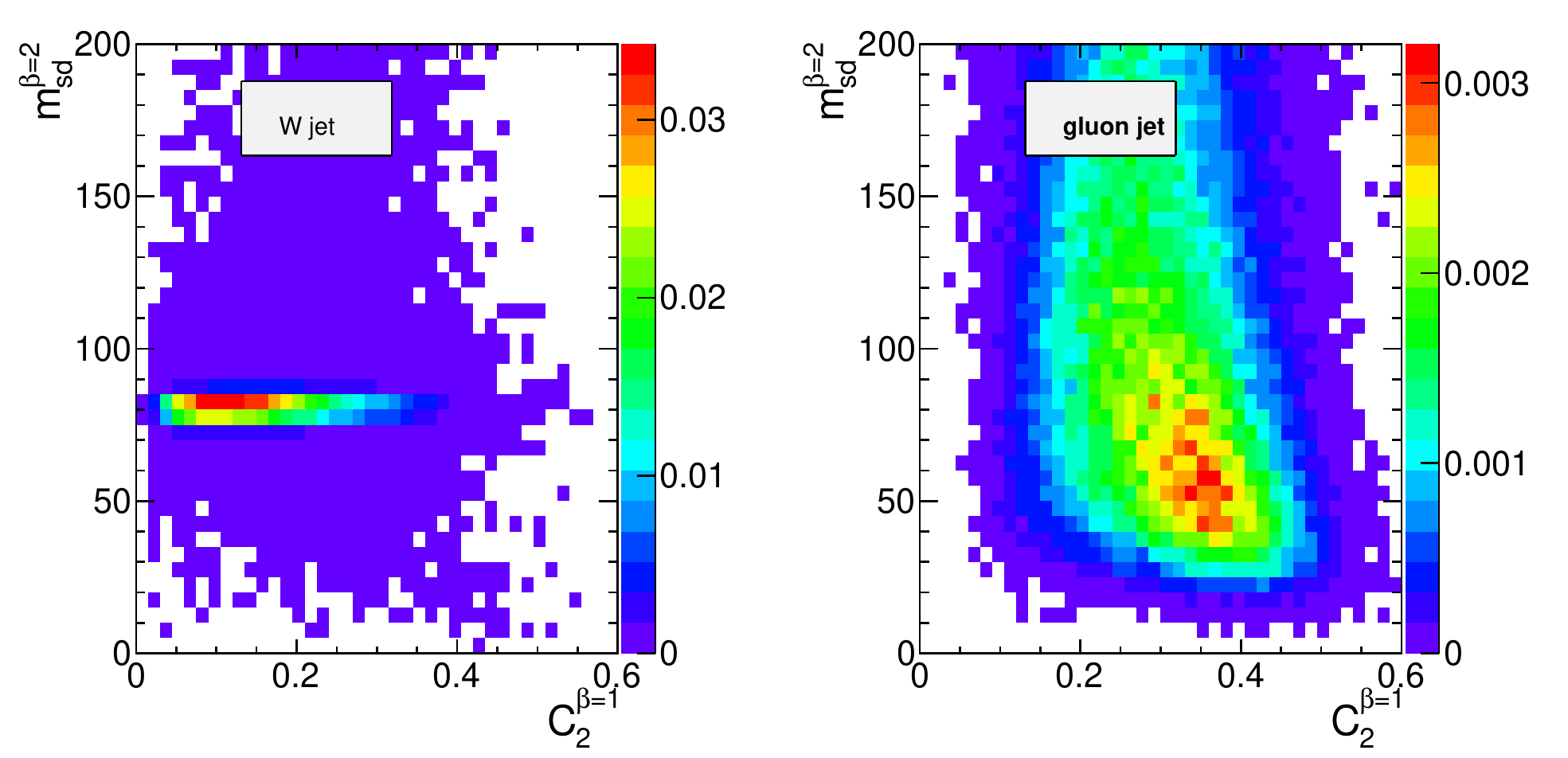}\label{fig:pt1000_2d_c2b1_AKt_R12}}
\caption{2-D histograms of $m_{sd}^{\beta=2}$ versus $C_2^{\beta=1}$
  for $R=0.4$, 0.8 and 1.2 jets in the \pt = 1.0-1.1 \TeV bin, shown separately for signal and background.}
\label{fig:2d_c2b1_pt1000}
\end{figure*}

\begin{figure*}
\centering
\subfigure[$R=0.4$]{\includegraphics[width=0.78\textwidth]{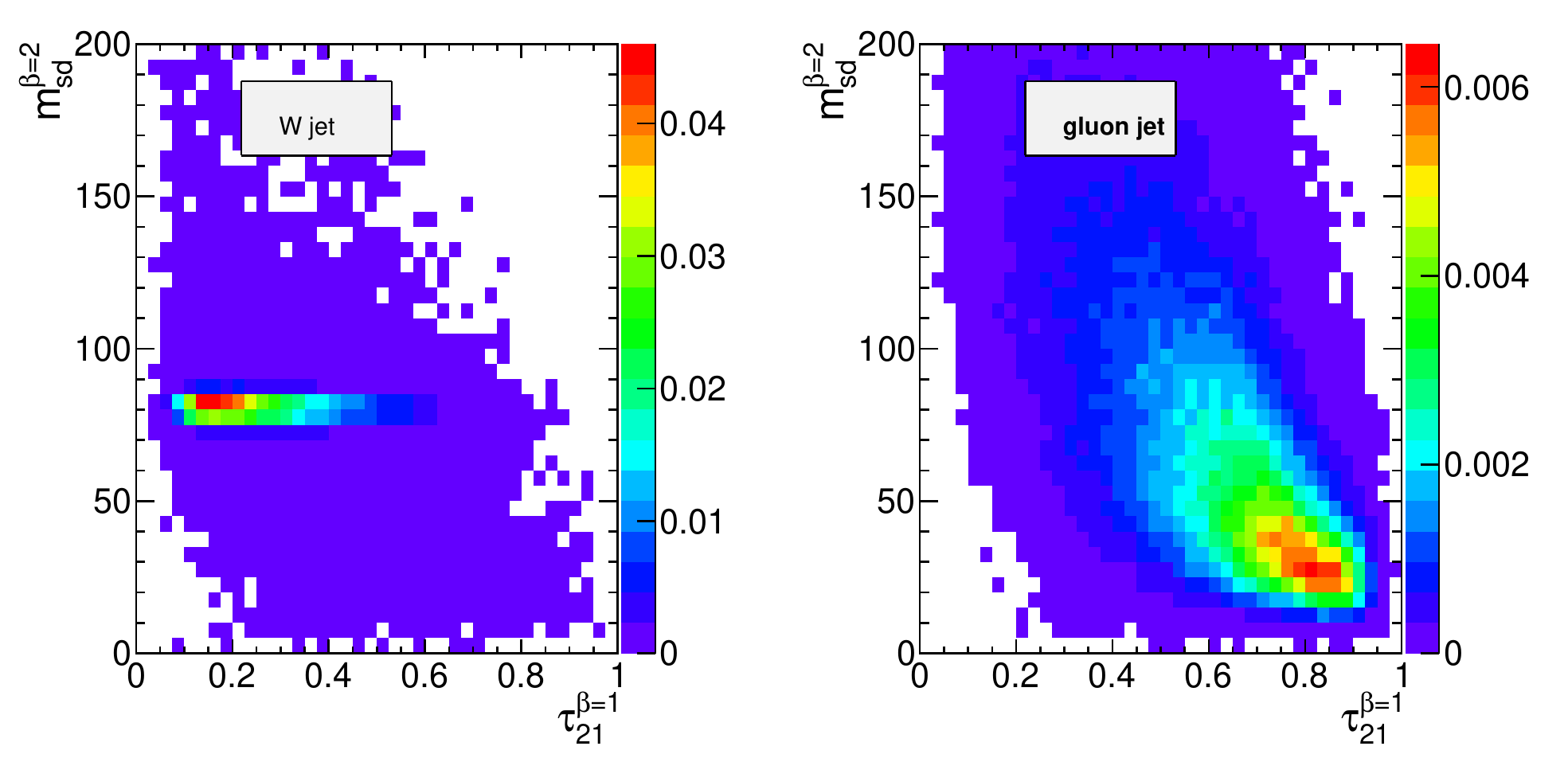}\label{fig:pt1000_2d_tau21_AKt_R04}}
\subfigure[$R=0.8$]{\includegraphics[width=0.78\textwidth]{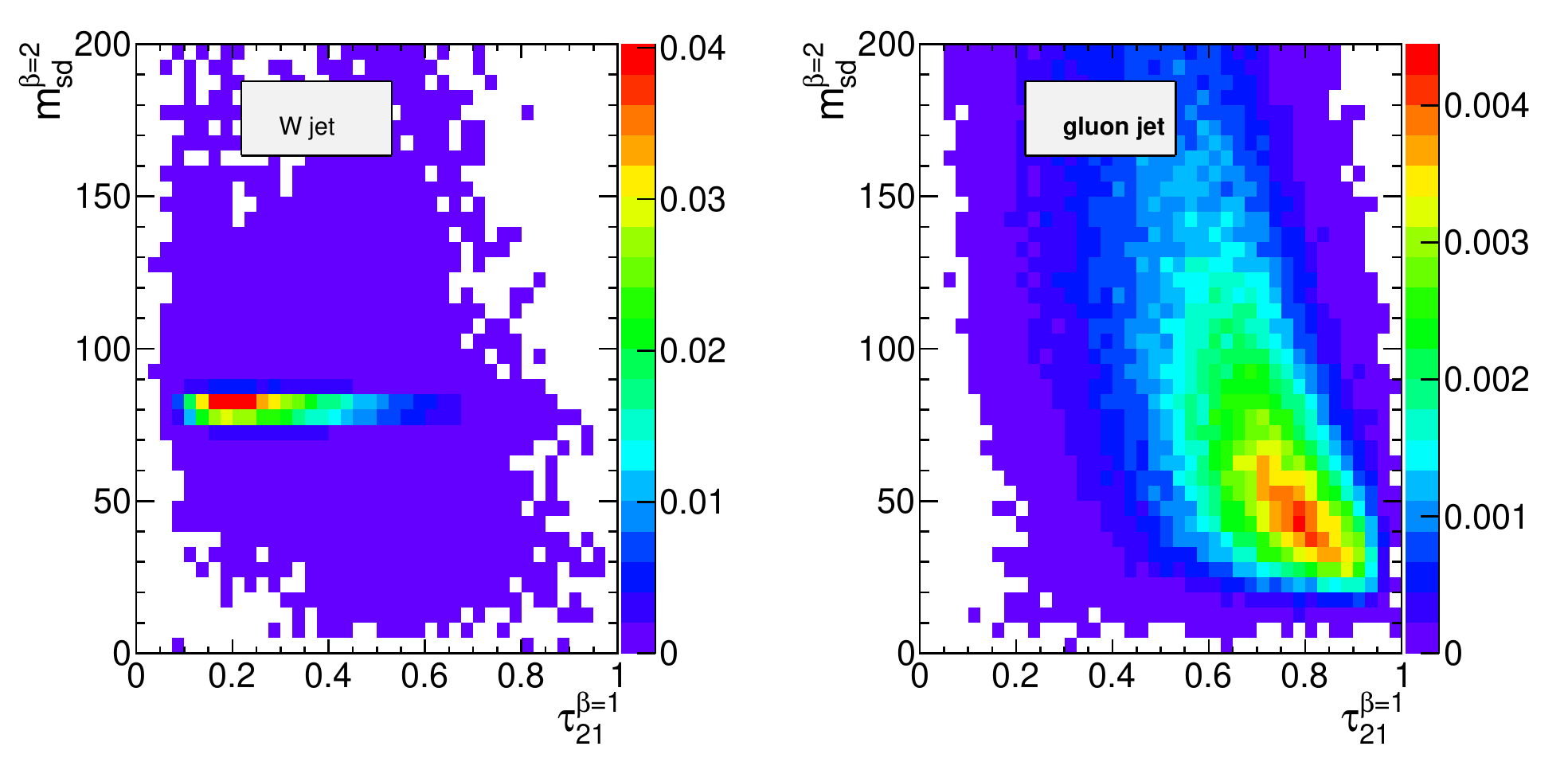}\label{fig:pt1000_2d_tau21_AKt_R08_b}}
\subfigure[$R=1.2$]{\includegraphics[width=0.78\textwidth]{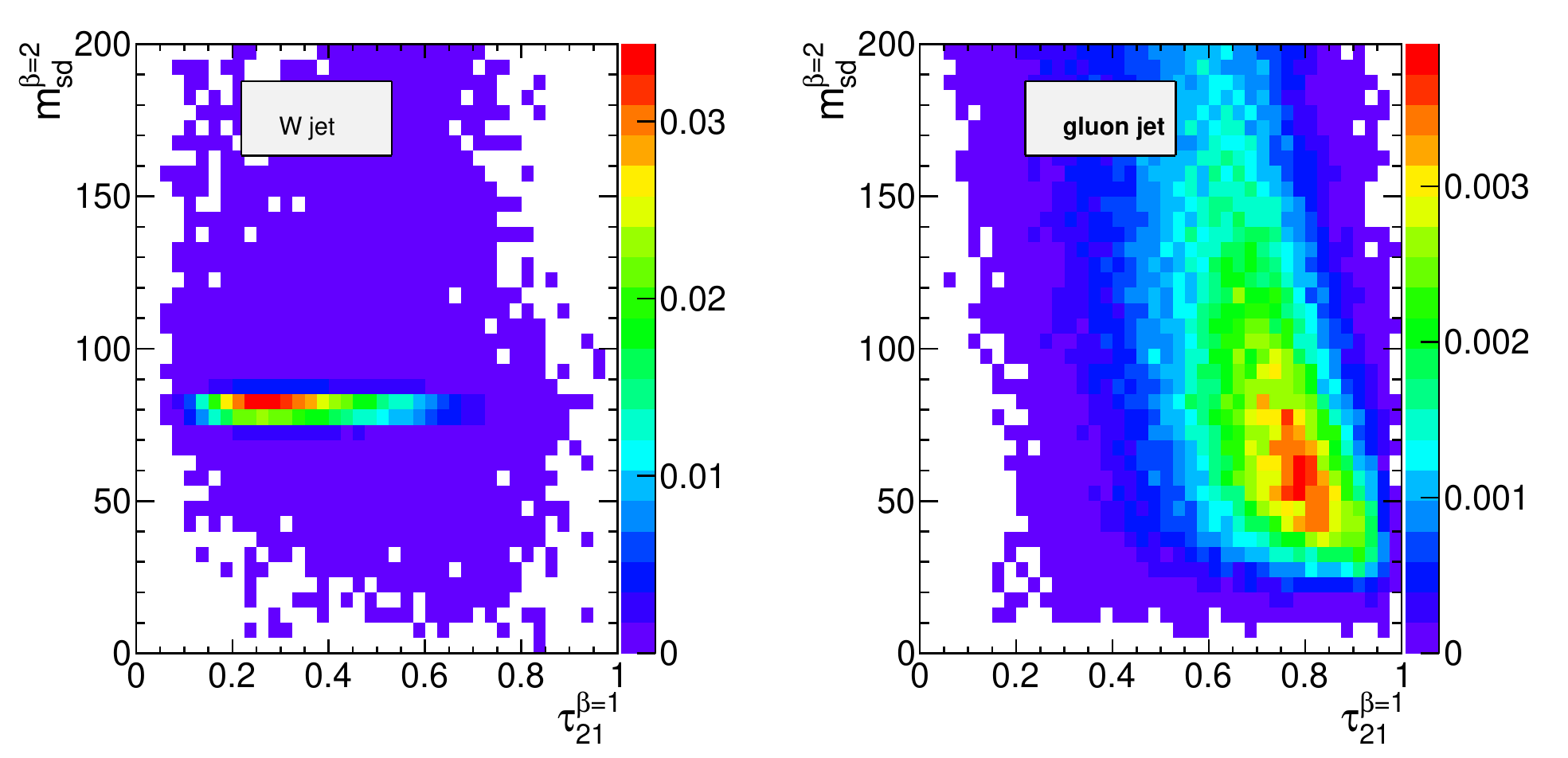}\label{fig:pt1000_2d_tau21_AKt_R12}}
\caption{2-D histograms of $m_{sd}^{\beta=2}$ versus $\tau_{21}^{\beta=1}$
  for $R=0.4$, 0.8 and 1.2 jets in the \pt = 1.0-1.1 \TeV bin, shown separately for signal and background.}
\label{fig:2d_tau21_pt1000}
\end{figure*}

However, when we switch to a jet radius of $R=1.2$ the picture for
$C_2^{\beta=1}$ combinations changes dramatically. These become
significantly less powerful, and the most powerful variable in groomed
mass combinations becomes $\tau_{21}^{\beta=1}$ for all jet
\pt considered. Figure~\ref{fig:2d_c2b1_pt1000} shows the correlation between
$m_{sd}^{\beta=2}$ and $C_2^{\beta=1}$ in the \pt = 1.0 - 1.1 \TeV bin for
the various jet radii considered. Figure~\ref{fig:2d_tau21_pt1000} is
the equivalent set of distributions for $m_{sd}^{\beta=2}$ and
$\tau_{21}^{\beta=1}$. One can see from
Figure~\ref{fig:2d_c2b1_pt1000} that, due to the sensitivity of the
observable to to soft, wide-angle radiation, as the jet radius increases
$C_2^{\beta=1}$ increases and becomes more and more smeared out for both signal and
background, leading to worse discrimination power. This does not
happen to the same extent for $\tau_{21}^{\beta=1}$. We can see from Figure~\ref{fig:2d_tau21_pt1000} that
the negative correlation between $m_{sd}^{\beta=2}$ and
$\tau_{21}^{\beta=1}$ that is clearly visible for $R=0.4$ decreases for
larger jet radius, such that the groomed mass and substructure variable
are far less correlated and $\tau_{21}^{\beta=1}$ offers improved
discrimination within a $m_{sd}^{\beta=2}$ mass window.

\subsubsection{Mass + Mass Performance}

\begin{figure*}
\centering
\subfigure[Pruned mass vs ungroomed mass]{\includegraphics[width=0.78\textwidth]{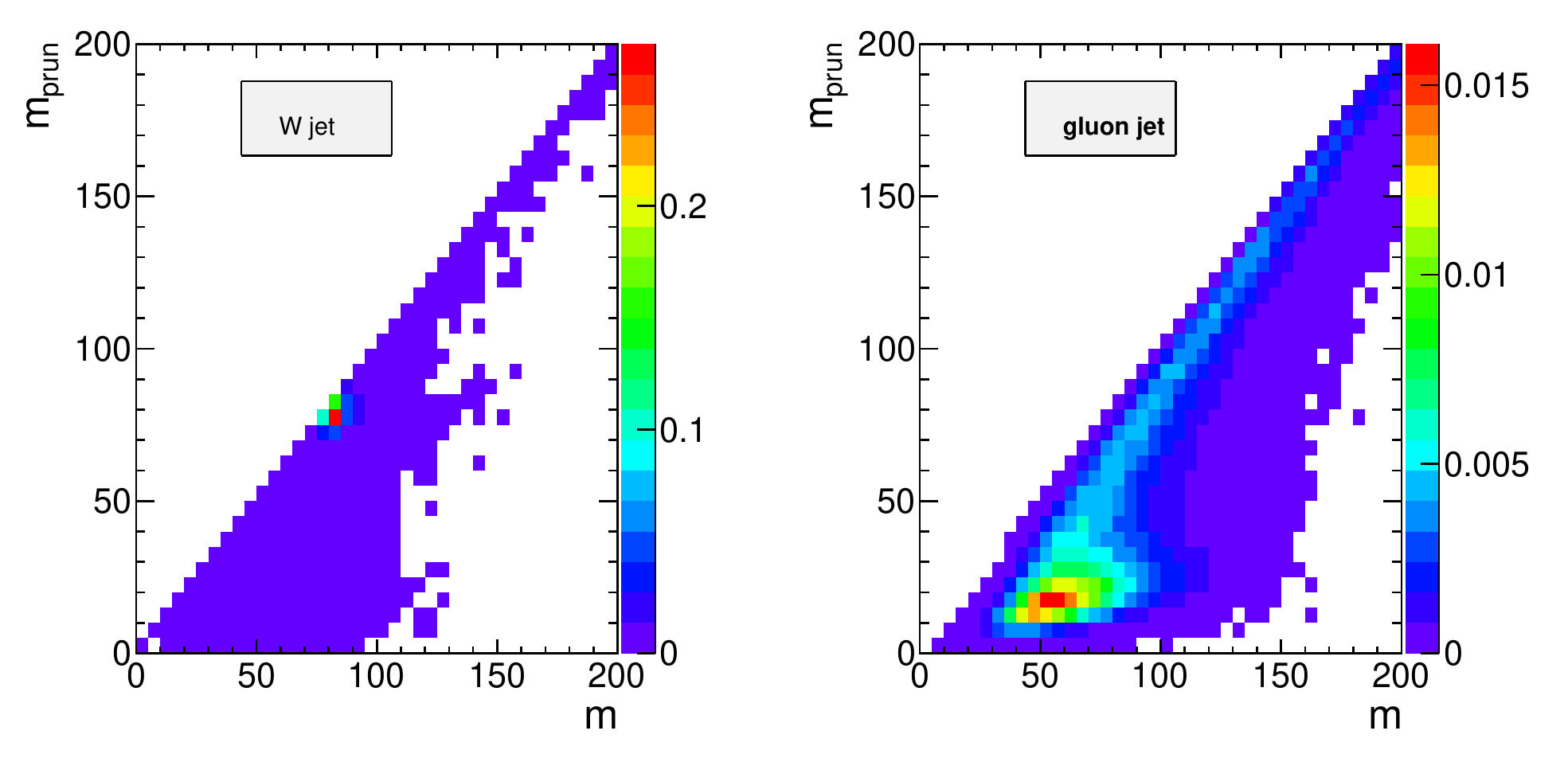}\label{fig:pt1000_2d_prun_AKt_R04}}
\subfigure[Trimmed mass vs ungroomed mass]{\includegraphics[width=0.78\textwidth]{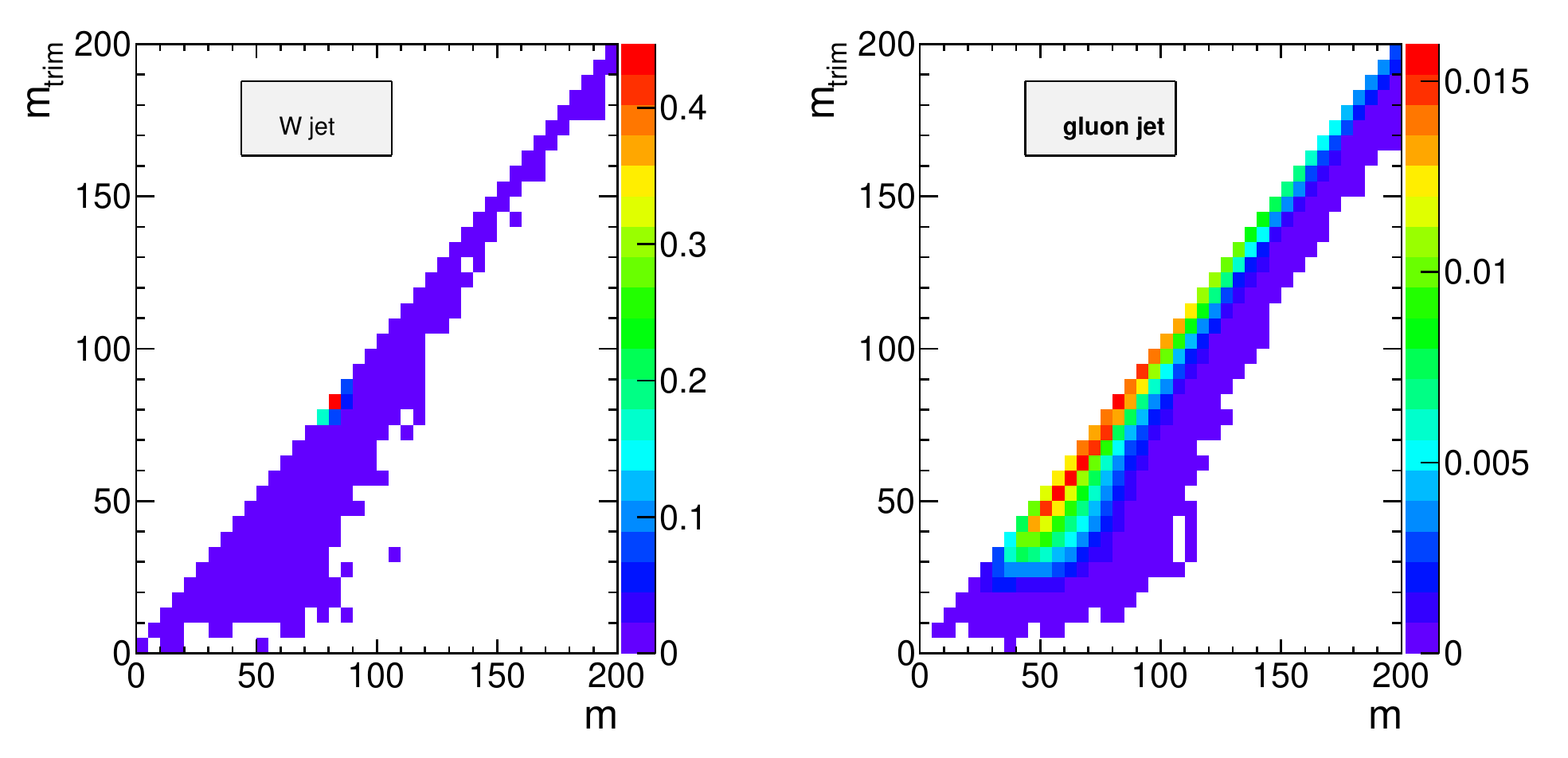}\label{fig:pt1000_2d_trim_AKt_R04}}
\caption{2-D histograms of groomed mass versus ungroomed mass in the \pt = 1.0-1.1 \TeV bin using the
  anti-\kT $R=0.4$ algorithm, shown separately for signal and background.}
\label{fig:pt1000_2d_masses_AKt_R04}
\end{figure*}

\begin{figure*}
\centering
\subfigure[Pruned mass vs ungroomed mass]{\includegraphics[width=0.78\textwidth]{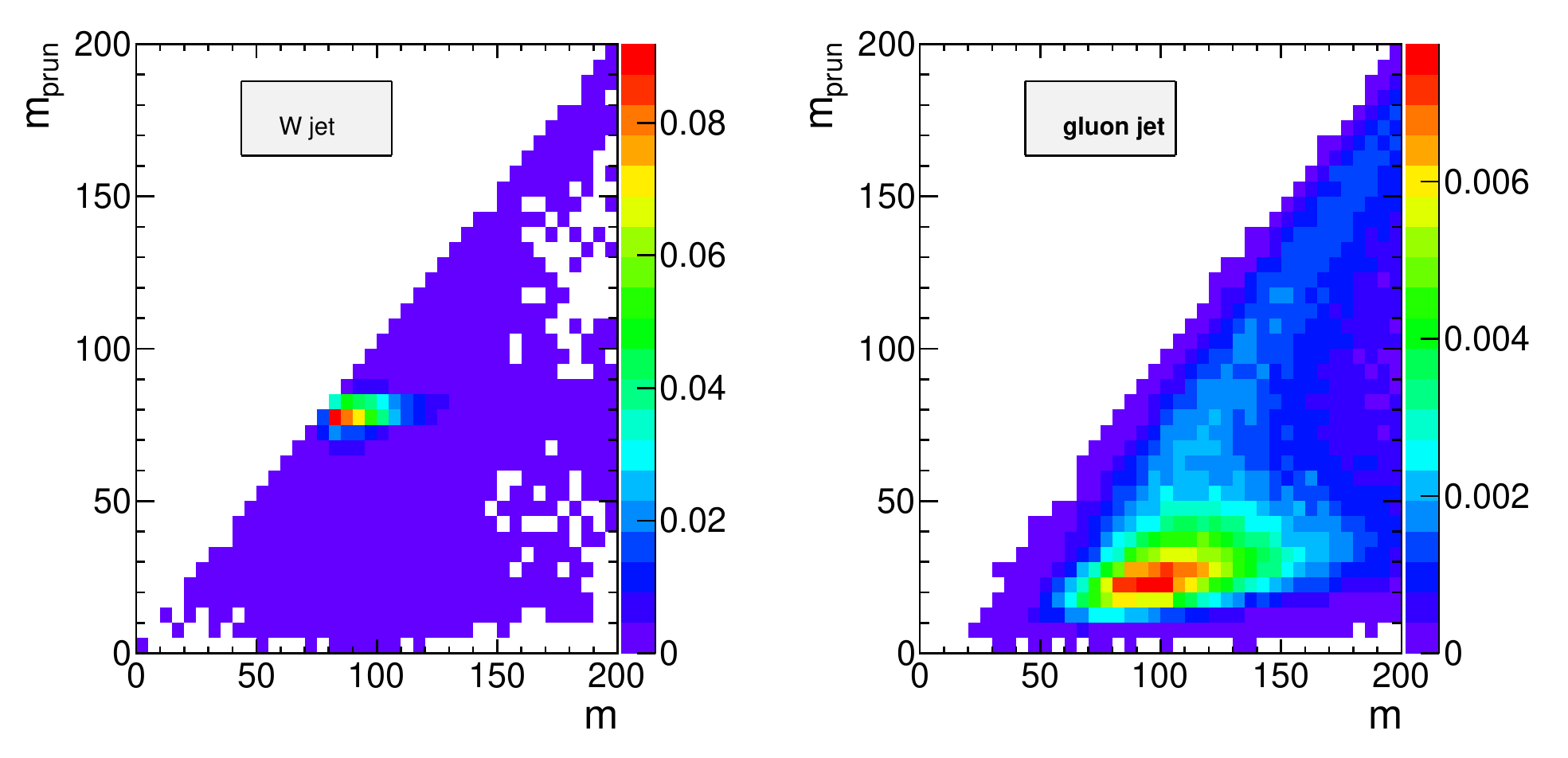}\label{fig:pt1000_2d_prun_AKt_R08}}
\subfigure[Trimmed mass vs ungroomed mass]{\includegraphics[width=0.78\textwidth]{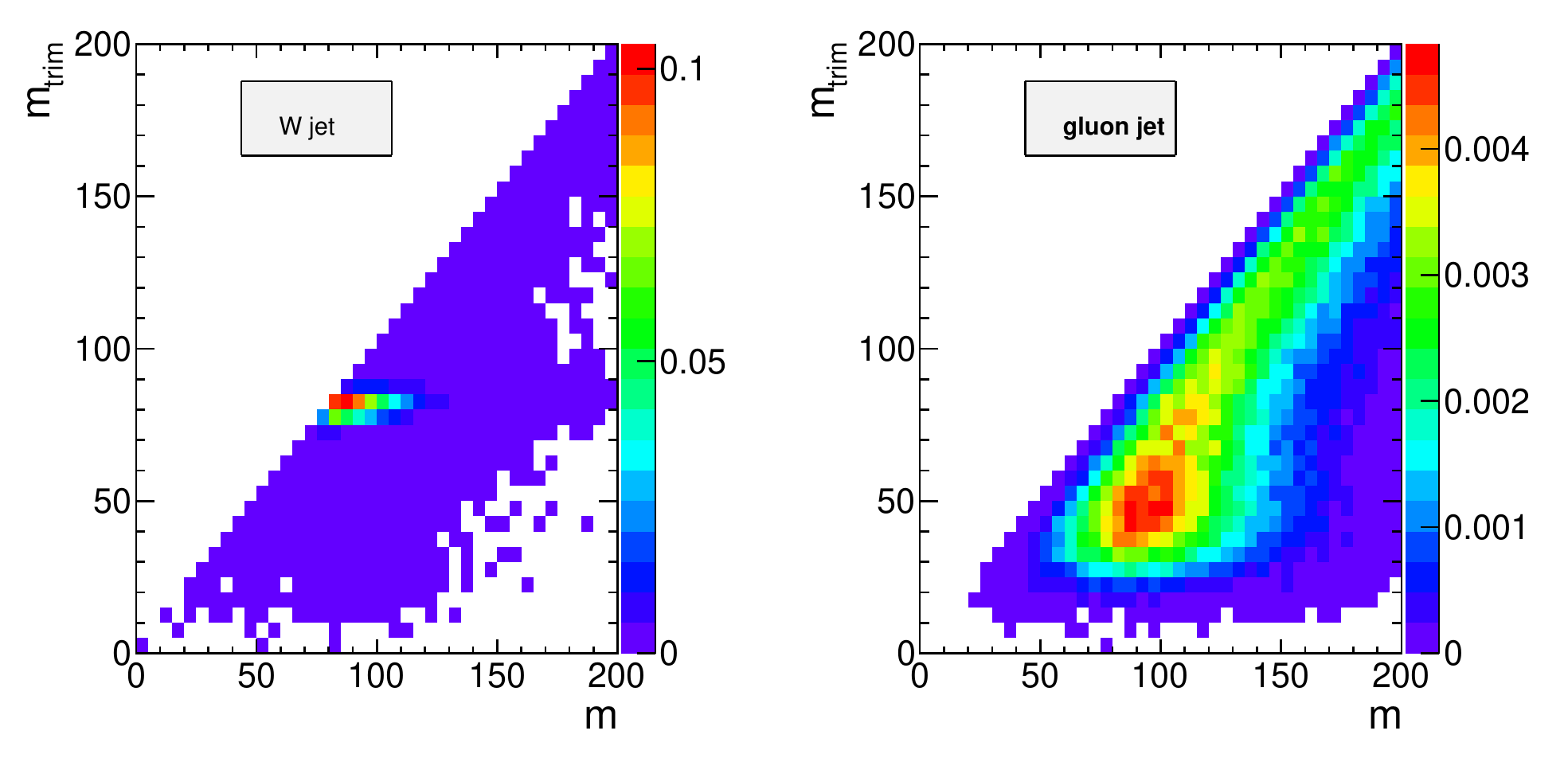}\label{fig:pt1000_2d_trim_AKt_R08}}
\caption{2-D histograms of groomed mass versus ungroomed mass in the \pt = 1.0-1.1 \TeV bin using the
  anti-\kT $R=0.8$ algorithm, shown separately for signal and background.}
\label{fig:pt1000_2d_masses_AKt_R08}
\end{figure*}

\begin{figure*}
\centering
\subfigure[Pruned mass vs ungroomed mass]{\includegraphics[width=0.78\textwidth]{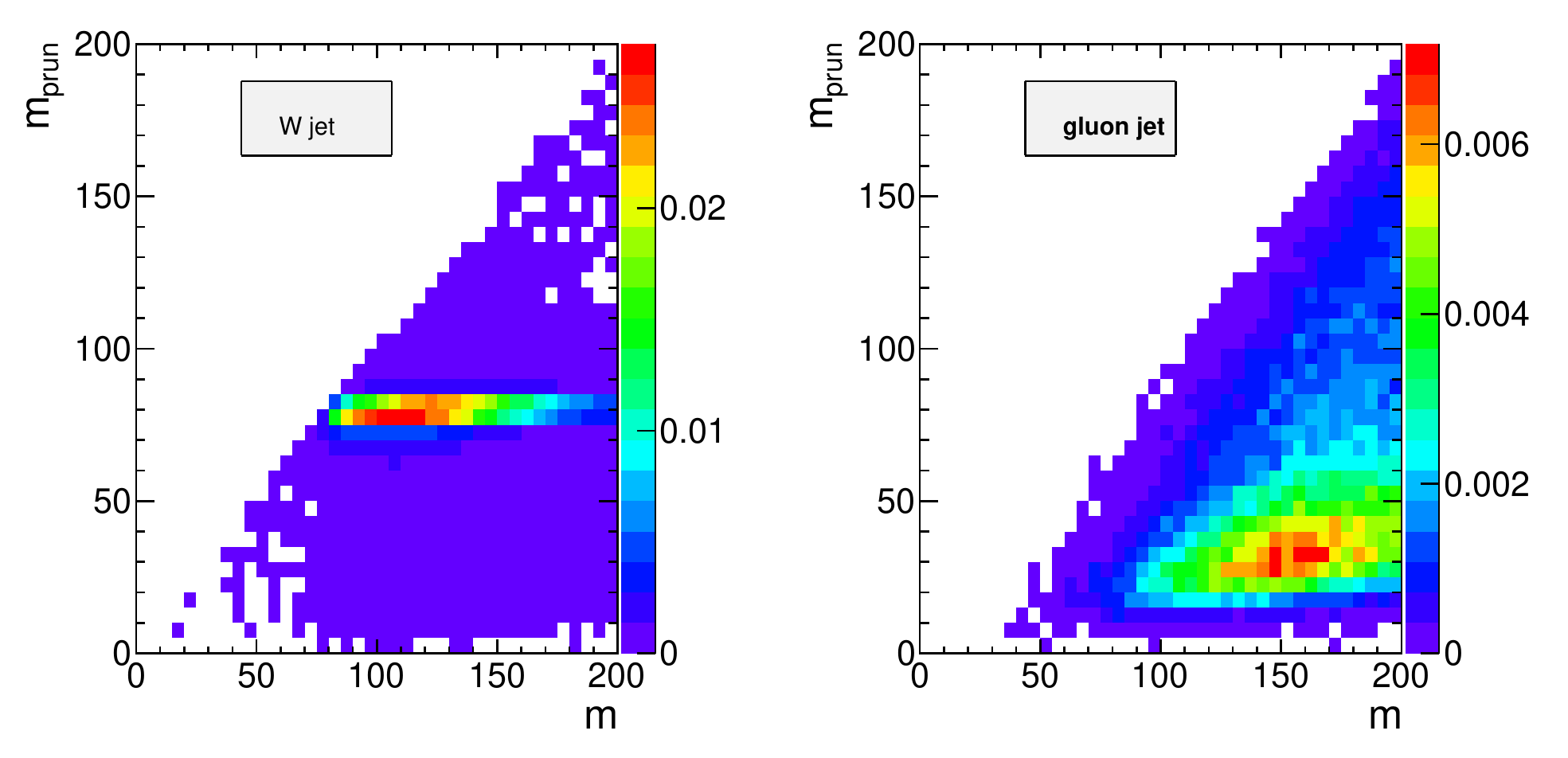}\label{fig:pt1000_2d_prun_AKt_R12}}
\subfigure[Trimmed mass vs ungroomed mass]{\includegraphics[width=0.78\textwidth]{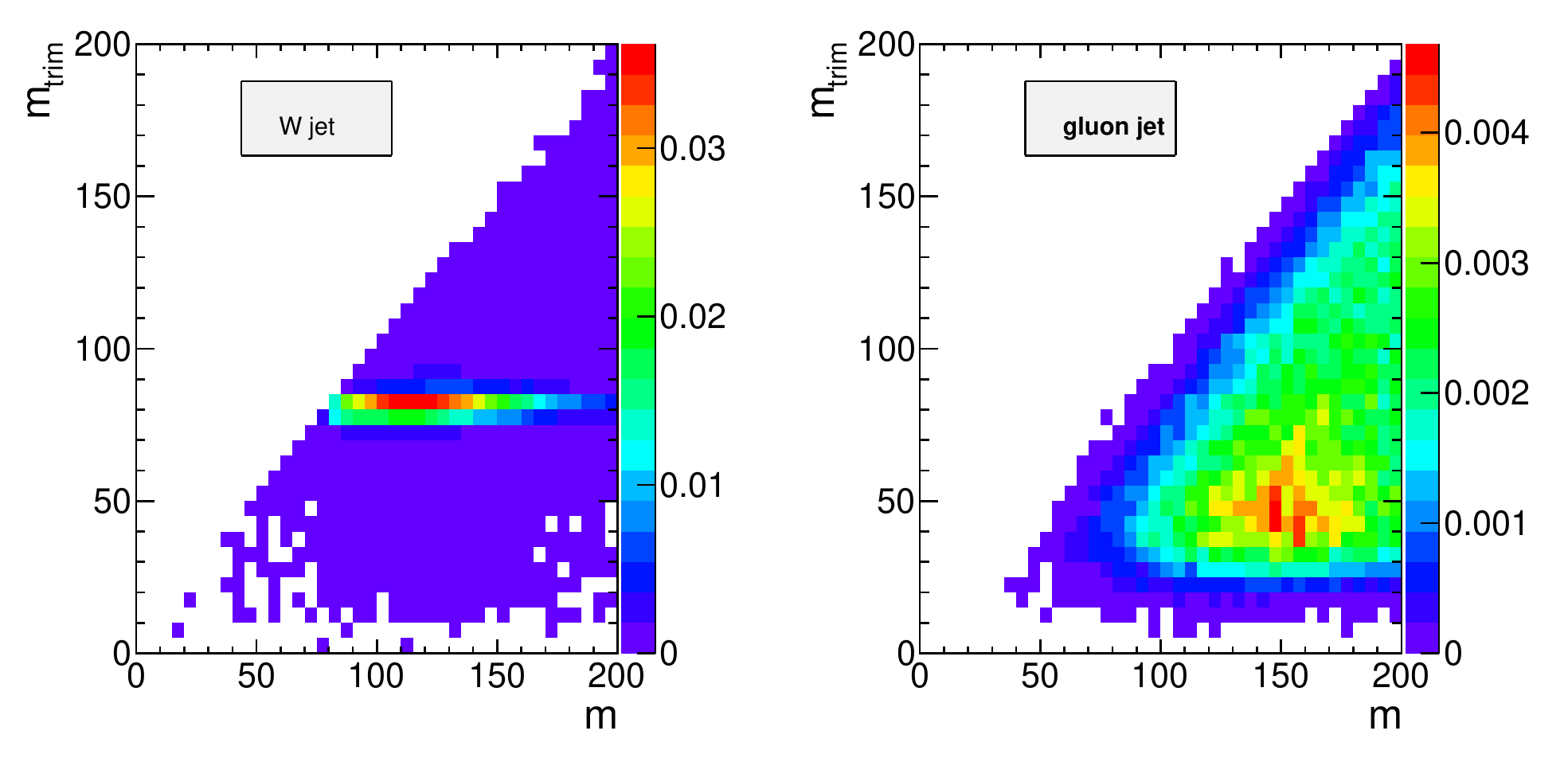}\label{fig:pt1000_2d_trim_AKt_R12}}
\caption{2-D histograms of groomed mass versus ungroomed mass in the \pt = 1.0-1.1 \TeV bin using the
  anti-\kT $R=1.2$ algorithm, shown separately for signal and background.}
\label{fig:pt1000_2d_masses_AKt_R12}
\end{figure*}

The different groomed masses and the ungroomed mass are of course not
fully correlated, and thus one can always see some kind of improvement
in the background rejection when two different mass variables are combined
in the BDT. However, in some cases the improvement can be dramatic,
particularly at higher \pt, and particularly for combinations with the
ungroomed mass. For example, in Figure~\ref{fig:pt1000_comb2D} we can
see that in the \pt =1.0-1.1 \TeV bin, the combination of pruned mass with
ungroomed mass produces a greater than eight-fold improvement in the
background rejection for $R=0.4$ jets, a greater than five-fold
improvement for $R=0.8$ jets, and a factor $\sim 2$ improvement for
$R=1.2$ jets. A similar behaviour can be seen for mMDT mass. 
In Figures~\ref{fig:pt1000_2d_masses_AKt_R04},~\ref{fig:pt1000_2d_masses_AKt_R08}
and~\ref{fig:pt1000_2d_masses_AKt_R12}, we show the 2-D correlation plots of the pruned mass versus the
ungroomed mass separately for the $WW$ signal and $gg$ background
samples in the \pt = 1.0-1.1 \TeV bin, for the various jet radii
considered. For comparison, the correlation of the trimmed mass with
the ungroomed mass, a combination that does not improve on the single
mass as dramatically, is shown. In all cases one can see that there is
a much smaller degree of correlation between the pruned mass and the
ungroomed mass in the backgrounds sample than for the trimmed mass and the ungroomed mass. This
is most obvious in Figure~\ref{fig:pt1000_2d_masses_AKt_R04}, where
the high degree of correlation between the trimmed and ungroomed mass
is expected, since with the parameters used (in particular
$R_{\rm trim} = 0.2$) we cannot expect trimming to have a significant
impact on an $R=0.4$ jet. The reduced correlation with ungroomed mass
for pruning in the background means that, once we have required that the
pruned mass is consistent with a $W$ (\emph{i.e.} $\sim 80 \GeV$), a relatively large difference
between signal and background in the ungroomed mass still remains, and
can be exploited to improve the background rejection further. In other
words, many of the background events which pass the pruned mass
requirement do so because they are shifted to lower mass (to be within
a signal mass window) by the grooming, but these events still have the
property that they look very much like background events before the
grooming. A  requirement on the groomed mass alone does not
exploit this property. Of course, the impact of pile-up, not considered in this
study, could limit the degree to which the ungroomed
mass could be used to improve discrimination in this way. 

\subsubsection{``All Variables'' Performance}\label{sec:Wtagallvars}

Figures~\ref{fig:pt300_comb2D},~\ref{fig:pt500_comb2D}
and~\ref{fig:pt1000_comb2D}  report the background rejection
achieved by a combination of all the variables considered into a
single BDT discriminant. In all cases, the rejection
power of this ``all variables'' BDT is significantly larger than the
best two-variable combination. This indicates
that, beyond the best two-variable combination, there is still
significant complementary information available in the remaining
observables  to improve the discrimination of signal and
background. How much complementary information is available appears to
be \pt dependent. In the lower \pt = 300-400 and 500-600 \GeV bins, the
background rejection of the ``all variables'' combination is a factor
$\sim 1.5$ greater than the best two-variable combination, but in the
highest \pt bin it is a factor $\sim 2.5$ greater. 

The final column in Figures ~\ref{fig:pt300_comb2D},~\ref{fig:pt500_comb2D}
and~\ref{fig:pt1000_comb2D} allows us to further explore the all variables
performance relative to the pair-wise performance. It shows the background rejection for 
three-variable BDT combinations of $m_{\rm sd}^{\beta=2} + C_2^{\beta=1} +
X$, where $X$ is the variable on the y-axis. For jets with $R=0.4$ and
$R=0.8$, the combination $m_{\rm sd}^{\beta=2} + C_2^{\beta=1}$
is (at least close to) the best performant
two-variable combination in every \pt bin considered. For $R=1.2$ this
is not the case, as $C_2^{\beta=1}$ is superseded by
$\tau_{21}^{\beta=1}$ in performance, as discussed earlier. Thus, in
considering the three-variable combination results, it is simplest to focus
on the $R=0.4$ and $R=0.8$ cases. Here we see that, for the lower \pt =
300-400 and 500-600 \GeV bins, adding the third variable to the best
two-variable combination brings us to within $\sim 15\%$ of the ``all
variables'' background rejection. However, in the highest \pt = 1.0-1.1 \TeV bin, whilst adding the third
variable does improve the performance considerably, we are still $\sim
40\%$ from the observed ``all variables'' background rejection, and
clearly adding a fourth or maybe even fifth variable would bring
considerable gains. In terms of which variable offers the best
improvement when added to the $m_{\rm sd}^{\beta=2} + C_2^{\beta=1}$
combination, it is hard to see an obvious pattern; the best third
variable changes depending on the \pt and $R$ considered.

It appears that there is a rich and
complex structure in terms of the degree to which the discriminatory
information provided by the set of variables considered overlaps, with
the degree of overlap apparently decreasing at higher \pt. This
suggests that in all \pt ranges, but especially at higher \pt, there
are substantial performance gains to be made by designing a more
complex multivariate $W$ tagger.

\subsection{Conclusions}

We have studied the performance, in terms of the separation of a
hadronically decaying $W$ boson from a gluon-initiated jet
background, of a number of groomed jet masses, substructure variables,
and BDT combinations of the above. We have used this to gain insight into
 how the discriminatory information contained in the
variables overlaps, and how this complementarity between the variables
changes with jet \pt and \antikt distance parameter $R$. 

In terms of the performance of individual variables, we find that, in
agreement with other studies~\cite{ATL-PHYS-PUB-2014-004}, the groomed masses generally perform best, 
with a background rejection
power that increases with larger \pt, but which is more consistent
with respect to changes in $R$. We have explained the dependence of the
groomed mass performance on \pt and $R$ using the understanding of the QCD mass distribution
developed in Section~\ref{sec:qg_mass}. Conversely, the performance of other
substructure variables, such as $C_2^{\beta=1}$ and
$\tau_{21}^{\beta=1}$, is more susceptible to changes in radius, with
background rejection power decreasing with increasing $R$. This is due
to the inherent sensitivity of these observables to soft, wide angle radiation.

The best two-variable performance is obtained by combining a groomed
mass with a substructure variable. Which particular substructure
variable works best in combination strongly depends on \pT and
$R$. The variable $C_2^{\beta=1}$ offers significant complementarity to groomed mass
for the smaller values of $R$ investigated ($R=0.4$ and 0.8), owing to the small degree of correlation between the
variables. 
However, the sensitivity of $C_2^{\beta=1}$ to soft,
wide-angle radiation leads to worse discrimination power at $R=1.2$,
where $\tau_{21}^{\beta=1}$ performs better in combination. The best
two-variable performance in each \pt bin examined is obtained for
$C_2^{\beta=1}$ in combination with a groomed mass, using $R=0.8$,
with a performance that is better at higher \pt.
Our
studies also demonstrate the potential for enhancing discrimination by
combining groomed and ungroomed mass information, although the use of
ungroomed mass in this may be limited in practice by the presence of
pile-up that is not considered in these studies.

By examining the performance of a BDT combination of all  variables
considered, it is clear that there are potentially substantial performance gains to be made by designing a more
complex multivariate $W$ tagger, especially at higher \pt.

%% file: toptagging.tex
In this section, we investigate the identification of boosted top quarks using jet substructure. Boosted top quarks result in large-radius jets with complex substructure, containing a $b$-subjet and a boosted $W$. As a consequence of the many kinematic differences between top and QCD jets, top taggers are typically complex, with a couple of input parameters necessary for any given algorithm. We study the variation in performance of top tagging techniques with respect to jet \pt and $R$, re-optimizing the tagger inputs for each kinematic range and jet radius considered. We also investigate the effects of combining dedicated top tagging algorithms with other jet substructure variables, giving insight into the correlations among top-tagging variables.

\subsection{Methodology}\label{sec:topmethod}

We use the top quark MC samples for each bin described in Section \ref{sec:top-samples}. The analysis  relies on \textsc{FastJet} 3.0.3 for jet clustering and
calculation of jet substructure variables. Jets are clustered using the \antikt algorithm, and only the leading jet is used in each analysis. To ensure similar $\pt$ spectra in each bin an upper and lower $\pt$ cut are applied to each sample after jet clustering. The bins in leading jet $\pt$
 for top tagging are 600-700 \GeV, 1-1.1 \TeV, and
1.5-1.6 TeV. Jets are clustered with radii $R=0.4$, 0.8, and 1.2; $R=0.4$ jets are only studied in the 1.5-1.6 TeV bin
because the top decay products are all contained within an $R=0.4$ jet for top quarks with this boost.

We study a number of top-tagging strategies, which can be divided into two distinct categories. In the first category are dedicated top-tagging algorithms, which aim to directly reconstruct the top and $W$ candidates in the top decay. In particular, we study:
\begin{enumerate}
\item HEPTopTagger
\item Johns Hopkins Tagger (JH)
\item Trimming with $W$-identification
\item Pruning with $W$-identification
\end{enumerate}
as described in Section~\ref{sec:taggers}. 
In the case of the HepTopTagger and JH tagger, the algorithms produce three output variables ($\topmass$, $\wmass$ and helicity angle) that can be used to discriminate top jets from QCD. The trimming and pruning algorithms as used here produce two outputs, $\topmass$ and $\wmass$. All of the above taggers and groomers incorporate a step to remove contributions from the underlying event and other soft radiation to the reconstructed $\topmass$ and $\wmass$, and also explicitly rejects jets that do not meet basic selection criteria, as explained in detail in Section~\ref{sec:taggers}.

In the second category are individual jet substructure variables that are sensitive to the radiation pattern within the jet, which we refer to as ``jet-shape variables''. While the most sensitive top-tagging variables are typically sensitive to three-pronged radiation, we also consider variables sensitive to two-pronged radiation in the limit where the $W$ is very boosted and its subjets overlap. The variables we consider are:
\begin{itemize}
\item The ungroomed jet mass.
\item $N$-subjettiness ratios \tautwoone and \tauthreetwo, using the ``winner-takes-all'' axes definition.
\item 2-point energy correlation function ratios $C_2^{\beta=1}$ and $C_3^{\beta=1}$.
\item The pruned Qjet mass volatility, $\Gamma_{\rm Qjet}$.
\end{itemize}
Several of these variables were also considered earlier for $q/g$-tagging and $W$-tagging.

To study the correlations amongst the above substructure variables and tagging algorithms,
we combine the relevant tagger output variables and/or jet shapes into a BDT\footnote{Similar studies were recently performed for the HepTopTagger in~\cite{Anders:2013oga,Kasieczka:2015jma}, in the context of trying to improve the tagger by combining it's outputs with $N$-subjettiness.}, as described in Section~\ref{sec:multivariate}. 
Additionally, because each tagger has two input parameters, we scan over reasonable values of the input parameters to determine the optimal value that gives the largest background rejection for each top tagging signal efficiency. This allows a direct comparison of the optimized version of each tagger. The input parameter values scanned for the various algorithms are:
\begin{itemize}
\item {\bf HEPTopTagger:} $m\in[30,100]$ \GeV, $\mu\in[0.5,1]$
\item {\bf JH Tagger:} $\delta_p\in[0.02,0.15]$, $\delta_R\in[0.07,0.2]$
\item {\bf Trimming:} $f_{\rm cut}\in[0.02,0.14]$, $R_{\rm trim}\in[0.1,0.5]$
\item {\bf Pruning:} $z_{\rm cut}\in[0.02,0.14]$, $R_{\rm cut}\in[0.1,0.6]$
\end{itemize}
We also investigate the degradation in performance of the top-tagging variables when moving away from the optimal parameter choice.

\subsection{Single Variable Performance}\label{sec:single_variable}
We begin by investigating the behaviour of individual jet substructure variables. Because of the rich, three-pronged structure of the top decay, it is expected that combinations of masses and jet shapes will far outperform single variables in identifying boosted tops. However, a study of the top-tagging performance of single variables facilitates a direct comparison with the $W$ tagging results in Section \ref{sec:wtagging}, and also allows a straightforward examination of the performance of each variable for different $\pt$ and jet radius.

\begin{figure*}
\centering
\subfigure[Jet shapes]{\includegraphics[width=0.49\textwidth]{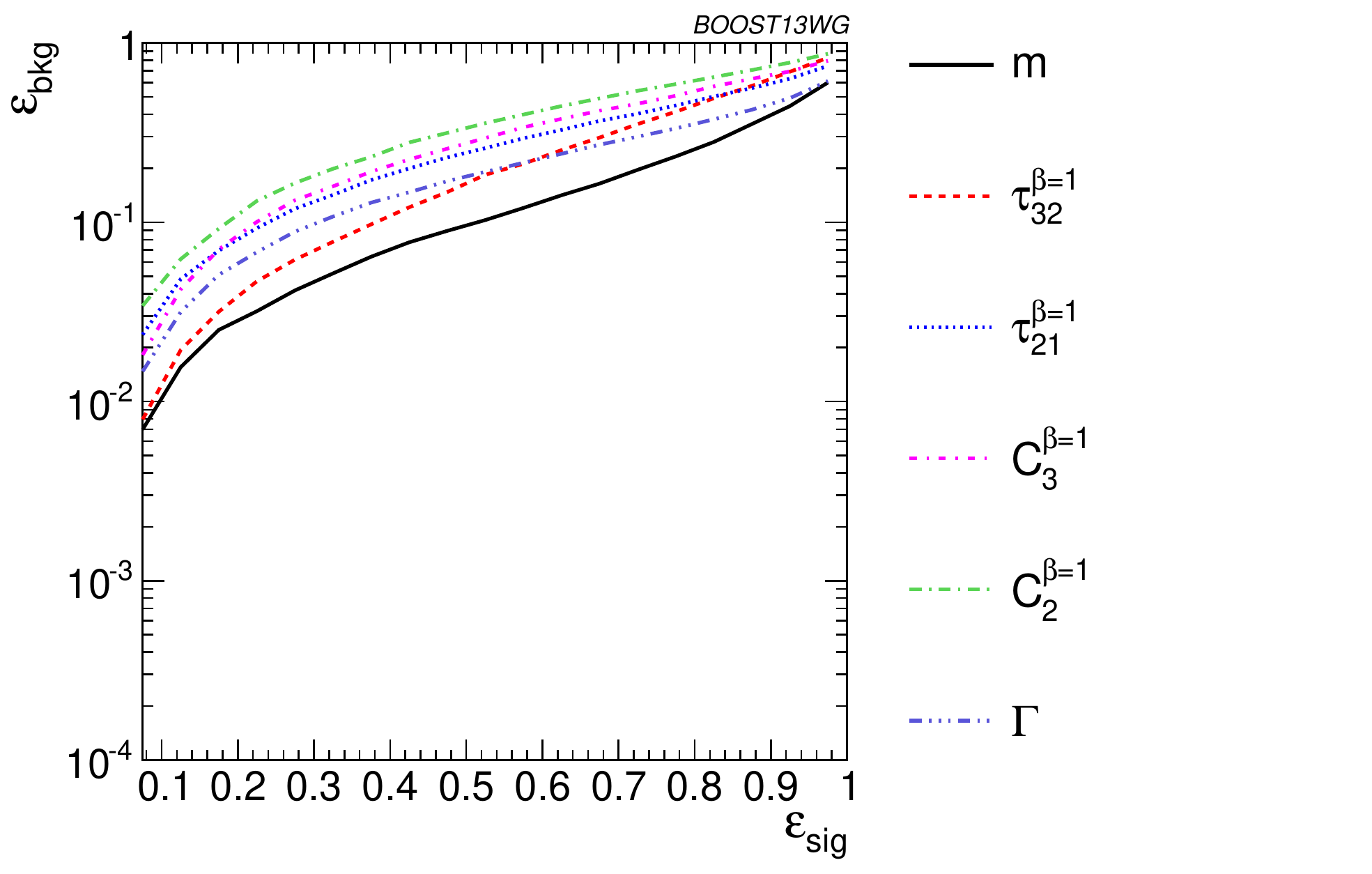}\label{fig:single_variable_ROC_shape}}
\subfigure[top mass]{\includegraphics[width=0.49\textwidth]{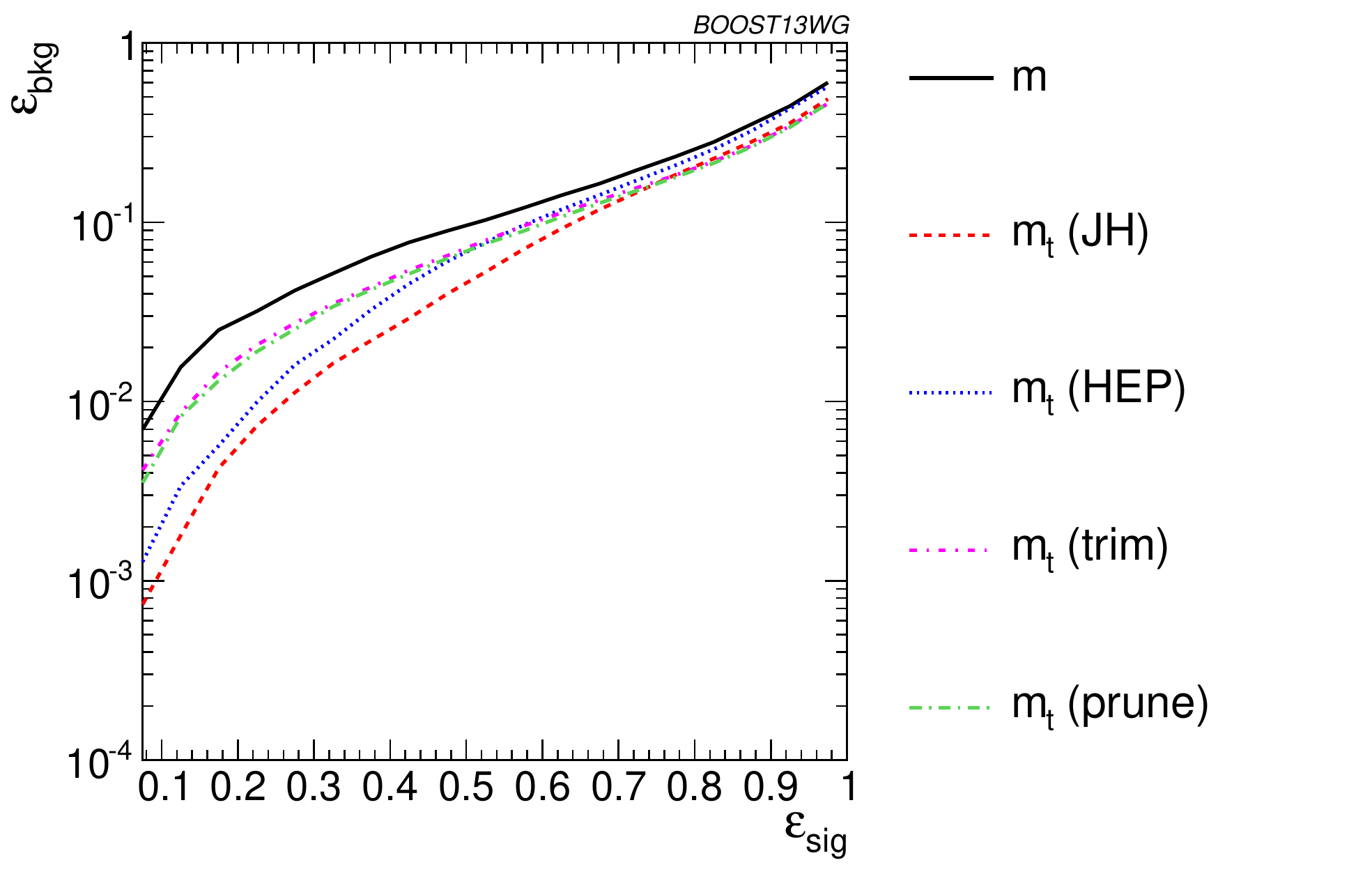}\label{fig:single_variable_ROC_topmass}}
\caption{Comparison of single-variable top-tagging performance in the $\pt= 1-1.1$ GeV bin using the anti-\kT, R=0.8 algorithm.}
\label{fig:single_variable_ROC}
\end{figure*}

Top-tagging performance is quantified using ROC curves. Figure~\ref{fig:single_variable_ROC} shows the ROC curves for each of the top-tagging variables, with the bare (ungroomed) jet mass also plotted for comparison. The jet-shape variables all perform substantially worse than ungroomed jet mass; this is in contrast with $W$ tagging, for which several variables are competitive with or perform better than ungroomed jet mass (see, for example, Figures~\ref{fig:pt500_comb2D_08}, \ref{fig:pt1000_comb2D_04} and~\ref{fig:pt1000_comb2D_08}).
To understand why this is the case, consider $N$-subjettiness:~the $W$ is two-pronged and the top is three-pronged, and so we expect $\tau_{21}$ and $\tau_{32}$ to be the best-performant $N$-subjettiness ratios, respectively. However, a cut selection small values of $\tau_{21}$ necessarily selects for events with large $\tau_1$, which is strongly correlated with jet mass, up to exponentially suppressed contributions. Therefore, $\tau_{21}$ applied to $W$-tagging indirectly incorporates some information about the jet mass in addition to shape information. By contrast, $\tau_{32}$ applied to top tagging does not include any information on the ungroomed jet mass information.  This likely accounts for why, relative to a cut on ungroomed mass, $\tau_{32}$ for top tagging performs substantially worse than $\tau_{21}$ for $W$-tagging.

\begin{figure*}
\centering
\subfigure[JH, $R=0.4$]{\includegraphics[width=0.245\textwidth]{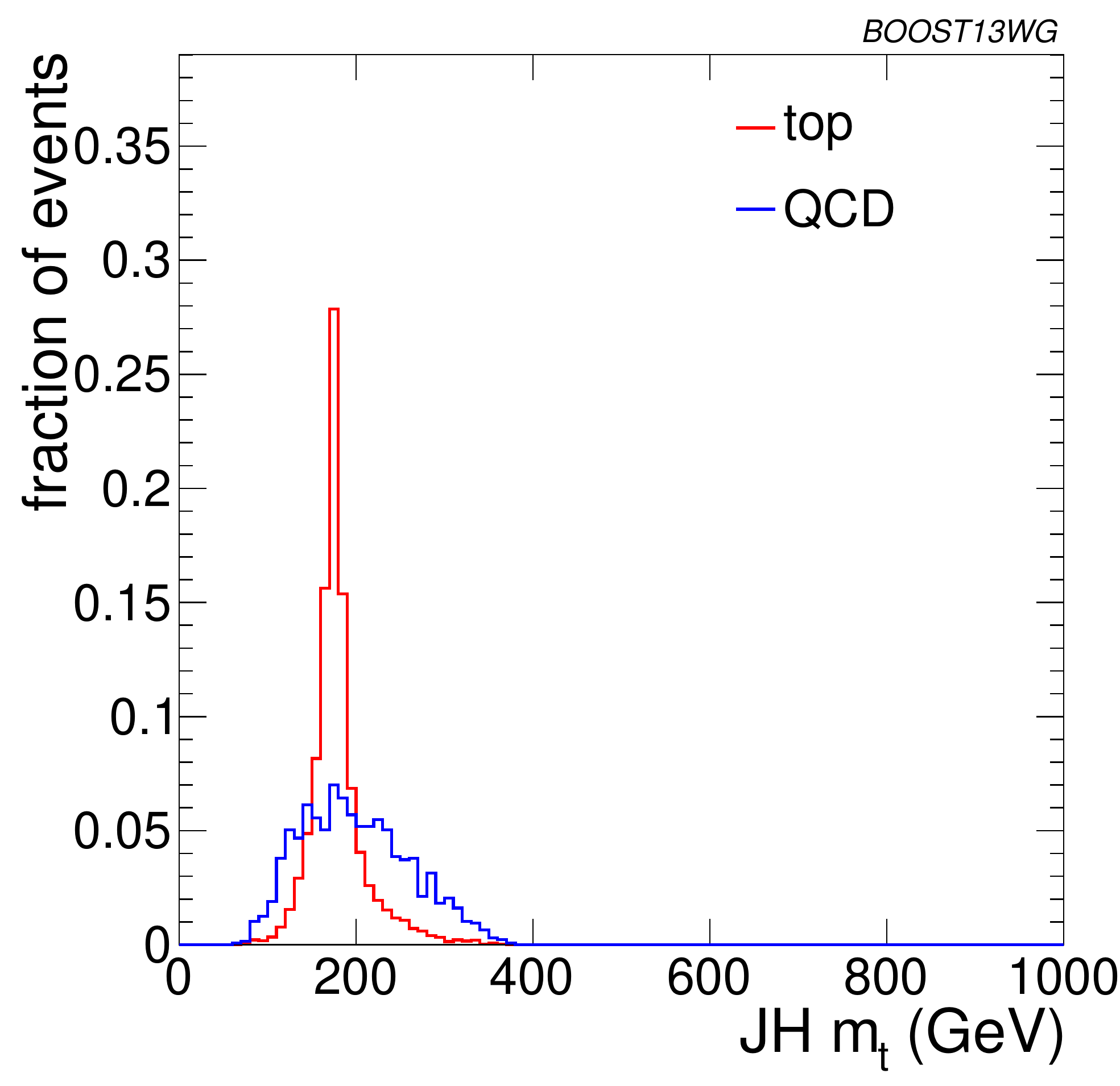}}
\subfigure[HEP, $R=0.4$]{\includegraphics[width=0.245\textwidth]{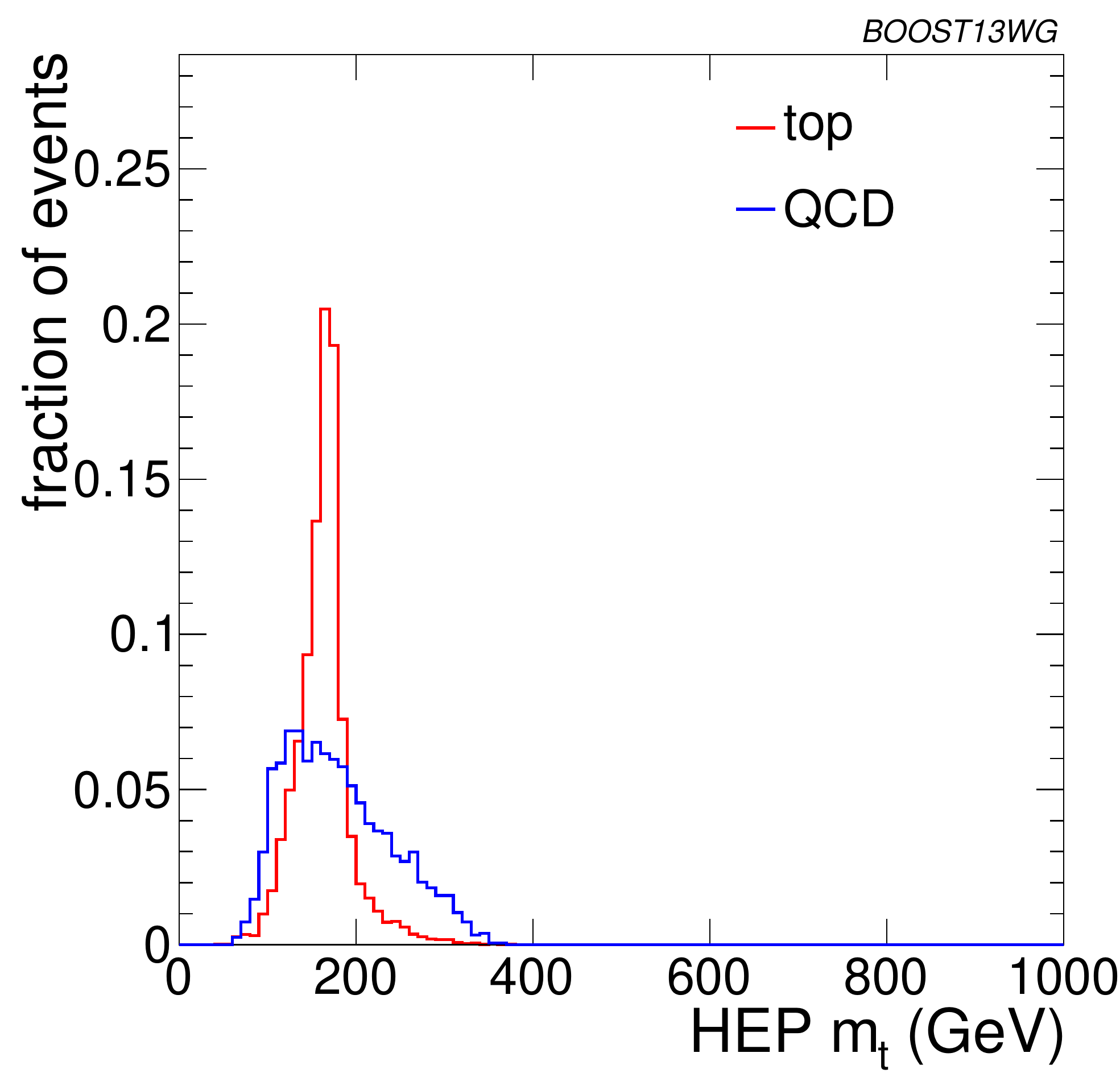}}
\subfigure[JH, $R=1.2$]{\includegraphics[width=0.245\textwidth]{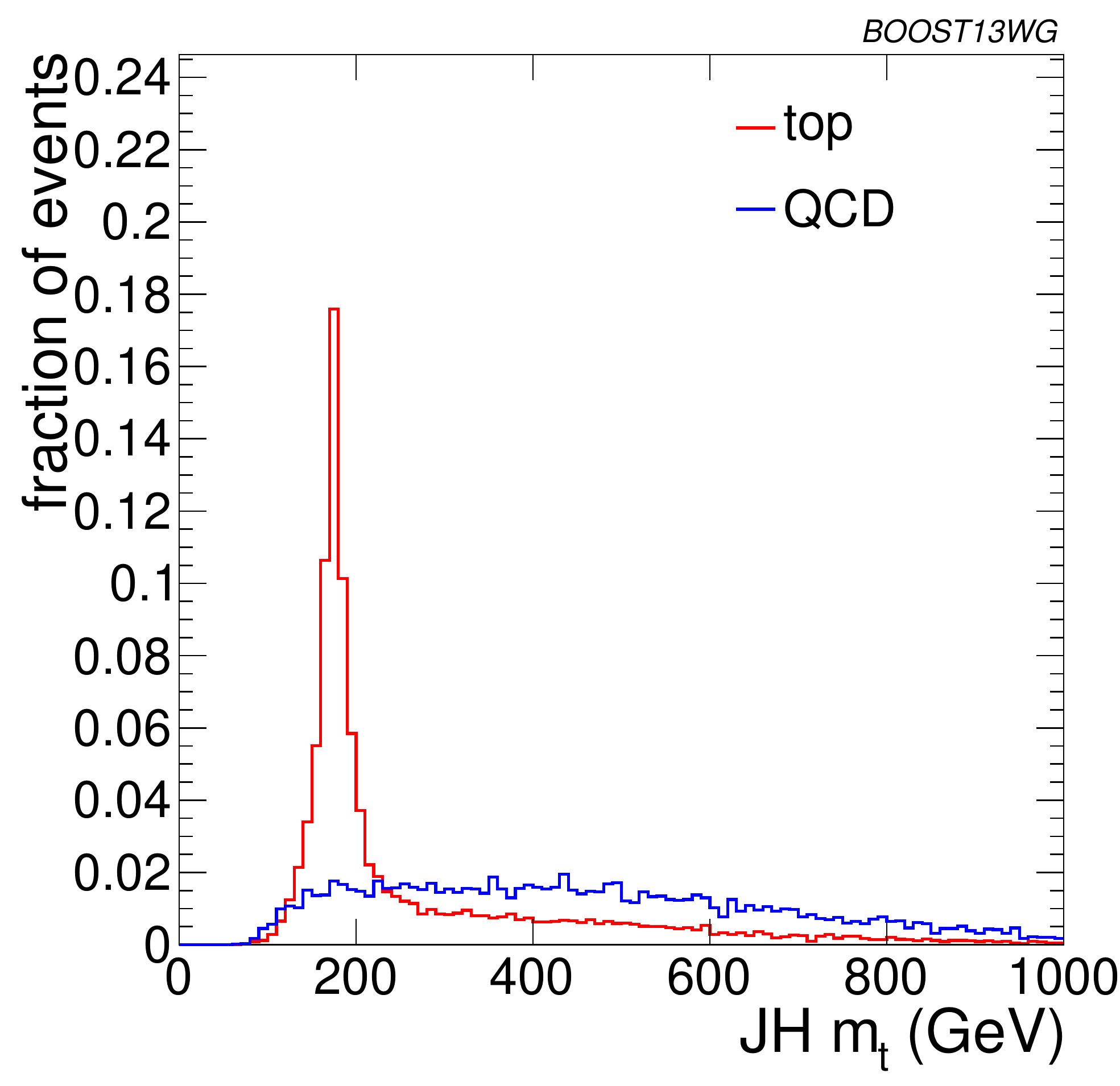}}
\subfigure[HEP, $R=1.2$]{\includegraphics[width=0.245\textwidth]{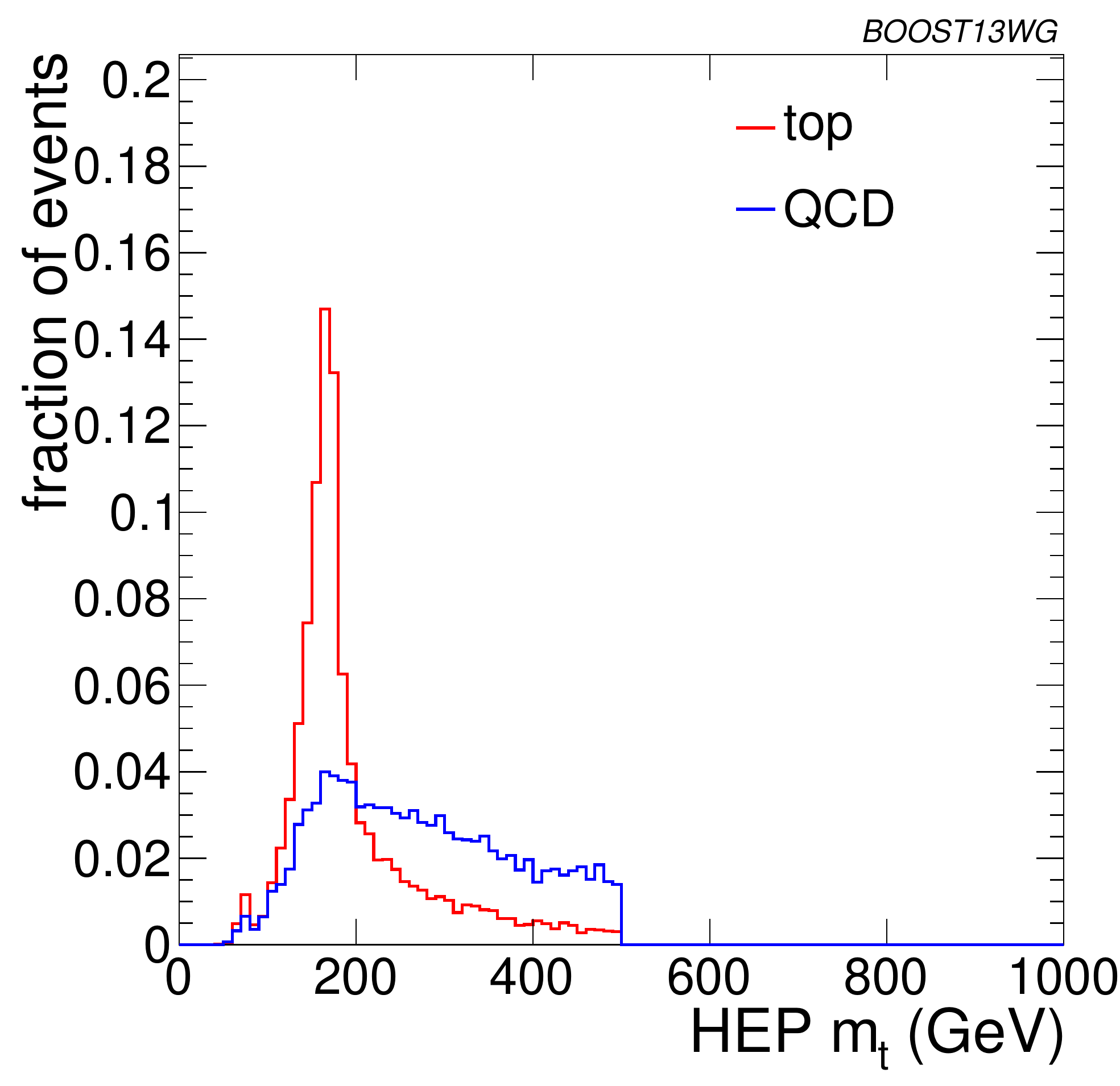}\label{fig:topmass_histogram_HEP_R12}}\\
\subfigure[prune, $R=0.4$]{\includegraphics[width=0.245\textwidth]{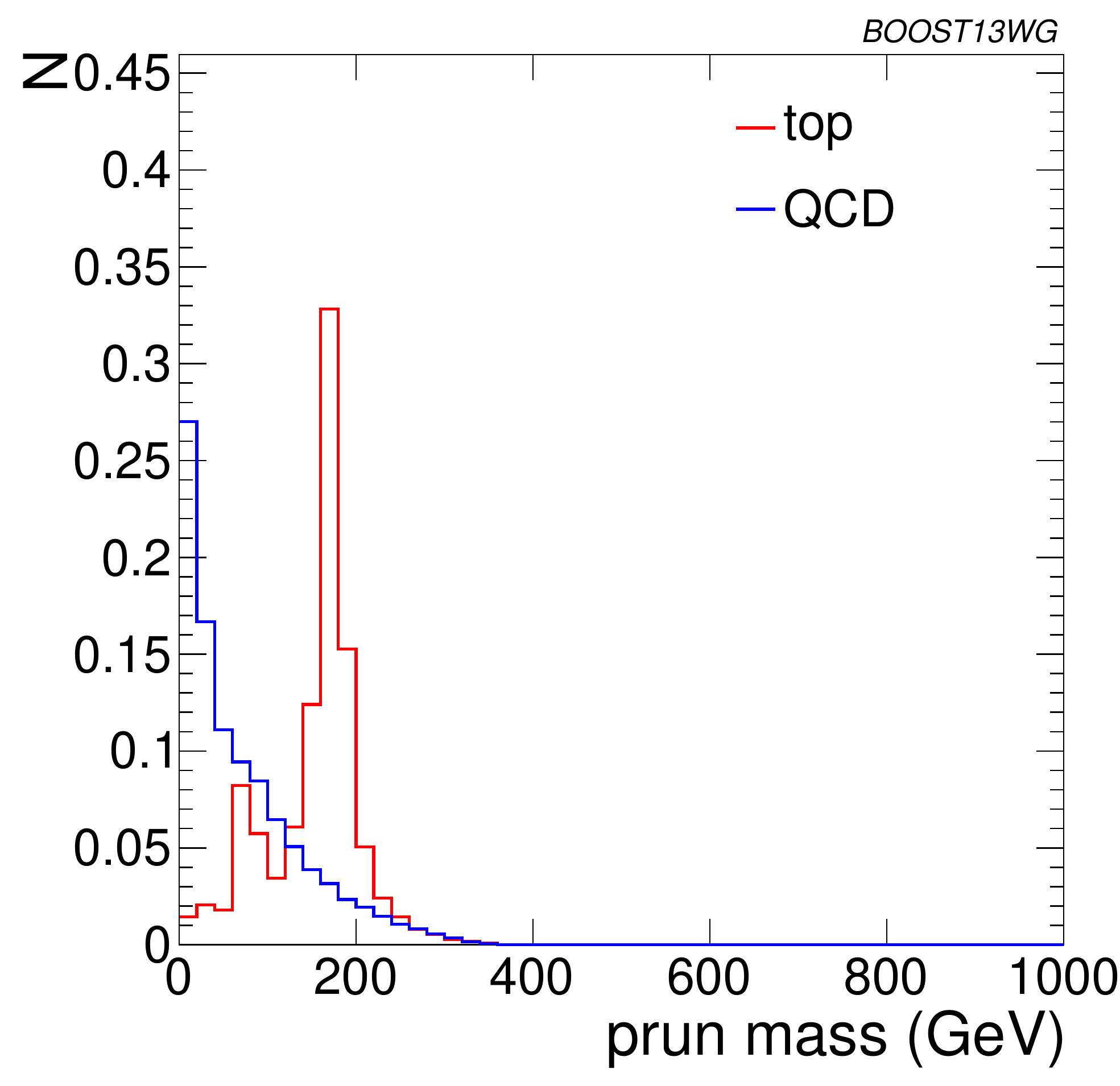}}
\subfigure[trim, $R=0.4$]{\includegraphics[width=0.245\textwidth]{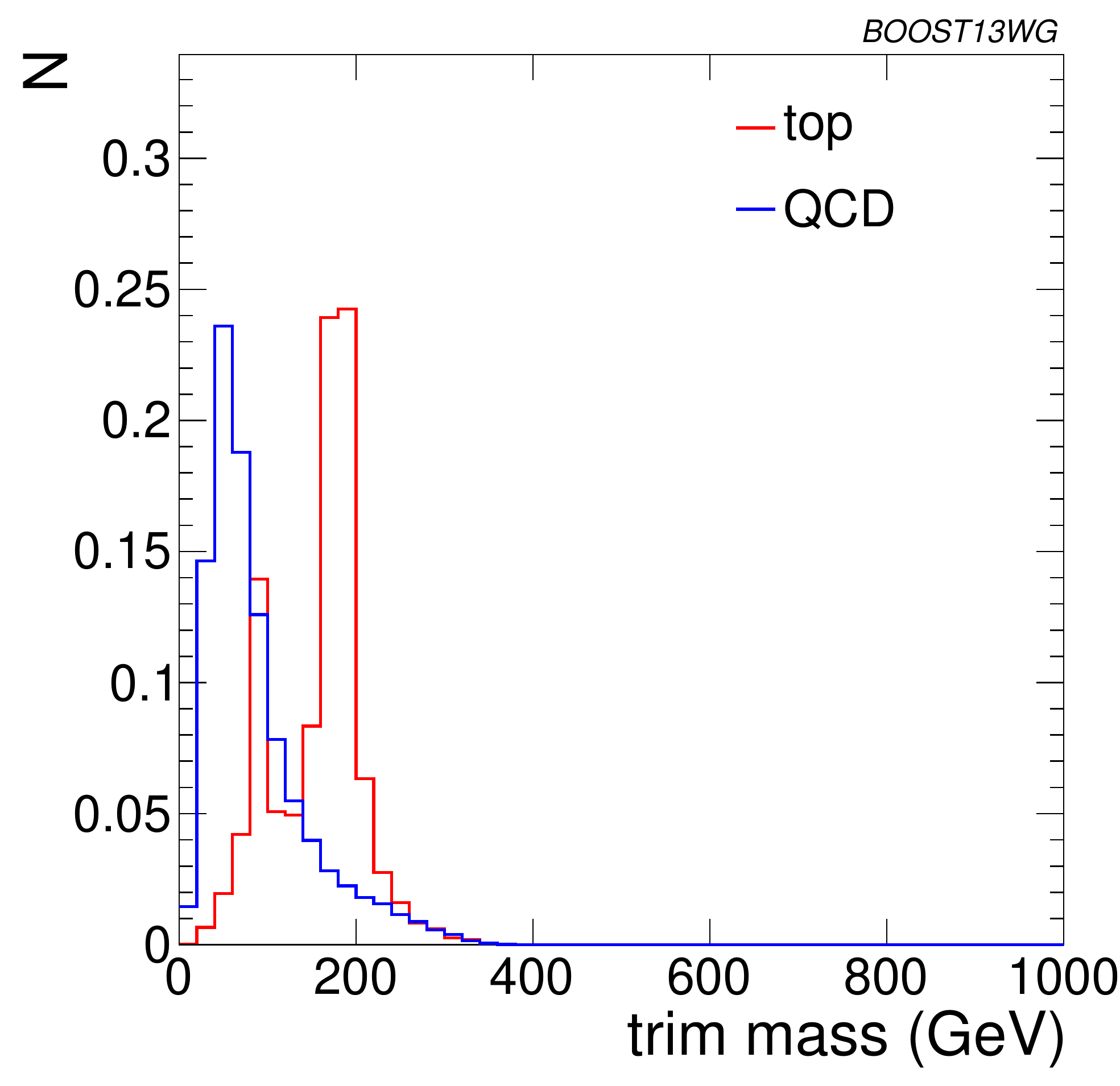}}
\subfigure[prune, $R=1.2$]{\includegraphics[width=0.245\textwidth]{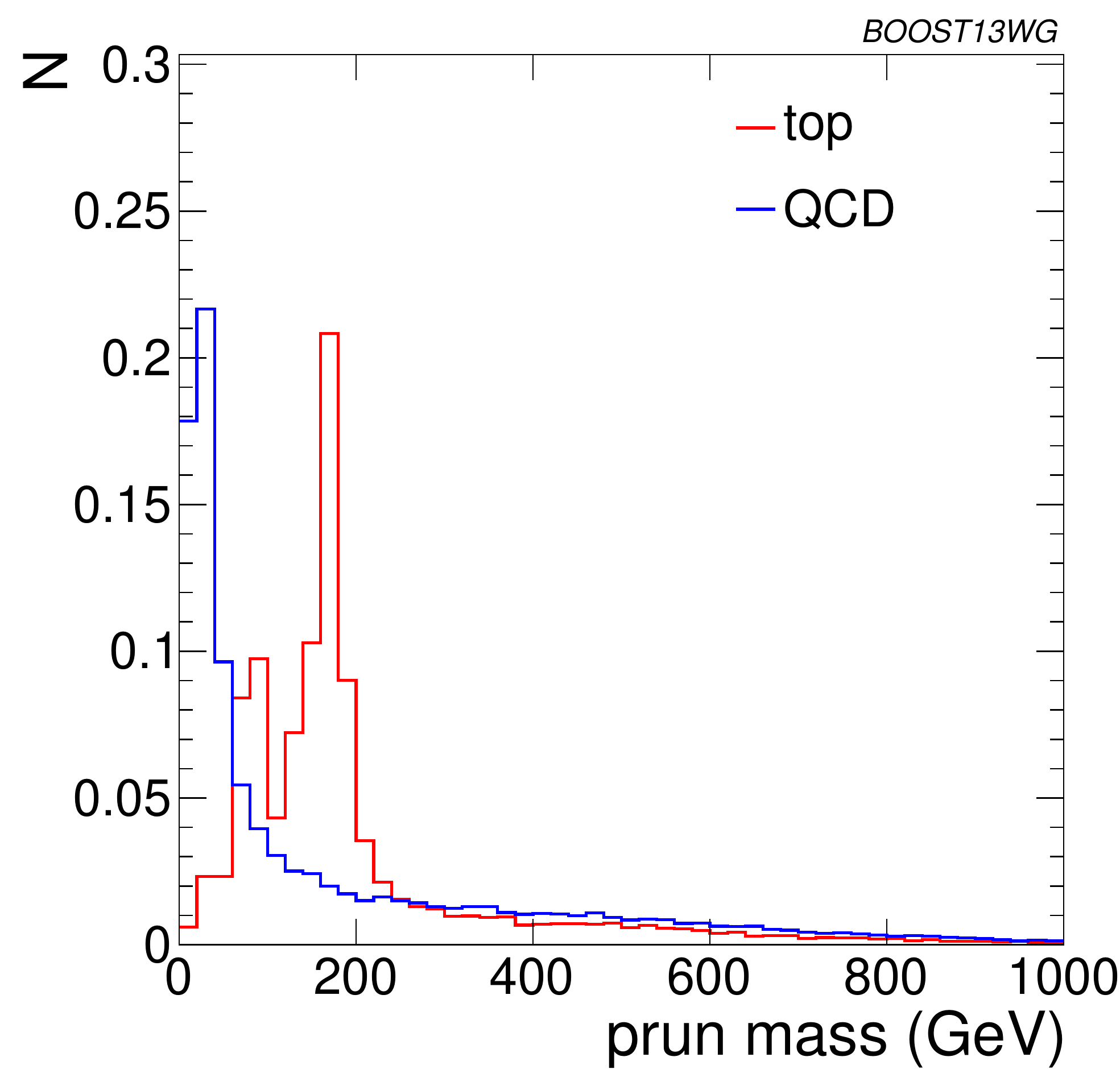}}
\subfigure[trim, $R=1.2$]{\includegraphics[width=0.245\textwidth]{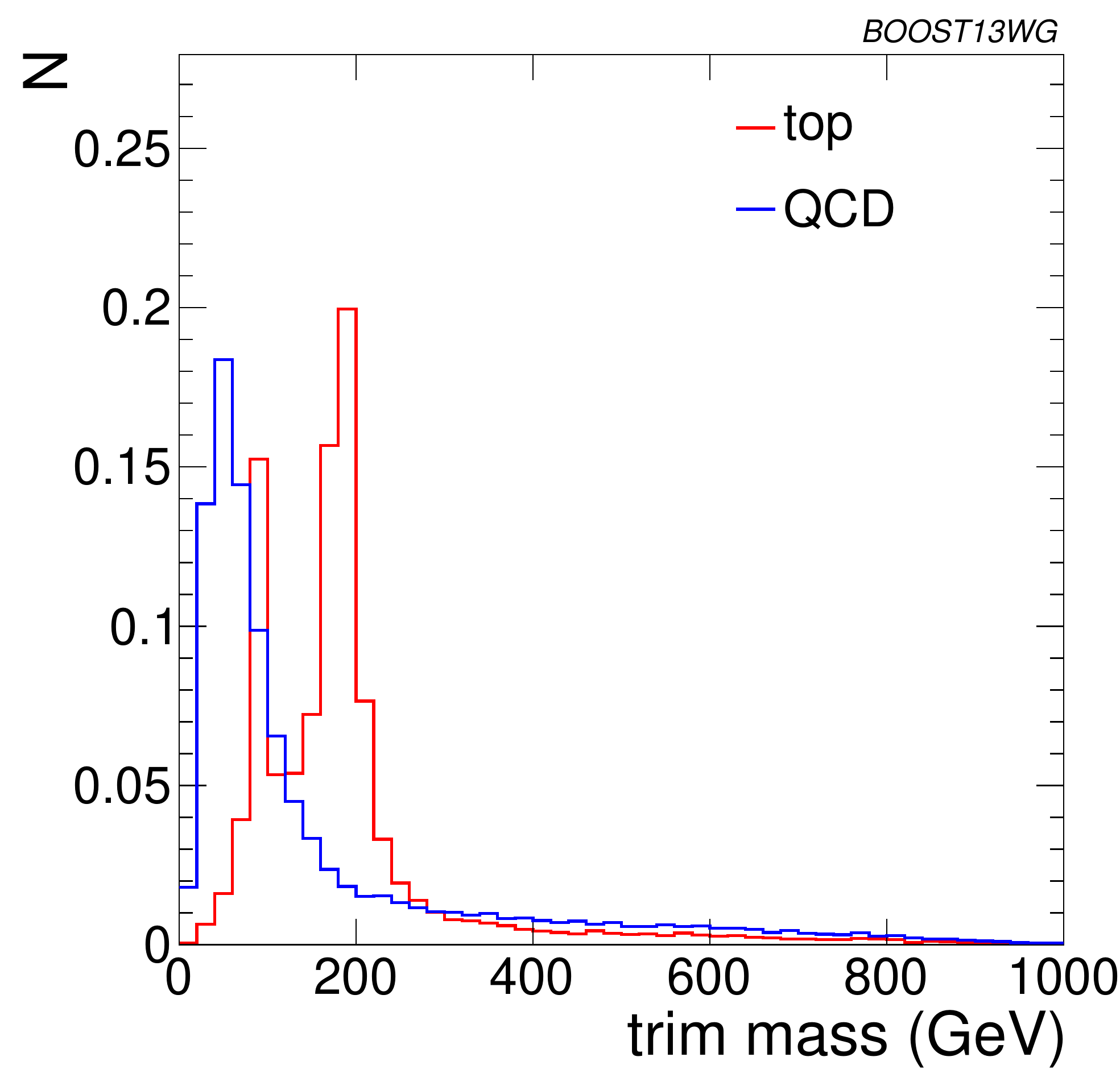}}
\caption{Comparison of top mass reconstruction with the Johns Hopkins (JH), HEPTopTaggers (HEP), pruning, and trimming at different $R$ using the \antikt algorithm in the \pt = 1.5-1.6 \TeV bin. Each histogram is shown for the working point optimized for best performance with $m_t$ in the $0.3$-$0.35$ signal efficiency bin, and is normalized to the fraction of events passing the tagger. In this and subsequent plots, the HEPTopTagger distribution cuts off at 500 \GeV because the tagger fails to tag jets with a larger mass.}
\label{fig:topmass_histogram_HEP_JH}
\end{figure*}

\begin{figure*}
\centering
\subfigure[JH, \pt = 600-700 \GeV]{\includegraphics[width=0.245\textwidth]{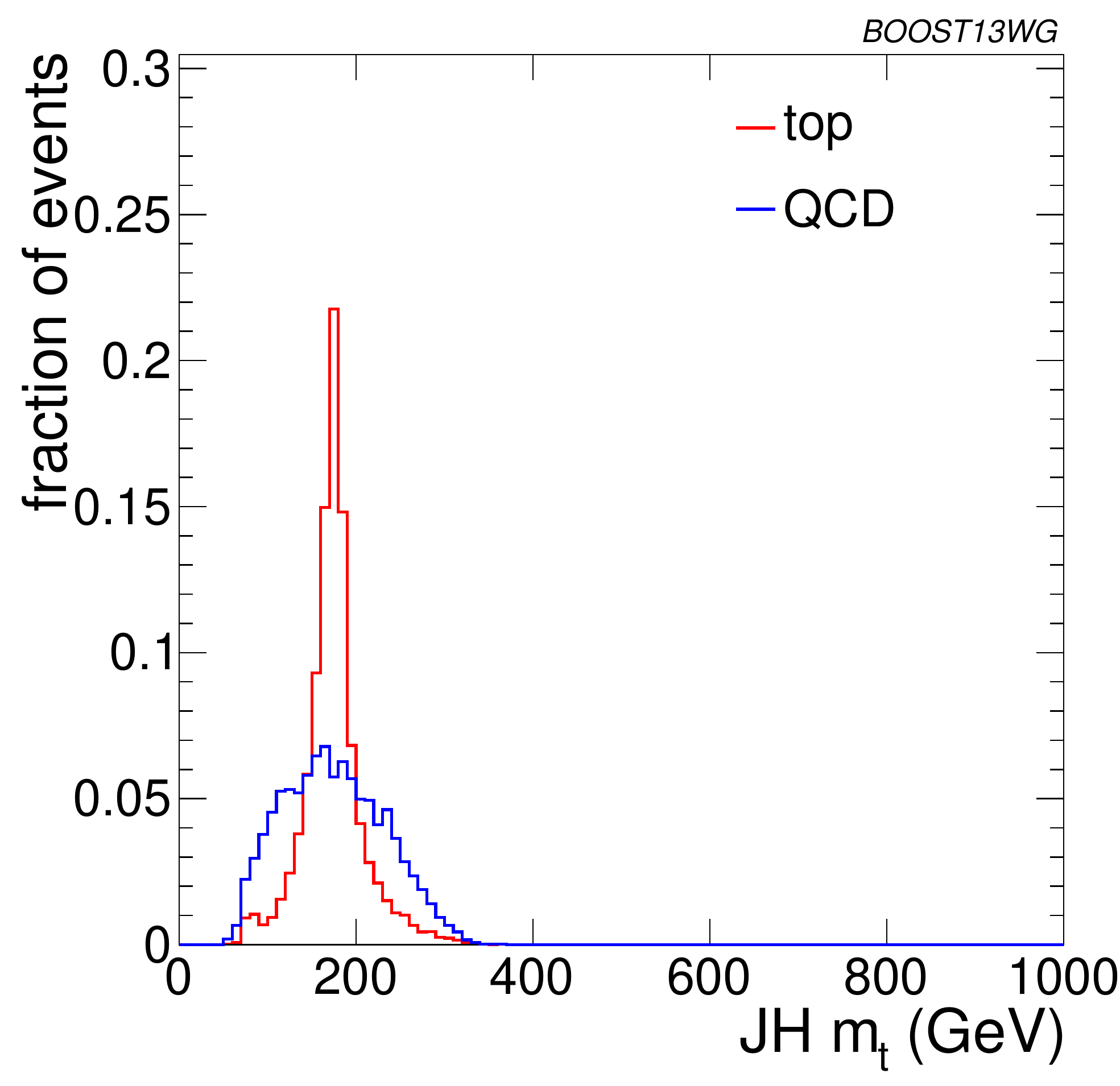}}
\subfigure[HEP, \pt = 600-700 \GeV]{\includegraphics[width=0.245\textwidth]{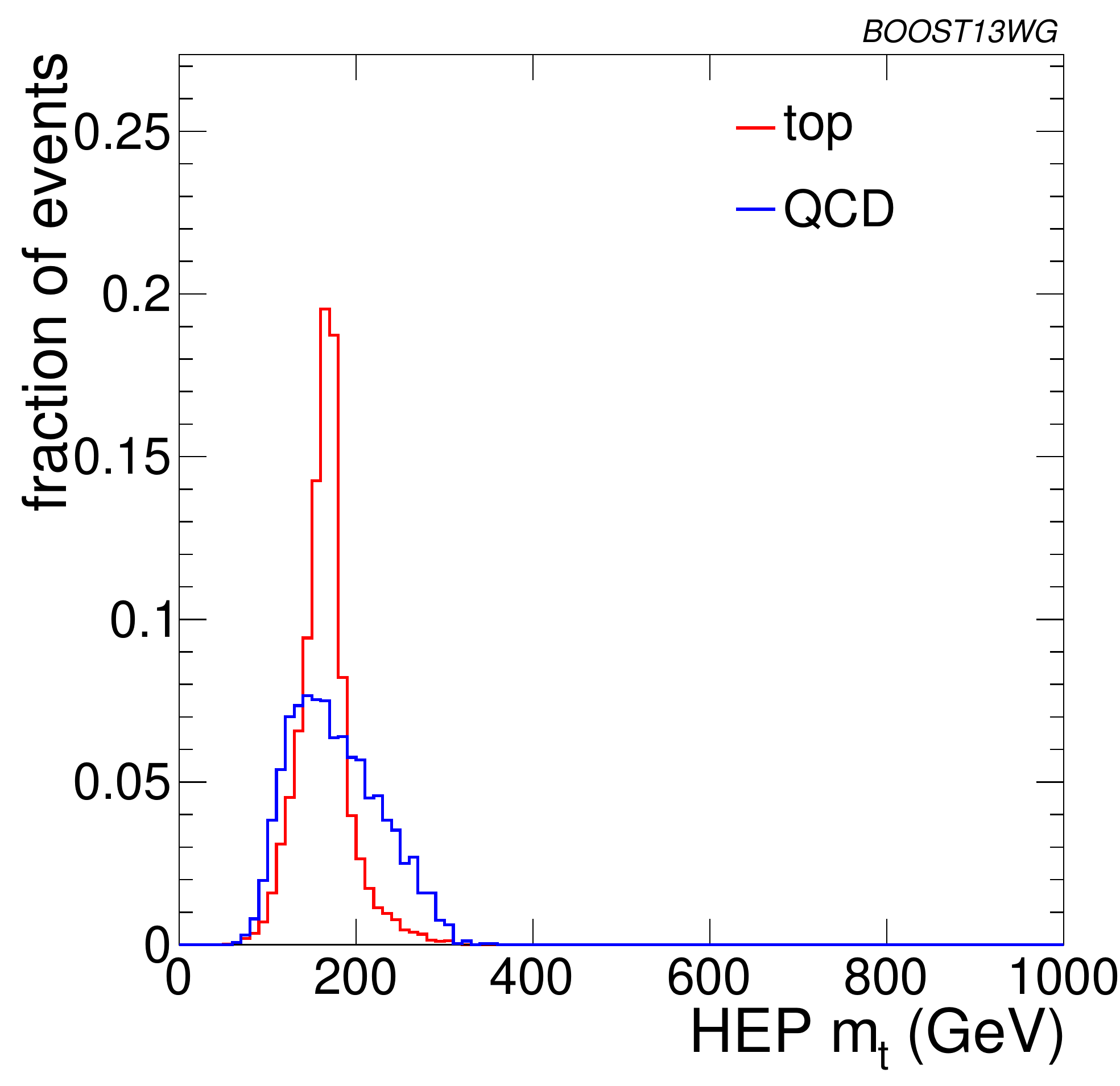}}
\subfigure[JH, \pt = 1.5-1.6 \TeV]{\includegraphics[width=0.245\textwidth]{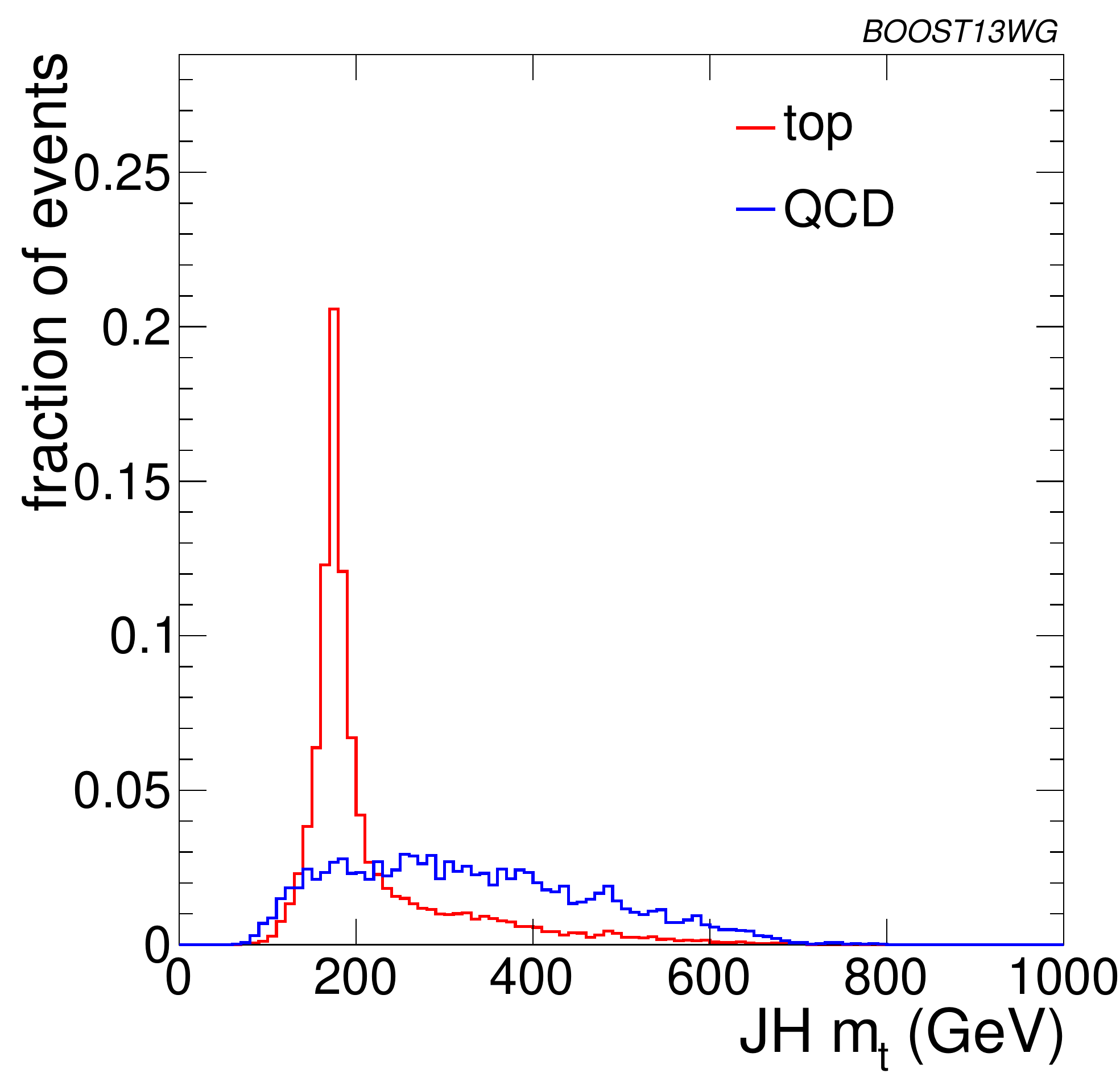}}
\subfigure[HEP, \pt = 1.5-1.6 \TeV]{\includegraphics[width=0.245\textwidth]{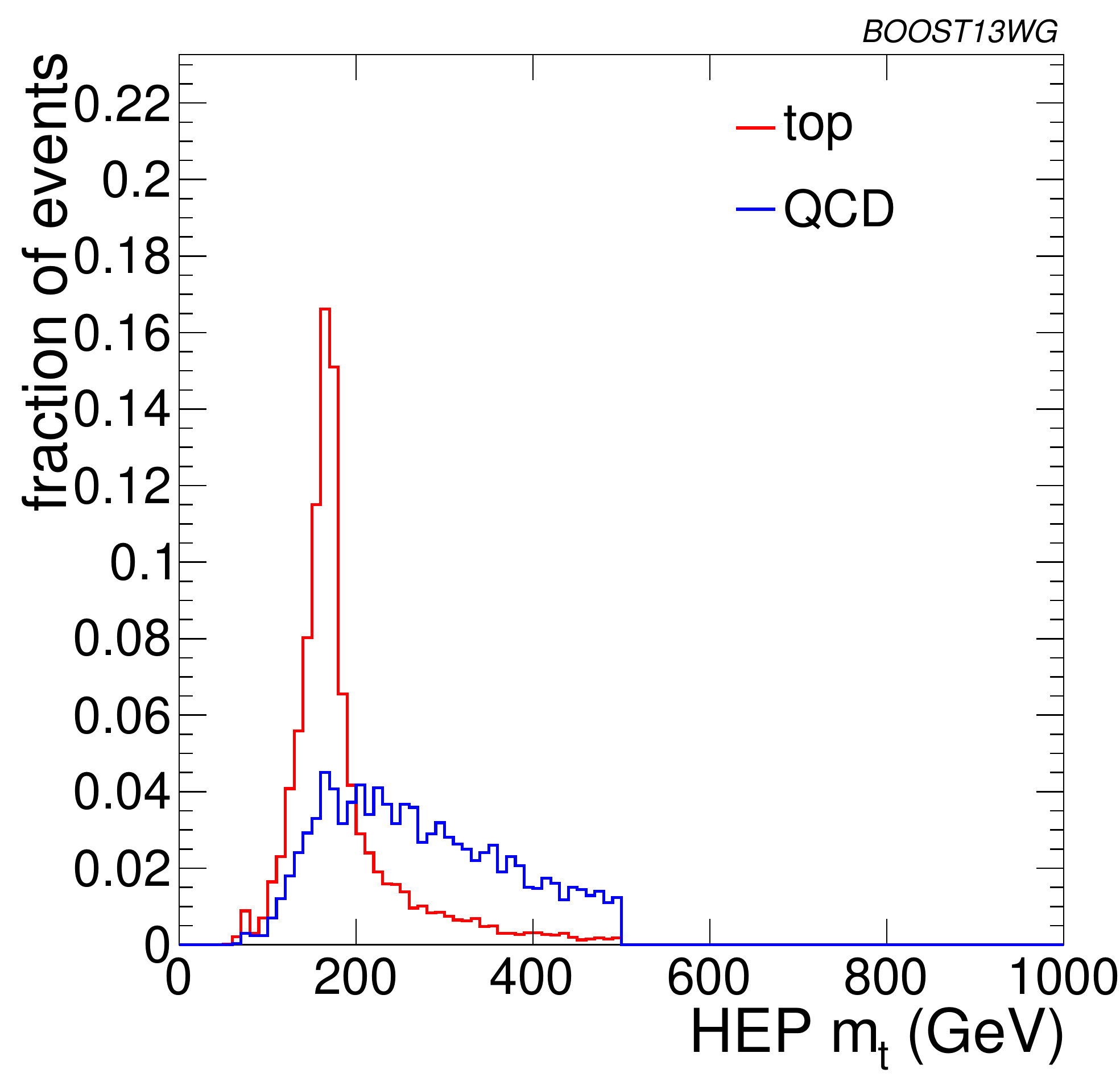}\label{fig:topmass_histogram_HEP_pT15}}\\
\subfigure[prune, \pt = 600-700 \GeV]{\includegraphics[width=0.245\textwidth]{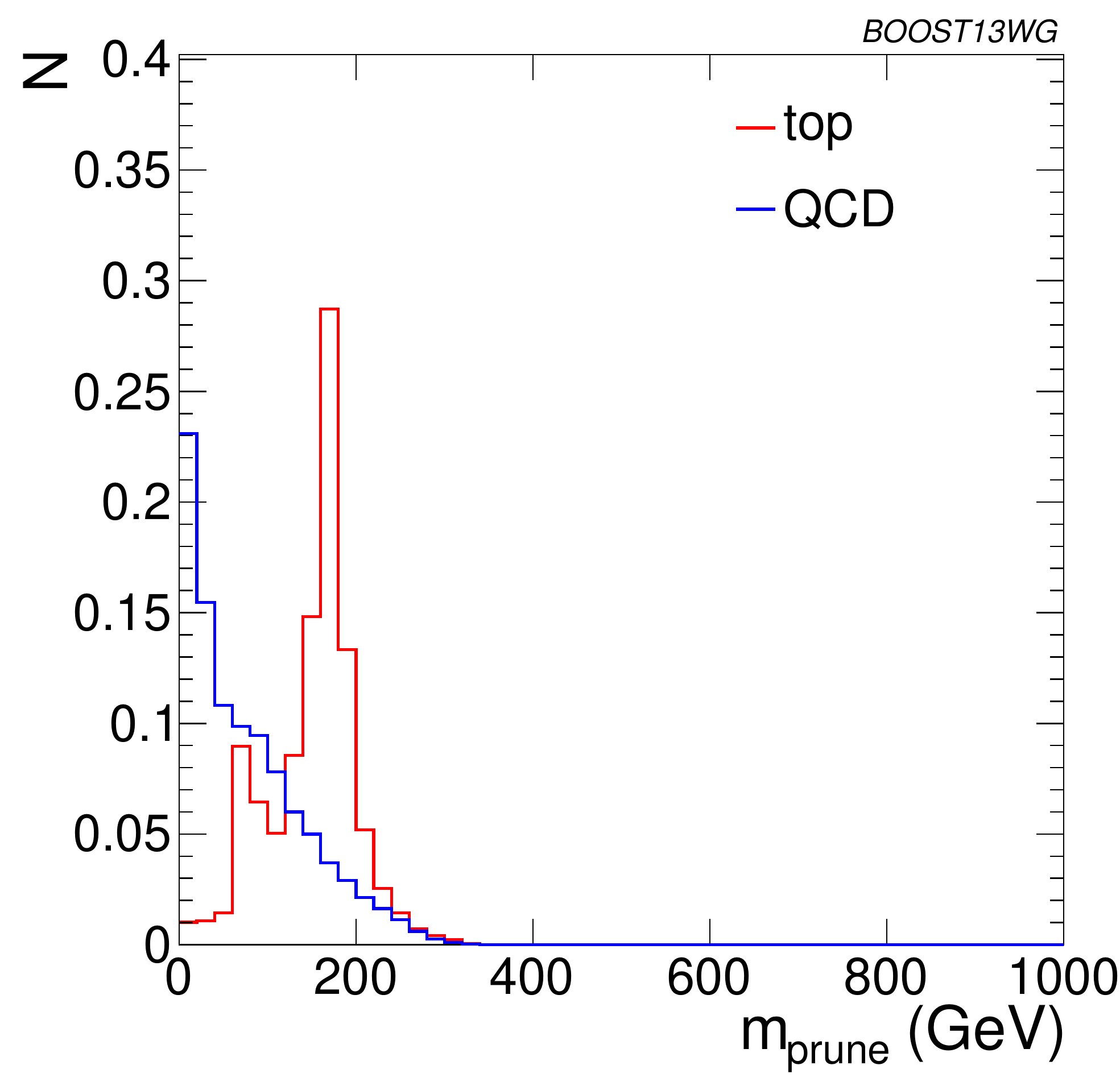}}
\subfigure[trim, \pt = 600-700 \GeV]{\includegraphics[width=0.245\textwidth]{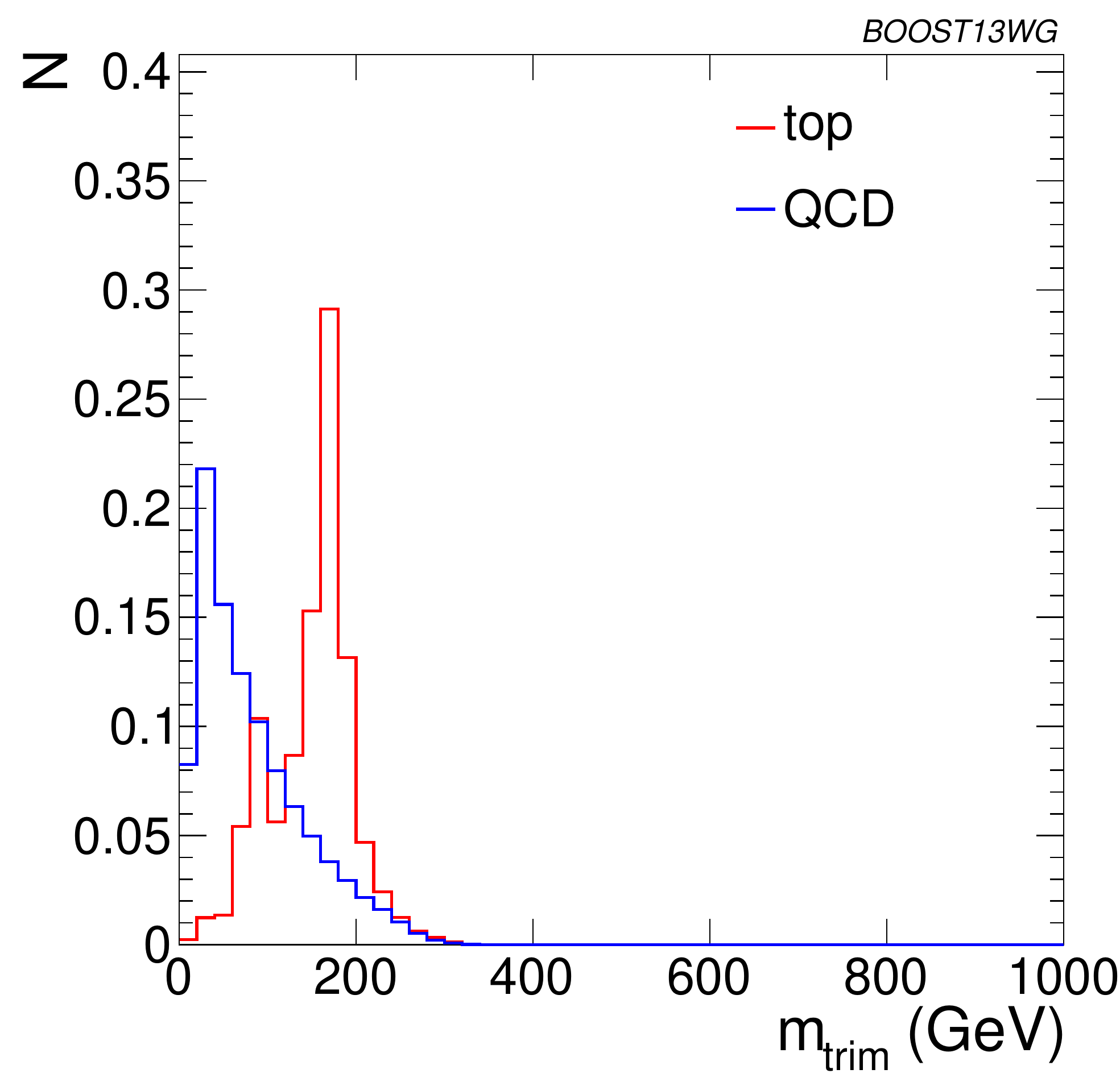}}
\subfigure[prune, \pt = 1.5-1.6 \TeV]{\includegraphics[width=0.245\textwidth]{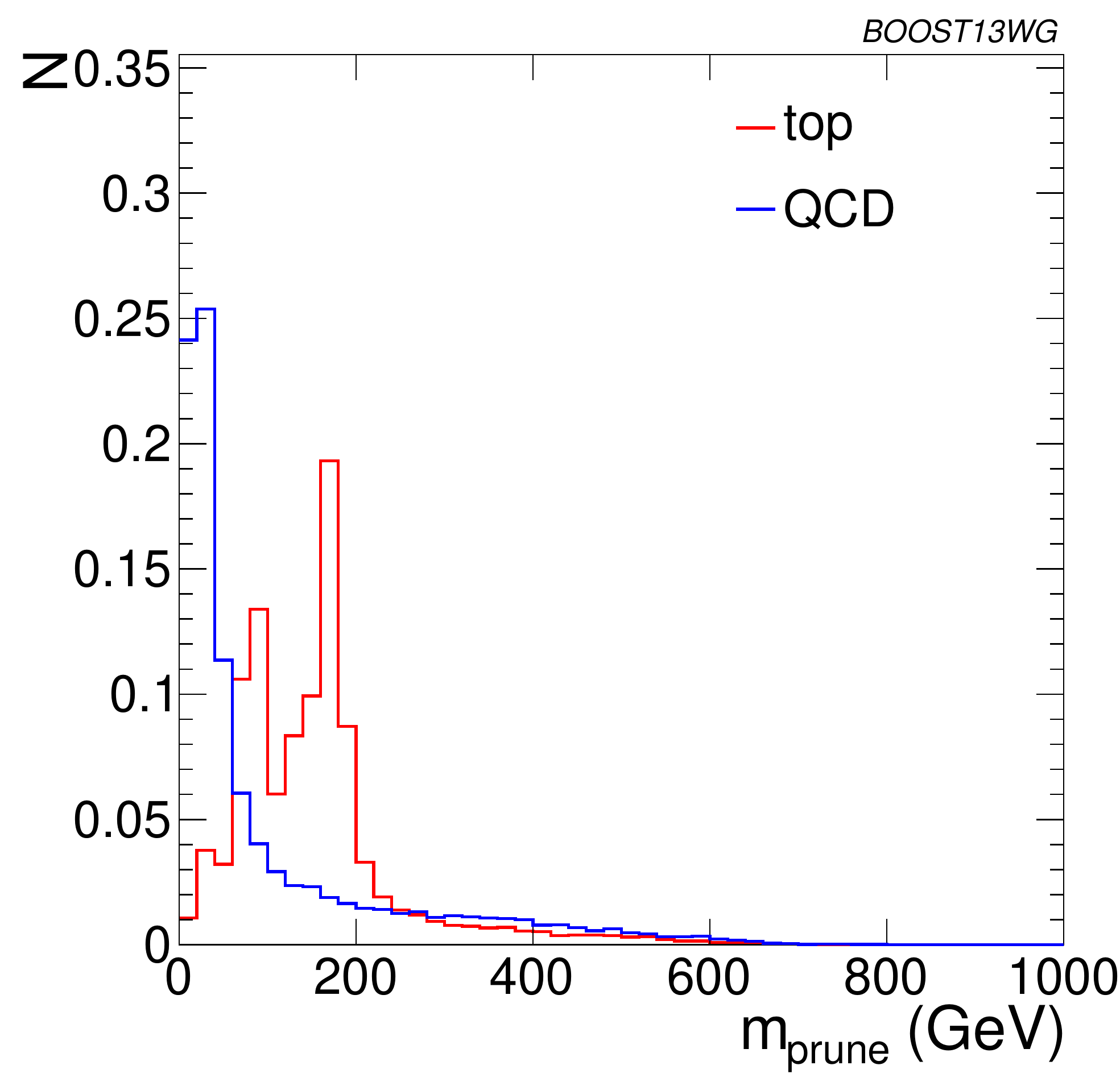}}
\subfigure[trim, \pt = 1.5-1.6 \TeV]{\includegraphics[width=0.245\textwidth]{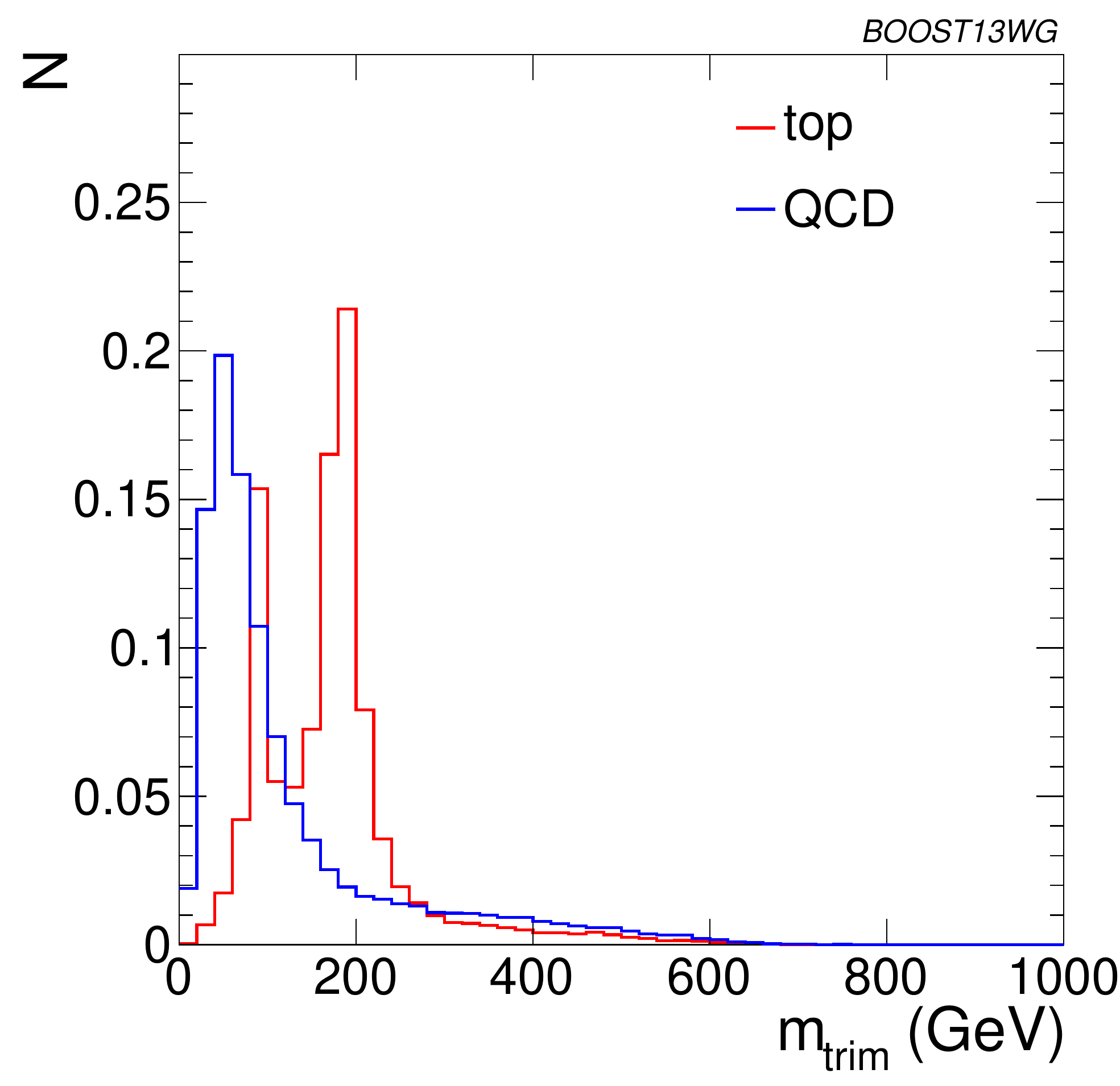}}
\caption{Comparison of top mass reconstruction with the Johns Hopkins (JH), HEPTopTaggers (HEP), pruning, and trimming at different \pt using the \antikt algorithm, $R=0.8$. Each histogram is shown for the working point optimized for best performance with $m_t$ in the $0.3$-$0.35$ signal efficiency bin, and is normalized to the fraction of events passing the tagger.}
\label{fig:topmass_histogram_HEP_JH_pT}
\end{figure*}

\begin{figure*}
\centering
\subfigure[HEPTopTagger $m_t$]{\includegraphics[width=0.48\textwidth]{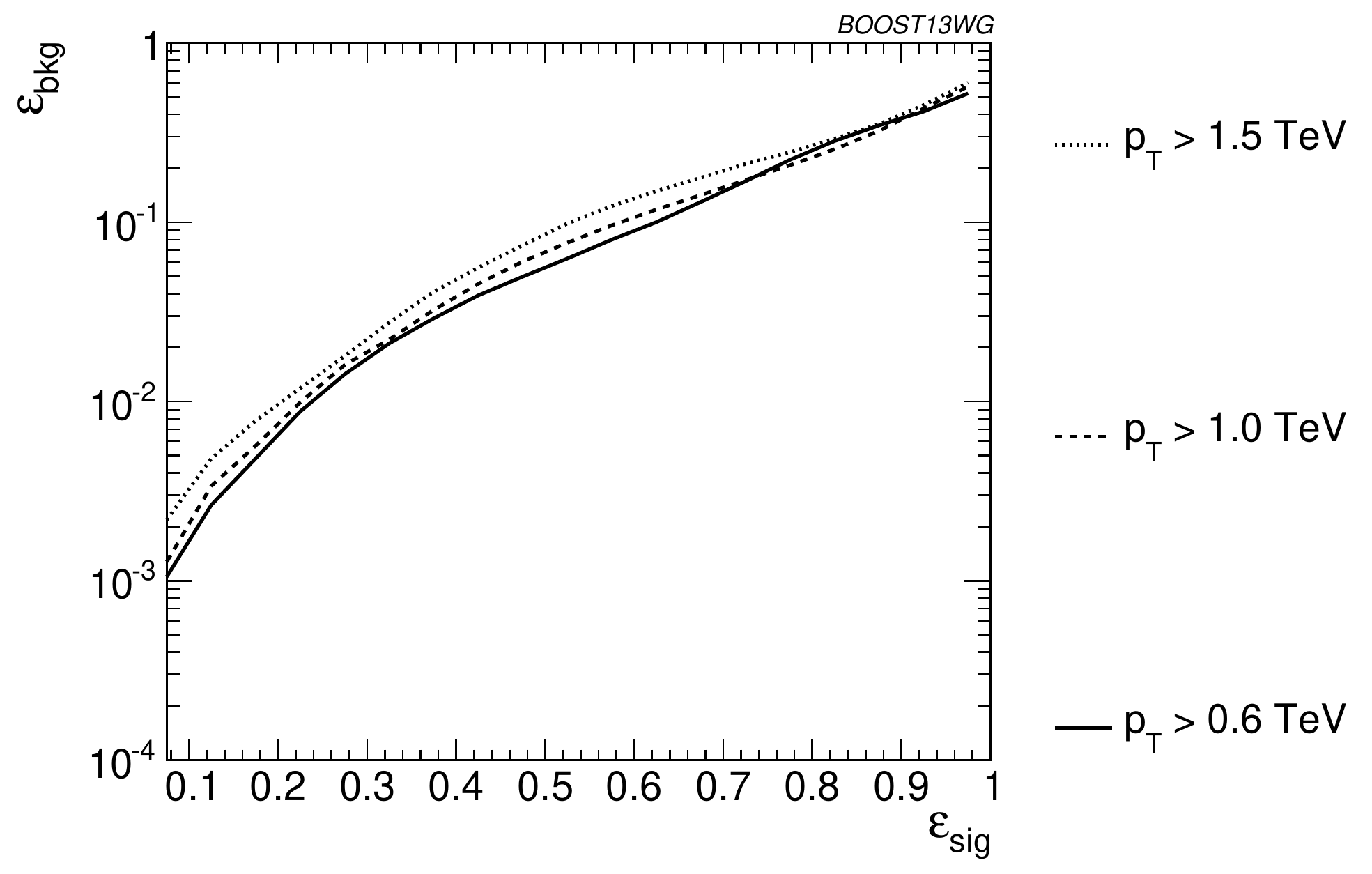}}
\subfigure[Johns Hopkins Tagger $m_t$]{\includegraphics[width=0.48\textwidth]{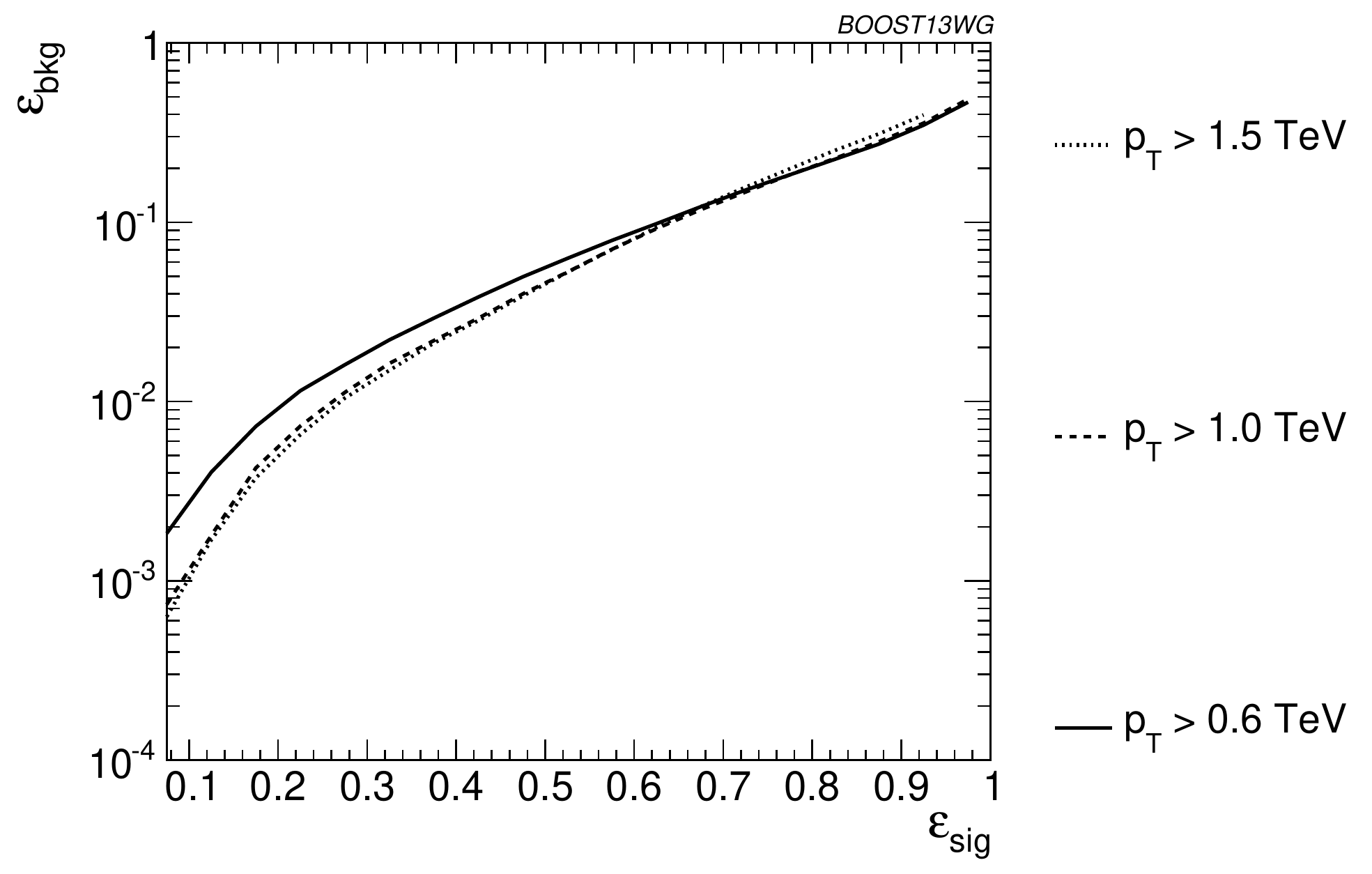}}
\subfigure[Pruning $m_t$]{\includegraphics[width=0.48\textwidth]{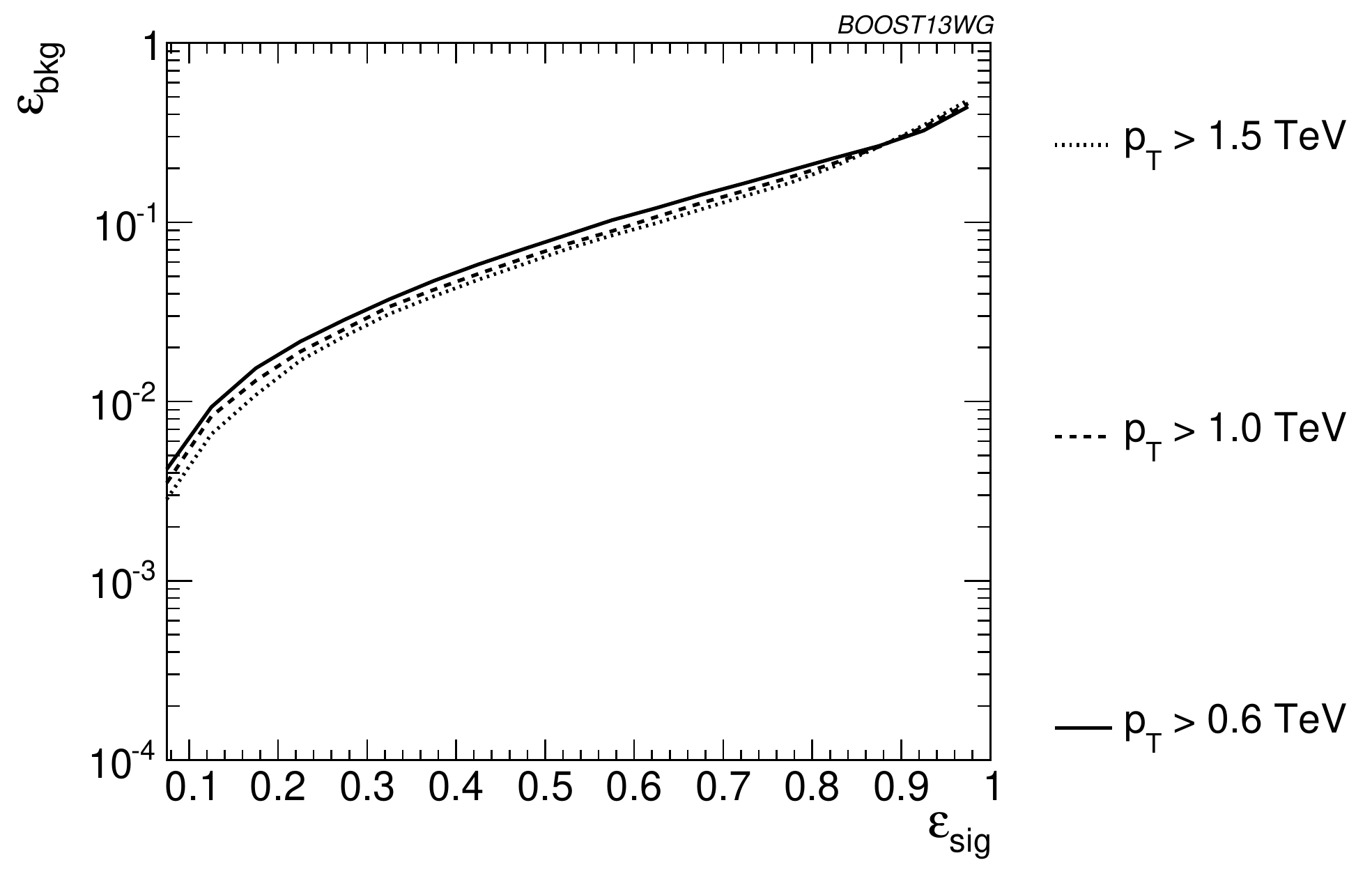}}
\subfigure[Trimming $m_t$]{\includegraphics[width=0.48\textwidth]{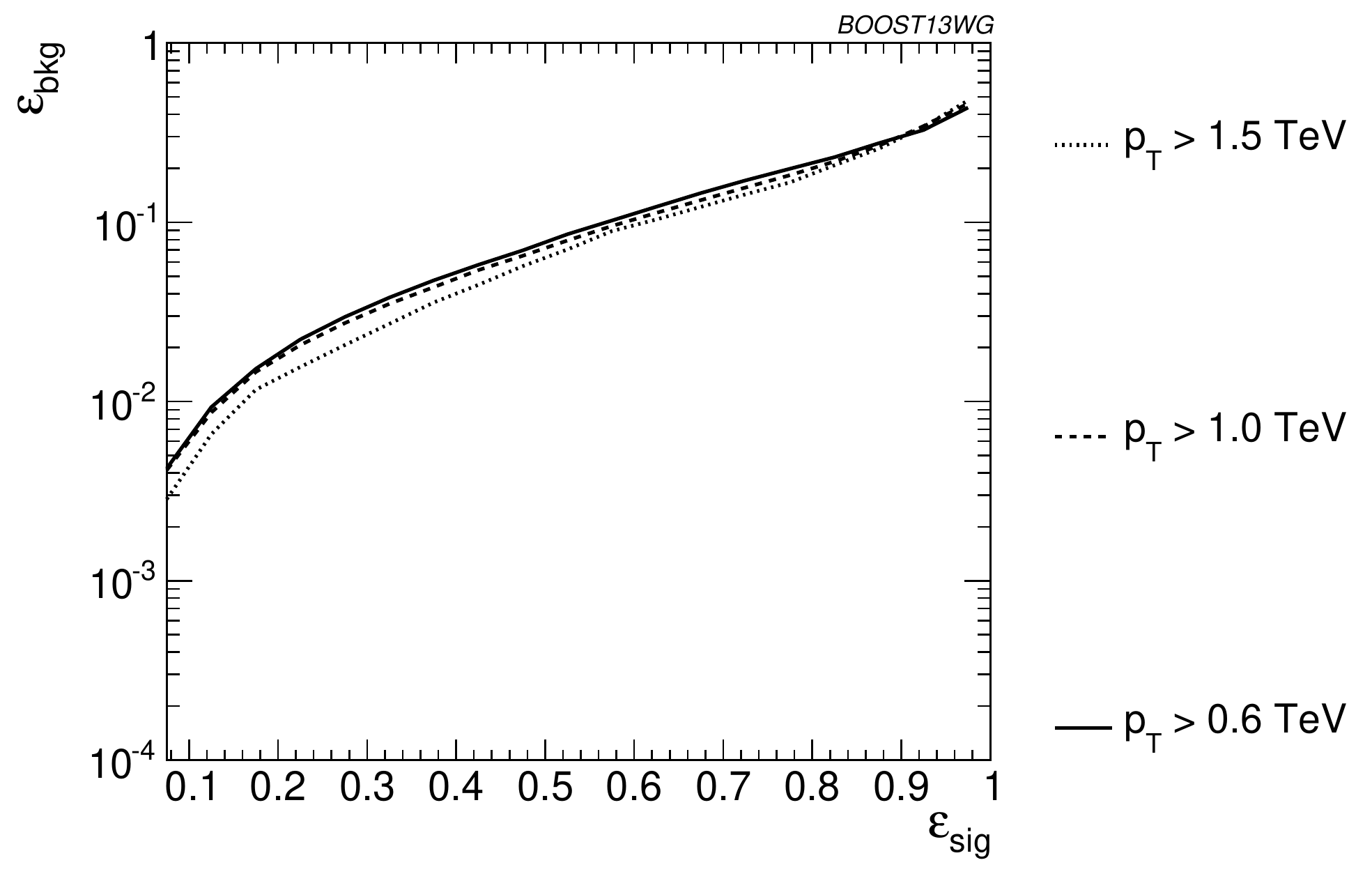}}
\caption{Comparison of top mass performance of different taggers at different \pt using the anti-\kT R=0.8 algorithm.}
\label{fig:ptcomparison_singletopmass_top}
\end{figure*}

\begin{figure*}
\centering
\subfigure[HEPTopTagger $m_t$]{\includegraphics[width=0.48\textwidth]{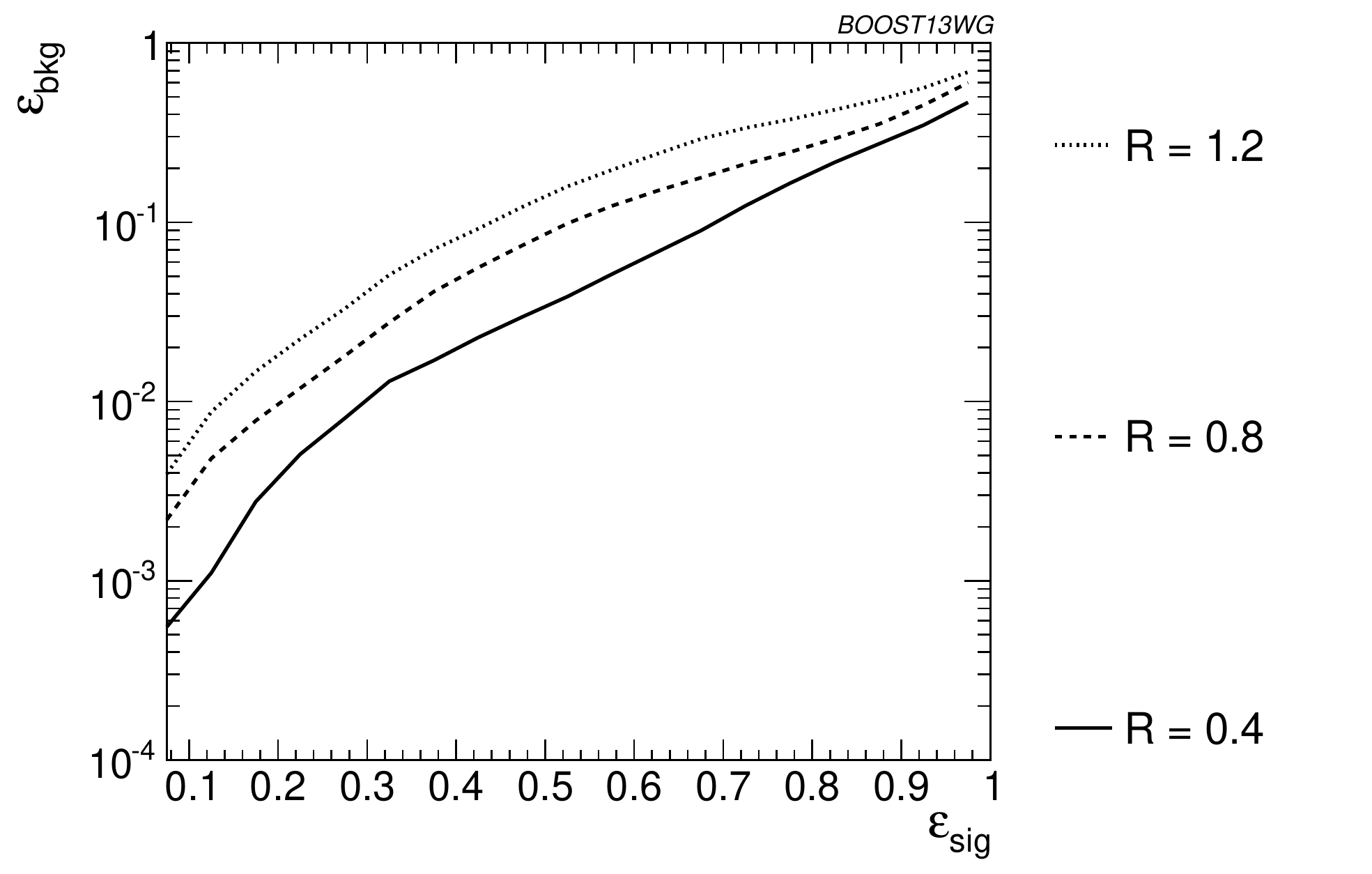}}
\subfigure[Johns Hopkins Tagger $m_t$]{\includegraphics[width=0.48\textwidth]{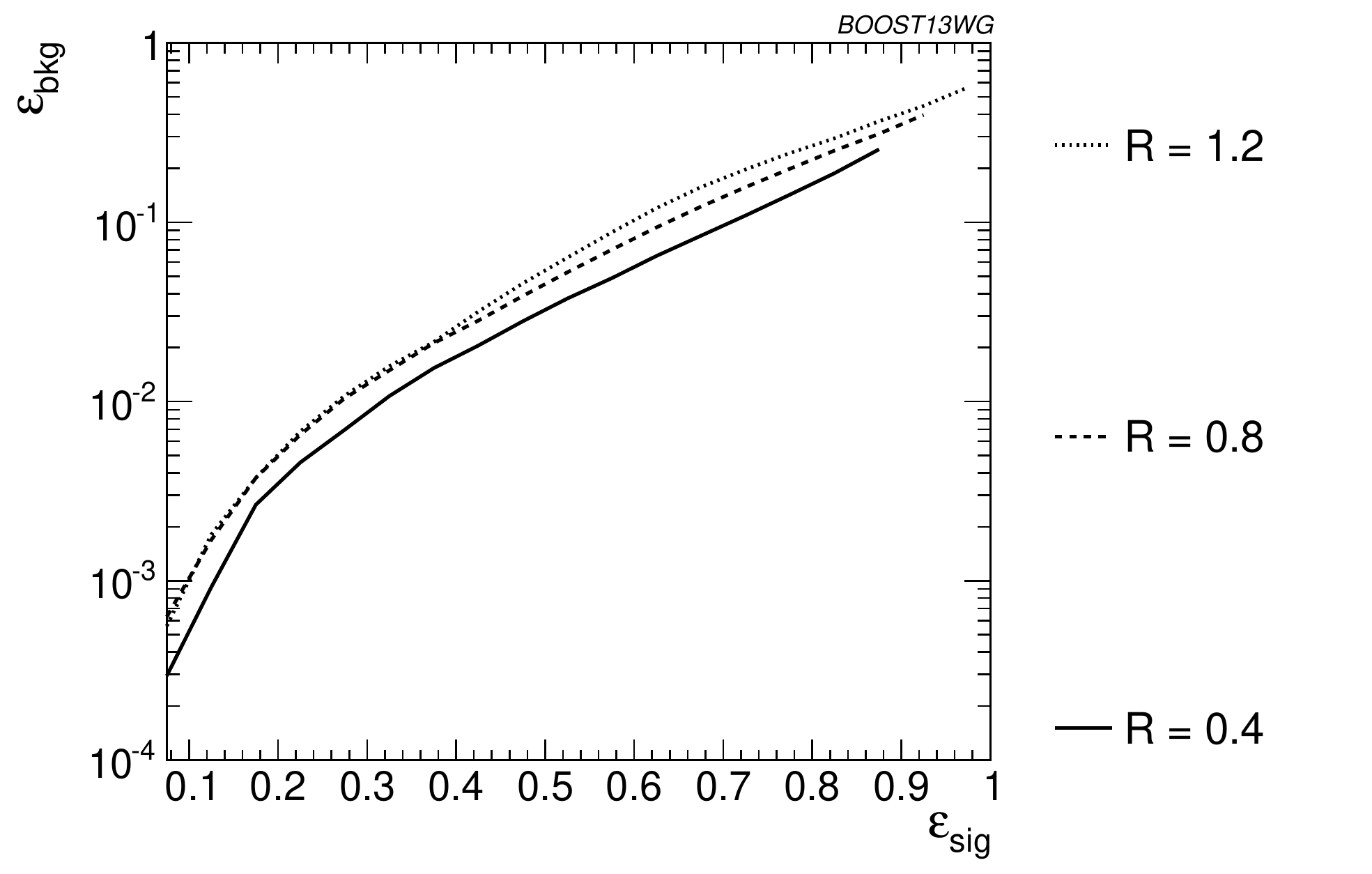}}
\subfigure[Pruning $m_t$]{\includegraphics[width=0.48\textwidth]{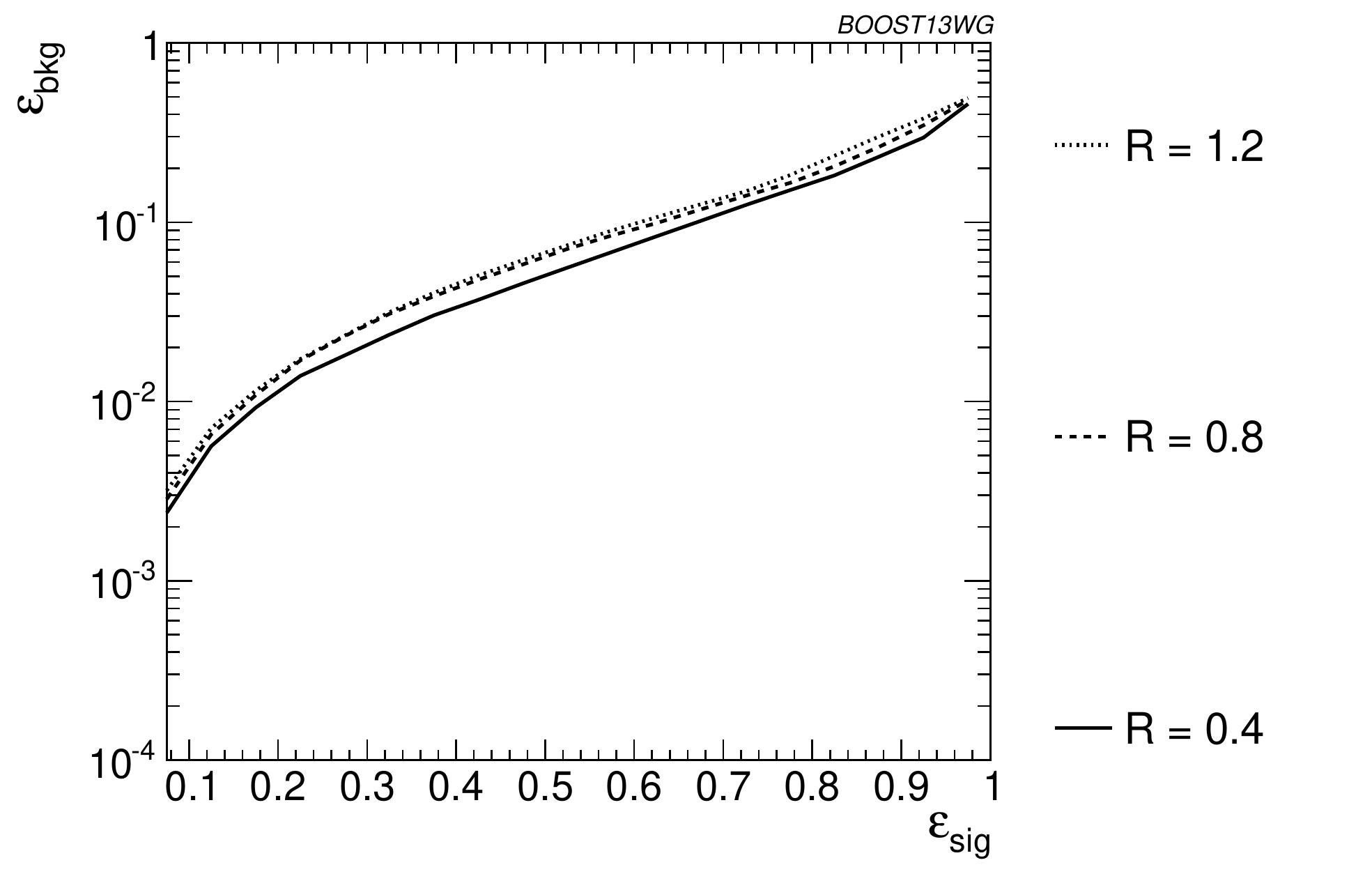}}
\subfigure[Trimming $m_t$]{\includegraphics[width=0.48\textwidth]{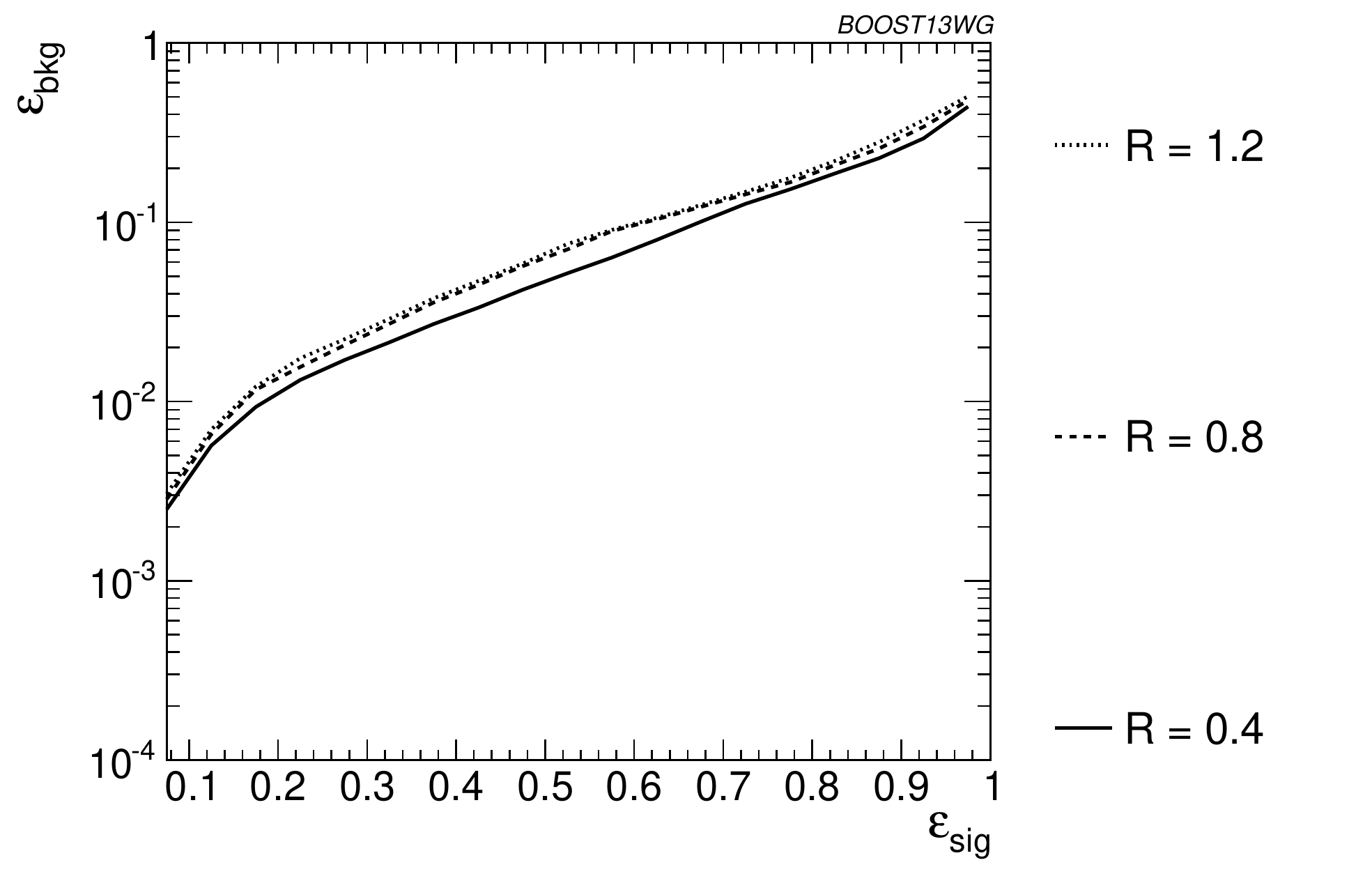}}
\caption{Comparison of top mass performance of different taggers at different $R$ in the \pt = 1.5-1.6 \TeV bin.}
\label{fig:Rcomparison_singletopmass_top}
\end{figure*}

Of the two top-tagging algorithms, it is apparent from Figure~\ref{fig:single_variable_ROC} that the Johns Hopkins tagger out-performs the HEPTopTagger in terms of its background rejection at fixed signal efficiency for both the top and $W$ candidate masses; this is expected, as the HEPTopTagger was designed to reconstruct moderate-\pt top jets in $ttH$ events (for a proposed high-\pt variant of the HEPTopTagger, see \cite{Schaetzel:2013vka}). In Figure~\ref{fig:topmass_histogram_HEP_JH}, we show the histograms for the top mass output from the JH and HEPTopTagger for different $R$ in the \pt =  1.5-1.6 TeV bin, and in Figure~\ref{fig:topmass_histogram_HEP_JH_pT} for different \pt at $R=0.8$, optimized at a signal efficiency of 30\%. 
A particular feature of the HepTopTagger algorithm is that, after the jet is filtered to select the five hardest subjets, the three subjets are chosen which most closely reconstruct the top mass. This requirement tends to shape a peak in the QCD background around $m_t$ for the HEPTopTagger, as can be seen from Figures~\ref{fig:topmass_histogram_HEP_R12} and~\ref{fig:topmass_histogram_HEP_pT15}; this is the likely reason for the better performance of the JH tagger, which has no such requirement. This effect is more pronounced  at higher $\pt$ and larger jet radius (see Figures~\ref{fig:ptcomparison_singletopmass_top} and \ref{fig:Rcomparison_singletopmass_top}). 
It has been proposed  \cite{Anders:2013oga,Kasieczka:2015jma} that performance of the HEPTopTagger may be improved by changing the selection criteria and/or performing a multivariate analysis with other variables. For example, the three subjets reconstructing the top should be selected only among those sets that pass the $W$ mass constraints, which reduces the shaping of the background. We indeed confirm below that combining the HEPTopTagger with other variables  reduces the discrepancy between the JH and the HEPTopTagger, and a preliminary study indicates that the new  ordering prescriptions  makes the tagger performances more comparable.  

We also see in Figure~\ref{fig:single_variable_ROC_topmass} that the top mass from the JH tagger and the HEPTopTagger has superior performance relative to either of the grooming algorithms; this is because  the pruning and trimming algorithms do not have inherent $W$-identification steps and are not optimized for this purpose. Indeed, because of the lack of a $W$-identification step, grooming algorithms are forced to strike a balance between under-grooming the jet, which broadens the signal peak due to underlying event contamination and features a larger background rate, and over-grooming the jet, which occasionally throws out the $b$-jet and preserves only the $W$ components inside the jet. We demonstrate this effect in Figures~\ref{fig:topmass_histogram_HEP_JH} and \ref{fig:topmass_histogram_HEP_JH_pT}, showing that with 30\% signal efficiency, the optimal performance of the tagger over-grooms a substantial fraction of the jets ($\sim20-30\%$), leading to a spurious second peak at \wmass. This effect is more pronounced at large $R$ and $\pt$, since more aggressive grooming is required in these limits to combat the increased contamination from underlying event and QCD radiation.

\begin{figure*}
\centering
\subfigure[$C_2^{\beta=1}$]{\includegraphics[width=0.48\textwidth]{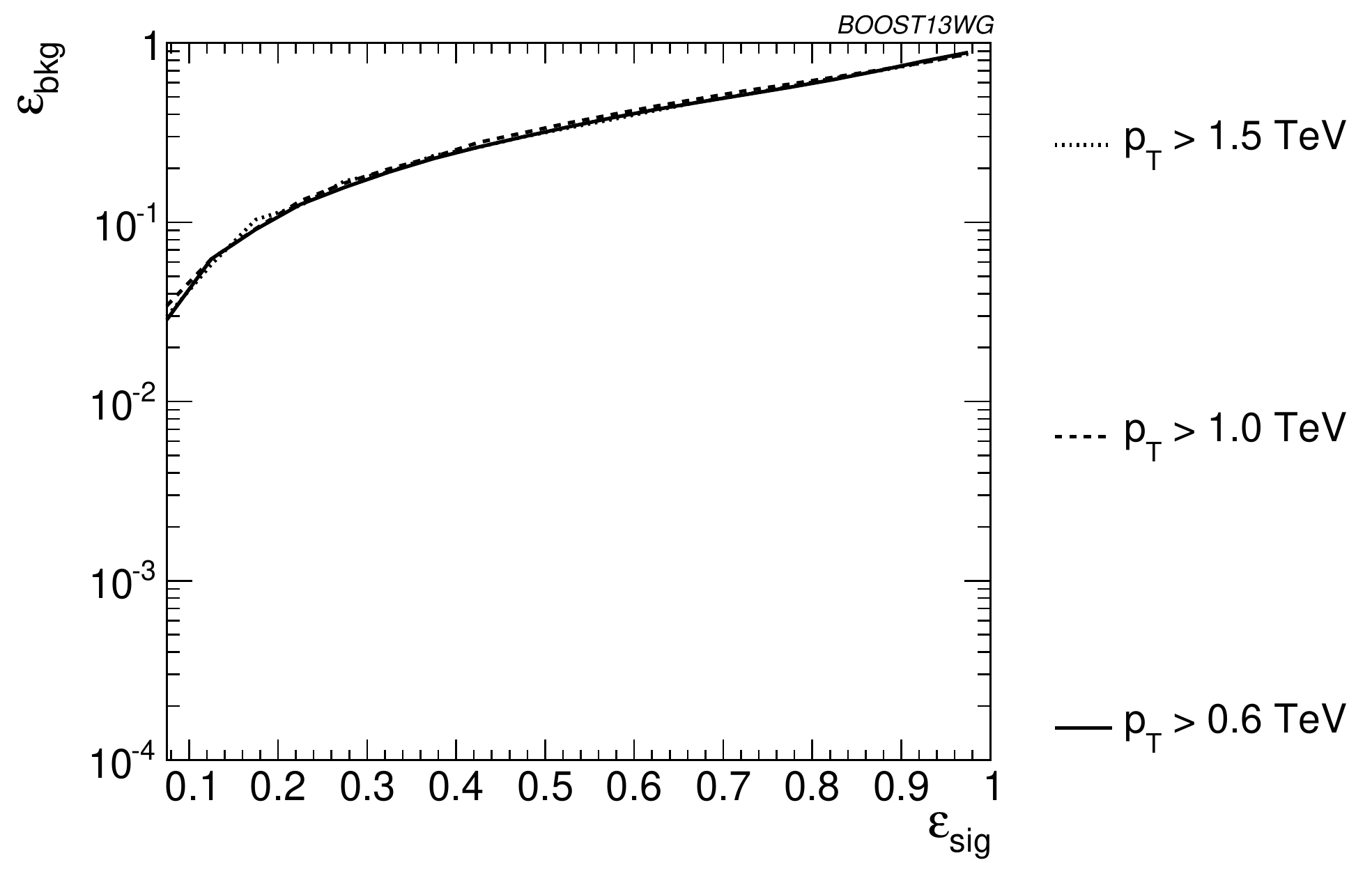}}
\subfigure[$C_3^{\beta=1}$]{\includegraphics[width=0.48\textwidth]{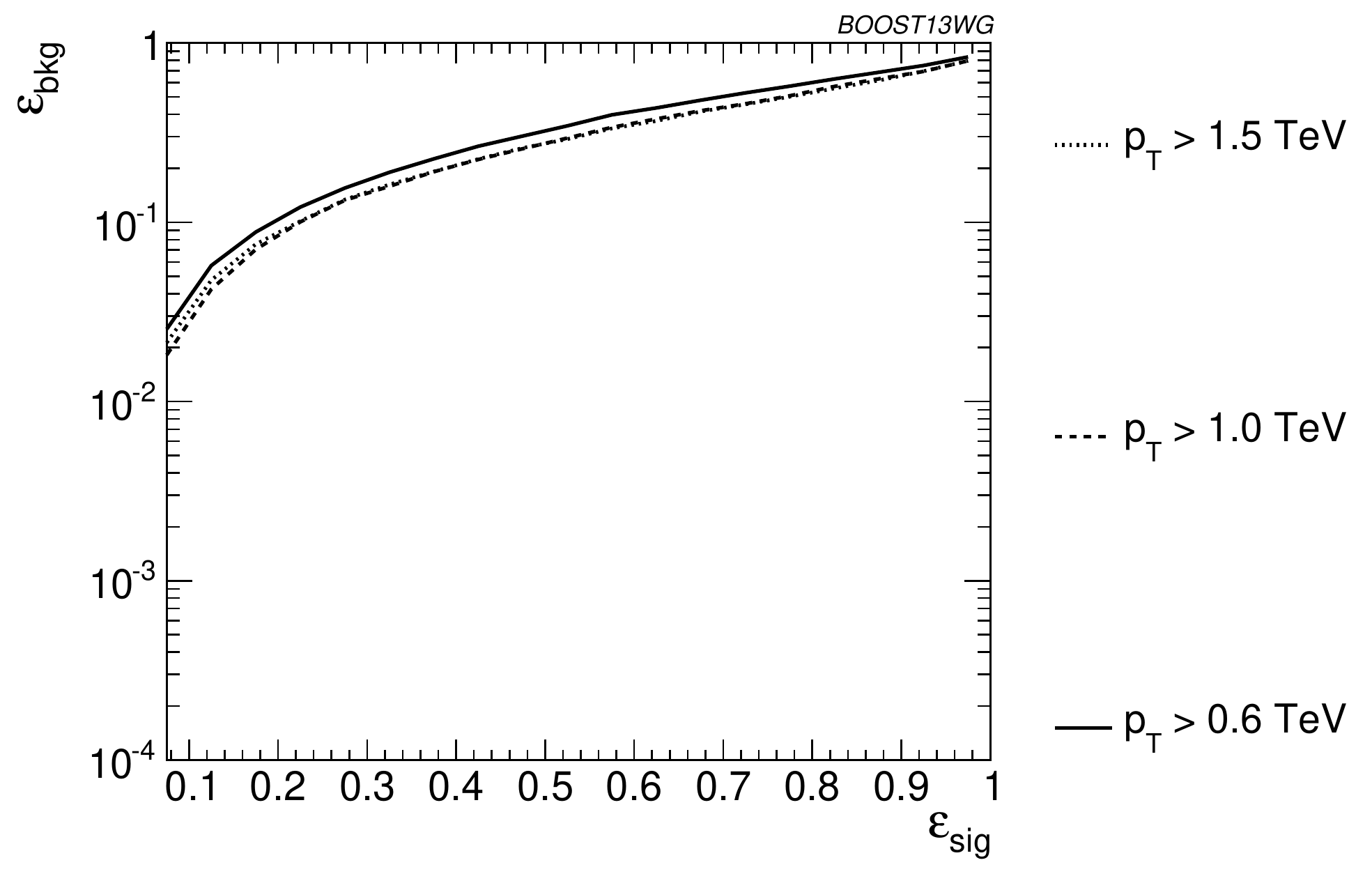}\label{fig:ptcomparison_singleshape_top_C3}}
\subfigure[\tautwoone]{\includegraphics[width=0.48\textwidth]{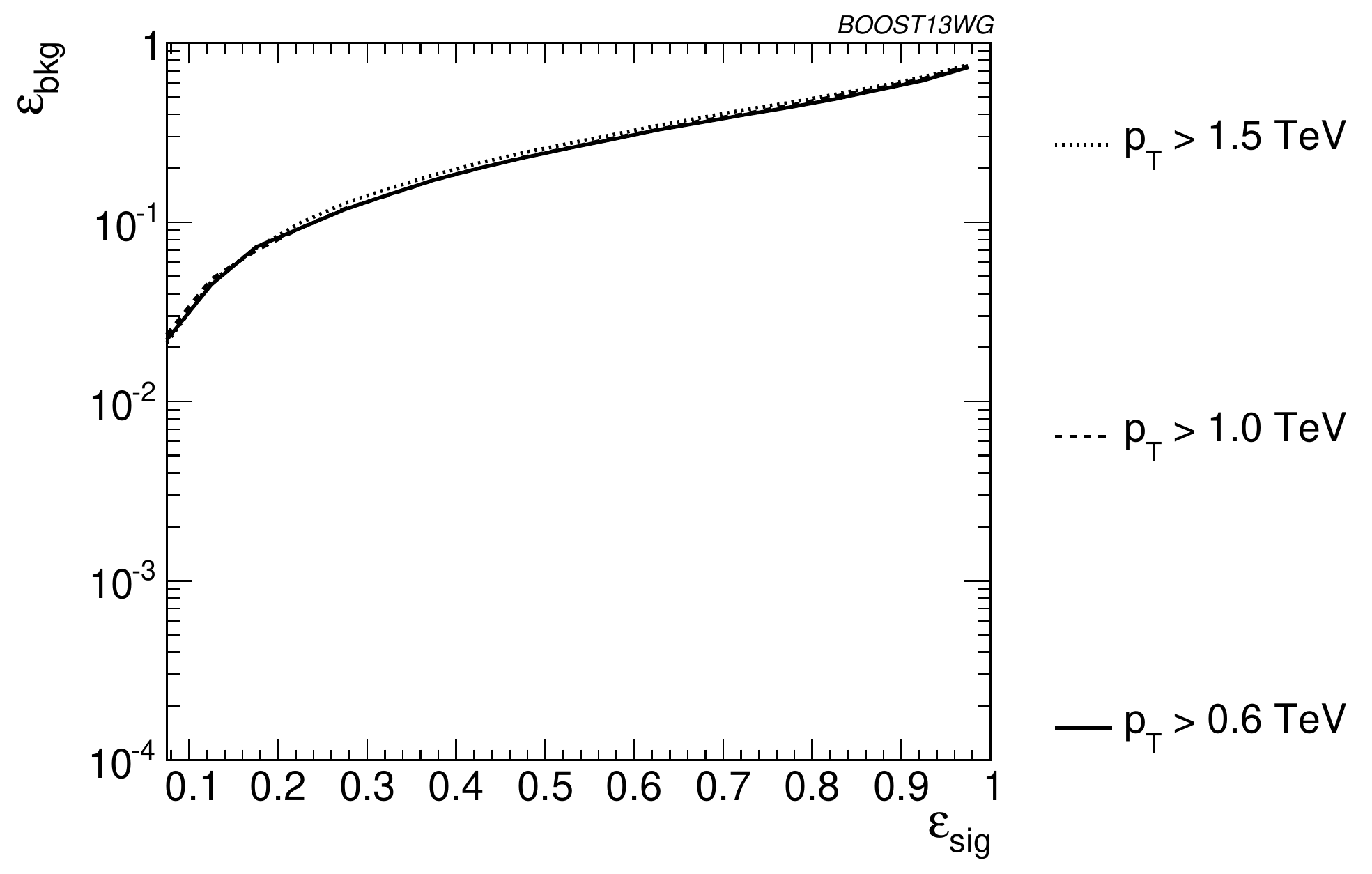}}
\subfigure[\tauthreetwo]{\includegraphics[width=0.48\textwidth]{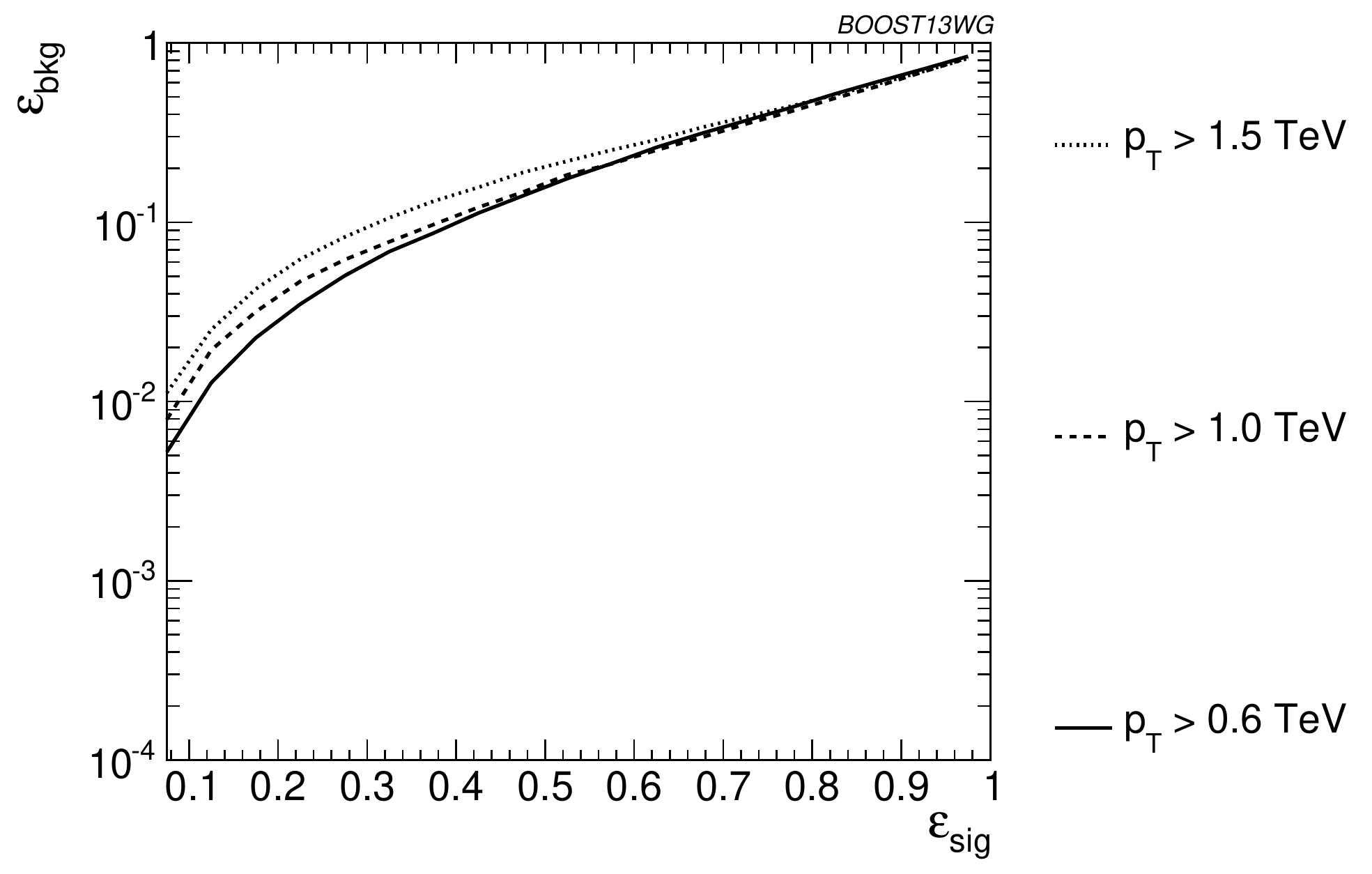}}
\subfigure[$\Gamma_{\rm Qjet}$]{\includegraphics[width=0.48\textwidth]{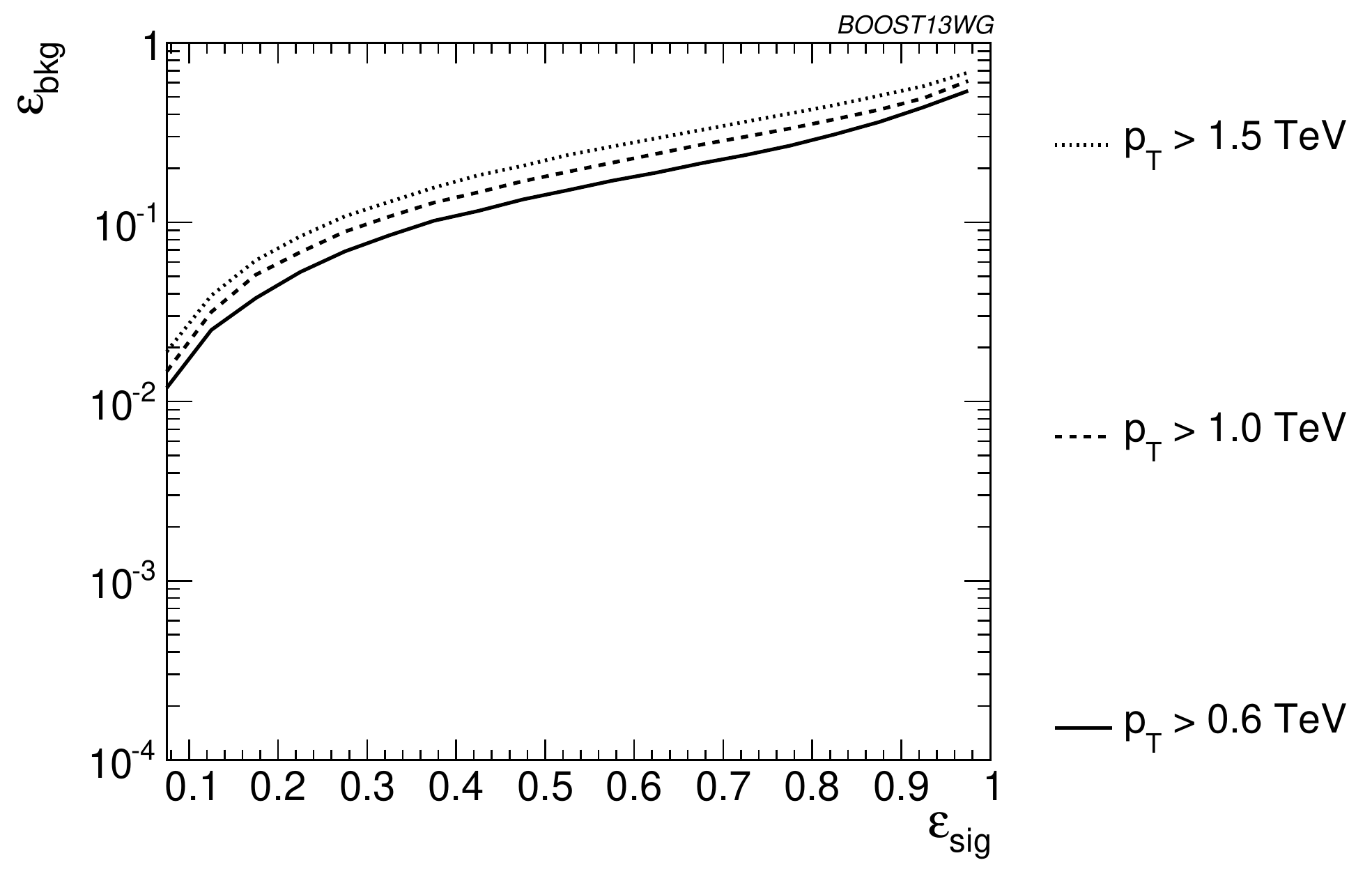}}
\caption{Comparison of individual jet shape performance at different \pt using the anti-\kT $R=0.8$ algorithm.}
\label{fig:ptcomparison_singleshape_top}
\end{figure*}

\begin{figure*}
\centering
\subfigure[$\Gamma_{\rm Qjet}$, \pt = 600-700 \GeV]{\includegraphics[width=0.245\textwidth]{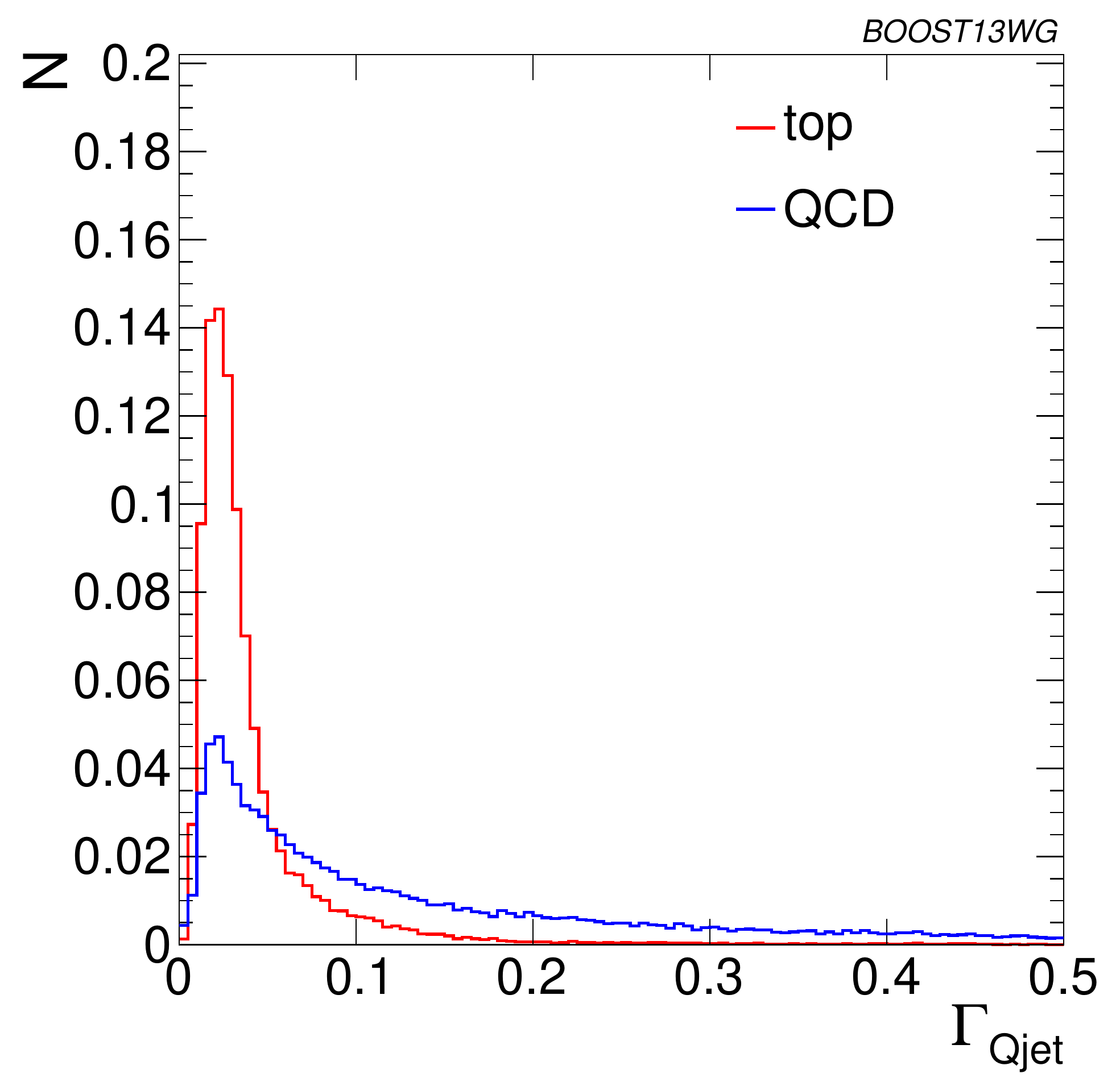}}
\subfigure[$\Gamma_{\rm Qjet}$, \pt = 1.5-1.6 \TeV]{\includegraphics[width=0.245\textwidth]{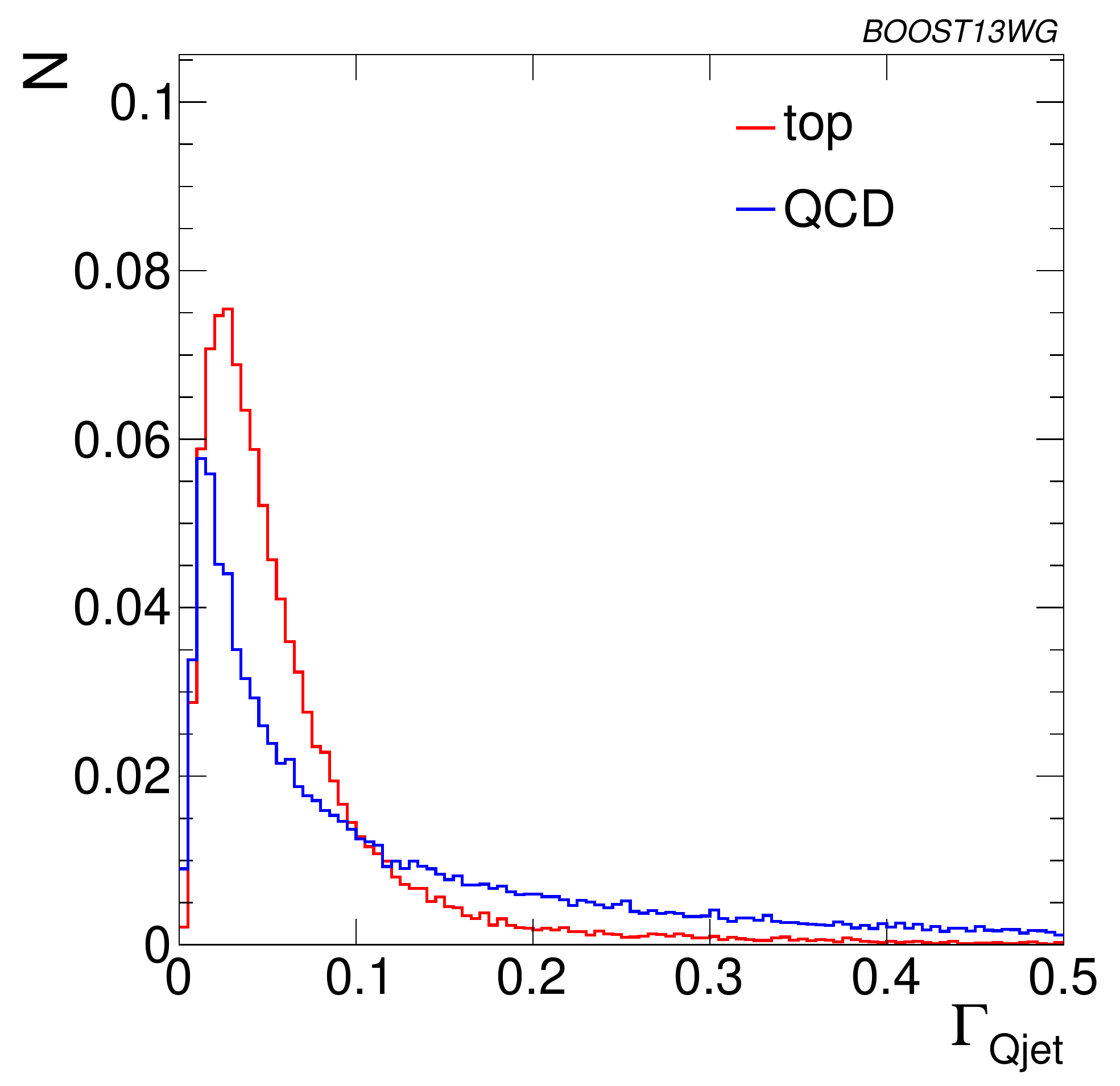}}
\subfigure[\tauthreetwo, \pt = 600-700 \GeV]{\includegraphics[width=0.245\textwidth]{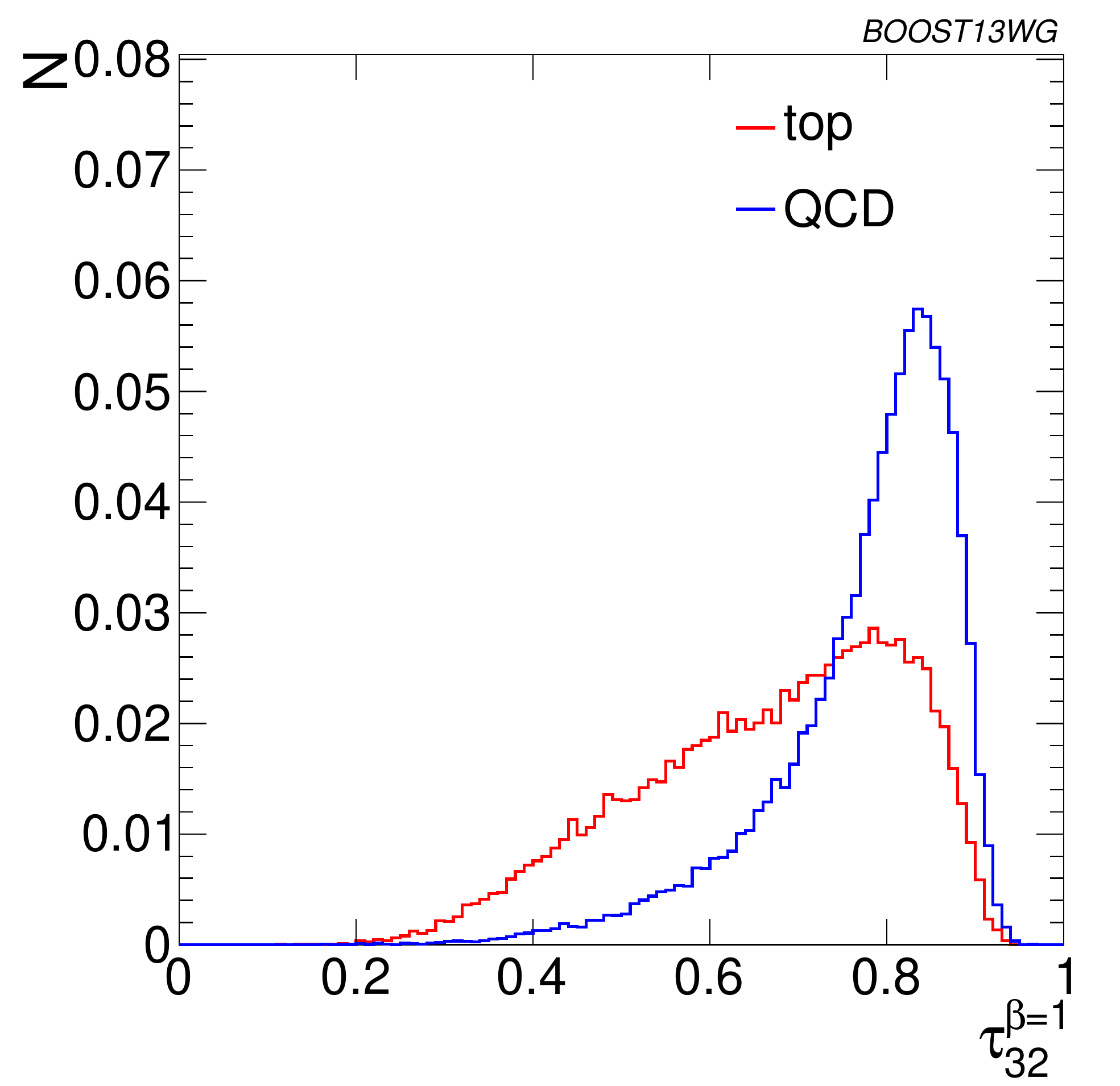}}
\subfigure[\tauthreetwo, \pt = 1.5-1.6 \TeV]{\includegraphics[width=0.245\textwidth]{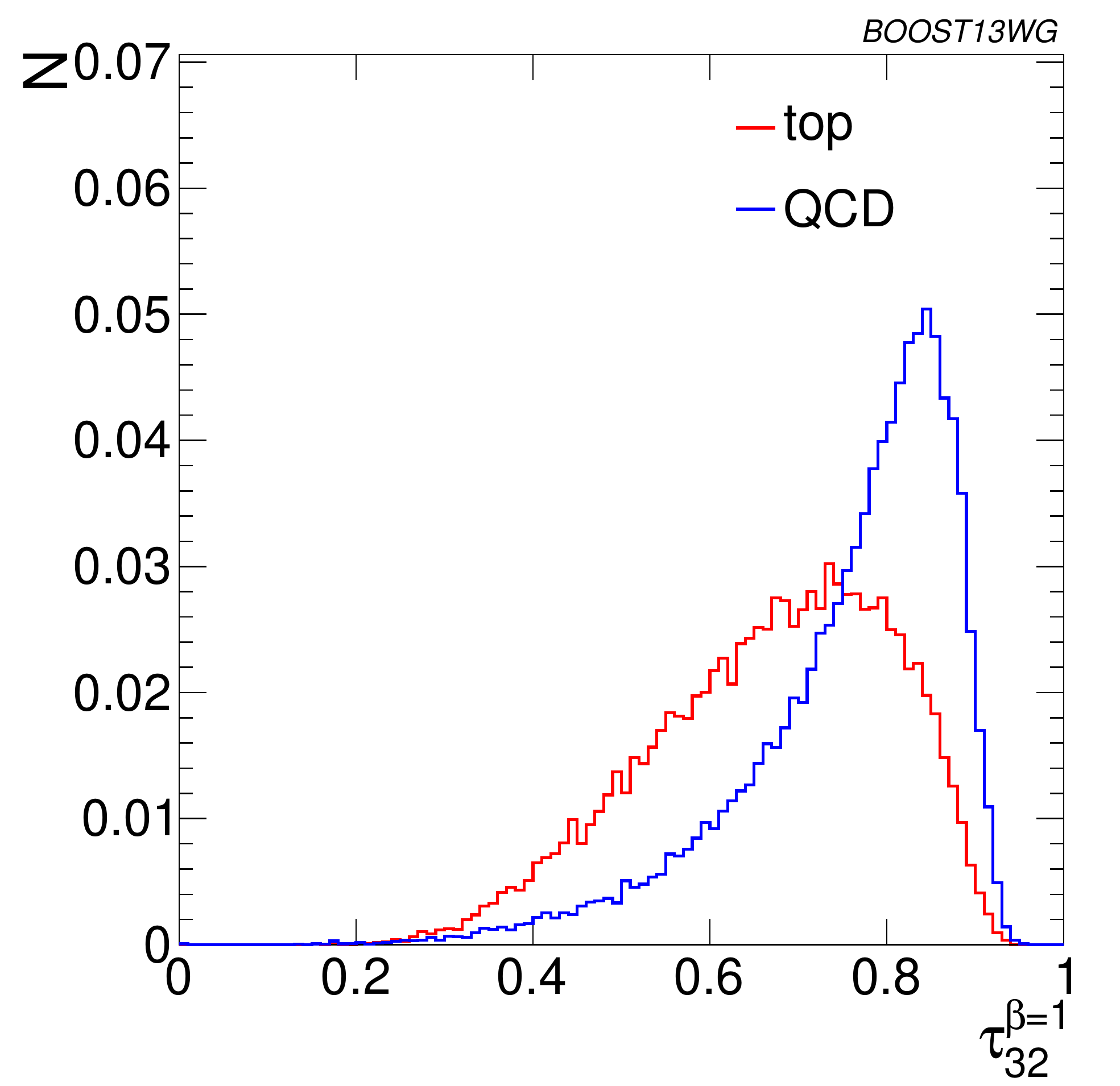}}
\caption{Comparison of $\Gamma_{\rm Qjet}$ and $\tau_{32}^{\beta=1}$ at $R=0.8$ and different values of the \pt. These shape variables are the most sensitive to varying \pt.}
\label{fig:Qjet_comparison_pT}
\end{figure*}

In Figures~\ref{fig:ptcomparison_singleshape_top} and~\ref{fig:ptcomparison_singletopmass_top} we directly compare ROC curves for jet-shape variable performance and top-mass performance, respectively, in three different \pt bins  whilst keeping the jet radius fixed at $R=0.8$. The input parameters of the taggers, groomers and shape variables are separately optimized in each \pt bin.  One can see from Figure~\ref{fig:ptcomparison_singleshape_top} that the tagging performance of jet shapes do not change substantially with $\pt$. The variables \tauthreetwo and $\Gamma_{\rm Qjet}$ have the most variation and tend to degrade with higher $\pt$, as can be seen in Figure~\ref{fig:Qjet_comparison_pT}. This was also observed in the $W$-tagging studies in Section~\ref{sec:wtagging}, and makes sense, as higher-$\pt$ QCD jets have more, harder emissions within the jet, giving rise to substructure that fakes the signal. 
For the variable $\Gamma_{\rm Qjet}$ (again as discussed in Section~\ref{sec:wtagging}) increasing \pt leads to QCD jets with a narrower volatility distribution
due to the enhanced contribution of the ``shoulder'' region, while for the signal (top) jets the increased amount of soft radiation with increasing \pt results
in a broader volatility distribution.  This with increasing \pt the signal and background jets exhibit more similar volatility distributions, as we see explicitly
in Figures~\ref{fig:Qjet_comparison_pT} (a) and (b).  Thus $\Gamma_{\rm Qjet}$ becomes less discriminant for top identification as \pt increases.
By contrast, from Figure~\ref{fig:ptcomparison_singletopmass_top} we can see that most of the top-mass variables have superior performance at higher $\pt$, due to the radiation from the top quark becoming more collimated. The notable exception is the HEPTopTagger, which degrades at higher $\pt$, likely in part due to the background-shaping effects studied above and which is at least partially mitigated by recent updates to the HEPTopTagger  \cite{Anders:2013oga,Kasieczka:2015jma}.

\begin{figure*}
\centering
\subfigure[$C_2^{\beta=1}$]{\includegraphics[width=0.48\textwidth]{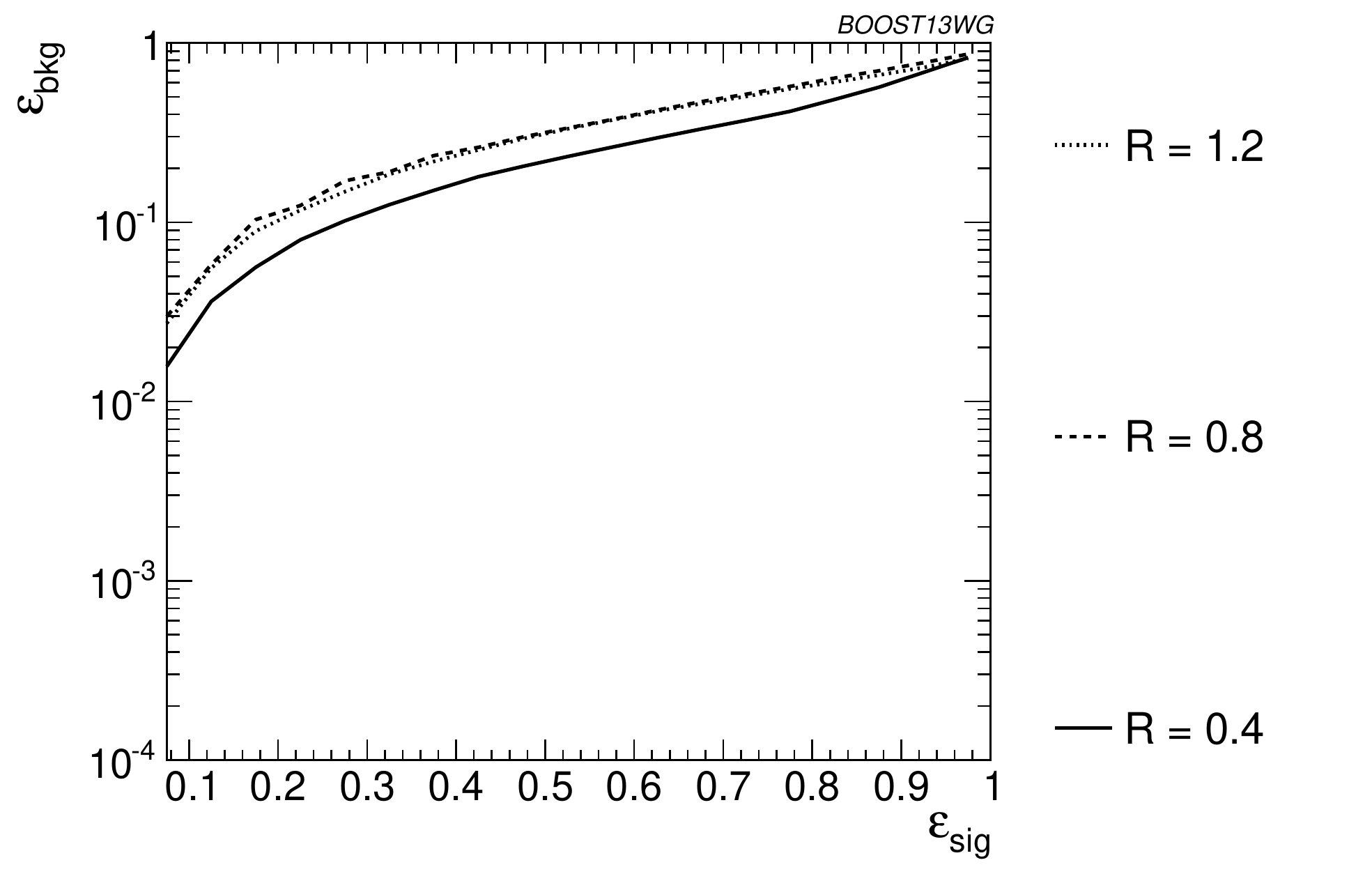}}
\subfigure[$C_3^{\beta=1}$]{\includegraphics[width=0.48\textwidth]{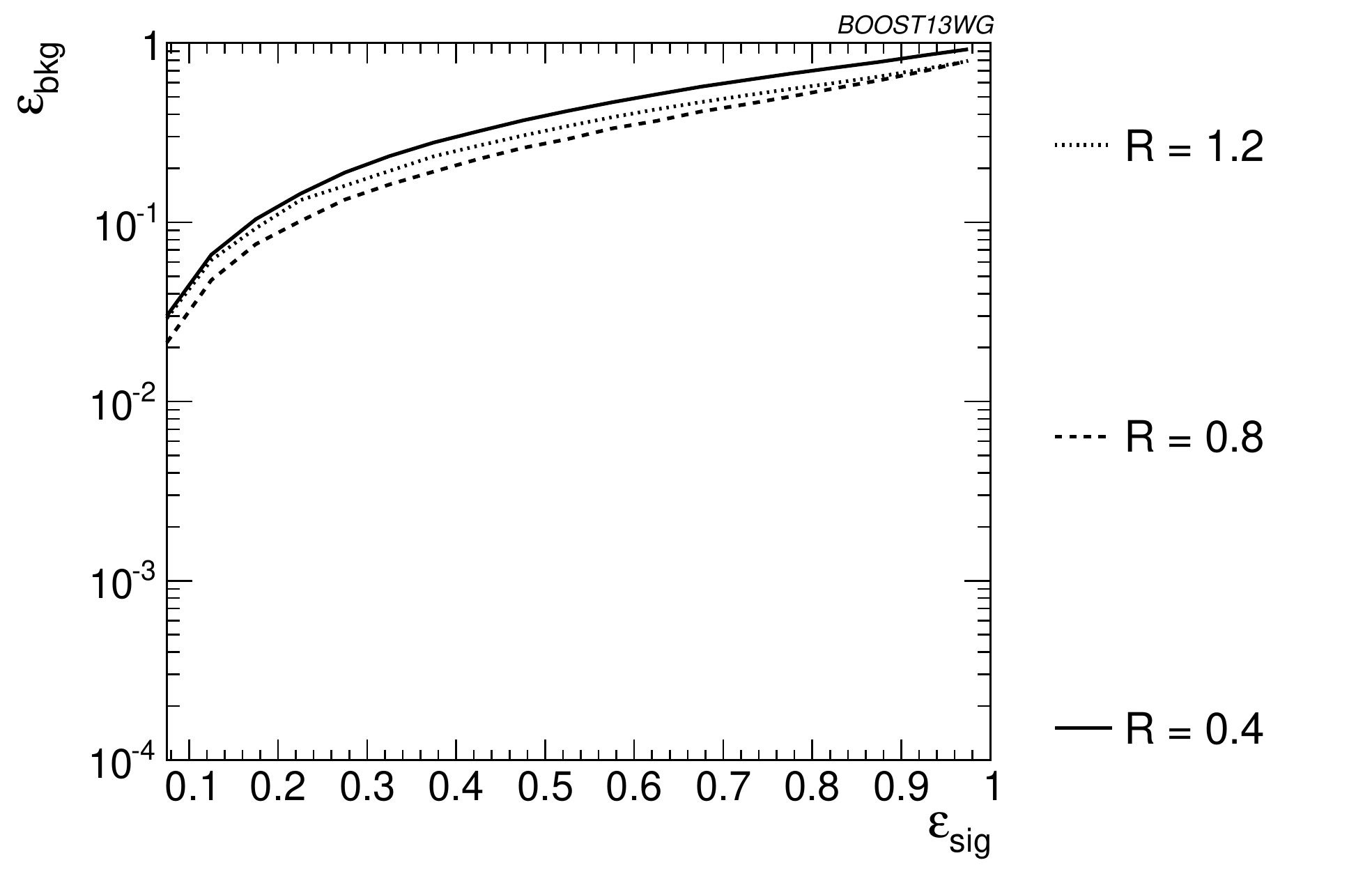}}
\subfigure[\tautwoone]{\includegraphics[width=0.48\textwidth]{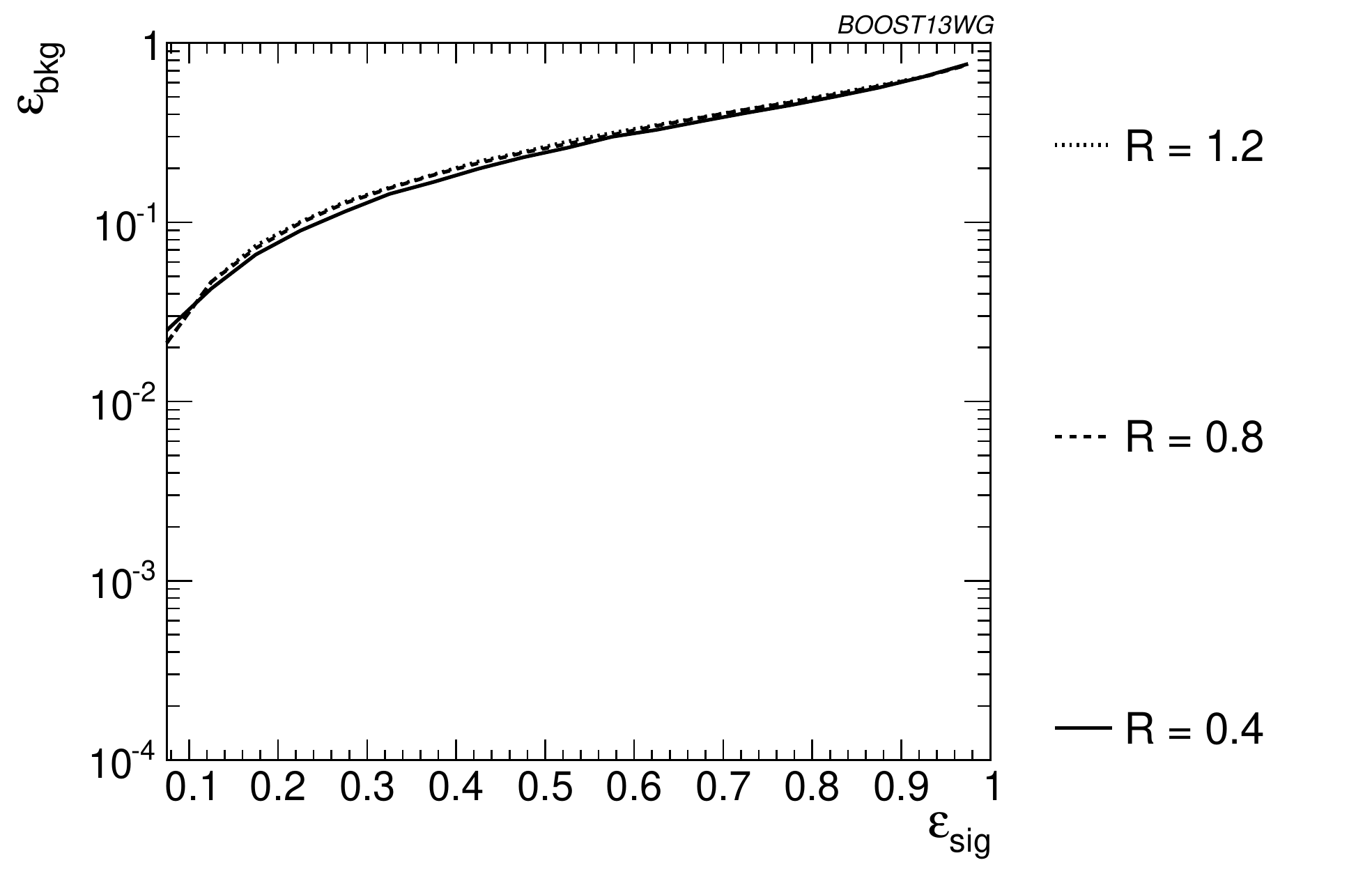}}
\subfigure[\tauthreetwo]{\includegraphics[width=0.48\textwidth]{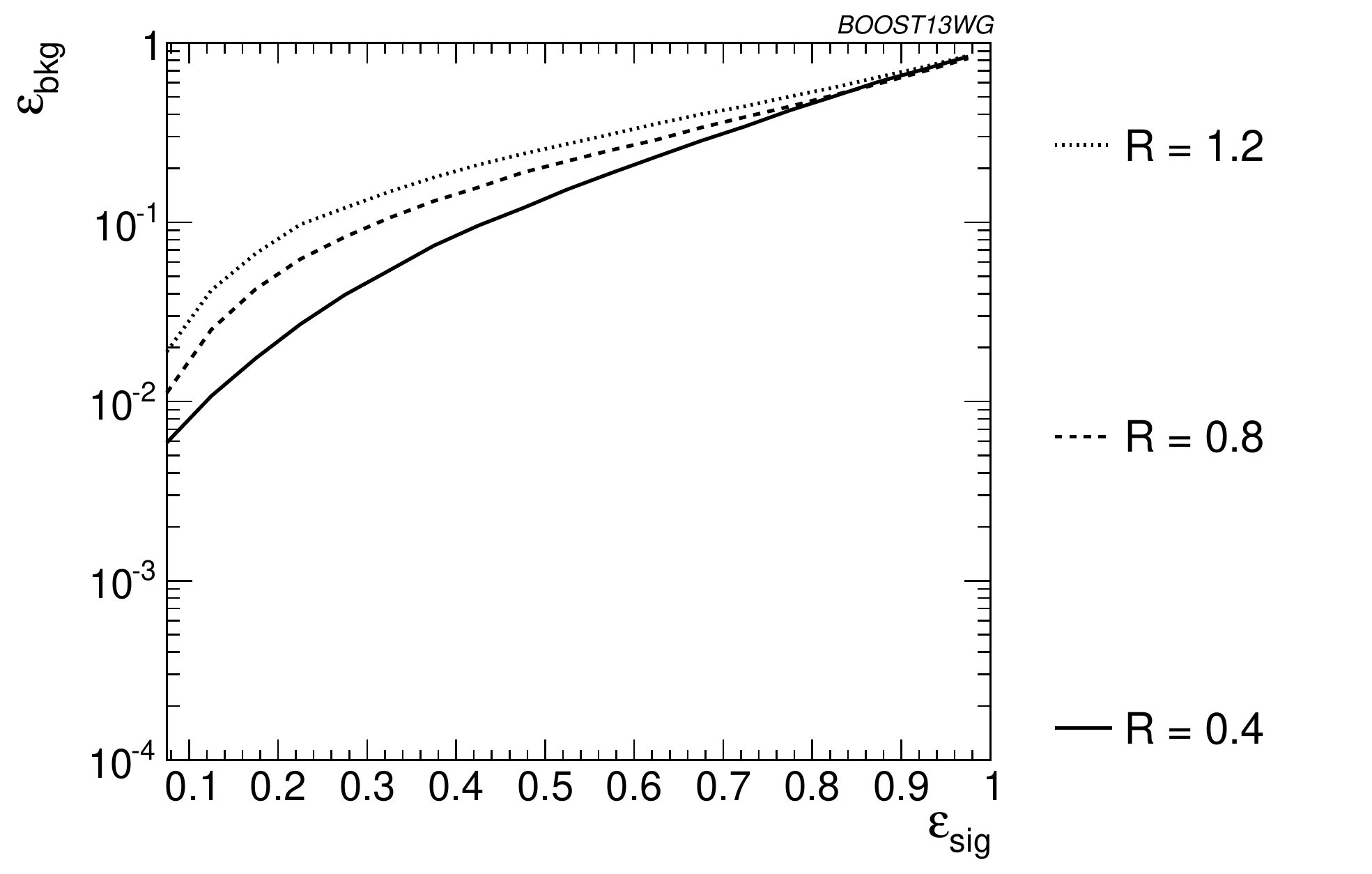}}
\subfigure[$\Gamma_{\rm Qjet}$]{\includegraphics[width=0.48\textwidth]{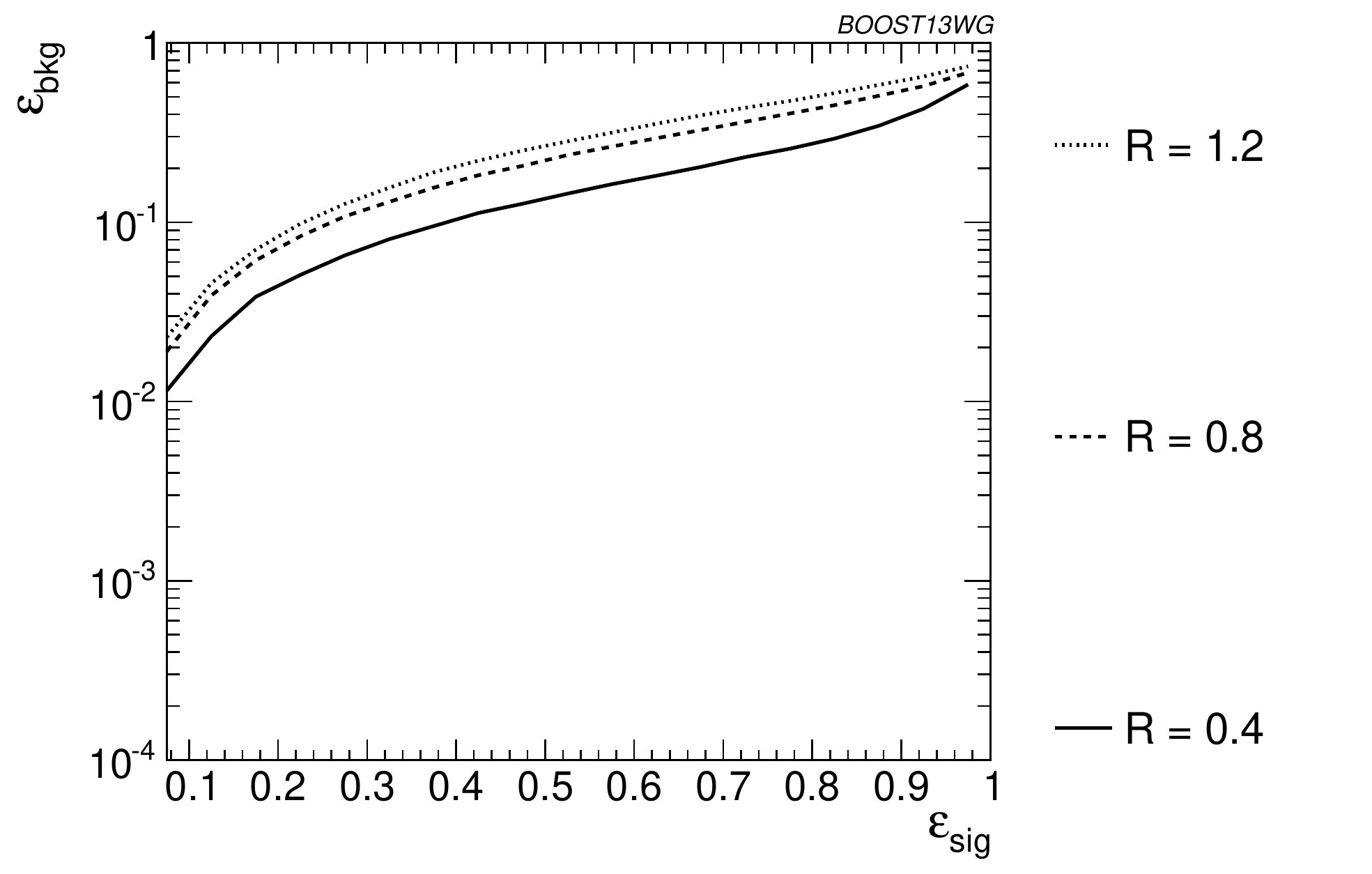}}
\caption{Comparison of individual jet shape performance at different $R$ in the \pt = 1.5-1.6 \TeV bin.}
\label{fig:Rcomparison_singleshape_top}
\end{figure*}

\begin{figure*}
\centering
\subfigure[$C_2^{\beta=1}$, $R=0.4$]{\includegraphics[width=0.245\textwidth]{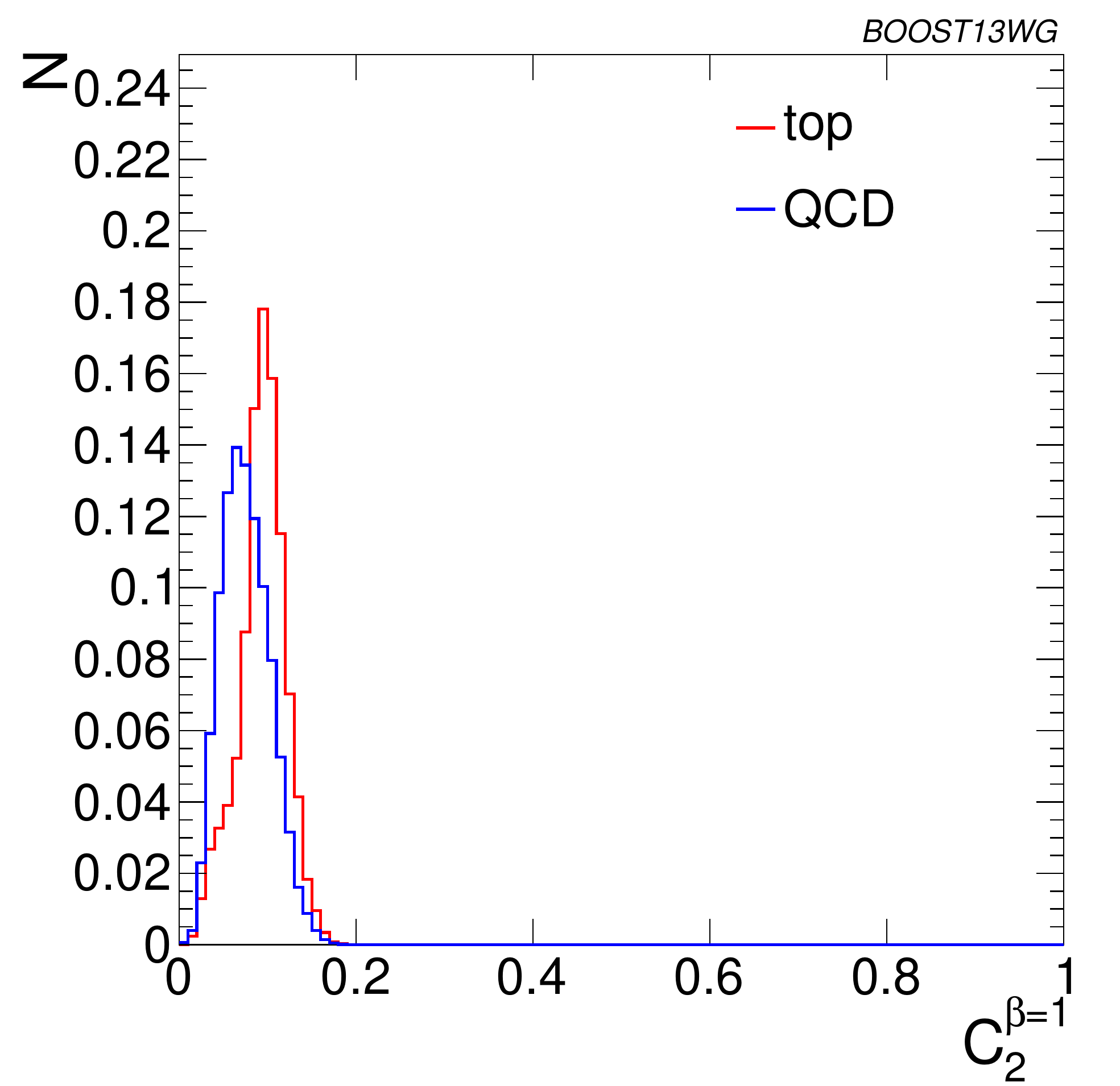}}
\subfigure[$C_2^{\beta=1}$, $R=1.2$]{\includegraphics[width=0.245\textwidth]{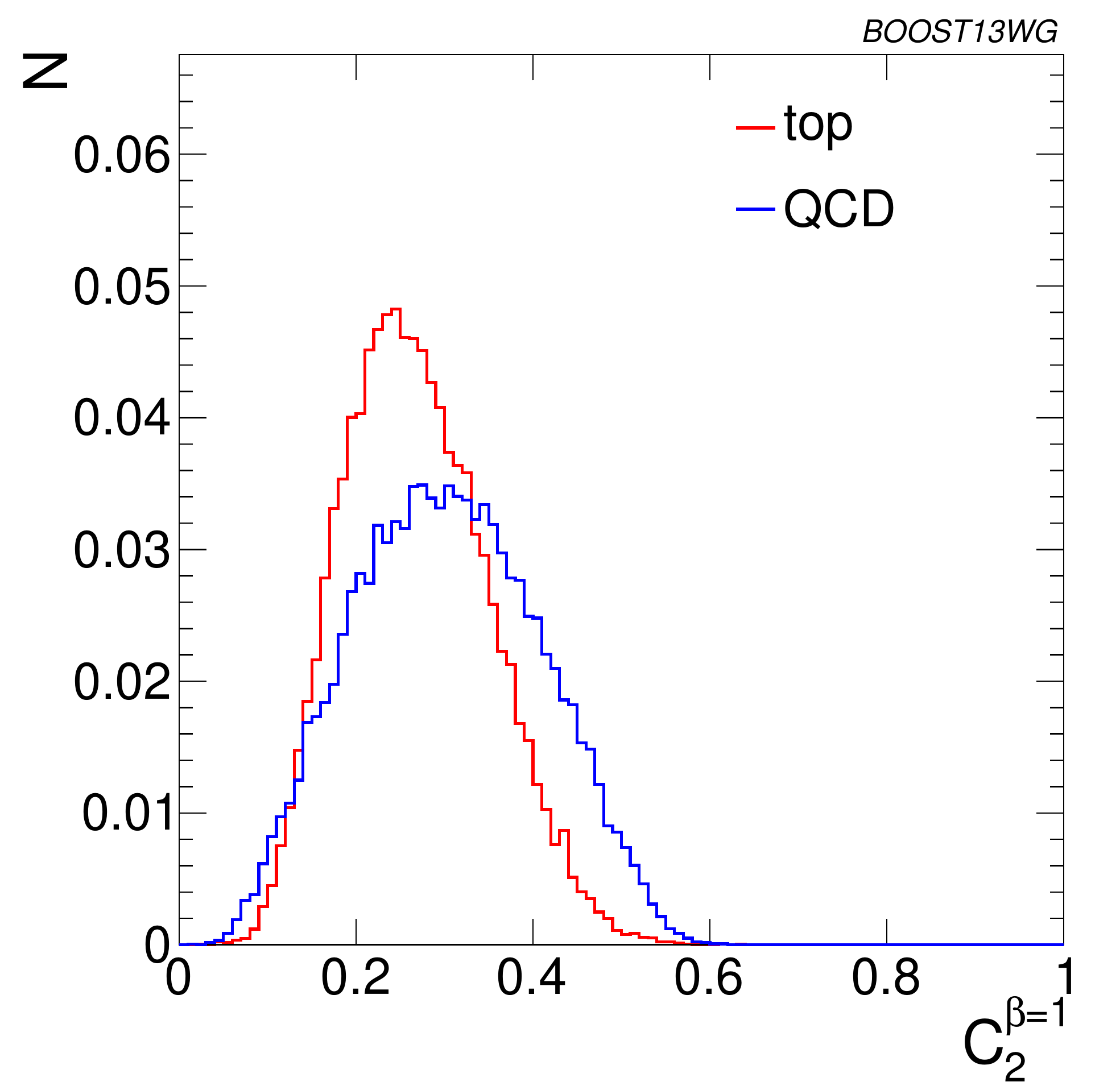}}
\subfigure[$C_3^{\beta=1}$, $R=0.4$]{\includegraphics[width=0.245\textwidth]{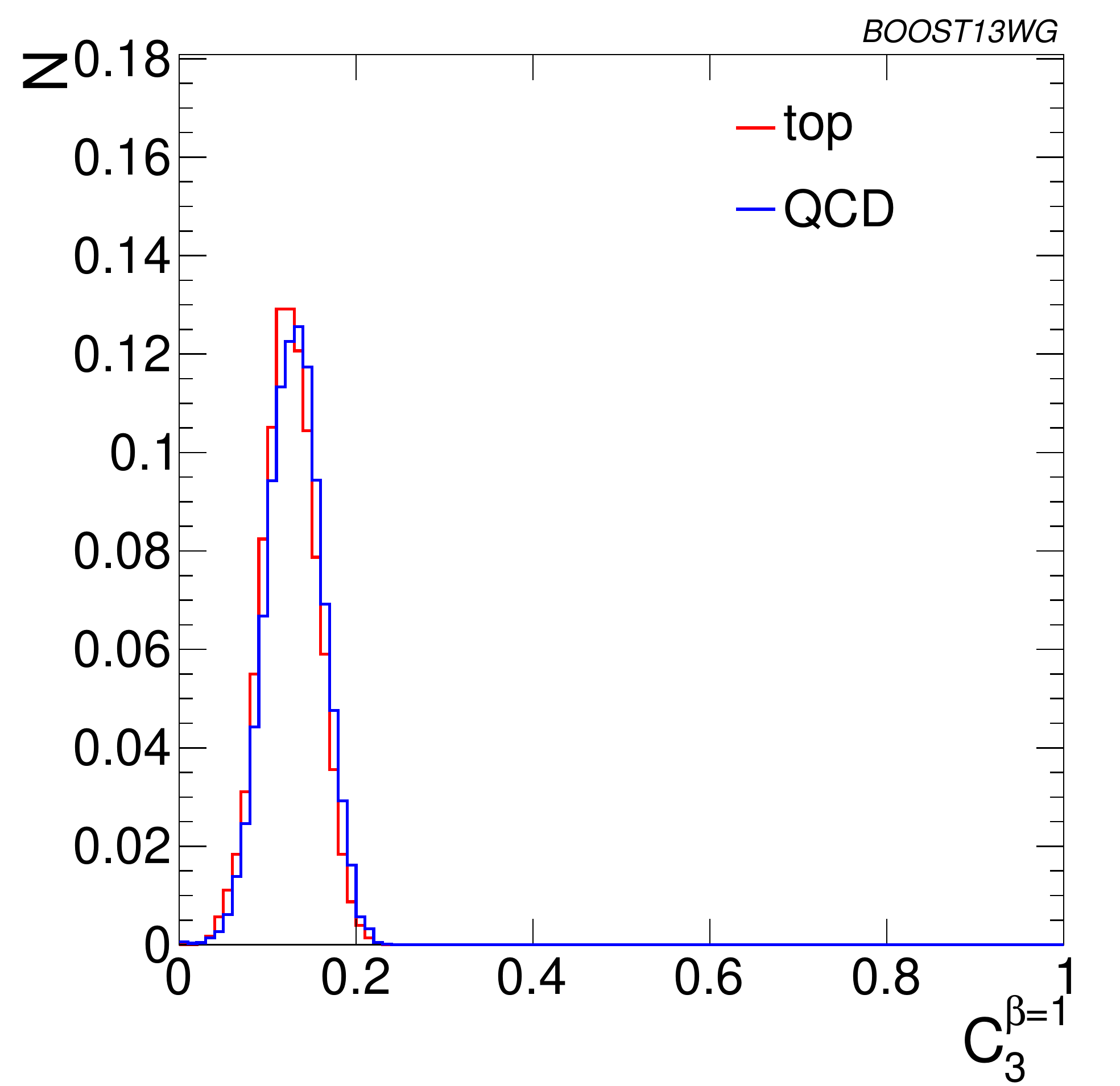}}
\subfigure[$C_3^{\beta=1}$, $R=1.2$]{\includegraphics[width=0.245\textwidth]{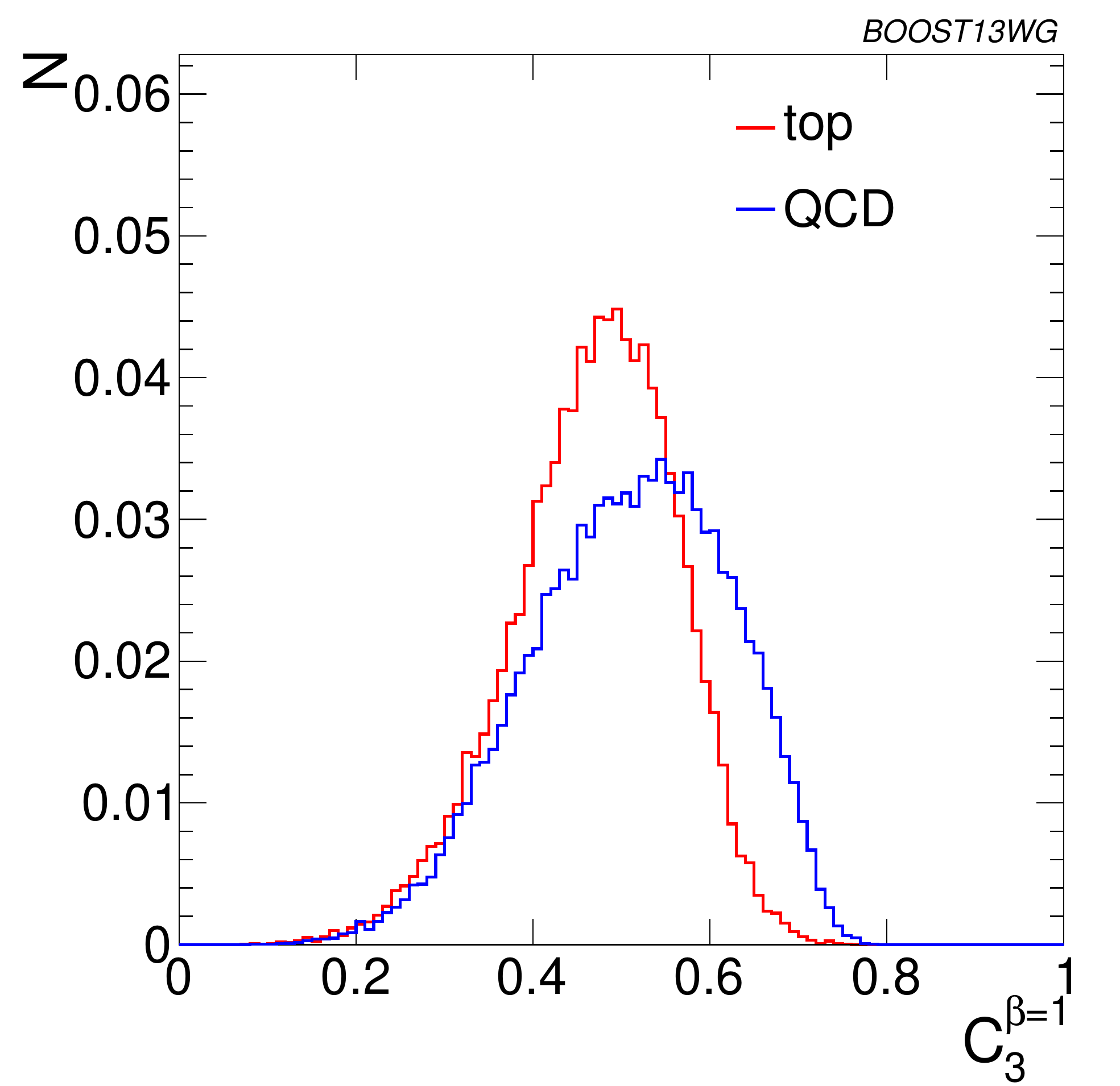}}\\
\subfigure[\tautwoone, $R=0.4$]{\includegraphics[width=0.245\textwidth]{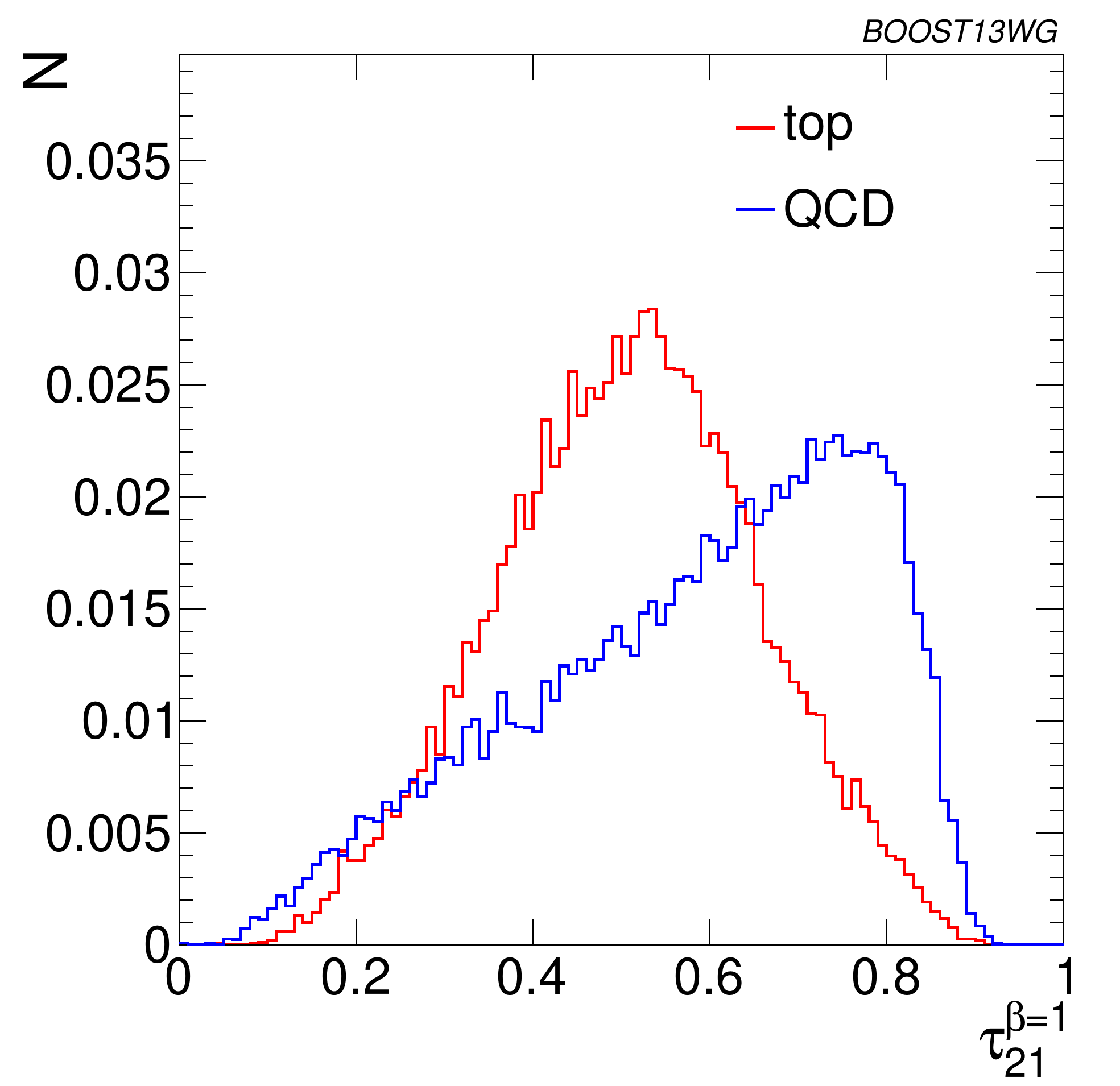}}
\subfigure[\tautwoone, $R=1.2$]{\includegraphics[width=0.245\textwidth]{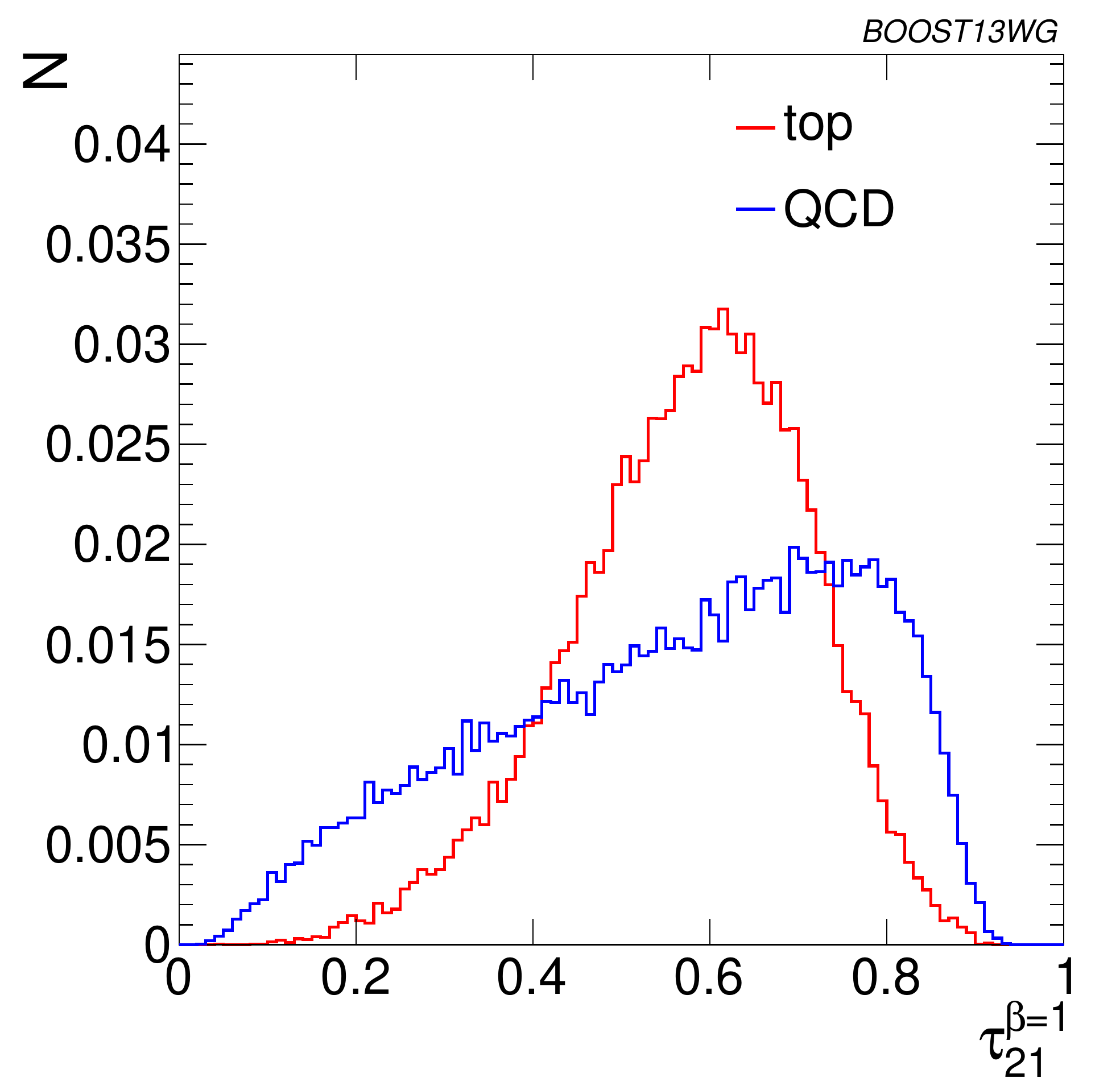}}
\subfigure[\tauthreetwo, $R=0.4$]{\includegraphics[width=0.245\textwidth]{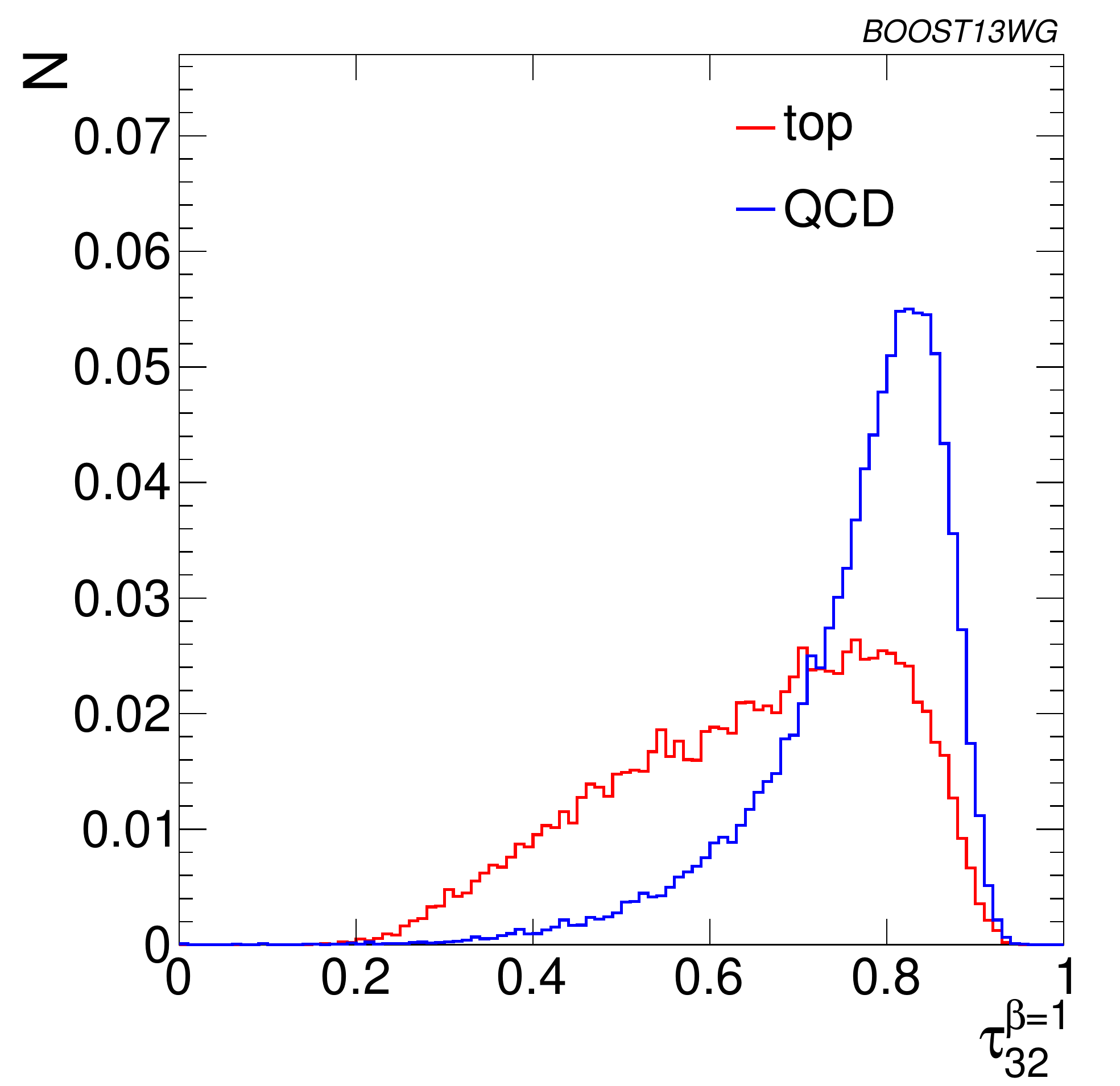}}
\subfigure[\tauthreetwo, $R=1.2$]{\includegraphics[width=0.245\textwidth]{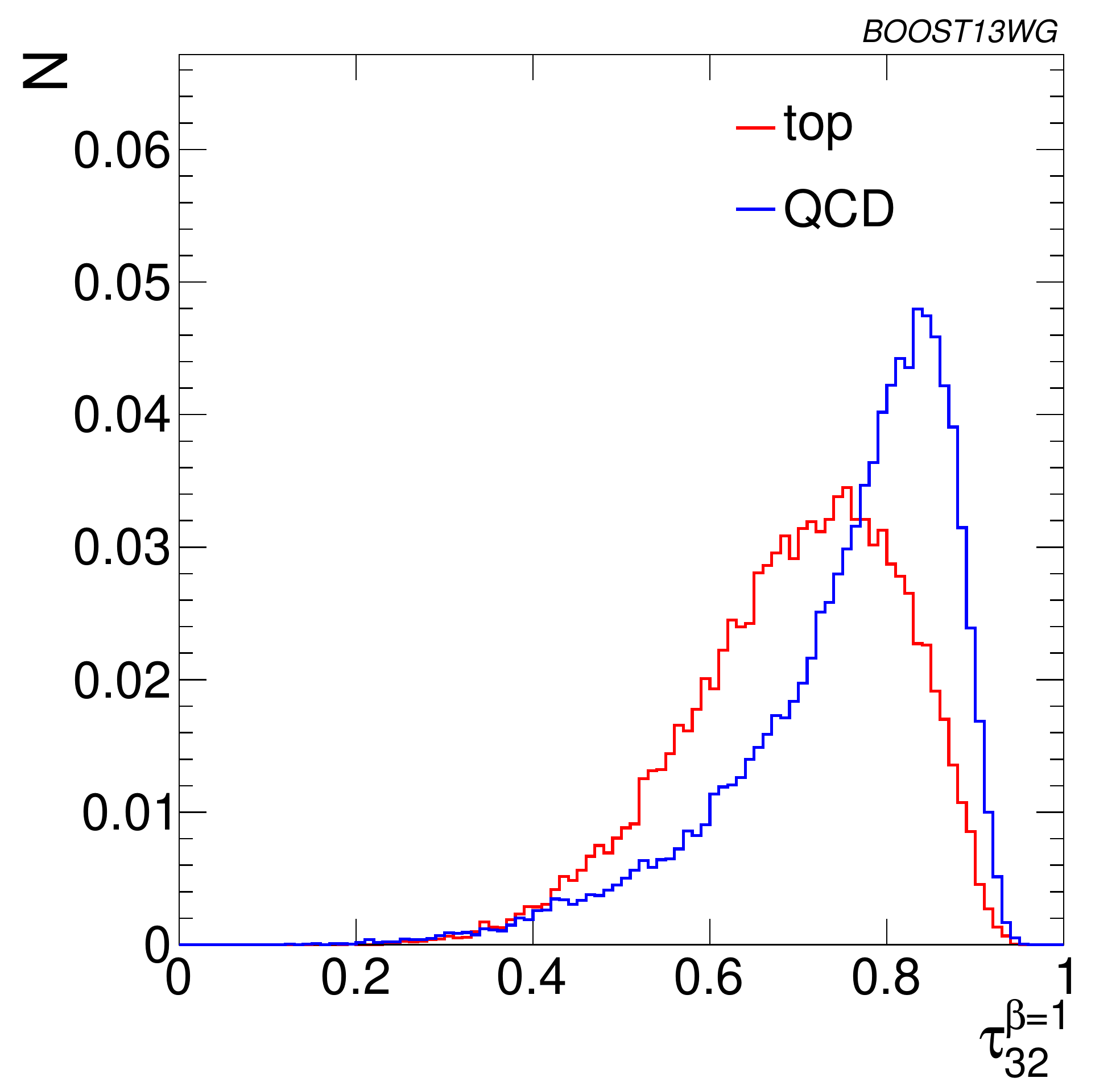}}\\
\subfigure[$\Gamma_{\rm Qjet}$, $R=0.4$]{\includegraphics[width=0.245\textwidth]{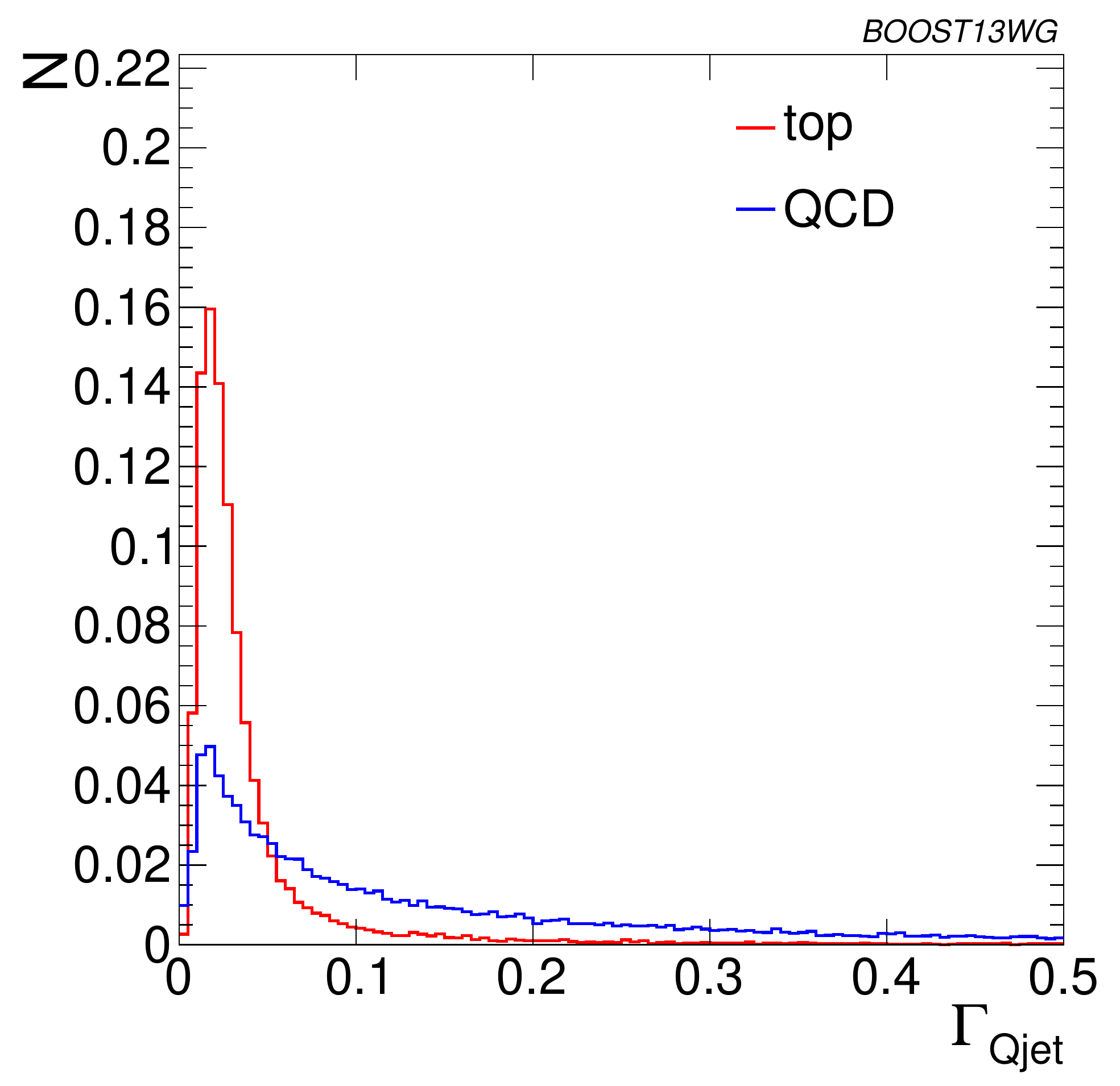}}
\subfigure[$\Gamma_{\rm Qjet}$, $R=1.2$]{\includegraphics[width=0.245\textwidth]{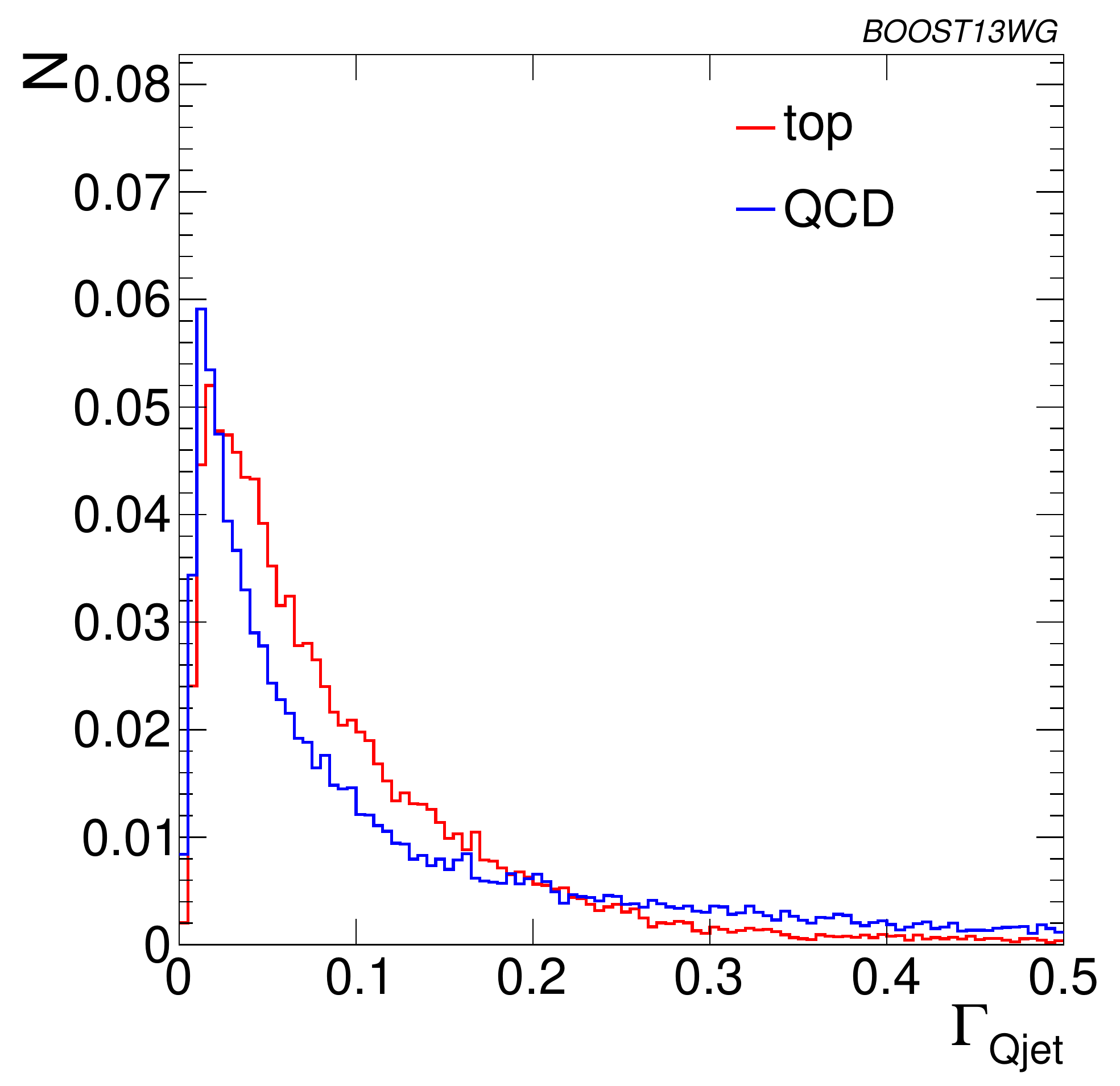}}
\caption{Comparison of various shape variables in the \pt = 1.5-1.6 \TeV bin and different values of the anti-$k_{\rm T}$ radius $R$.}
\label{fig:C_comparison_R}
\end{figure*}

In Figures~\ref{fig:Rcomparison_singleshape_top} and~\ref{fig:Rcomparison_singletopmass_top} we directly compare ROC curves for jet-shape variable performance and top-mass performance, respectively, for three different jet radii  within the \pt = 1.5-1.6 TeV bin. Again, the input parameters of the taggers, groomers and shape variables are separately optimized for each jet radius. We can see from these figures that most of the top-tagging variables, both shape and reconstructed top mass, perform best for smaller radius, as was generally observed in the case of $W$-tagging in Section~\ref{sec:wtagging}. This is likely because, at such high \pt, most of the radiation from the top quark is confined within $R=0.4$, and having a larger jet radius makes the variable more susceptible to contamination from the underlying event and other uncorrelated radiation. In Figure~\ref{fig:C_comparison_R}, we compare the individual top signal and QCD background distributions for each shape variable considered in the \pt = 1.5-1.6 TeV bin for the various jet radii. 
 In Figures~\ref{fig:C_comparison_R} (a) to (h) the distributions for both signal and background broaden with increasing $R$, degrading the discriminating power. 
For $C_2^{\beta=1}$ and $C_3^{\beta=1}$, the background distributions are shifted to larger values as well. 
For the variable $\Gamma_{\rm Qjet}$, as already discussed for increasing \pt (and in  Section~\ref{sec:wtagging}) the behavior with increasing $R$ is a bit
more complicated, with the QCD jets becoming less volatile and the signal jets more volatile, \textit{i.e.}, the two volatility distributions become more
similar as we move from Figure~\ref{fig:C_comparison_R} (i) to Figure~\ref{fig:C_comparison_R} (j).  So again 
the discriminating power decreases with increasing $R$. 
The main exception is   for $C_3^{\beta=1}$, which performs optimally at $R=0.8$; in this case, the signal and background coincidentally 
happen to have the same distribution around $R=0.4$, and so $R=0.8$ gives better discrimination.

\subsection{Performance of Multivariable Combinations}
We now consider various BDT combinations of the single variables considered in the last section, using the techniques described in Section~\ref{sec:multivariate}. In particular, we consider the performance of individual taggers such as the JH tagger and HEPTopTagger, which output information about the top and $W$ candidate masses and the helicity angle; for each tagger, all three output variables are combined in a BDT. For trimming and pruning, the output candidate $\wmass$ and $\topmass$ are combined in a BDT. Finally, we consider the combination of the full set of outputs of each of the above taggers/groomers with the shape variables, as well also a combination of the outputs of the HEPTopTagger and JH tagger. This allows us to determine the degree of complementary information in taggers/groomers and shape variables, as well as between the top tagging algorithms themselves. For all variables with tuneable input parameters, we scan and optimize over realistic values of such parameters, as described in Section~\ref{sec:topmethod}.

\begin{figure*}
\centering
{\includegraphics[width=0.68\textwidth]{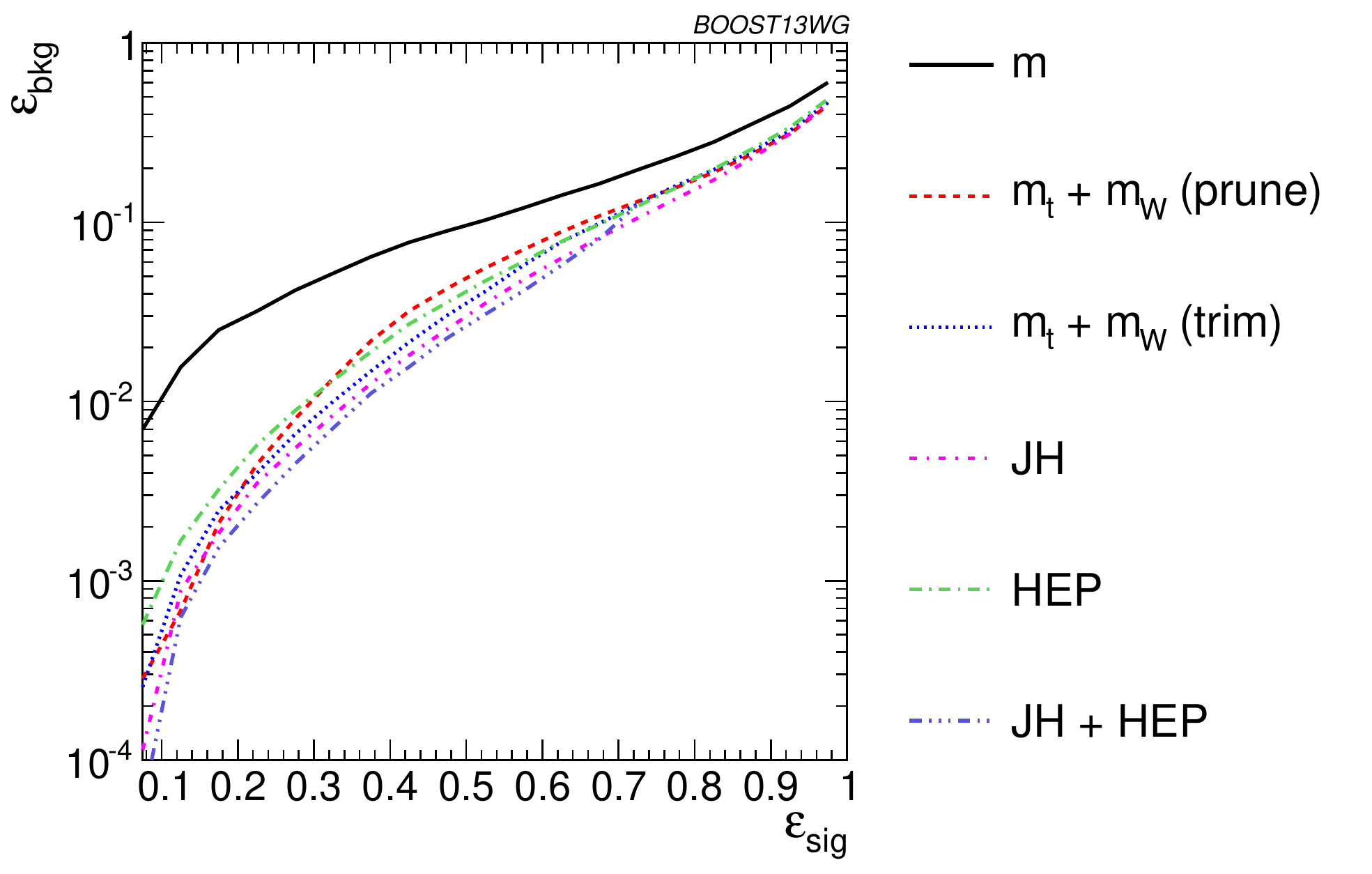}}
\caption{The performance of the various taggers in the $\pt = 1-1.1$ TeV bin using the anti-\kT R=0.8 algorithm. For the groomers a BDT combination of the reconstructed \topmass~and \wmass~are used. Also shown is a multivariable combination of all of the JH and HEPTopTagger outputs. The ungroomed mass performance is shown for comparison.}
\label{fig:pt1000_allcompare_AKt_R08_TG}
\end{figure*}

In Figure~\ref{fig:pt1000_allcompare_AKt_R08_TG}, we directly compare the performance of the HEPTopTagger, the JH tagger, trimming, and pruning, in the $\pt = 1-1.1$ TeV bin with $R=0.8$, where both \topmass~and \wmass~are used in the groomers. Generally, we find that pruning, which does not naturally incorporate subjets into the algorithm, does not perform as well as the others. Interestingly, trimming, which does include a subjet-identification step, performs comparably to the standard HEPTopTagger over much of the range, possibly due to the background-shaping observed in Section \ref{sec:single_variable}, although this can change with recent proposed updates to the HEPTopTagger  \cite{Anders:2013oga,Kasieczka:2015jma}. By contrast, the JH tagger outperforms the other standard algorithms. To determine whether there is complementary information in the mass outputs from different top taggers, we also consider in Figure~\ref{fig:pt1000_allcompare_AKt_R08_TG} a multivariable combination of all of the JH and HEPTopTagger outputs. The maximum efficiency of the combined JH and HEPTopTaggers is limited, as some fraction of signal events inevitably fails either one or other of the taggers. We do see a 20-50\% improvement in performance when combining all outputs, which suggests that the different algorithms used to identify the top and $W$ for different taggers contains complementary information.

\begin{figure*}
\centering
\subfigure[HEPTopTagger + Shape]{\includegraphics[width=0.48\textwidth]{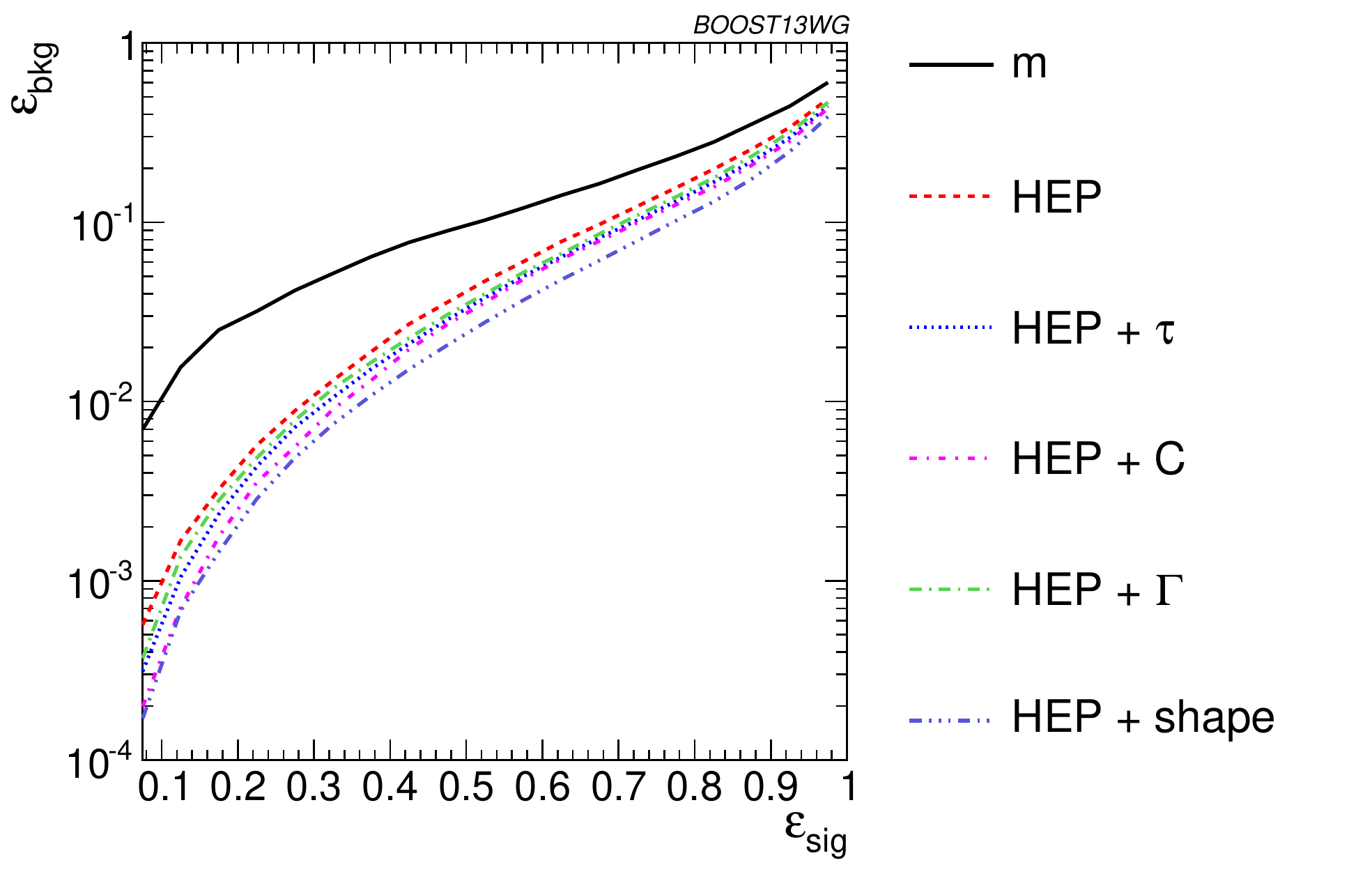}}
\subfigure[Johns Hopkins Tagger + shape]{\includegraphics[width=0.48\textwidth]{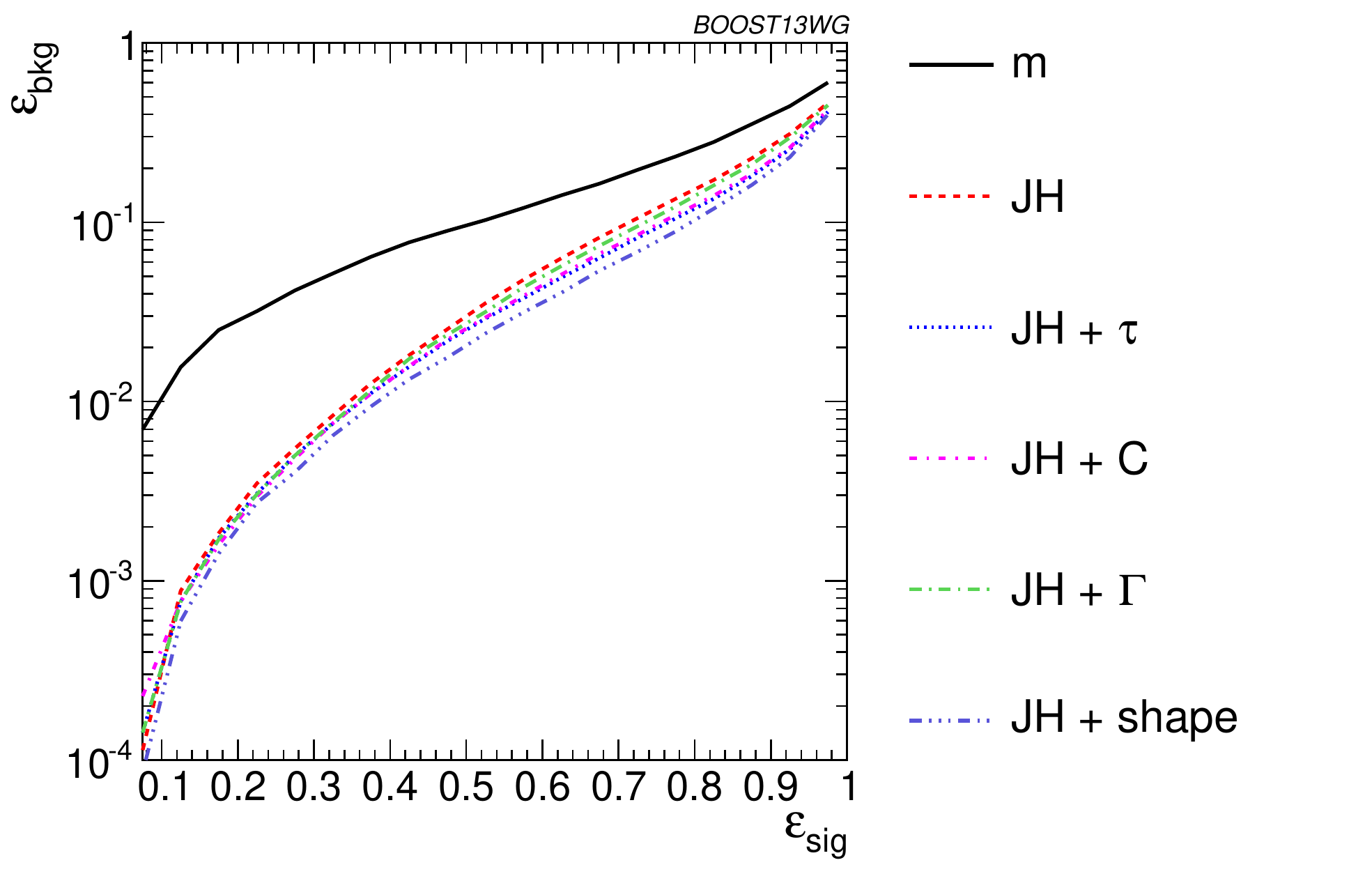}}
\subfigure[HEP vs.~JH comparison (incl. shape)]{\includegraphics[width=0.48\textwidth]{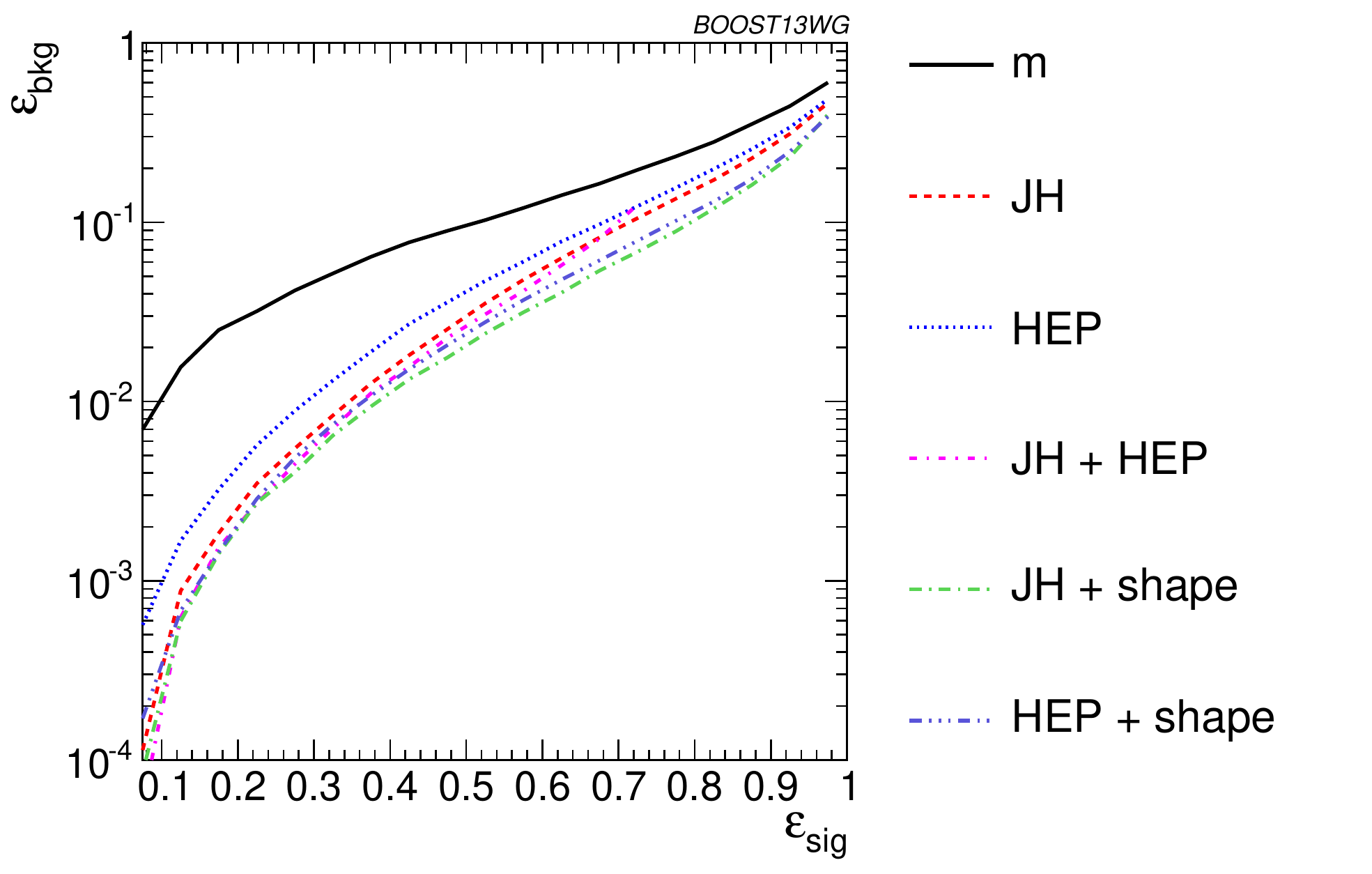}\label{fig:pt1000_allcompare_AKt_R08_TagSh_Comp}}
\caption{The performance of BDT combinations of the JH and HepTopTagger outputs with various shape variables in the $\pt = 1-1.1$ TeV bin using the anti-\kT $R=0.8$ algorithm. Taggers are combined with the following shape variables: $\tau_{21}^{\beta=1}+\tau_{32}^{\beta=1}$, $C_{2}^{\beta=1}+C_{3}^{\beta=1}$, $\Gamma_{\rm Qjet}$, and all of the above (denoted ``shape'').}
\label{fig:pt1000_allcompare_AKt_R08_TagSh}
\end{figure*}

In Figure~\ref{fig:pt1000_allcompare_AKt_R08_TagSh} we present the results for multivariable combinations of the top tagger outputs with and without shape variables. We see that, for both the HEPTopTagger and the JH tagger, the shape variables contain additional information uncorrelated with the masses and helicity angle, and give on average a factor 2-3 improvement in signal discrimination. We see that, when combined with the tagger outputs, both the energy correlation functions $C_2+C_3$ and the $N$-subjettiness ratios $\tau_{21}+\tau_{32}$ give comparable performance, while $\Gamma_{\rm Qjet}$ is slightly worse; this is unsurprising, as Qjets accesses shape information in a more indirect way from other shape variables. Combining all shape variables with a single top tagger provides even greater enhancement in discrimination power. We directly compare the performance of the JH and HEPTopTaggers in Figure~\ref{fig:pt1000_allcompare_AKt_R08_TagSh_Comp}. Combining the taggers with shape information nearly erases the difference between the tagging methods observed in Figure~\ref{fig:pt1000_allcompare_AKt_R08_TG}; this indicates that combining the shape information with the HEPTopTagger identifies the differences between signal and background missed by the standard tagger alone. This also suggests that further improvement to discriminating power may be minimal, as various multivariable combinations converge to within a factor of 20\% or so.

\begin{figure*}
\centering
\subfigure[Pruning + Shape]{\includegraphics[width=0.48\textwidth]{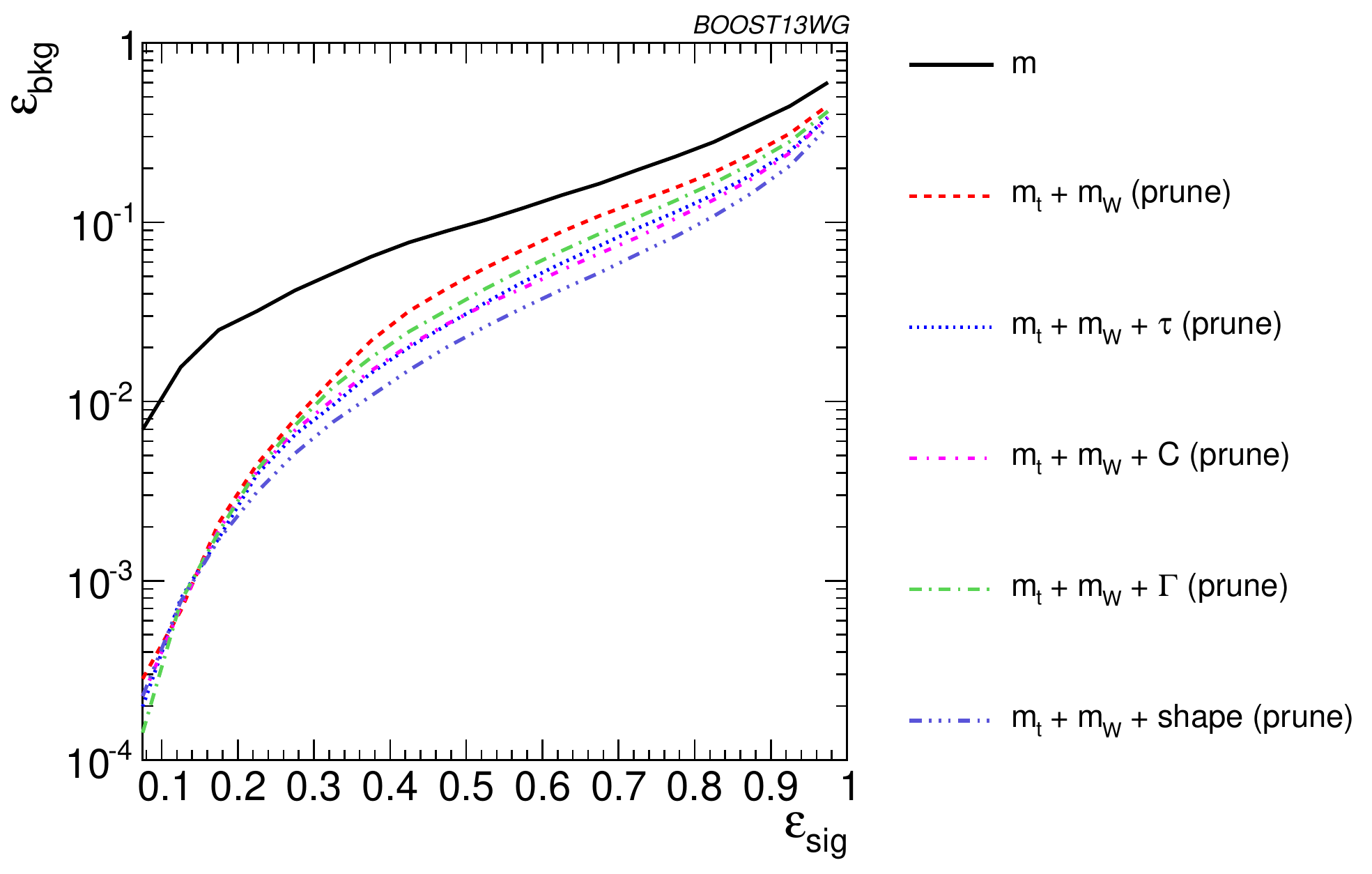}}
\subfigure[Trimming + Shape]{\includegraphics[width=0.48\textwidth]{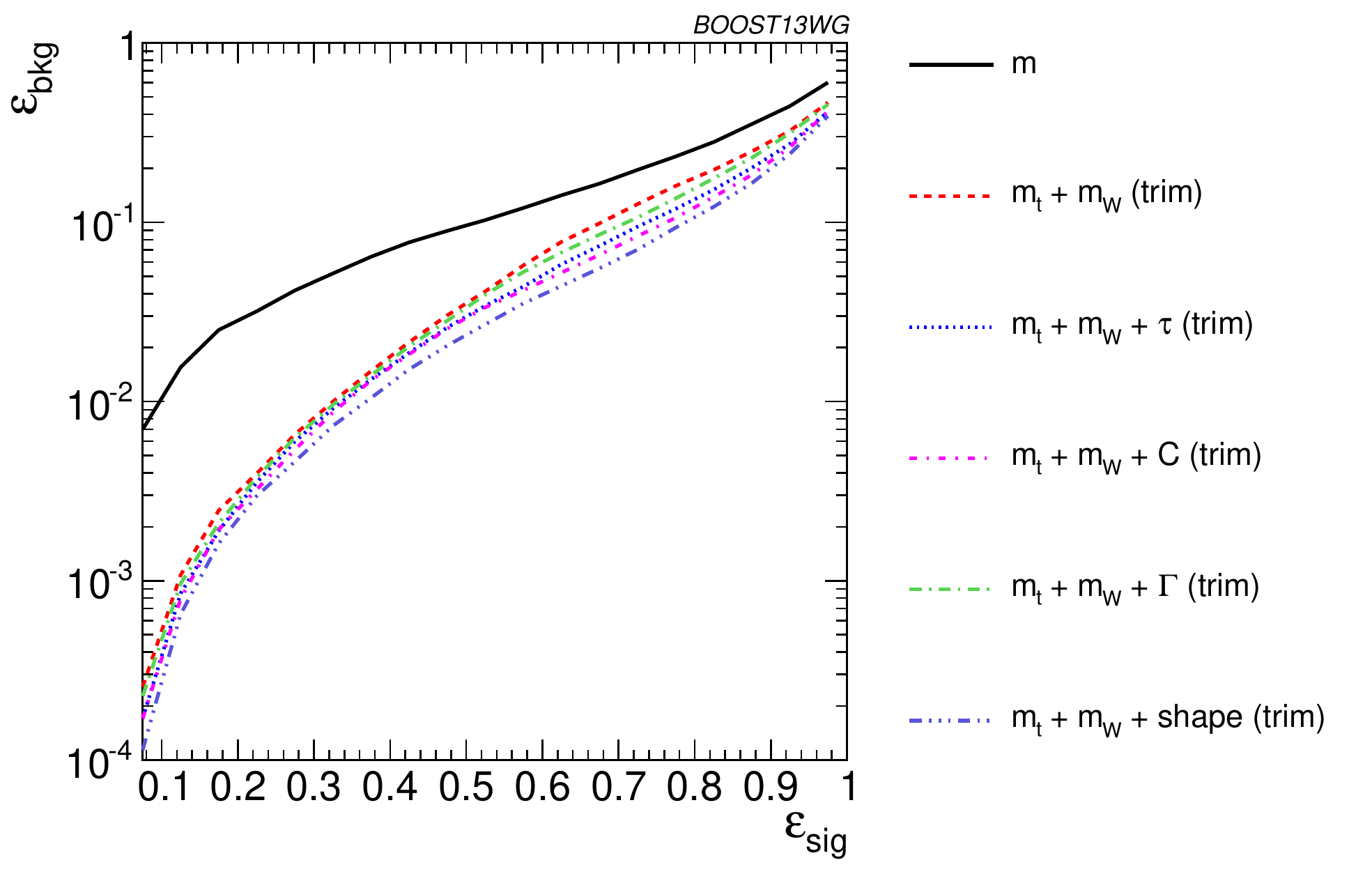}}
\subfigure[Trim vs.~Prune comparison (incl. shape)]{\includegraphics[width=0.48\textwidth]{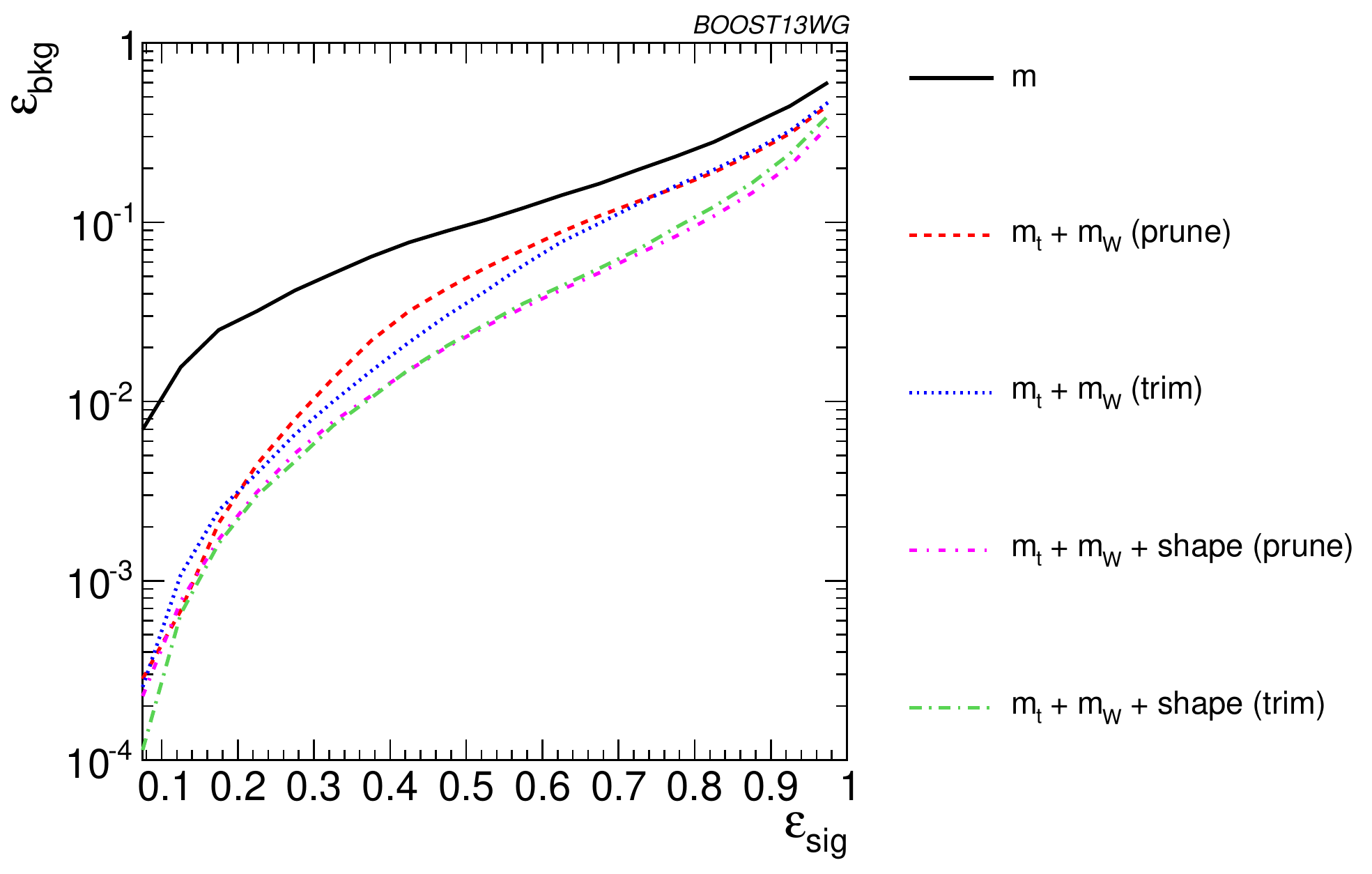}\label{fig:pt1000_allcompare_AKt_R08_GroomSh_Comp}}
\caption{The performance of the BDT combinations of the trimming and pruning outputs with various shape variables in the $\pt = 1-1.1$ TeV bin using the anti-\kT $R=0.8$ algorithm. Groomer mass outputs are combined with the following shape variables: $\tau_{21}^{\beta=1}+\tau_{32}^{\beta=1}$, $C_{2}^{\beta=1}+C_{3}^{\beta=1}$, $\Gamma_{\rm Qjet}$, and all of the above (denoted ``shape'').}
\label{fig:pt1000_allcompare_AKt_R08_GroomSh}
\end{figure*}

In Figure~\ref{fig:pt1000_allcompare_AKt_R08_GroomSh} we present the results for multivariable combinations of groomer outputs with and without shape variables. As with the tagging algorithms, combinations of groomers with shape variables improves their discriminating power; combinations with $\tau_{32}+\tau_{21}$ perform comparably to those with $C_3+C_2$, and both of these are superior to combinations with the mass volatility, $\Gamma_{\rm Qjet}$. Substantial further improvement is possible by combining the groomers with all shape variables. Not surprisingly, the taggers that lag behind in performance enjoy the largest gain in signal-background discrimination with the addition of shape variables. Once again, in Figure~\ref{fig:pt1000_allcompare_AKt_R08_GroomSh_Comp}, we find that the differences between pruning and trimming are erased when combined with shape information.

\begin{figure*}
\centering
{\includegraphics[width=0.68\textwidth]{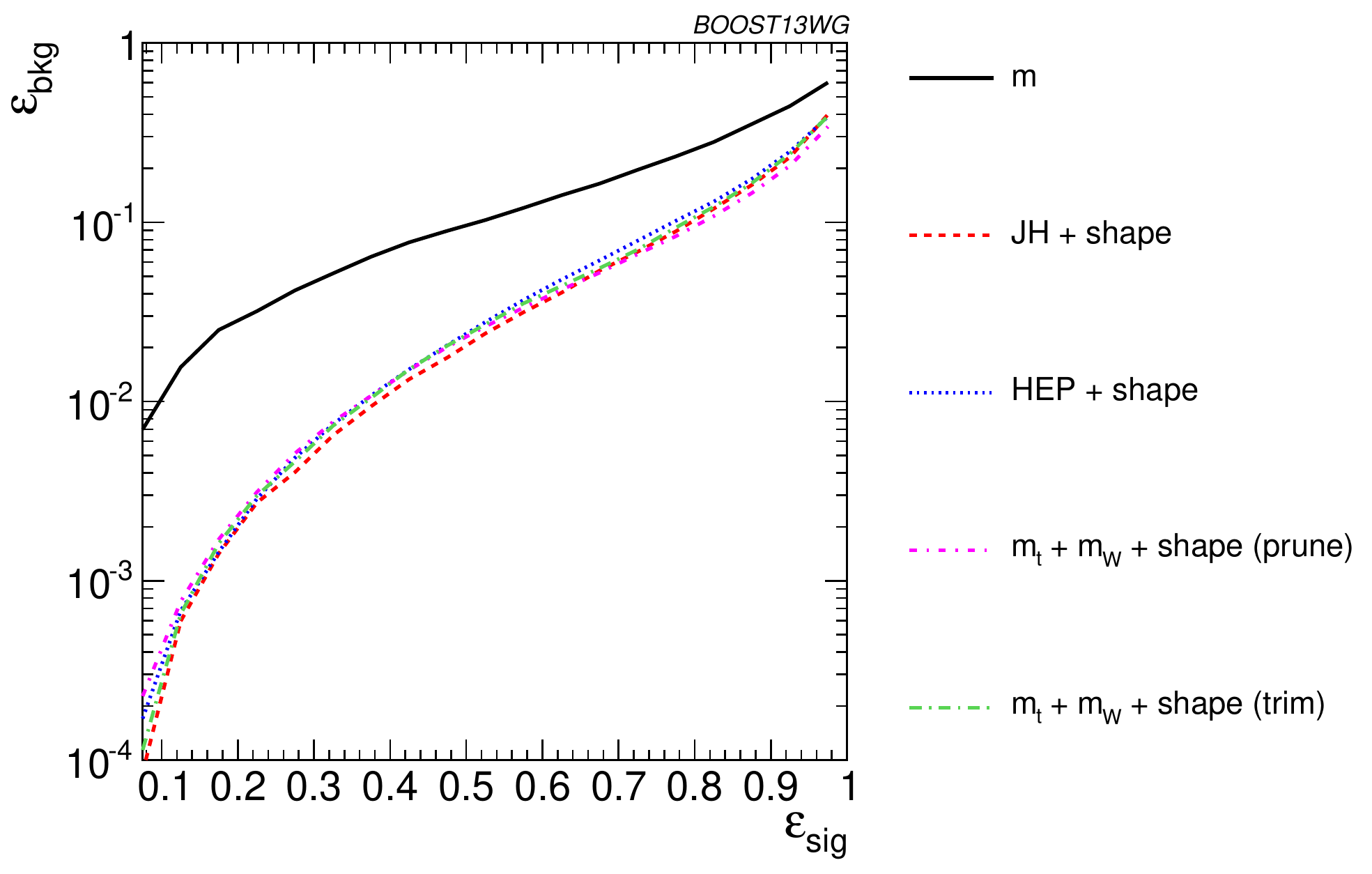}}
\caption{Comparison of the performance of the BDT combinations of all the groomer/tagger outputs with all the available shape variables in the $\pt = 1-1.1$ TeV bin using the anti-\kT R=0.8 algorithm. Tagger/groomer outputs are combined with all of the following shape variables: $\tau_{21}^{\beta=1}+\tau_{32}^{\beta=1}$, $C_{2}^{\beta=1}+C_{3}^{\beta=1}$, $\Gamma_{\rm Qjet}$.}
\label{fig:pt1000_allcompare_AKt_R08_Final}
\end{figure*}

Finally, in Figure~\ref{fig:pt1000_allcompare_AKt_R08_Final}, we compare the performance of each of the tagger/groomers when their outputs are combined with all of the shape variables considered. One can see that the discrepancies between the performance of the different taggers/groomers all but vanishes, suggesting perhaps that we are here utilising all available signal-background discrimination information, and that this is the optimal top tagging performance that could be achieved in these conditions. 

\begin{figure*}
\centering
\subfigure[HEPTopTagger]{\includegraphics[width=0.48\textwidth]{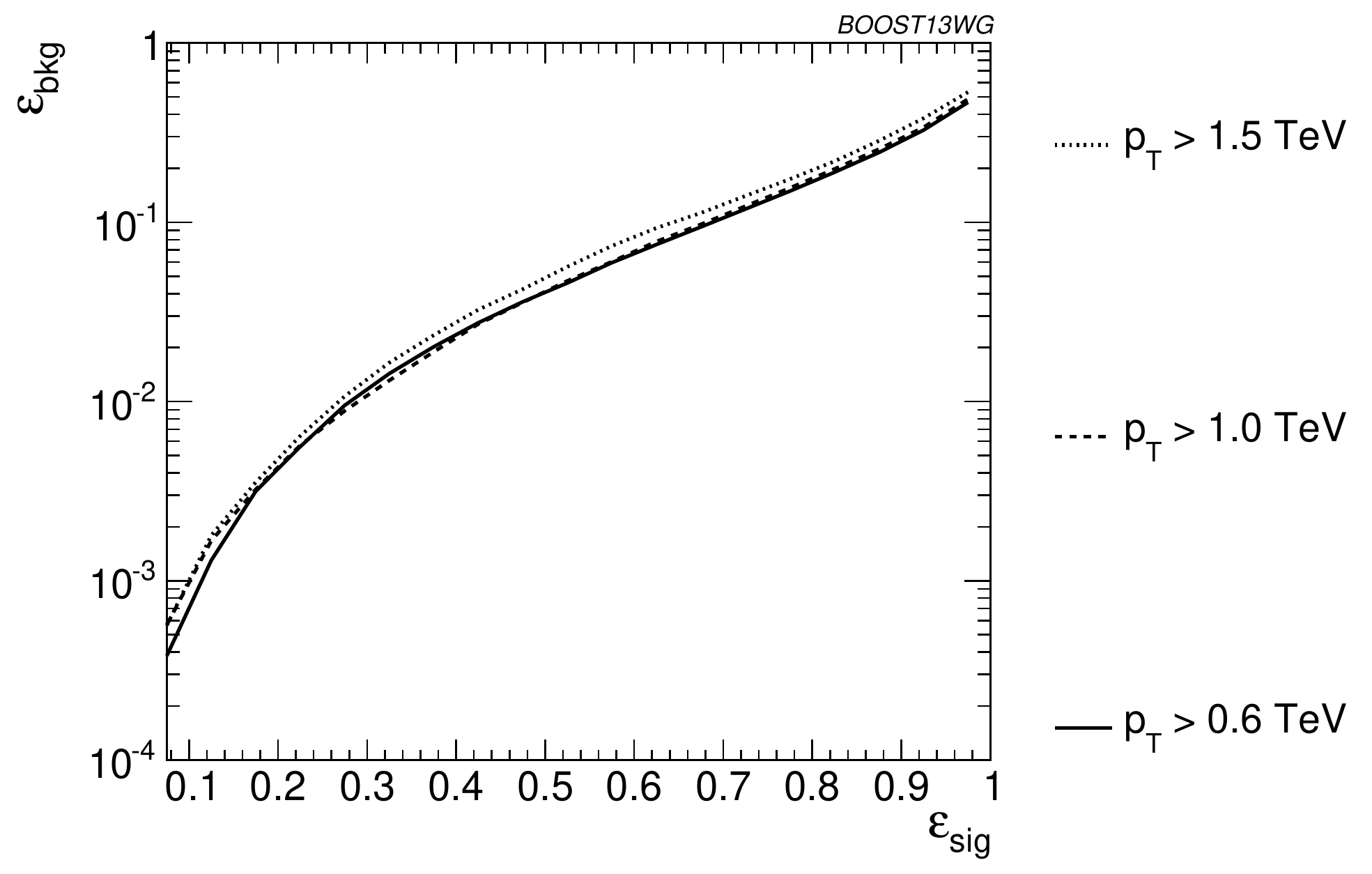}}
\subfigure[Johns Hopkins Tagger]{\includegraphics[width=0.48\textwidth]{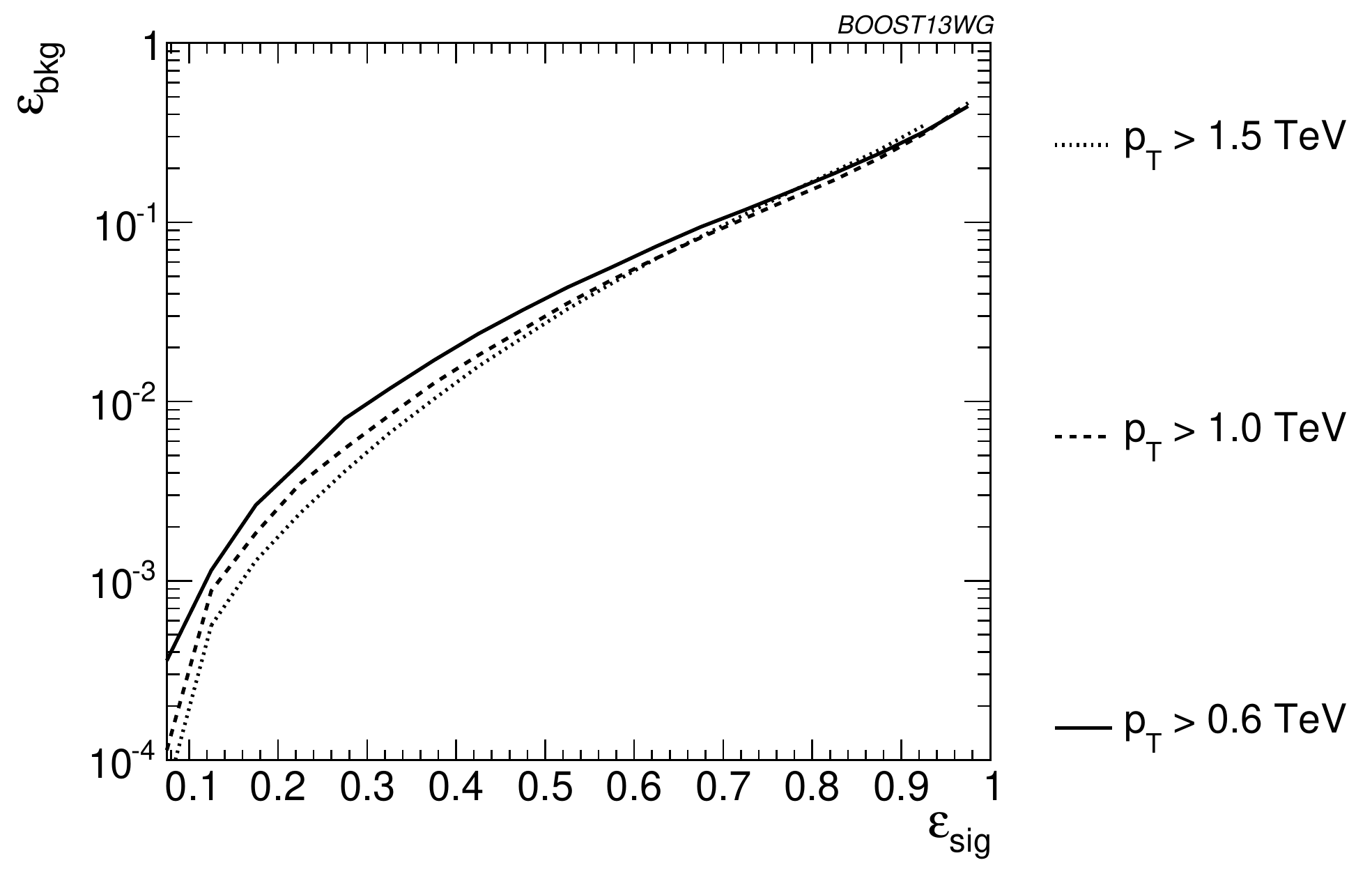}\label{fig:ptcomparison_top_JH}}
\subfigure[Trimming]{\includegraphics[width=0.48\textwidth]{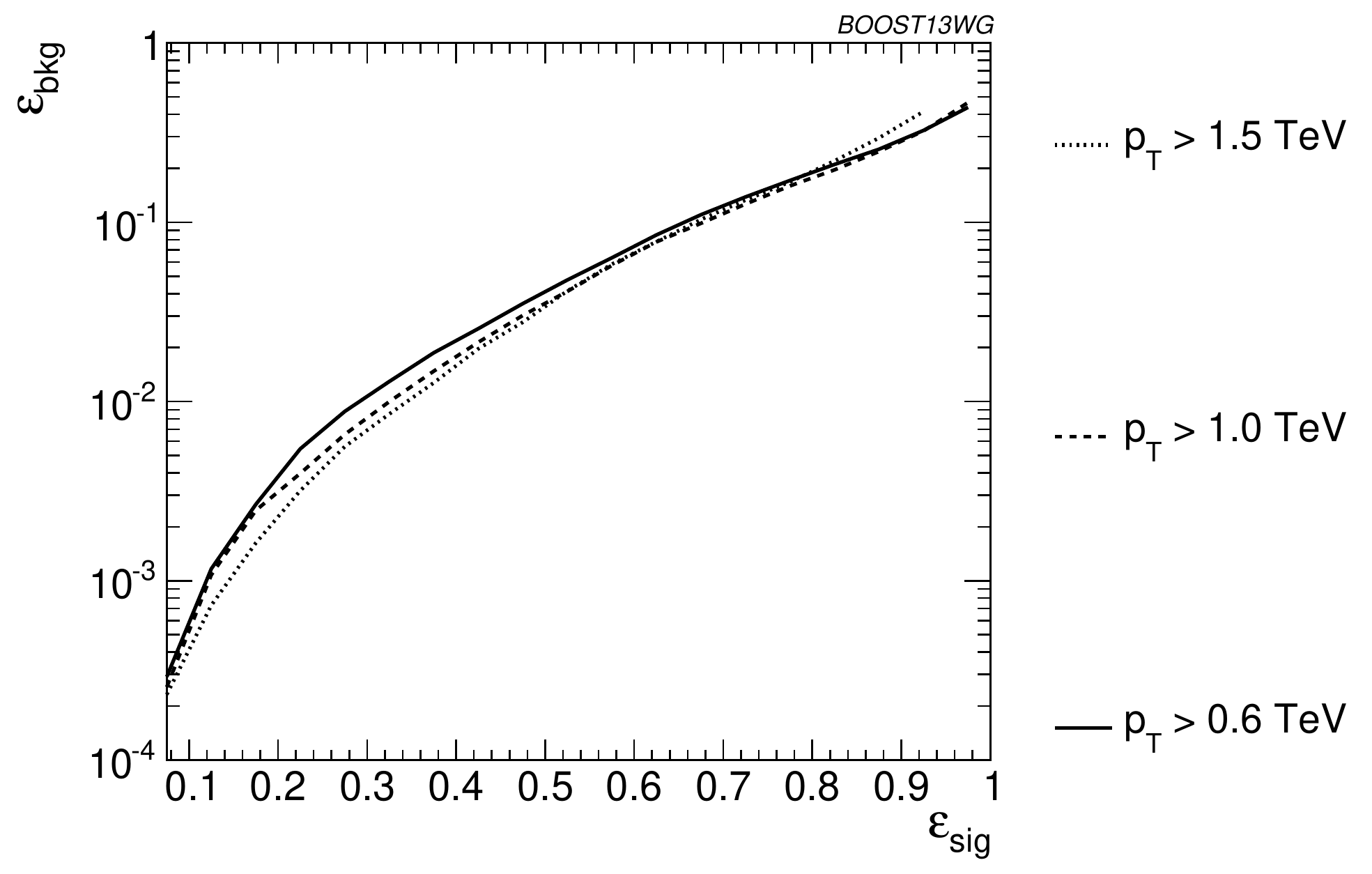}}
\subfigure[Pruning]{\includegraphics[width=0.48\textwidth]{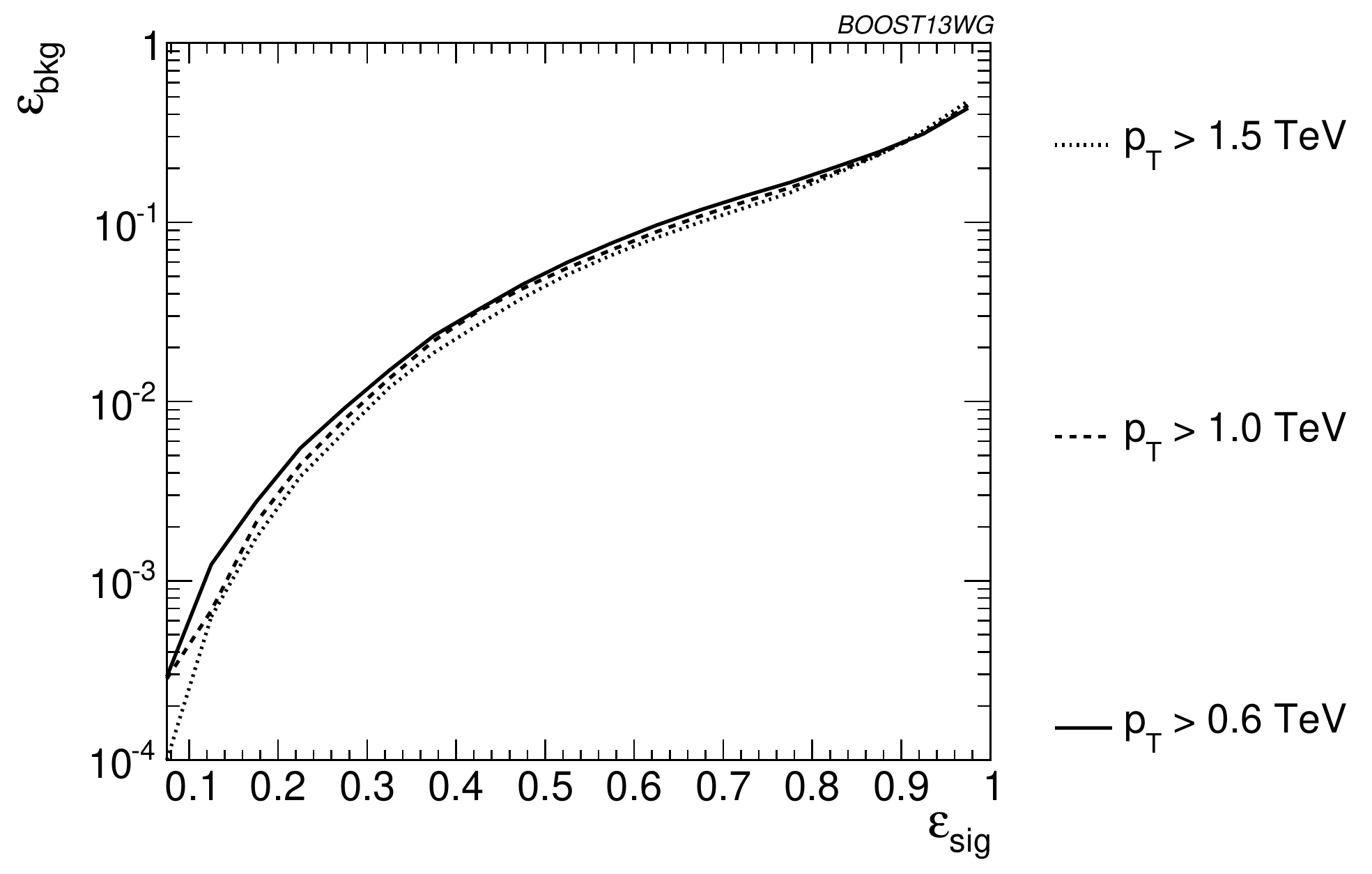}}
\caption{Comparison at different \pt of the performance of various top tagging/grooming algorithms using the anti-\kT $R=0.8$ algorithm. For each tagger/groomer, all output variables are combined in a BDT. }
\label{fig:ptcomparison_top}
\end{figure*}

Up to this point, we have  considered only the combined multivariable performance in the \pt =  1.0-1.1 TeV bin with jet radius $R=0.8$. We now compare the BDT combinations of tagger outputs, with and without shape variables, at different $\pt$. The taggers are optimized over all input parameters for each choice of \pt and signal efficiency. As with the single-variable study, we consider \antikt jets clustered with $R=0.8$ and compare the outcomes in the \pt = 500-600 \GeV, \pt = 1-1.1 \TeV, and \pt = 1.5-1.6 \TeV bins. The comparison of the taggers/groomers is shown in Figure~\ref{fig:ptcomparison_top}. The behaviour with \pt is qualitatively similar to the behaviour of the \topmass~variable for each tagger/groomer shown in Figure~\ref{fig:ptcomparison_singletopmass_top}; this suggests that the \pt behaviour of the taggers is dominated by the top-mass reconstruction. As before, the standard HEPTopTagger performance degrades slightly with increased \pt due to the background shaping effect (which may be mitigated by recently proposed updates), while the JH tagger and groomers modestly improve in performance.

\begin{figure*}
\centering
\subfigure[JH+$C_2^{\beta=1}$+$C_3^{\beta=1}$]{\includegraphics[width=0.48\textwidth]{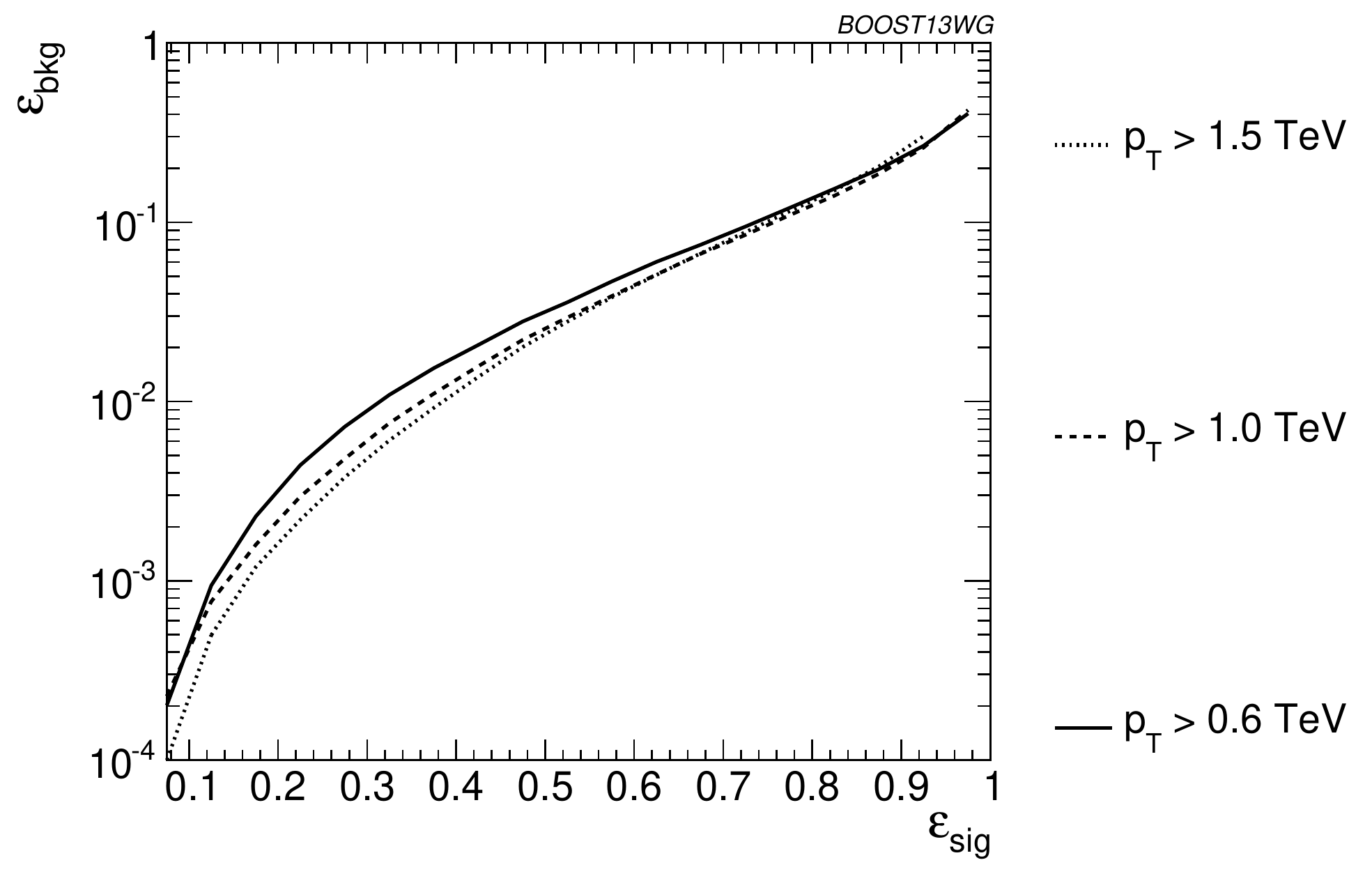}}
\subfigure[JH+\tautwoone+\tauthreetwo]{\includegraphics[width=0.48\textwidth]{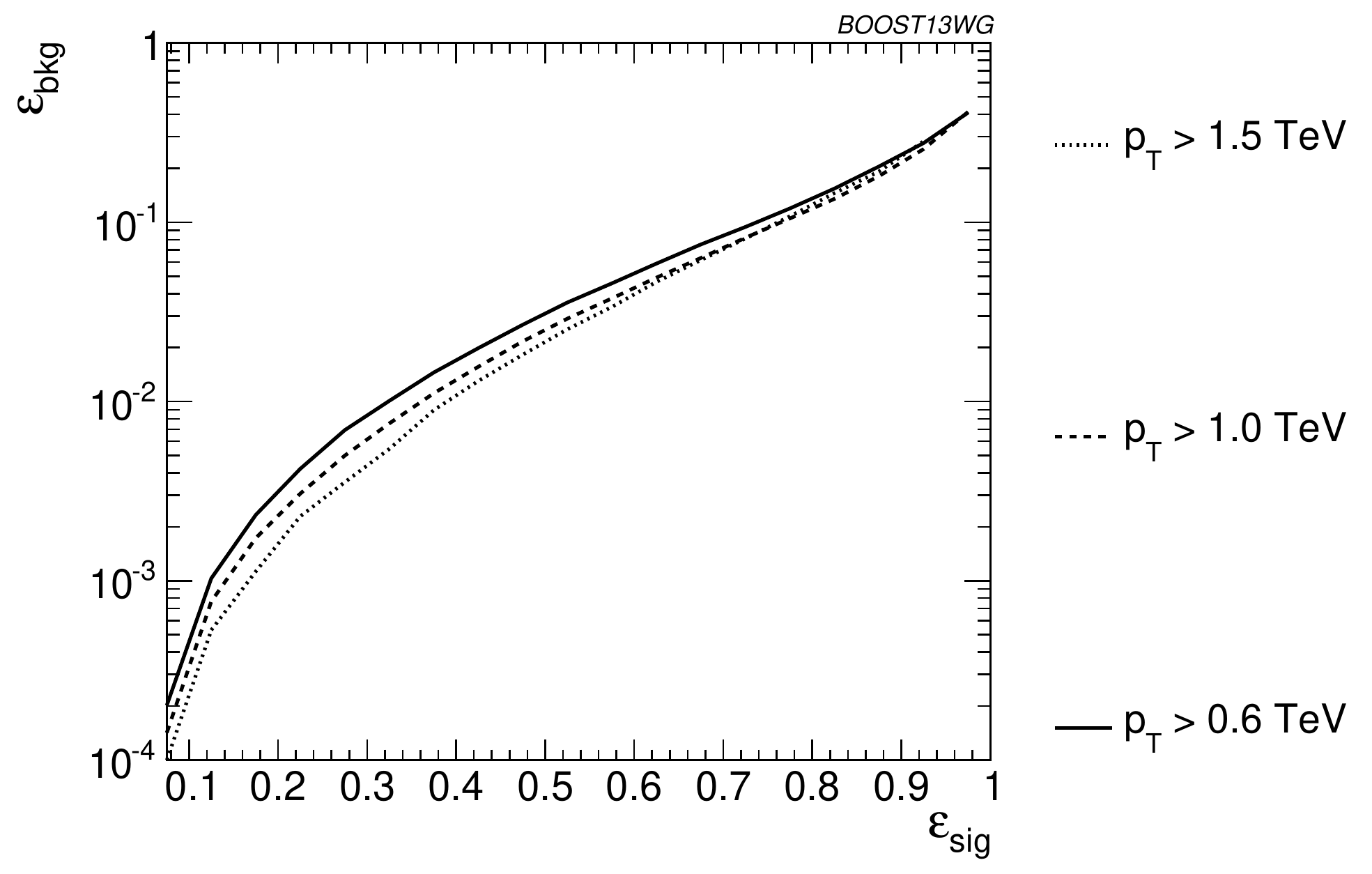}}
\subfigure[JH + $\Gamma_{\rm Qjet}$]{\includegraphics[width=0.48\textwidth]{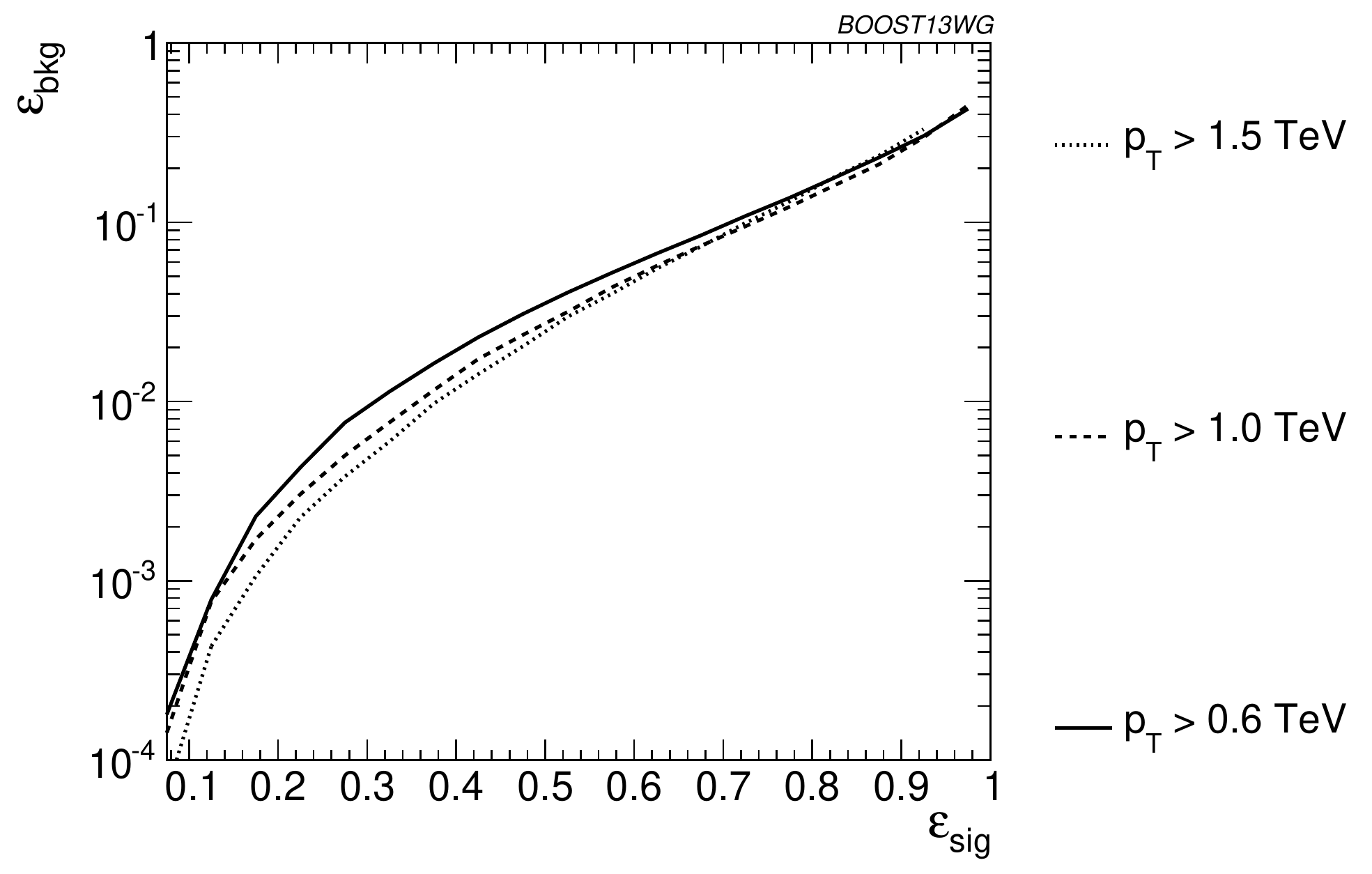}}
\subfigure[JH + all]{\includegraphics[width=0.48\textwidth]{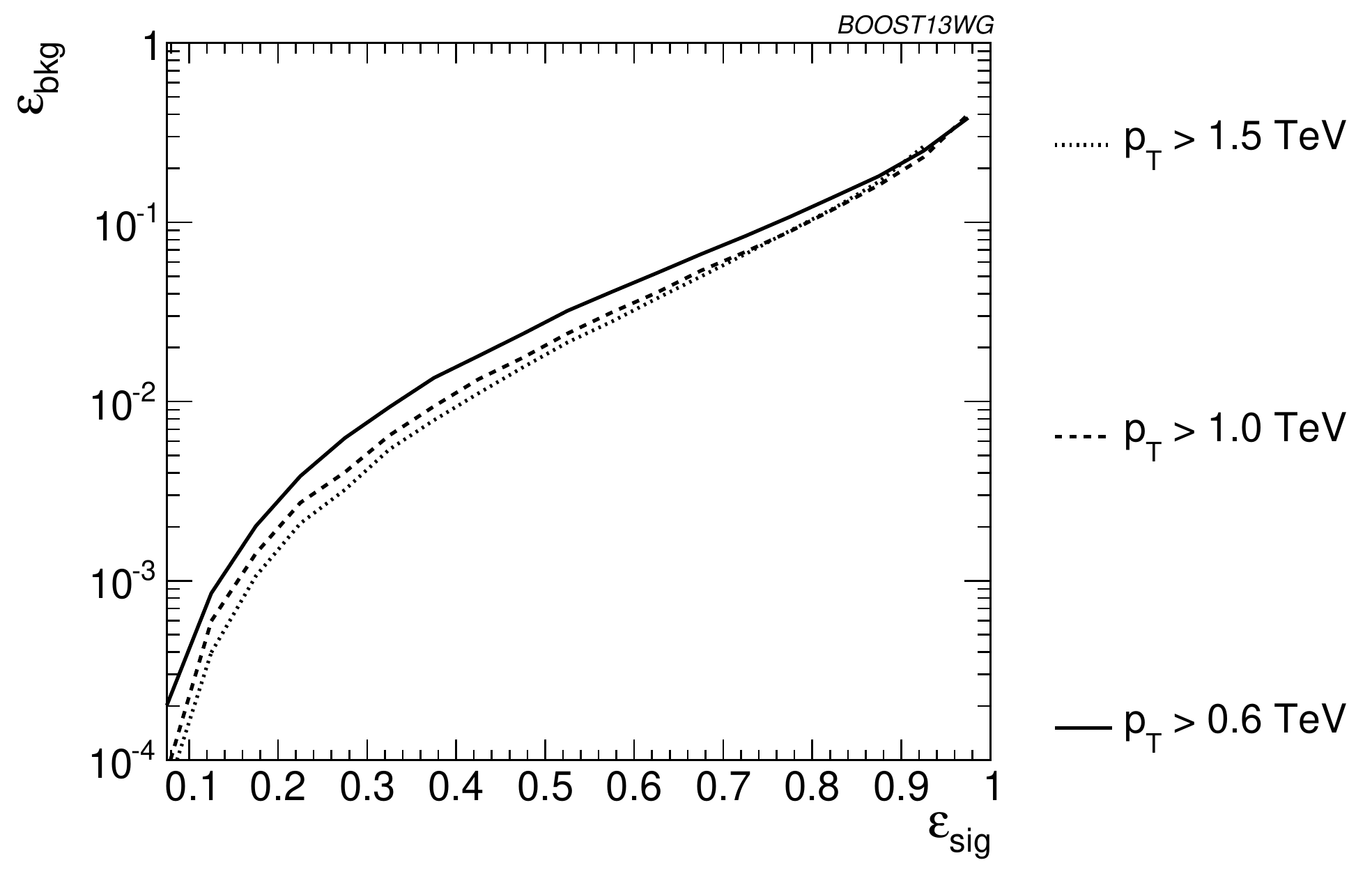}}
\caption{Comparison at different \pt of the performance of the JH tagger using the anti-\kT $R=0.8$ algorithm, where all tagger output variables are combined in a BDT with various shape variables.}
\label{fig:ptcomparison_JH_shape}
\end{figure*}

\begin{figure*}
\centering
\subfigure[HEP+$C_2^{\beta=1}$+$C_3^{\beta=1}$]{\includegraphics[width=0.48\textwidth]{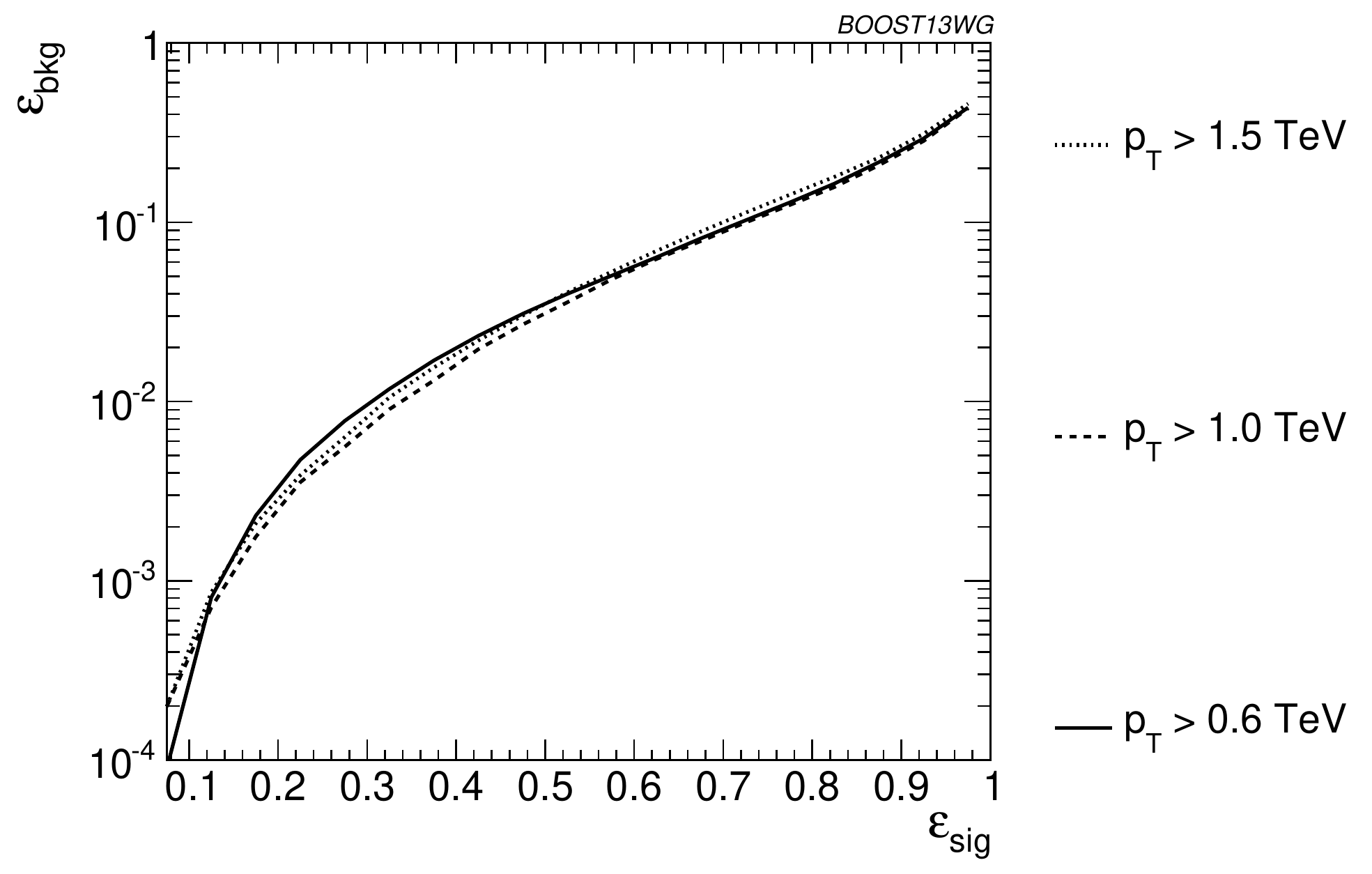}}
\subfigure[HEP+\tautwoone+\tauthreetwo]{\includegraphics[width=0.48\textwidth]{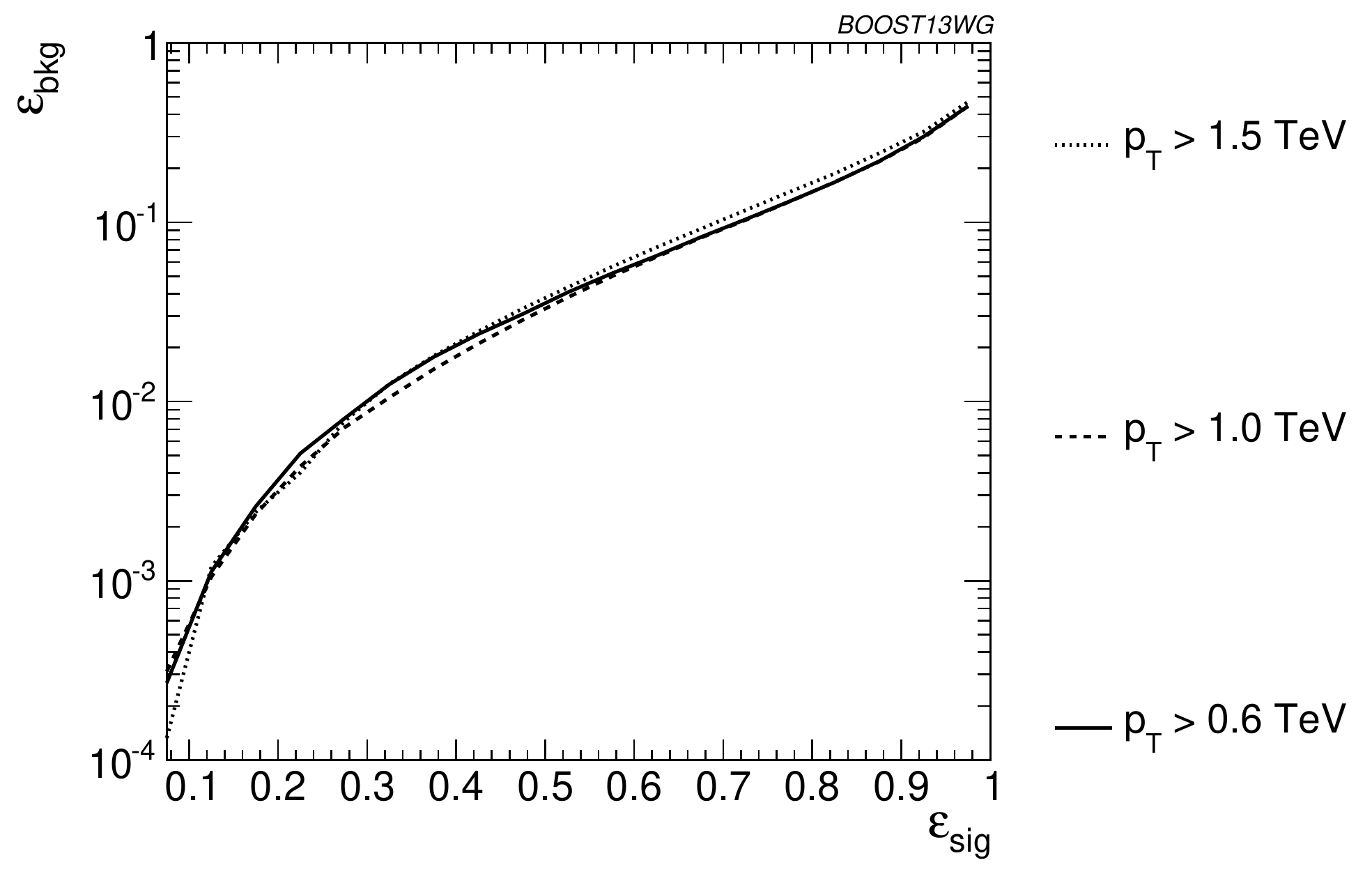}}
\subfigure[HEP + $\Gamma_{\rm Qjet}$]{\includegraphics[width=0.48\textwidth]{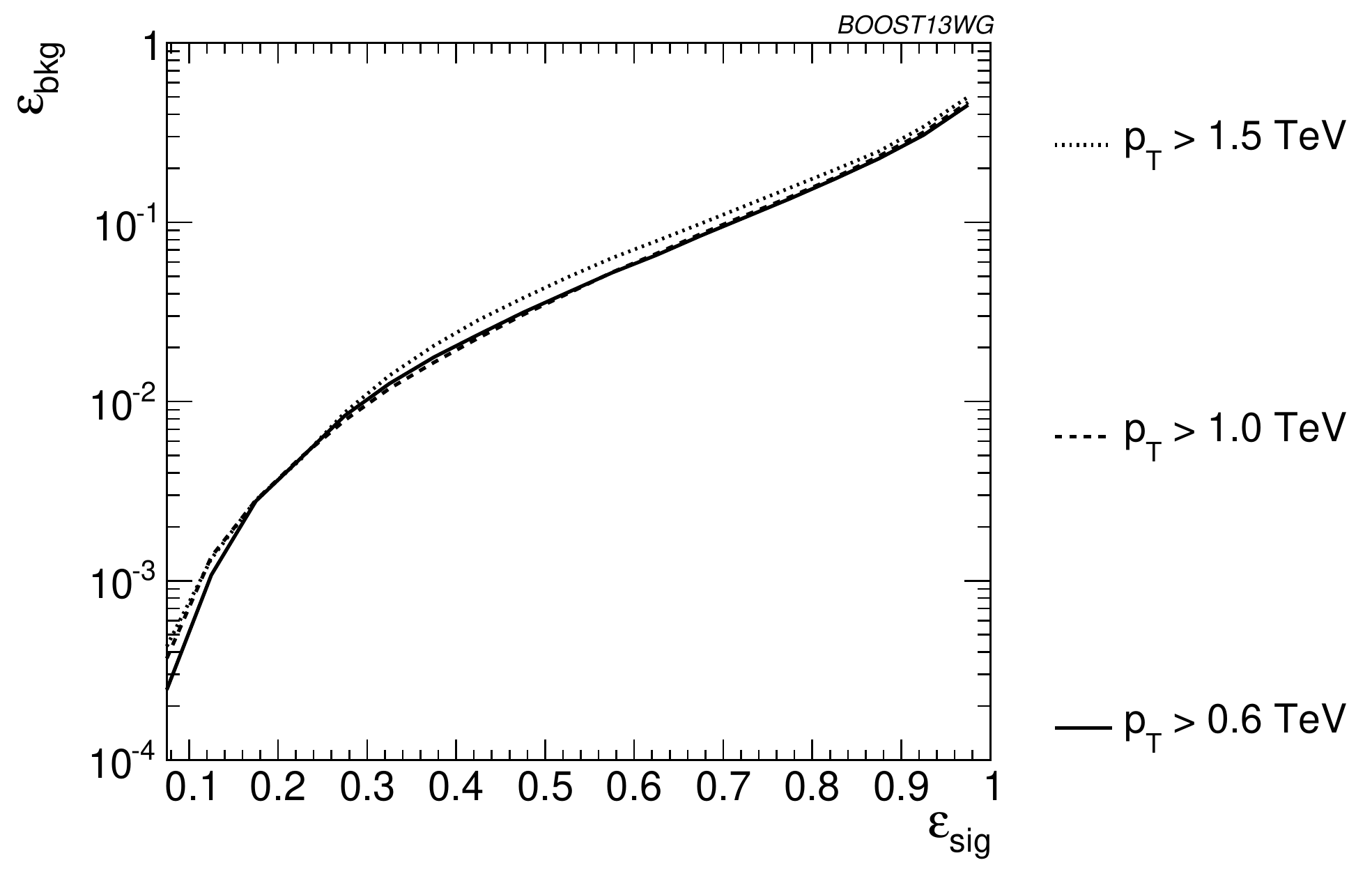}}
\subfigure[HEP + all]{\includegraphics[width=0.48\textwidth]{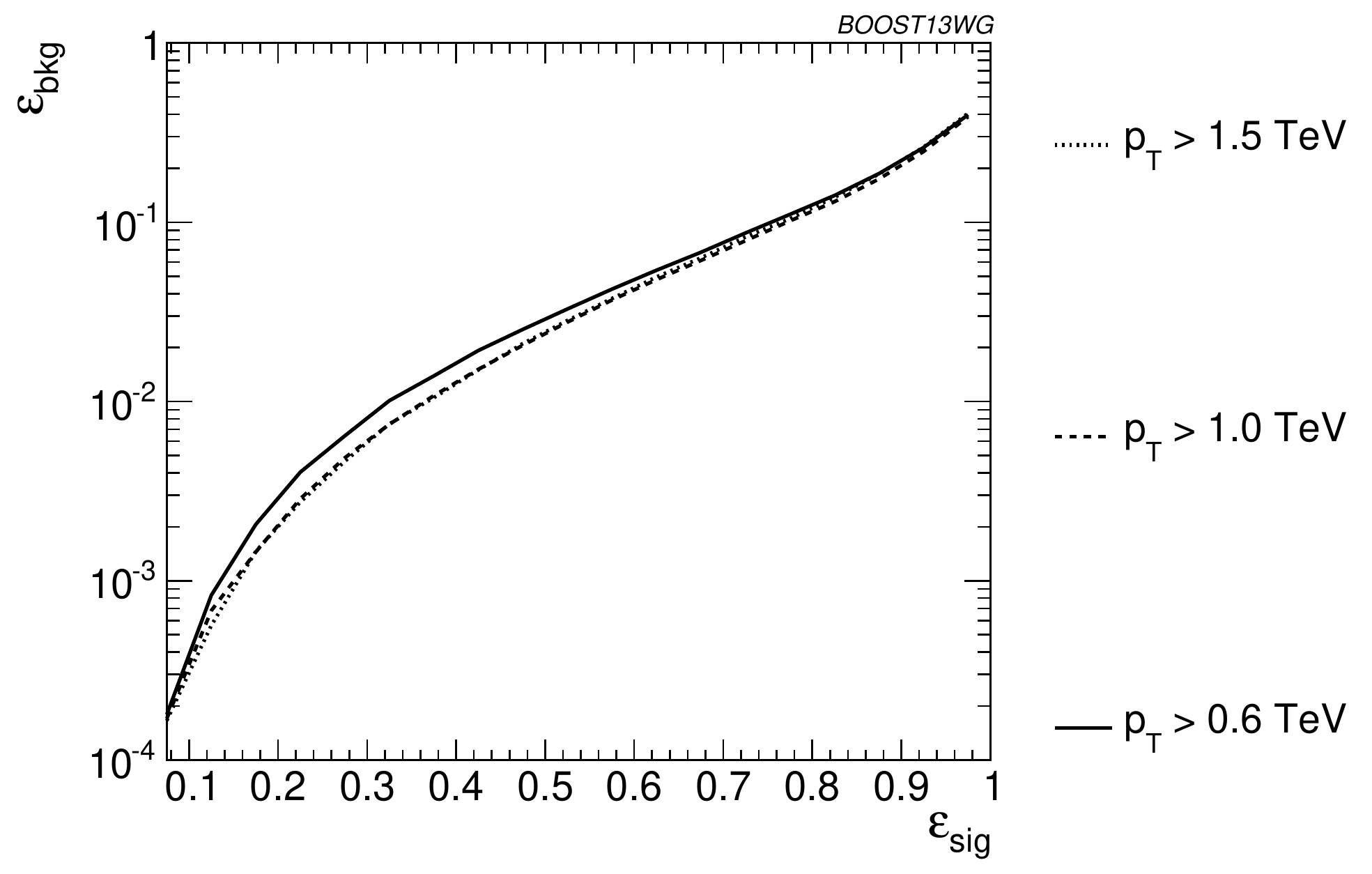}}
\caption{Comparison at different \pt of the performance of the HEPTopTagger using the anti-\kT $R=0.8$ algorithm,  where all tagger output variables are combined in a BDT with various shape variables.}
\label{fig:ptcomparison_HEP_shape}
\end{figure*}

In Figure~\ref{fig:ptcomparison_JH_shape}, we show the $\pt$-dependence of BDT combinations of the JH tagger output combined with shape variables. In terms of \pt dependence, we find that the curves look nearly identical to Figure~\ref{fig:ptcomparison_top_JH}:~the \pt dependence is again dominated by the top-mass reconstruction, and combining the tagger outputs with different shape variables does not substantially change this behavior. Although not shown here, the same behavior is observed for trimming and pruning. By contrast,  the \pt dependence of the HEPTopTagger ROC curves, shown in Figure~\ref{fig:ptcomparison_HEP_shape}, does change somewhat when combined with different shape variables; due to the suboptimal performance of the HEPTopTagger at high \pt in the conventional configuration, we find that combining the HEPTopTagger with $C_3^{\beta=1}$, which in Figure~\ref{fig:ptcomparison_singleshape_top_C3} is seen to have some modest improvement at high \pt, can improve its performance. Combining the standard HEPTopTagger with multiple shape variables gives the maximum improvement in performance at high \pt relative to at low \pt.\\

\begin{figure*}
\centering
\subfigure[HEPTopTagger]{\includegraphics[width=0.48\textwidth]{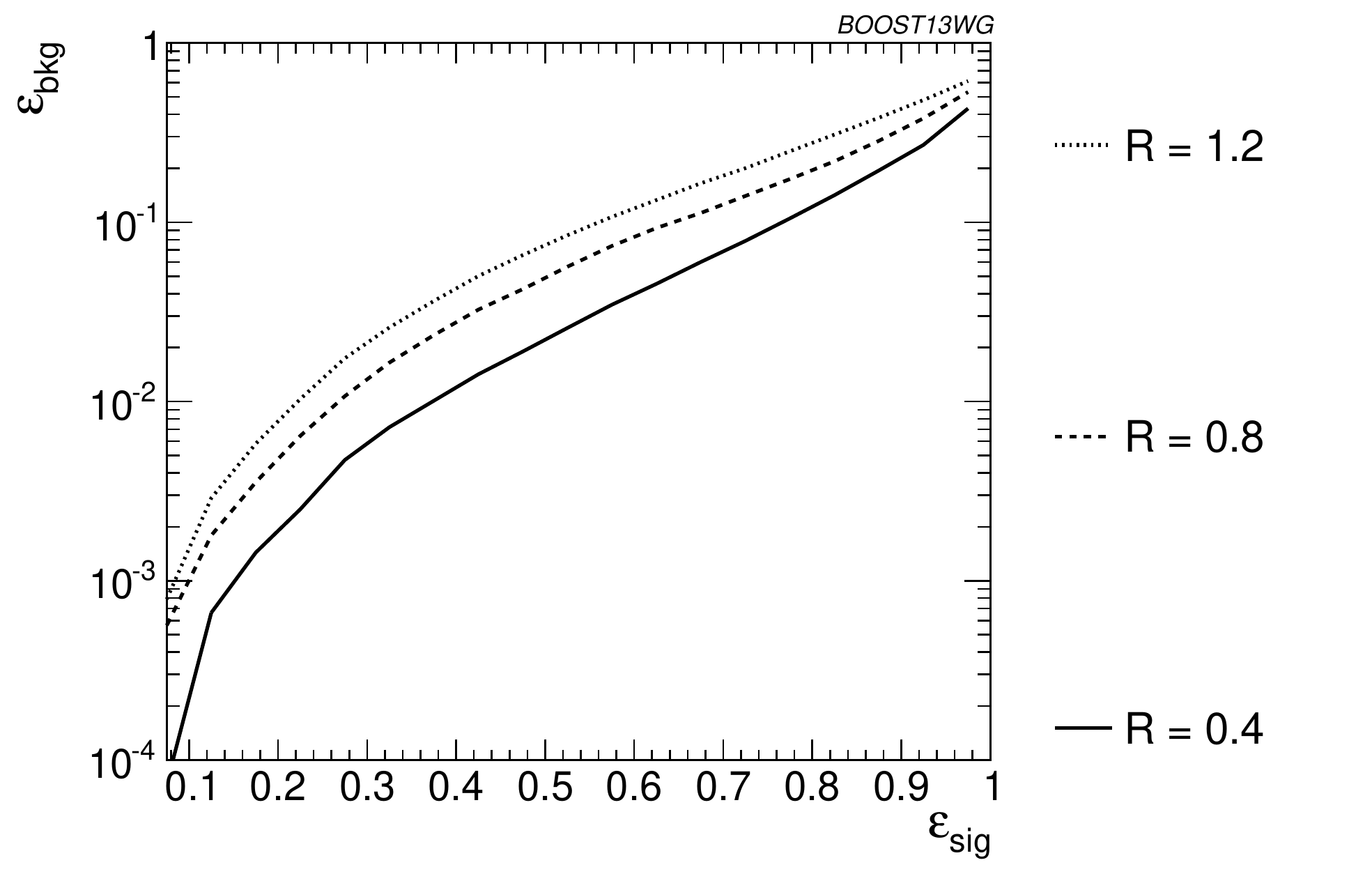}}
\subfigure[Johns Hopkins Tagger]{\includegraphics[width=0.48\textwidth]{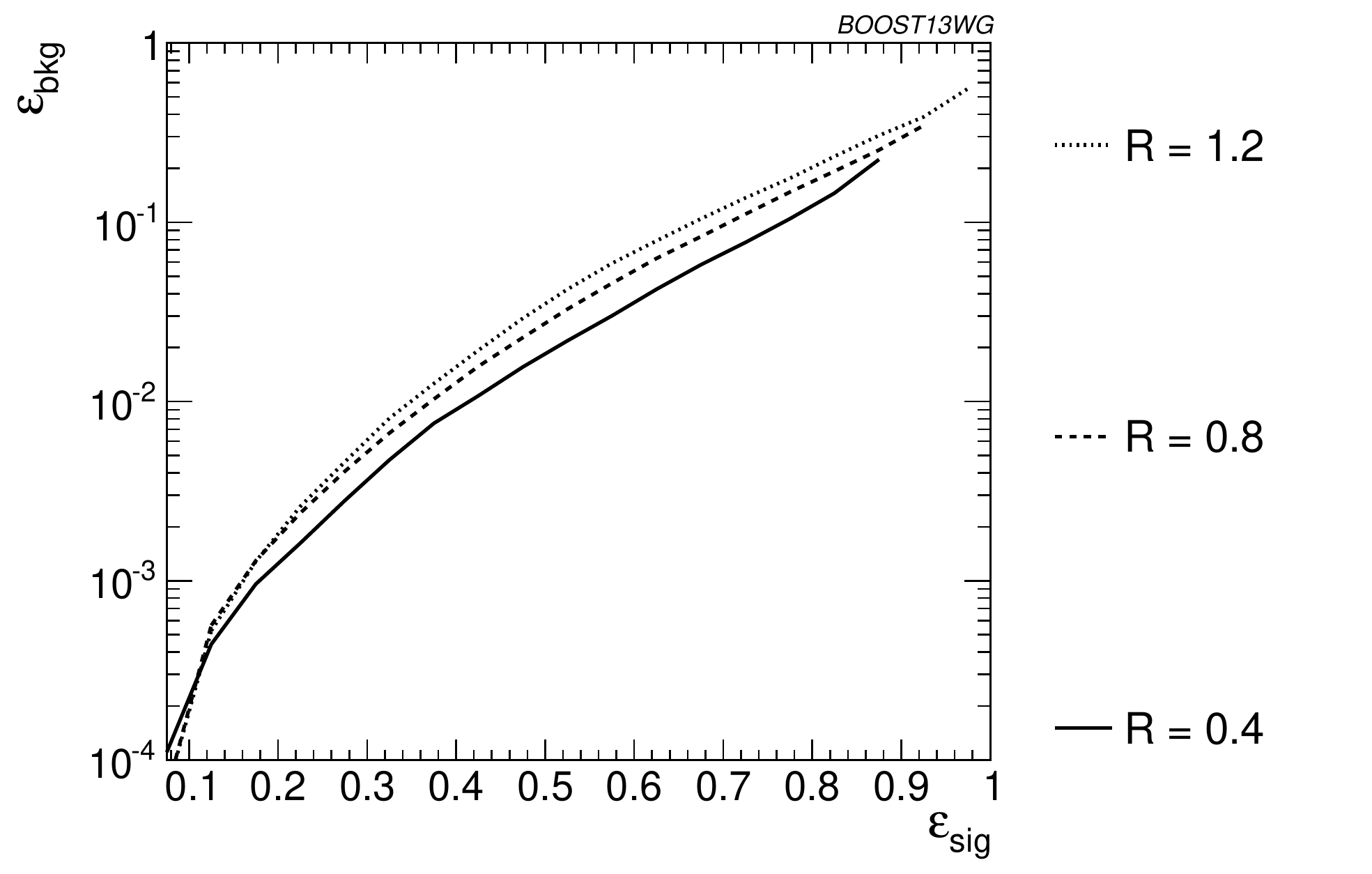}}
\subfigure[Trimming]{\includegraphics[width=0.48\textwidth]{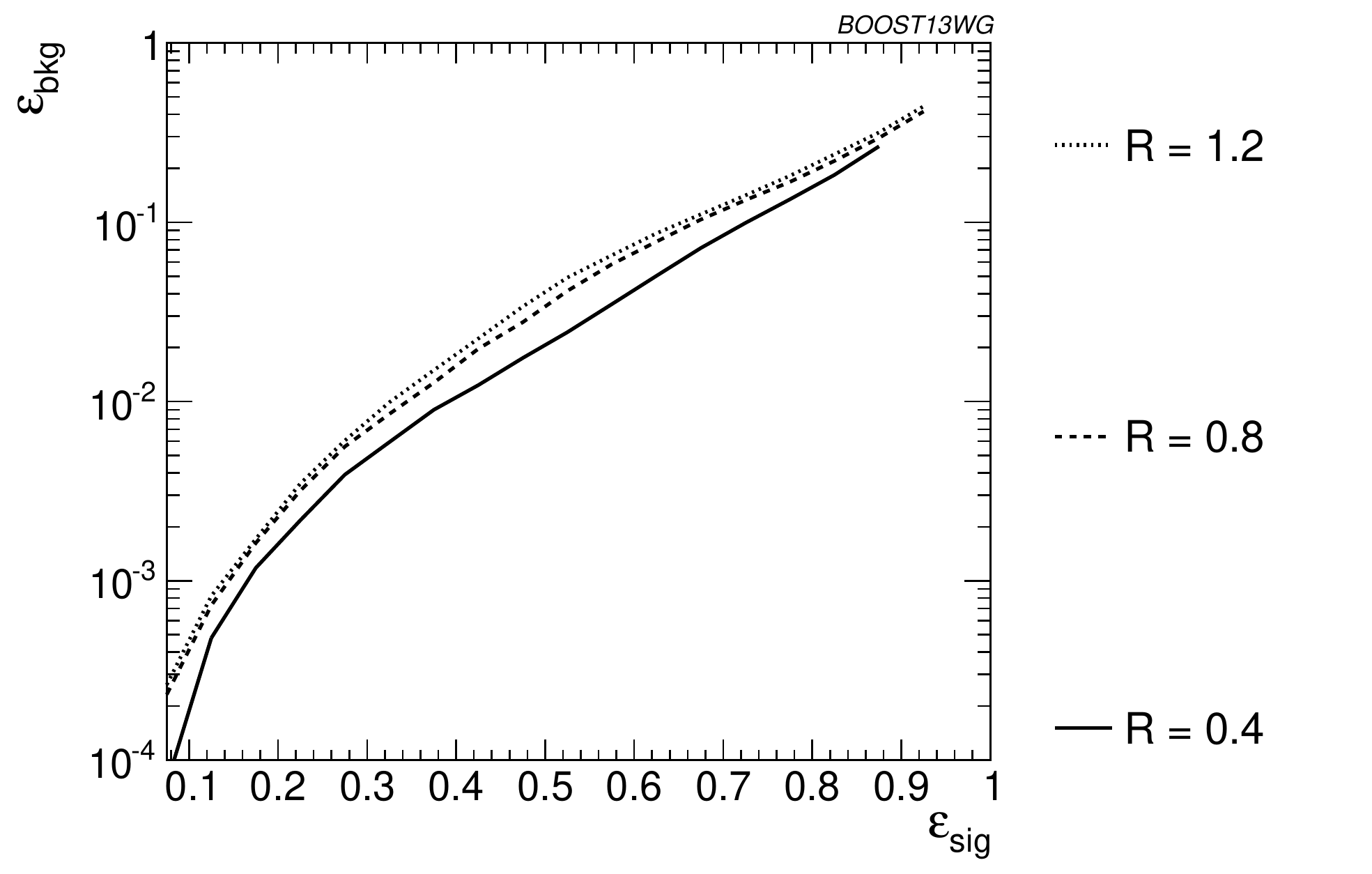}}
\subfigure[Pruning]{\includegraphics[width=0.48\textwidth]{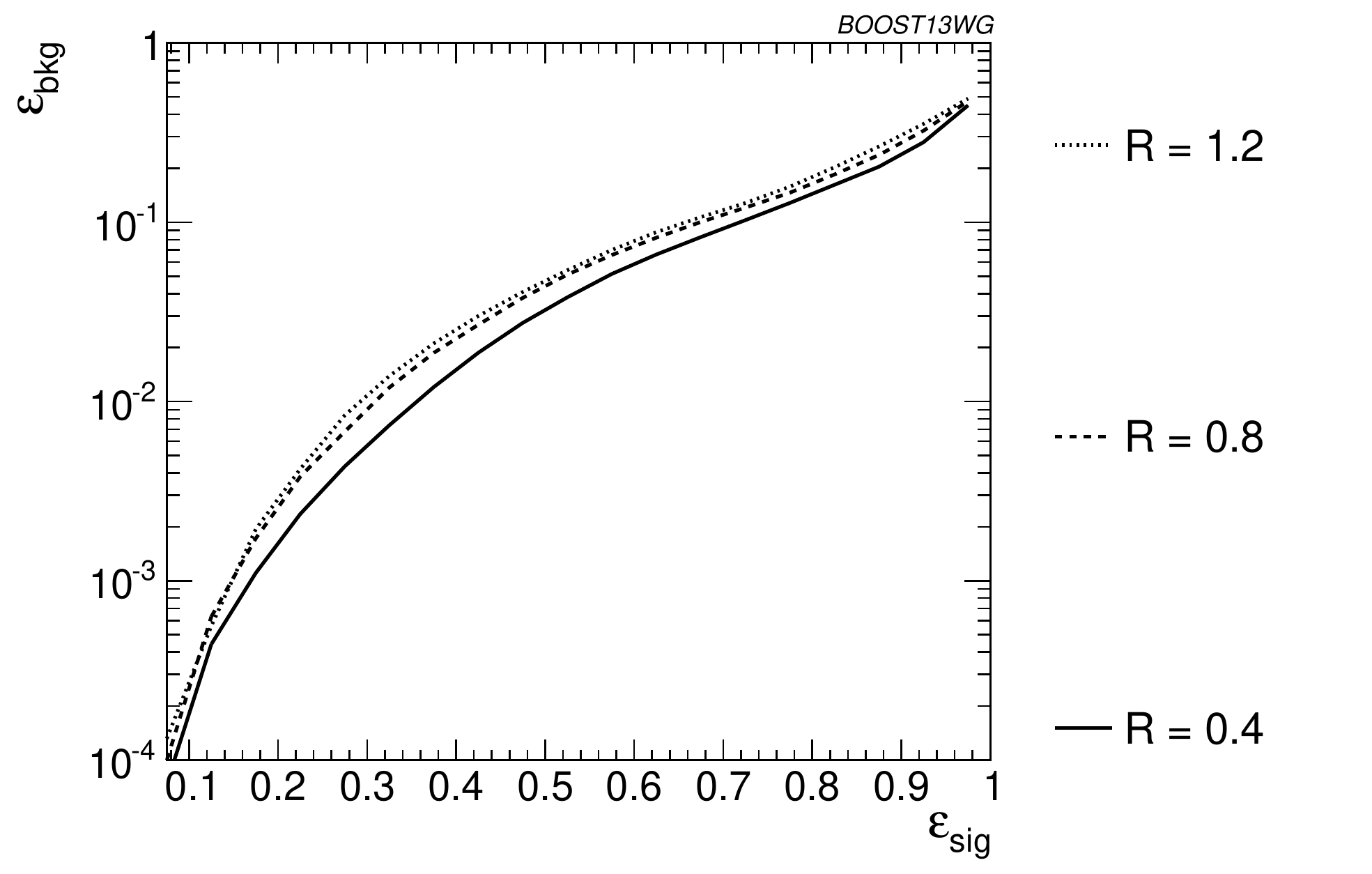}}
\caption{Comparison at different radii of the performance of various top tagging/grooming algorithms with \pt = 1.5-1.6 TeV. For each tagger/groomer, all output variables are combined in a BDT.}
\label{fig:Rcomparison_top}
\end{figure*}

\begin{figure*}
\centering
\subfigure[JH+$C_2^{\beta=1}$+$C_3^{\beta=1}$]{\includegraphics[width=0.48\textwidth]{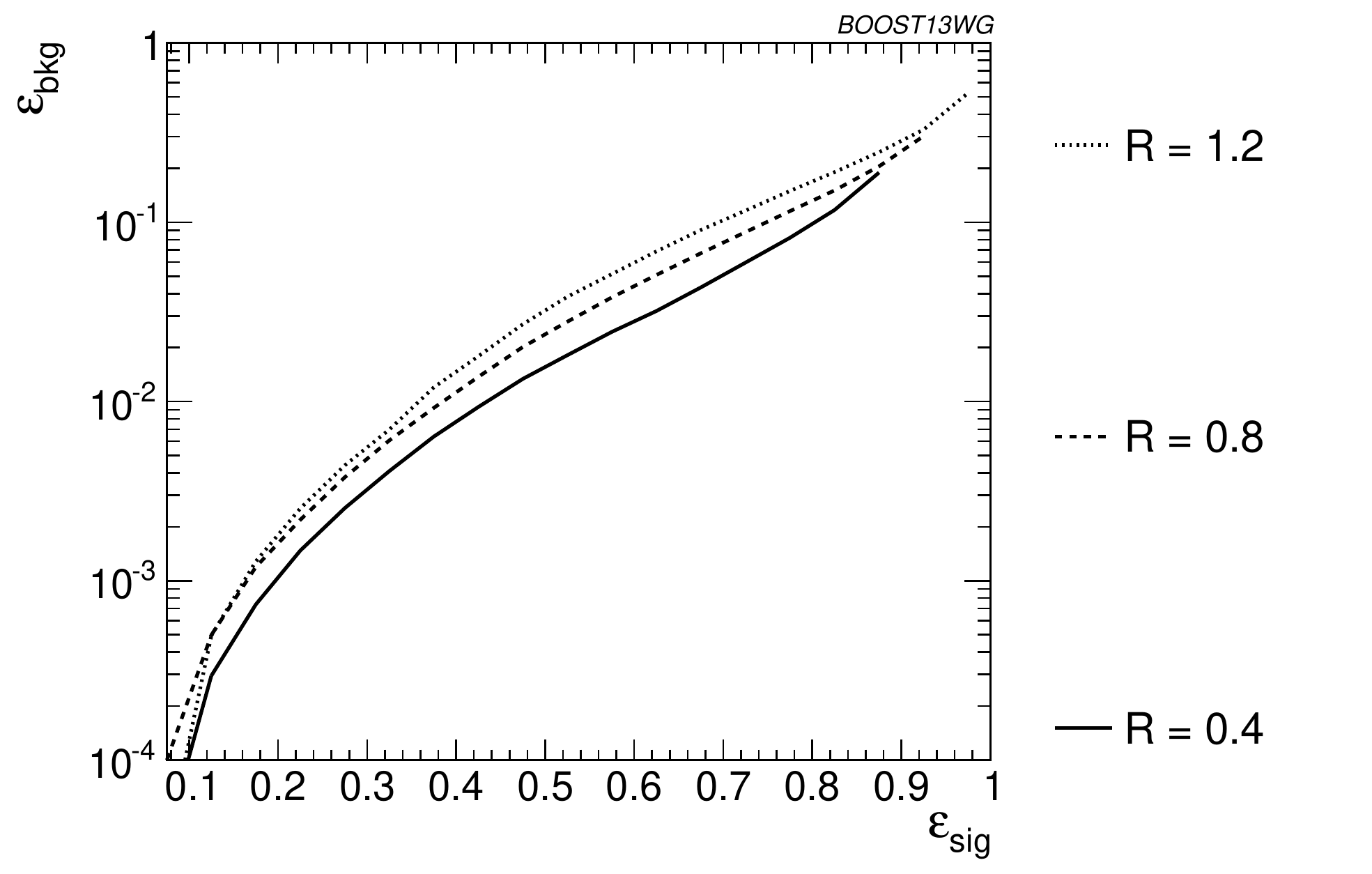}}
\subfigure[JH+\tautwoone+\tauthreetwo]{\includegraphics[width=0.48\textwidth]{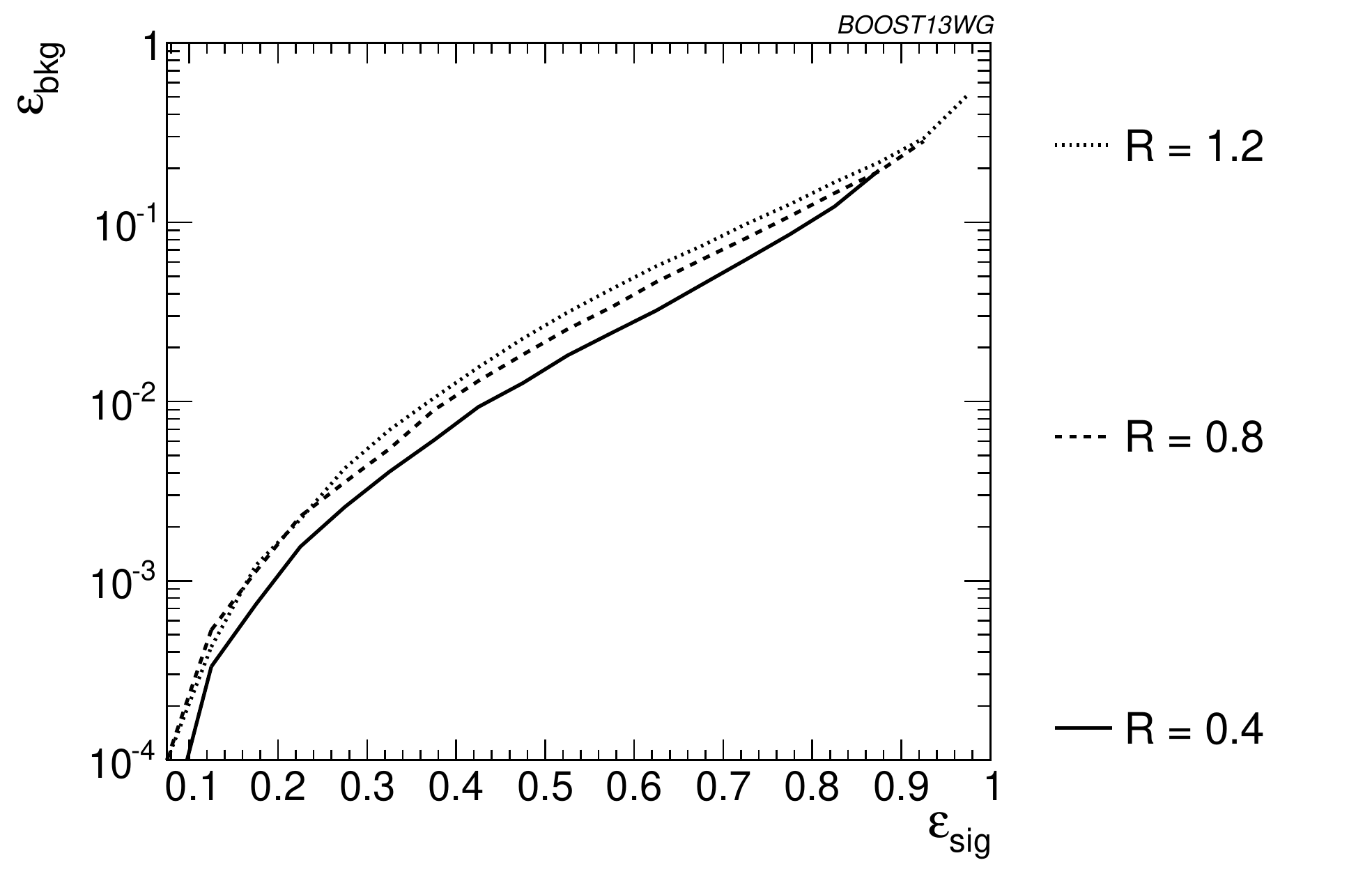}}
\subfigure[JH + $\Gamma_{\rm Qjet}$]{\includegraphics[width=0.48\textwidth]{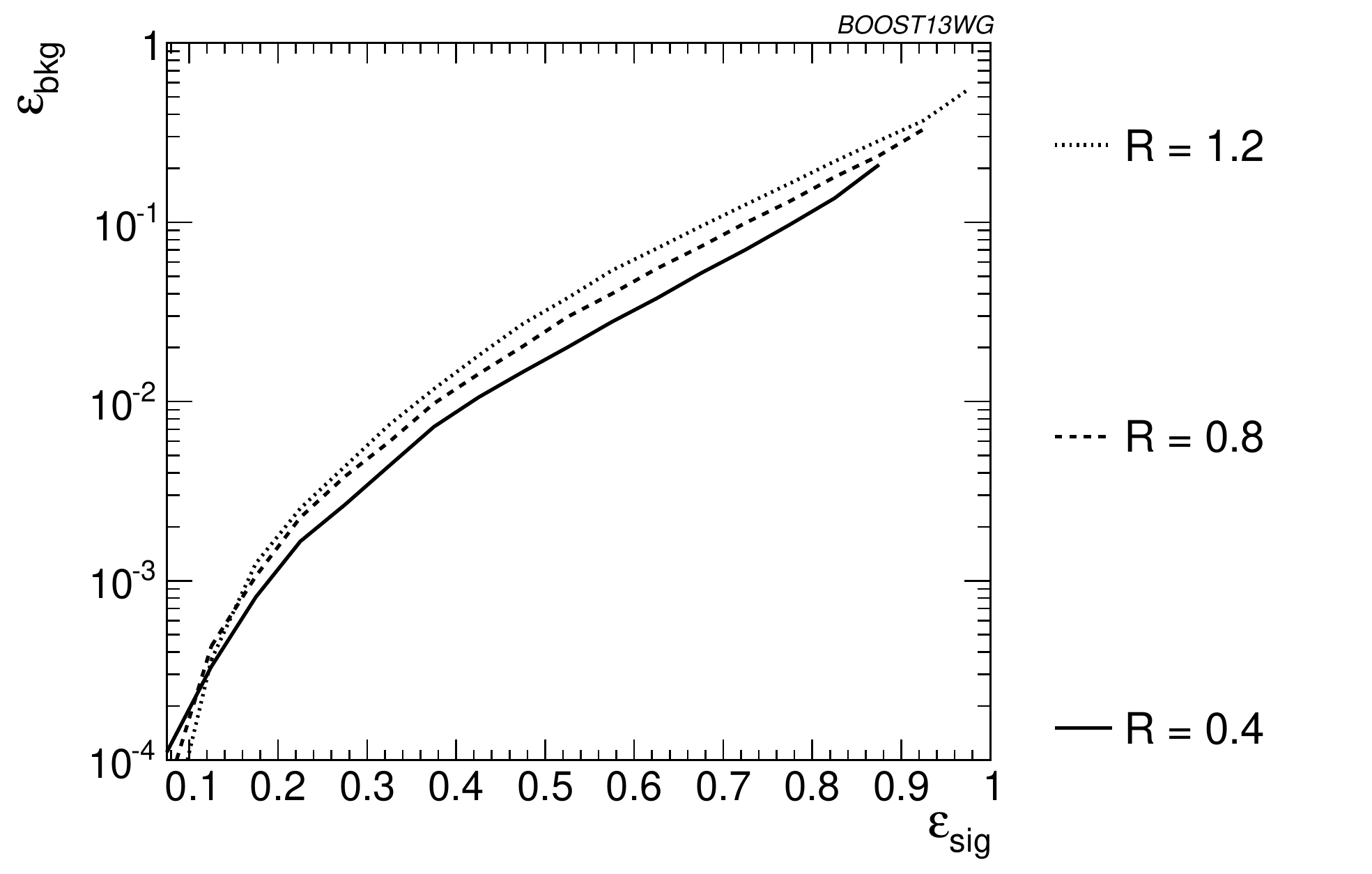}}
\subfigure[JH + all]{\includegraphics[width=0.48\textwidth]{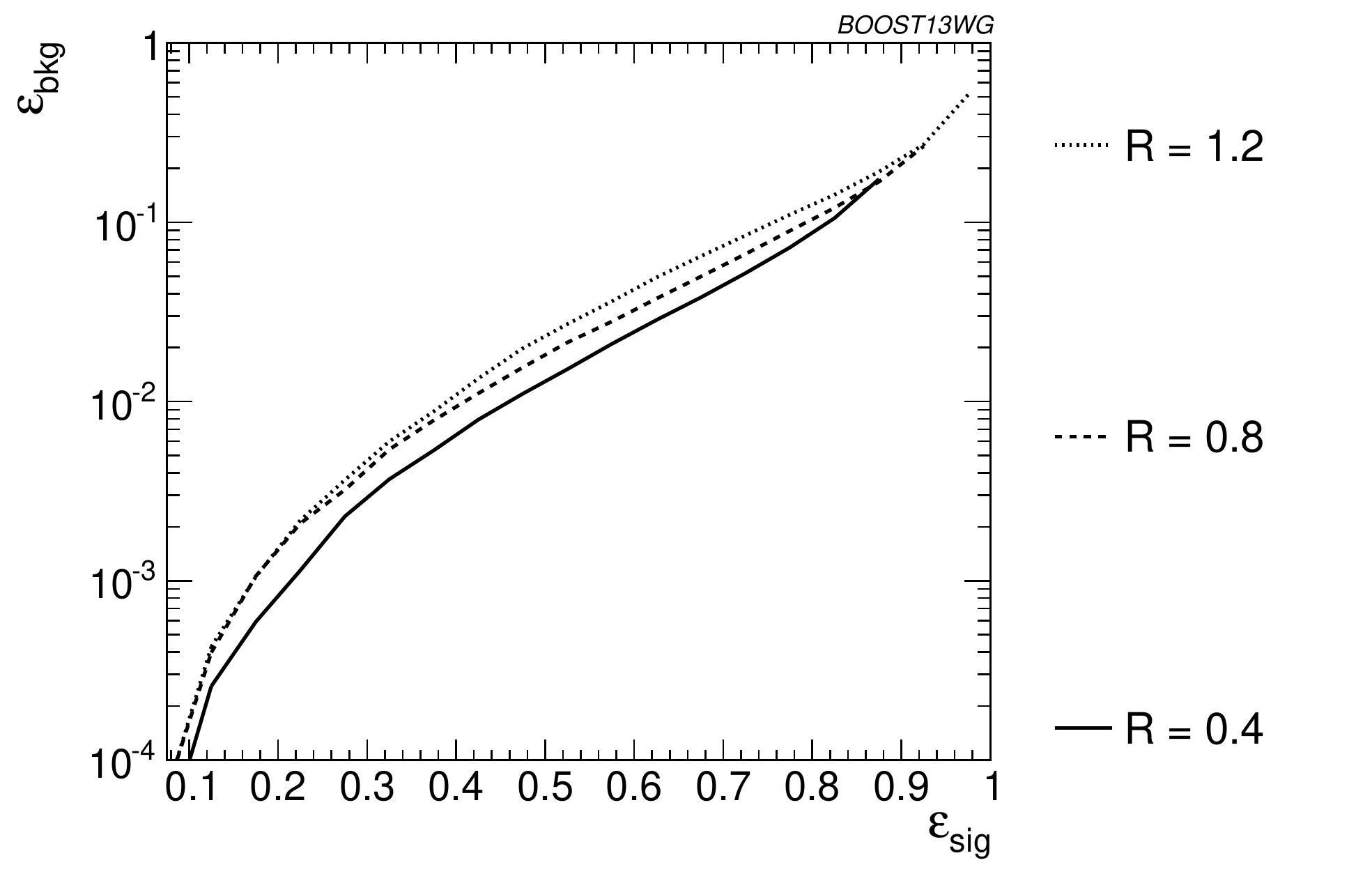}}
\caption{Comparison at different radii of the performance of the JH tagger in the \pt = 1.5-1.6 TeV bin, where all tagger output variables are combined in a BDT  with various shape variables}
\label{fig:Rcomparison_JH_shape}
\end{figure*}

In Figure ~\ref{fig:Rcomparison_top} we compare the BDT combinations of tagger outputs, with and without shape variables, at different jet radius $R$ in the \pt = 1.5-1.6 \TeV bin. The taggers are optimized over all input parameters for each choice of $R$ and signal efficiency. We find that, for all taggers and groomers, the performance is always best at small $R$; the choice of $R$ is sufficiently large to admit the full top quark decay at such high \pt, but is small enough to suppress contamination from additional radiation. This is not altered when the taggers are combined with shape variables. For example, in Figure~\ref{fig:Rcomparison_JH_shape} is shown the dependence on $R$ of the JH tagger when combined with shape variables, where one can see that the $R$-dependence is identical for all combinations. The same holds true for the HEPTopTagger, trimming, and pruning.

\subsection{Performance at Sub-Optimal Working Points}

Up until now, we have re-optimized our tagger and groomer parameters for each $\pt$, $R$, and signal efficiency working point. In reality, experiments will choose a finite set of working points to use. When this is taken into account, how will the top-tagging performance compare to the optimal results already shown? To address this concern, we replicate our analyses, but optimize the top taggers only for a single $\pt$ bin, single jet radius $R$, or single signal efficiency, and subsequently apply the same parameters to other scenarios. This allows us to determine the extent to which re-optimization is necessary to maintain the high signal-to-background discrimination power seen in the top-tagging algorithms we studied. In this section, we focus on the taggers and groomers, and their combination with shape variables, as the shape variables alone typically do not have any input parameters to optimize.

\begin{figure*}
\centering
\subfigure[HEPTopTagger $m_t$]{\includegraphics[width=0.48\textwidth]{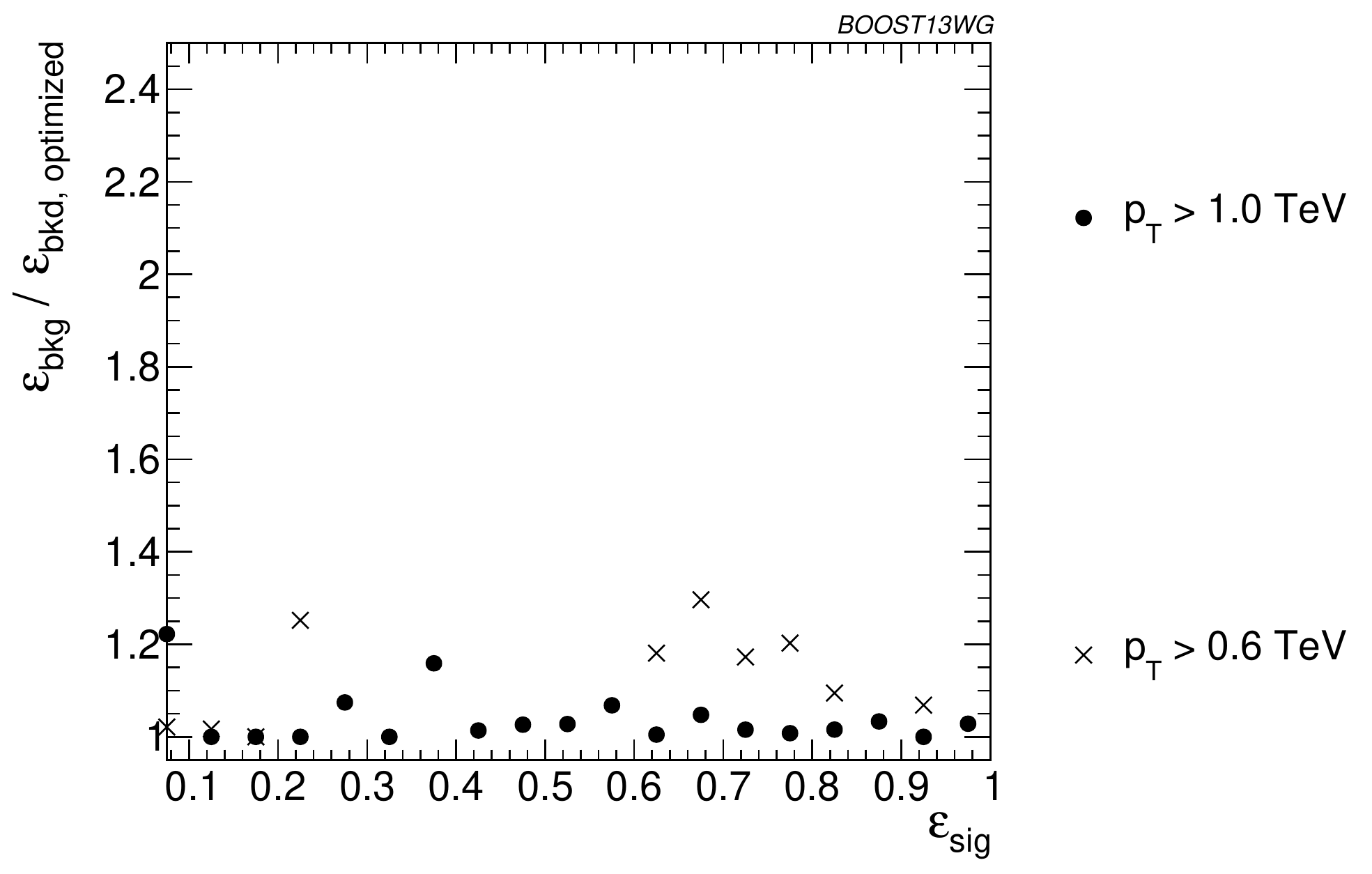}}
\subfigure[Johns Hopkins Tagger $m_t$]{\includegraphics[width=0.48\textwidth]{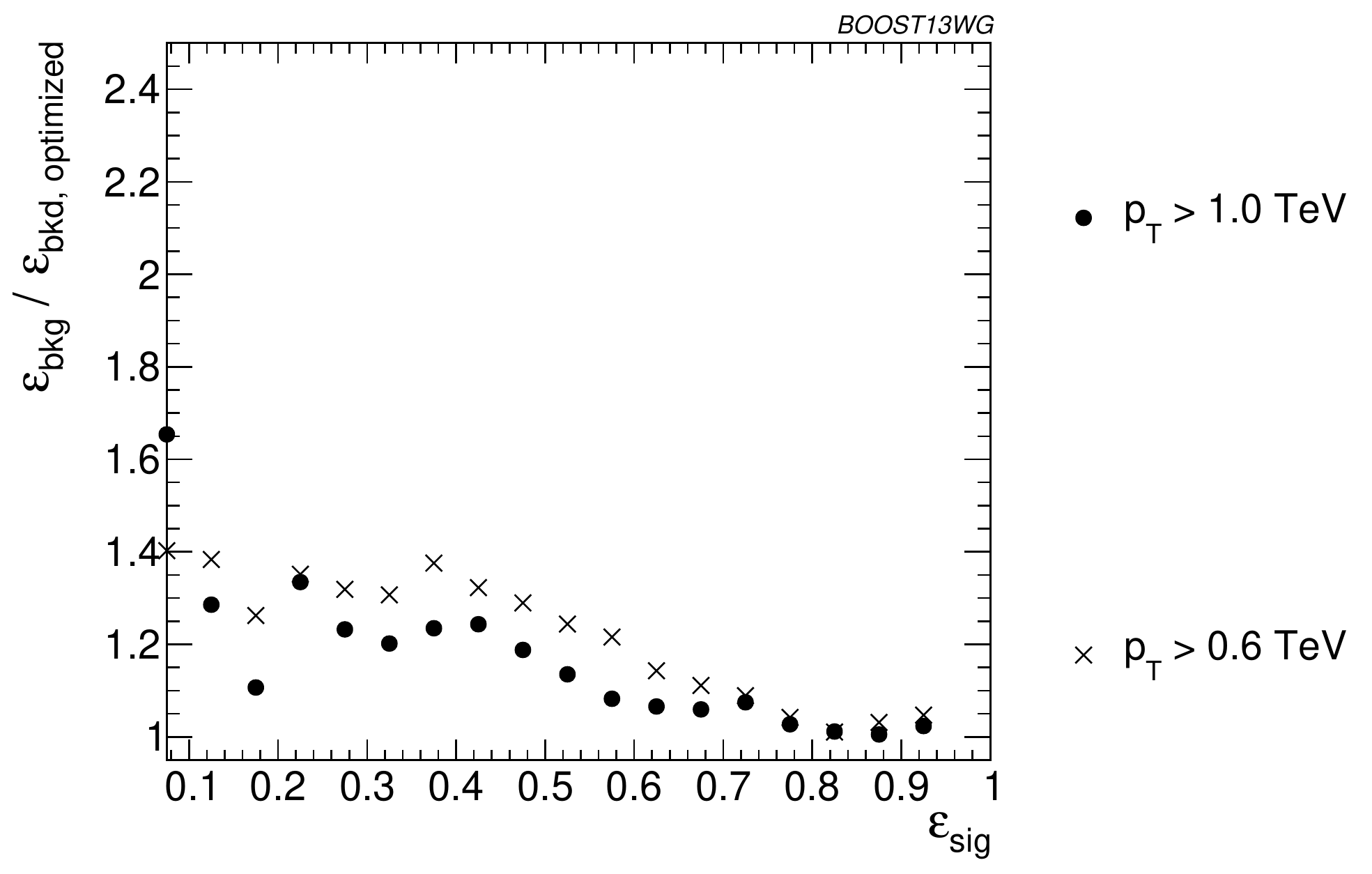}}
\subfigure[Pruning $m_t$]{\includegraphics[width=0.48\textwidth]{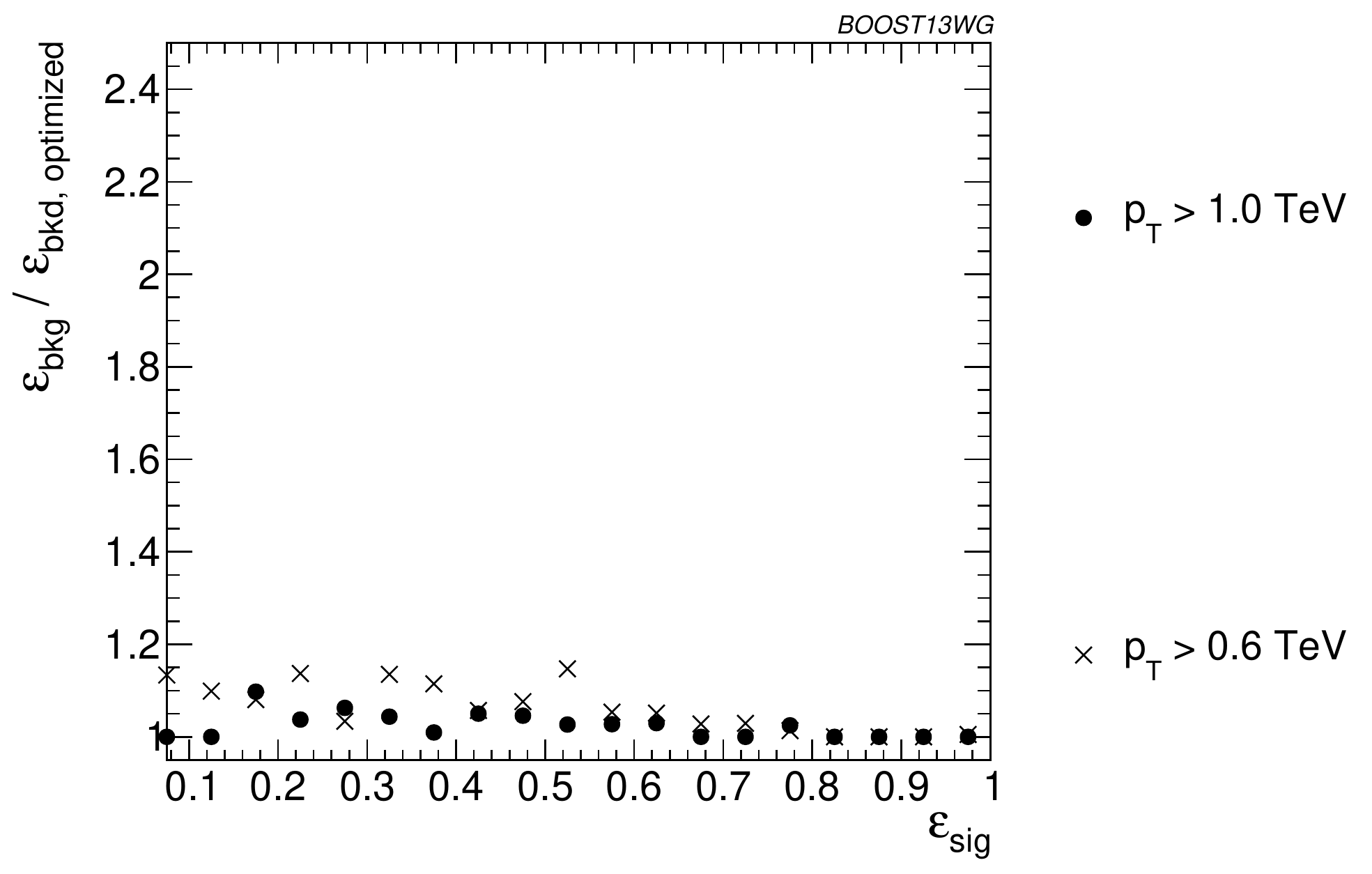}}
\subfigure[Trimming $m_t$]{\includegraphics[width=0.48\textwidth]{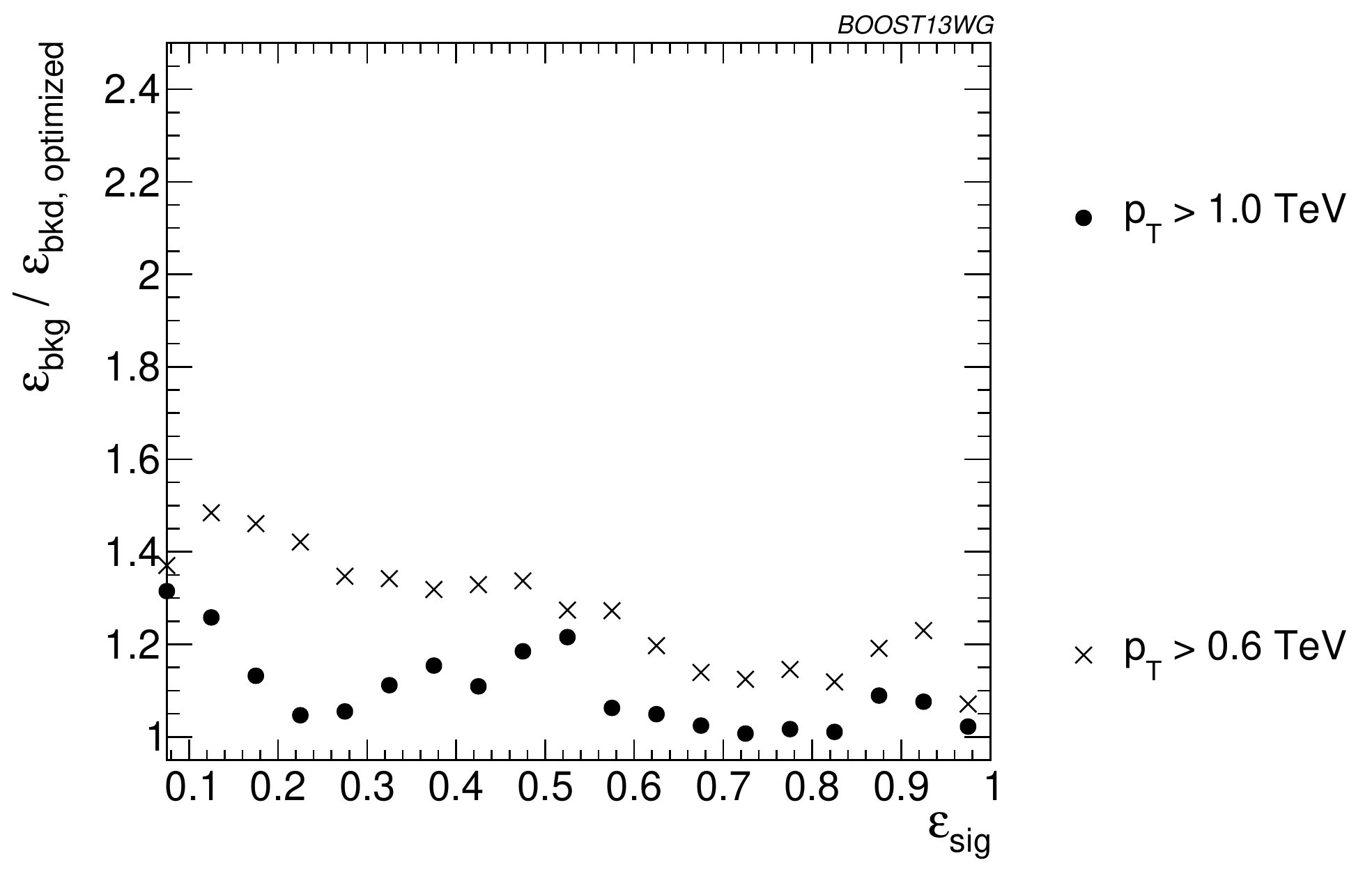}}
\caption{Comparison of the top mass performance of different taggers at different \pt using the anti-\kT $R=0.8$ algorithm. The tagger inputs are set to the optimum value for \pt = 1.5-1.6 \TeV, and the performance is normalized to the performance using the optimized tagger inputs at each \pt.}
\label{fig:ptcomparison_singletopmass_top_optOnce}
\end{figure*}

\noindent {\bf Optimizing at a single \pt:}
We show in Figure~\ref{fig:ptcomparison_singletopmass_top_optOnce} the performance of the reconstructed top mass for the \pt = 0.6-0.7 \TeV and \pt = 1.0-1.1 \TeV bins, with all input parameters optimized to the \pt = 1.5-1.6 \TeV bin (and $R=0.8$ throughout). This is normalized to the performance using the optimized tagger inputs at each \pt. The performance degradation is at the level of 20-30\% (at maximum 50\%) when the high-\pt optimized inputs are used at other momenta,
with trimming and the Johns Hopkins tagger degrading the most. The jagged behaviour of the points is due to the finite resolution of the scan. We also observe a particular effect associated with using suboptimal taggers:~since taggers sometimes fail to return a top candidate, parameters optimized for a particular signal efficiency $\varepsilon_{\rm sig}$ at \pt = 1.5-1.6 \TeV may not return enough signal candidates to reach the same efficiency at a different $\pt$. Consequently, no point appears for that $\pt$ value. This is not often a practical concern, as the largest gains in signal discrimination and significance are for smaller values of $\varepsilon_{\rm sig}$, but it may be an important effect to consider when selecting benchmark tagger parameters and signal efficiencies.

\begin{figure*}
\centering
\subfigure[HEPTopTagger]{\includegraphics[width=0.48\textwidth]{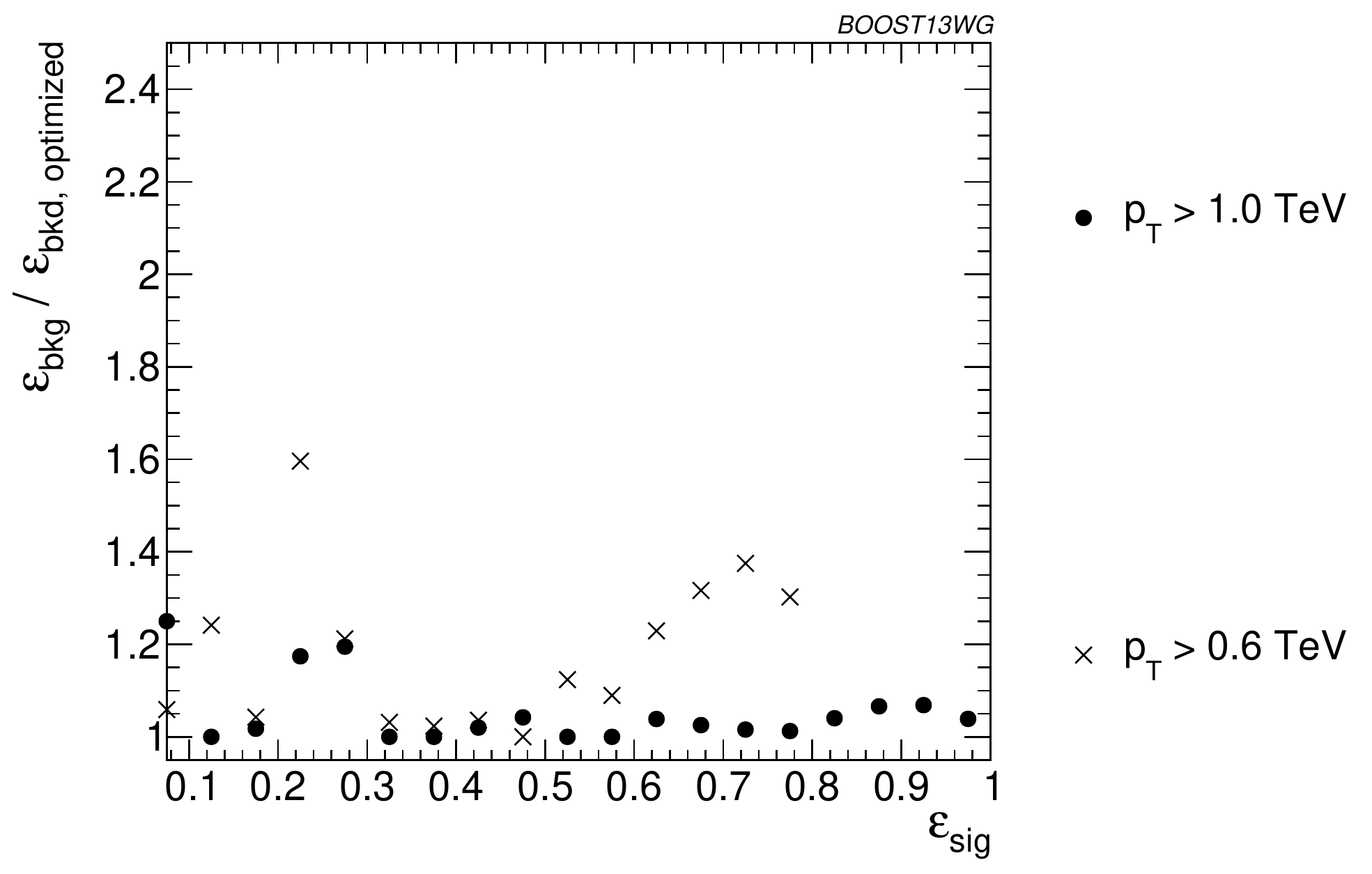}}
\subfigure[Johns Hopkins Tagger]{\includegraphics[width=0.48\textwidth]{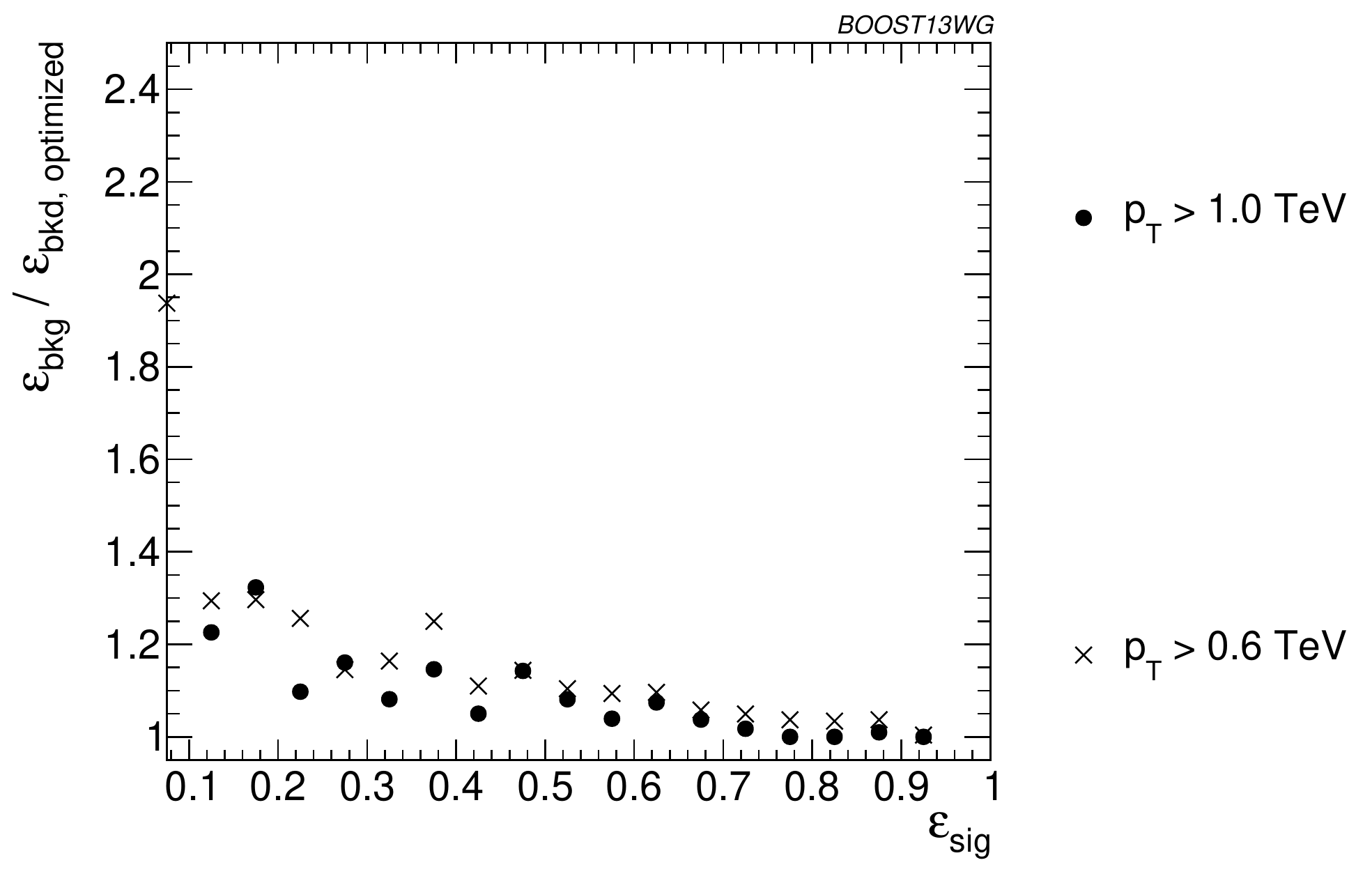}}
\subfigure[Pruning]{\includegraphics[width=0.48\textwidth]{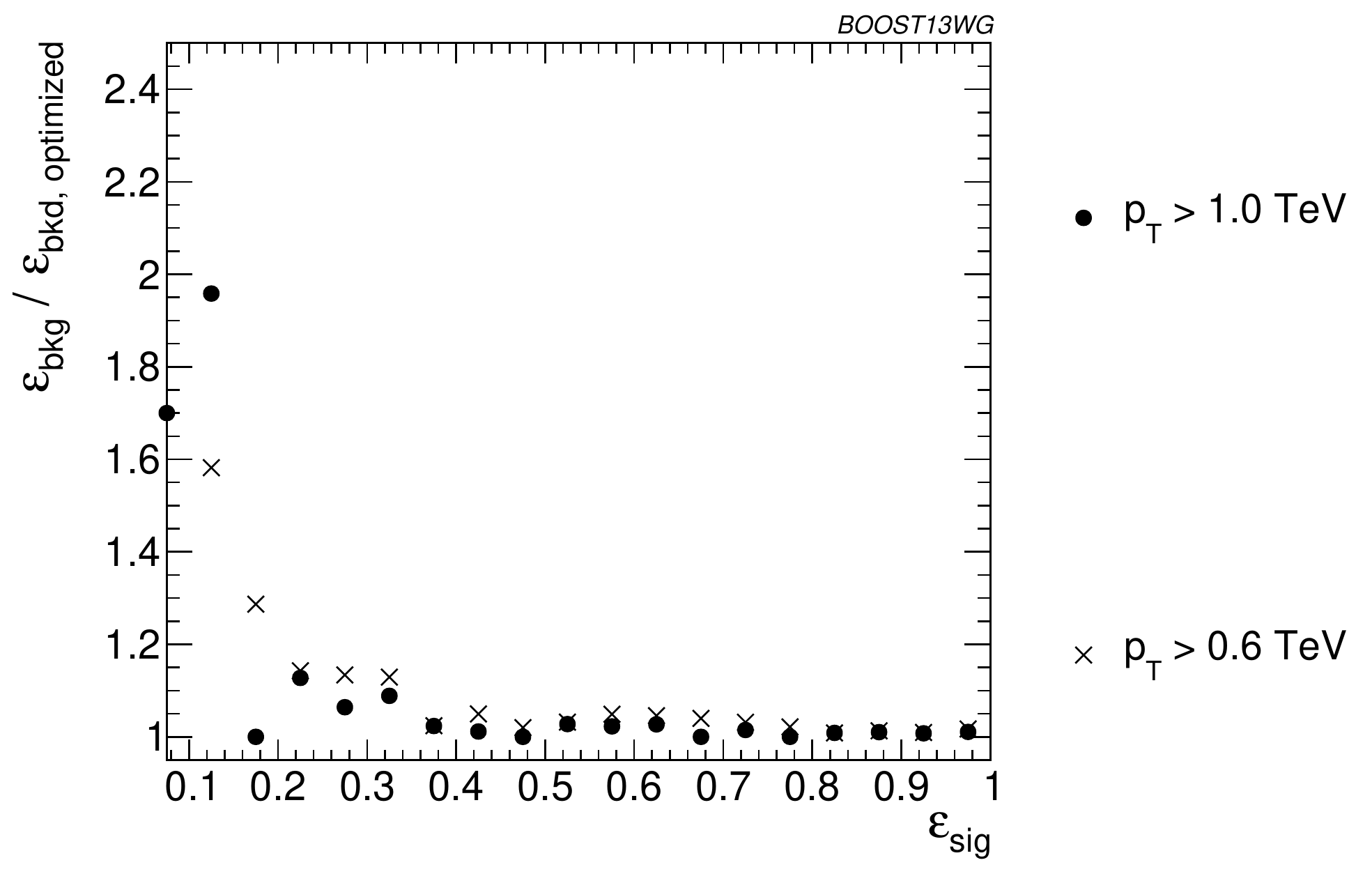}}
\subfigure[Trimming]{\includegraphics[width=0.48\textwidth]{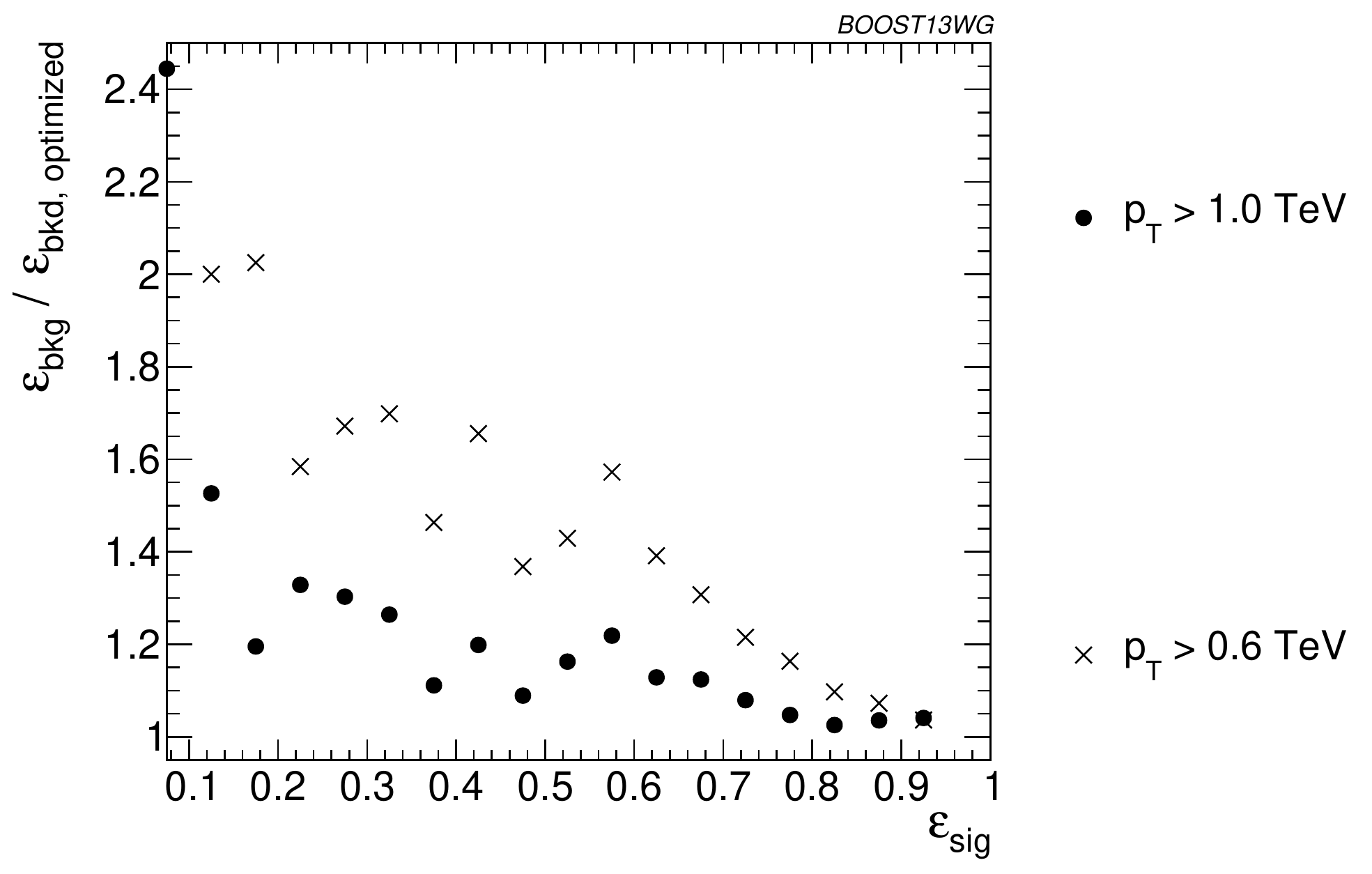}}
\caption{Comparison of tagger performance at different \pt using the \antikt $R=0.8$ algorithm. For each tagger/groomer, all output variables are combined in a BDT, and the tagger inputs are set to the optimum value for \pt = 1.5-1.6 \TeV. The performance is normalized to the performance using the optimized tagger inputs at each \pt.}
\label{fig:ptcomparison_top_optOnce}
\end{figure*}

The degradation in performance is more pronounced for the BDT combinations of the full tagger outputs, shown in Figure~\ref{fig:ptcomparison_top_optOnce}. This is true particularly at very low signal efficiency, where the optimization of inputs picks out a cut on the tail of some distribution that depends precisely on the $\pt$/$R$ of the jet. Once again, trimming and the Johns Hopkins tagger degrade more markedly.  Similar behavior holds for the BDT combinations of tagger outputs plus all shape variables.\\

\begin{figure*}
\centering
\subfigure[HEPTopTagger $m_t$]{\includegraphics[width=0.48\textwidth]{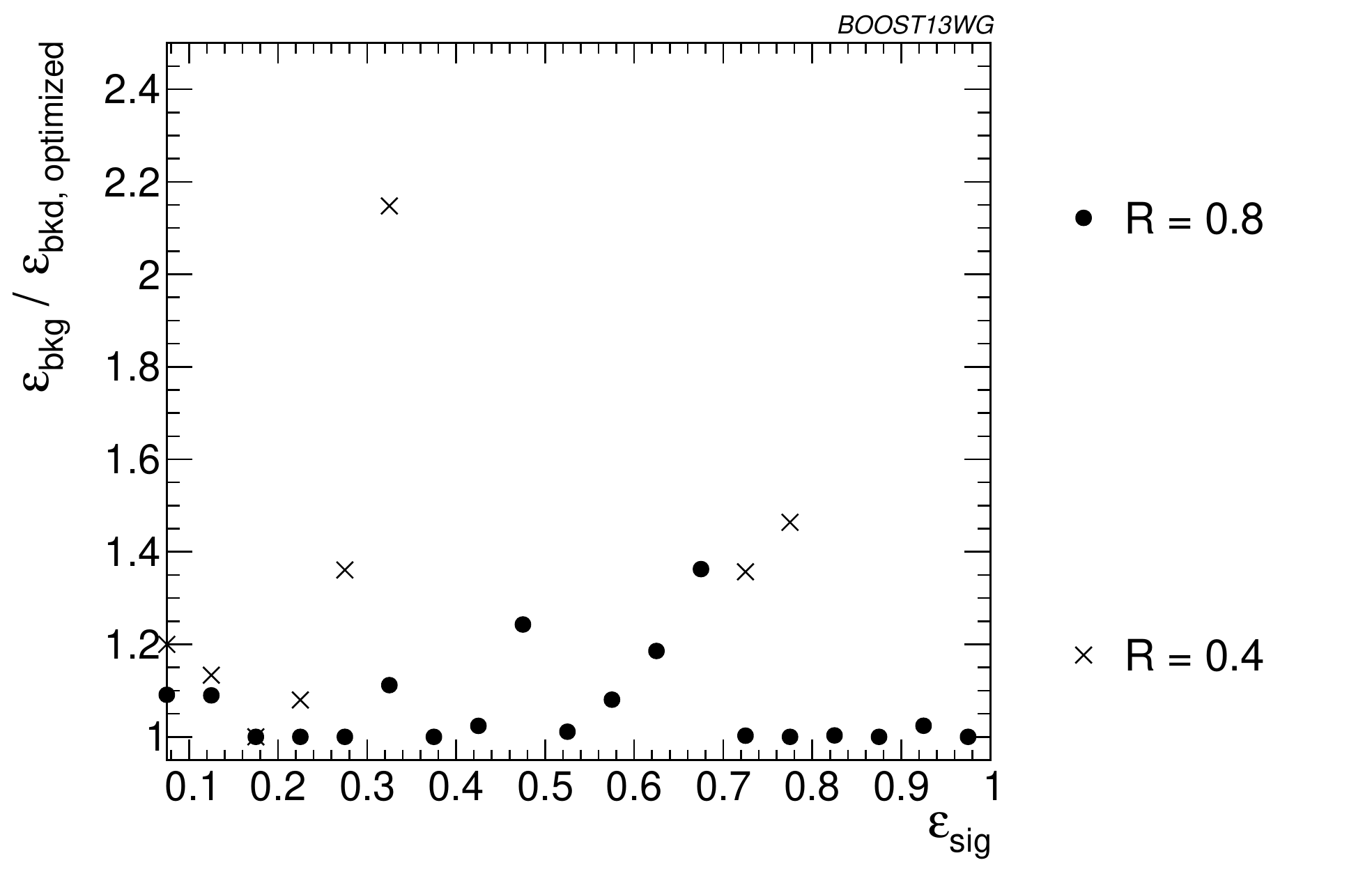}}
\subfigure[Johns Hopkins Tagger $m_t$]{\includegraphics[width=0.48\textwidth]{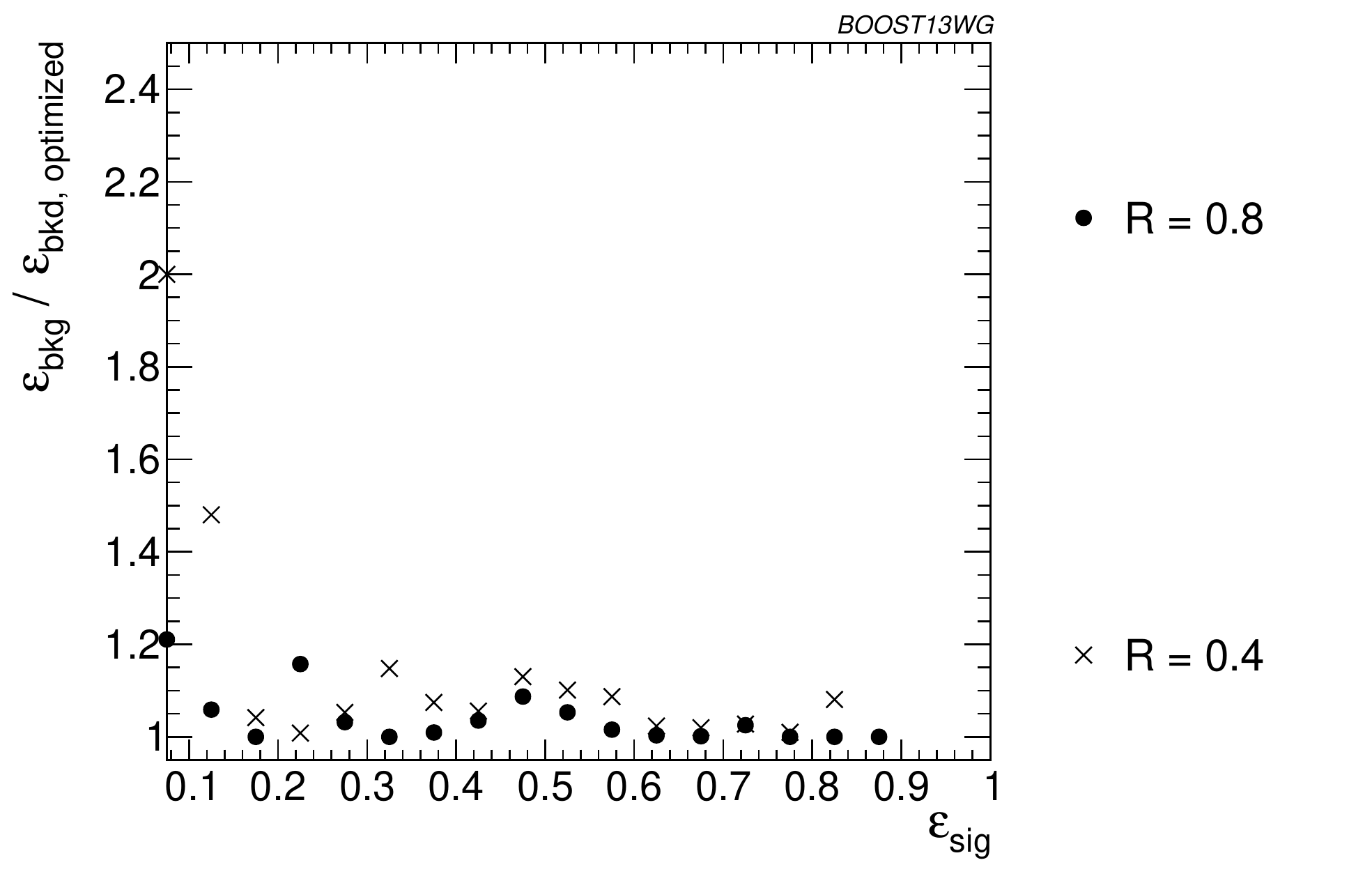}}
\subfigure[Pruning $m_t$]{\includegraphics[width=0.48\textwidth]{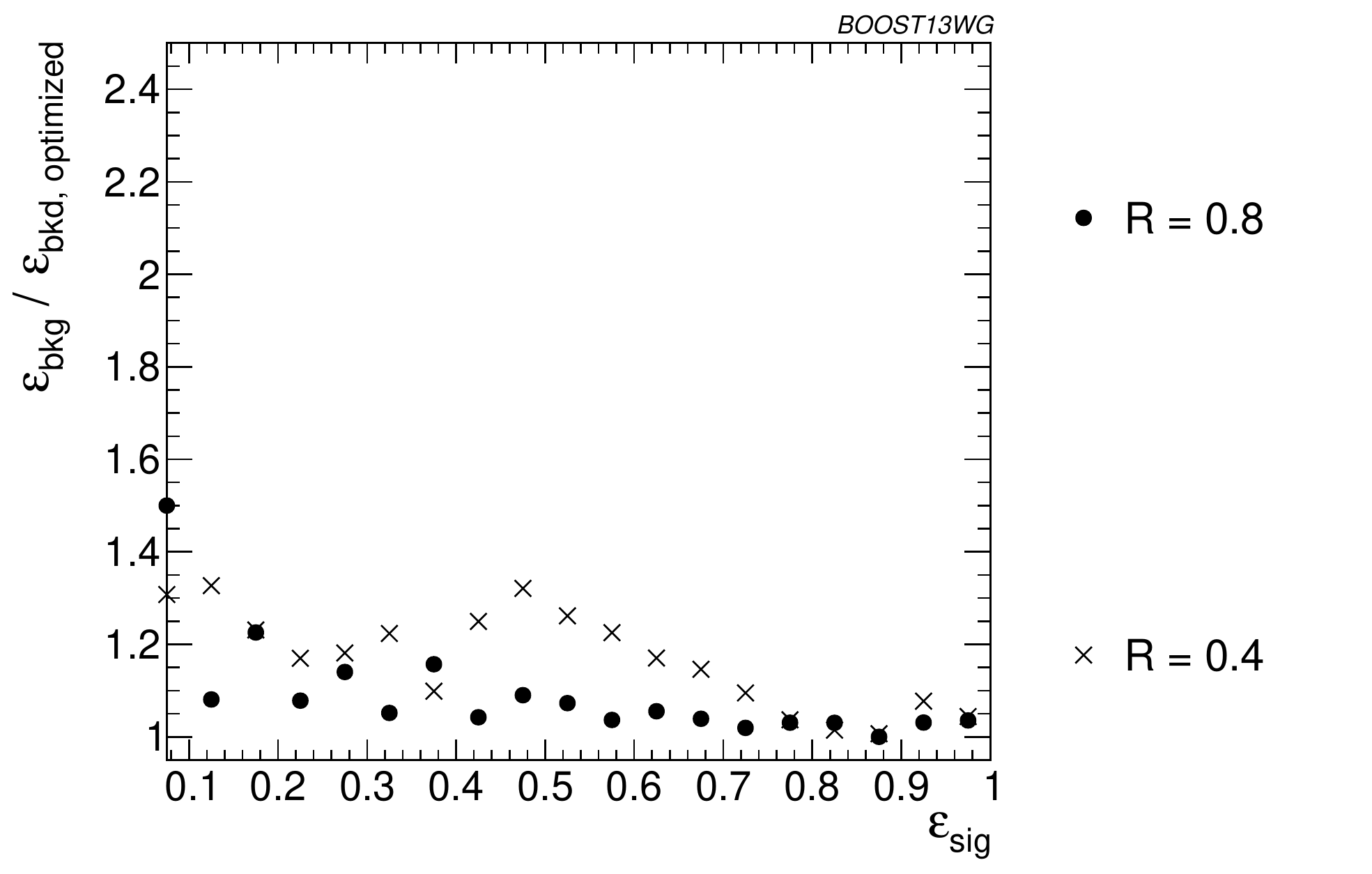}}
\subfigure[Trimming $m_t$]{\includegraphics[width=0.48\textwidth]{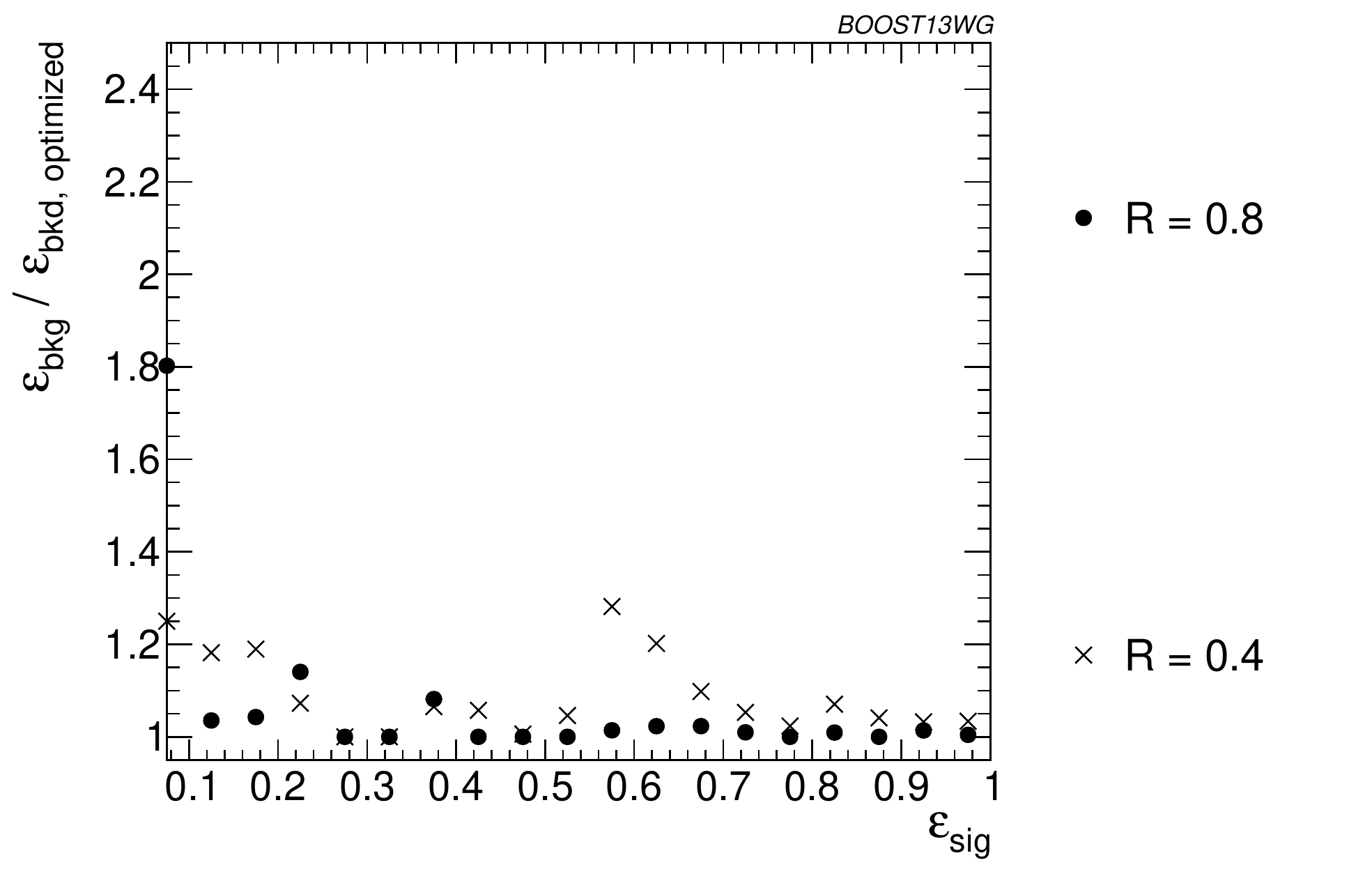}}
\caption{Comparison of the top mass performance of different taggers at different $R$ in the \pt = 1.5-1.6 \TeV bin. The tagger inputs are set to the optimum value for $R=1.2$, and the performance is normalized to the performance using the optimized tagger inputs at each $R$.}
\label{fig:Rcomparison_singletopmass_top_optOnce}
\end{figure*}

\noindent {\bf Optimizing at a single $R$:}
In Figure~\ref{fig:Rcomparison_singletopmass_top_optOnce}, we show the performance of the reconstructed top mass for $R=0.4$ and 0.8, with all input parameters optimized to $R=1.2$ TeV bin (and \pt = 1.5-1.6 \TeV throughout). This is normalized to the performance using the optimized tagger inputs at each $R$. While the performance of each variable degrades at small $\varepsilon_{\rm sig}$ compared to the optimized search, the HEPTopTagger fares the worst. It is not surprising that a tagger whose top mass reconstruction is susceptible to background-shaping at large $R$ and $\pt$ would require a more careful optimization of parameters to obtain the best performance; recent updates to the tagger algorithm  \cite{Anders:2013oga,Kasieczka:2015jma} may mitigate the need for this more careful optimization.

\begin{figure*}
\centering
\subfigure[HEPTopTagger]{\includegraphics[width=0.48\textwidth]{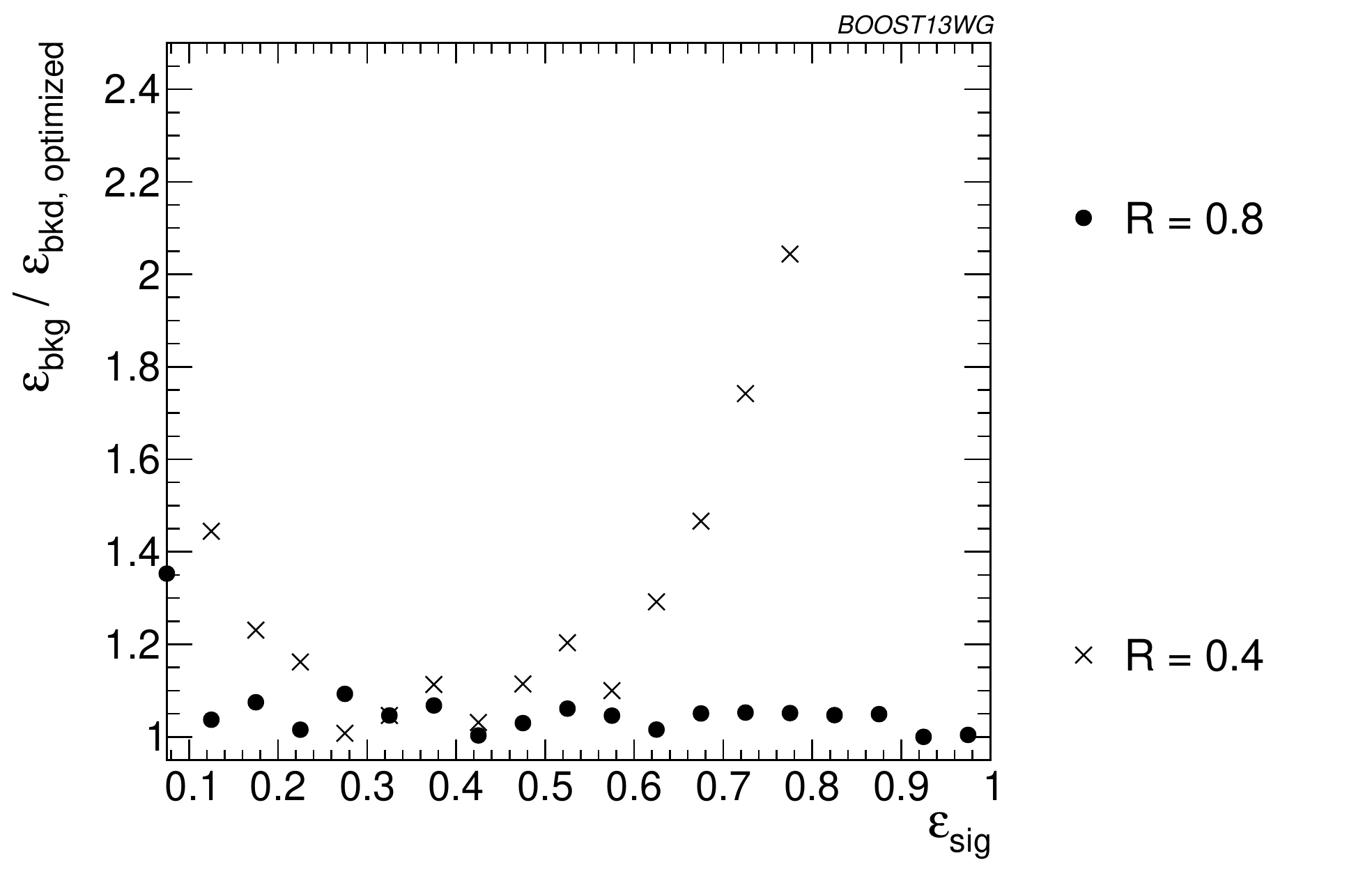}}
\subfigure[Johns Hopkins Tagger]{\includegraphics[width=0.48\textwidth]{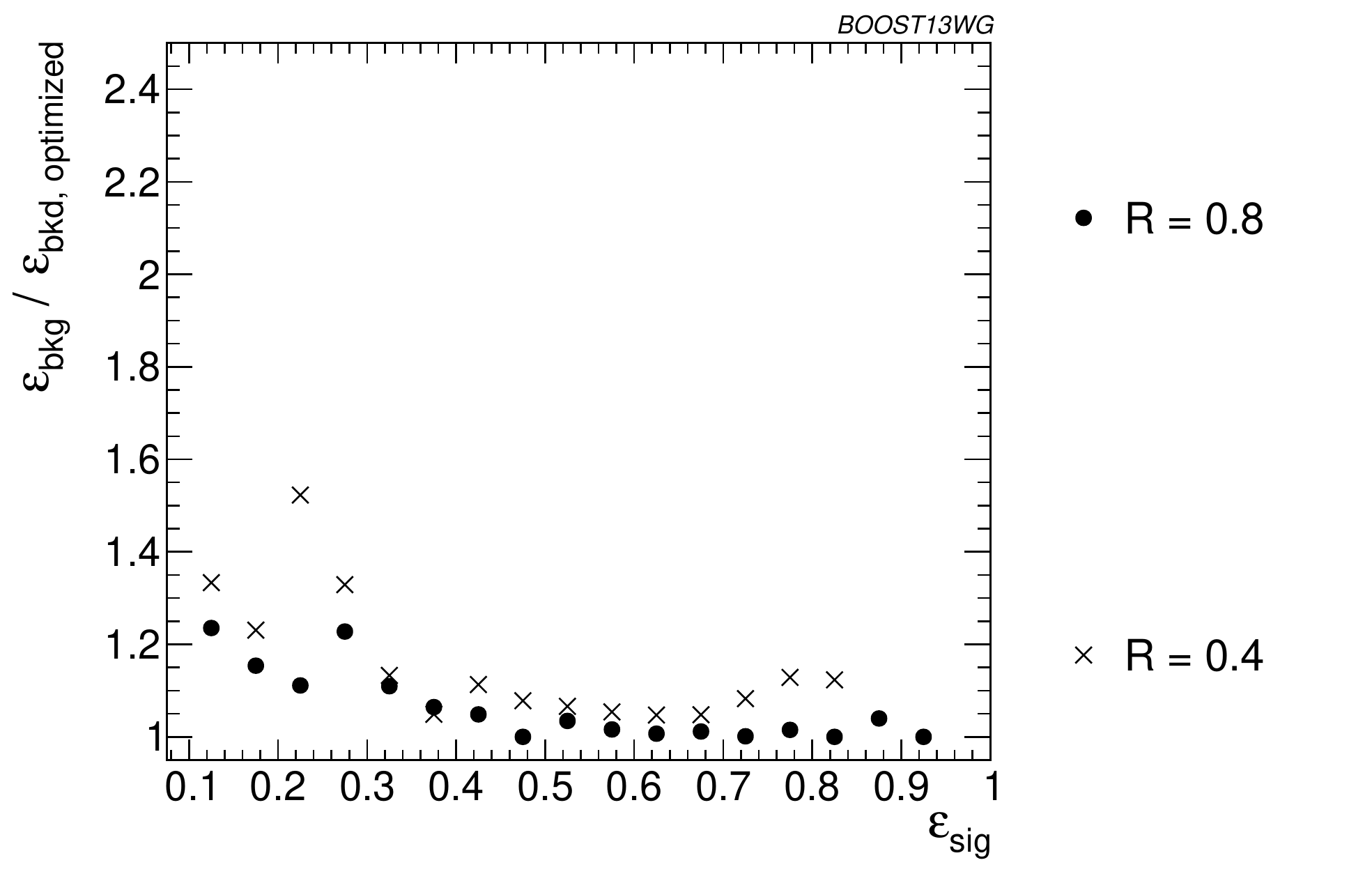}}
\subfigure[Pruning]{\includegraphics[width=0.48\textwidth]{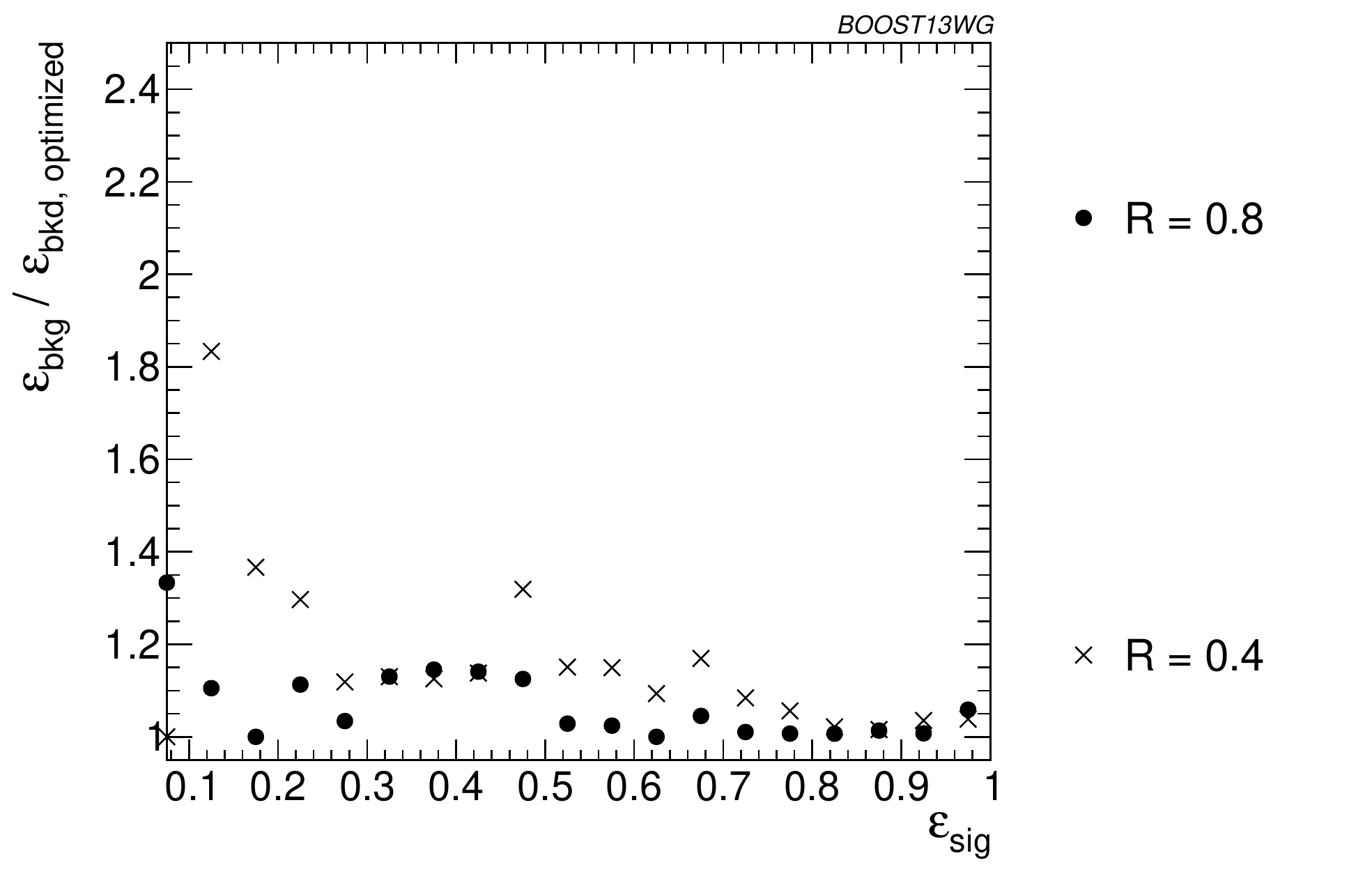}}
\subfigure[Trimming]{\includegraphics[width=0.48\textwidth]{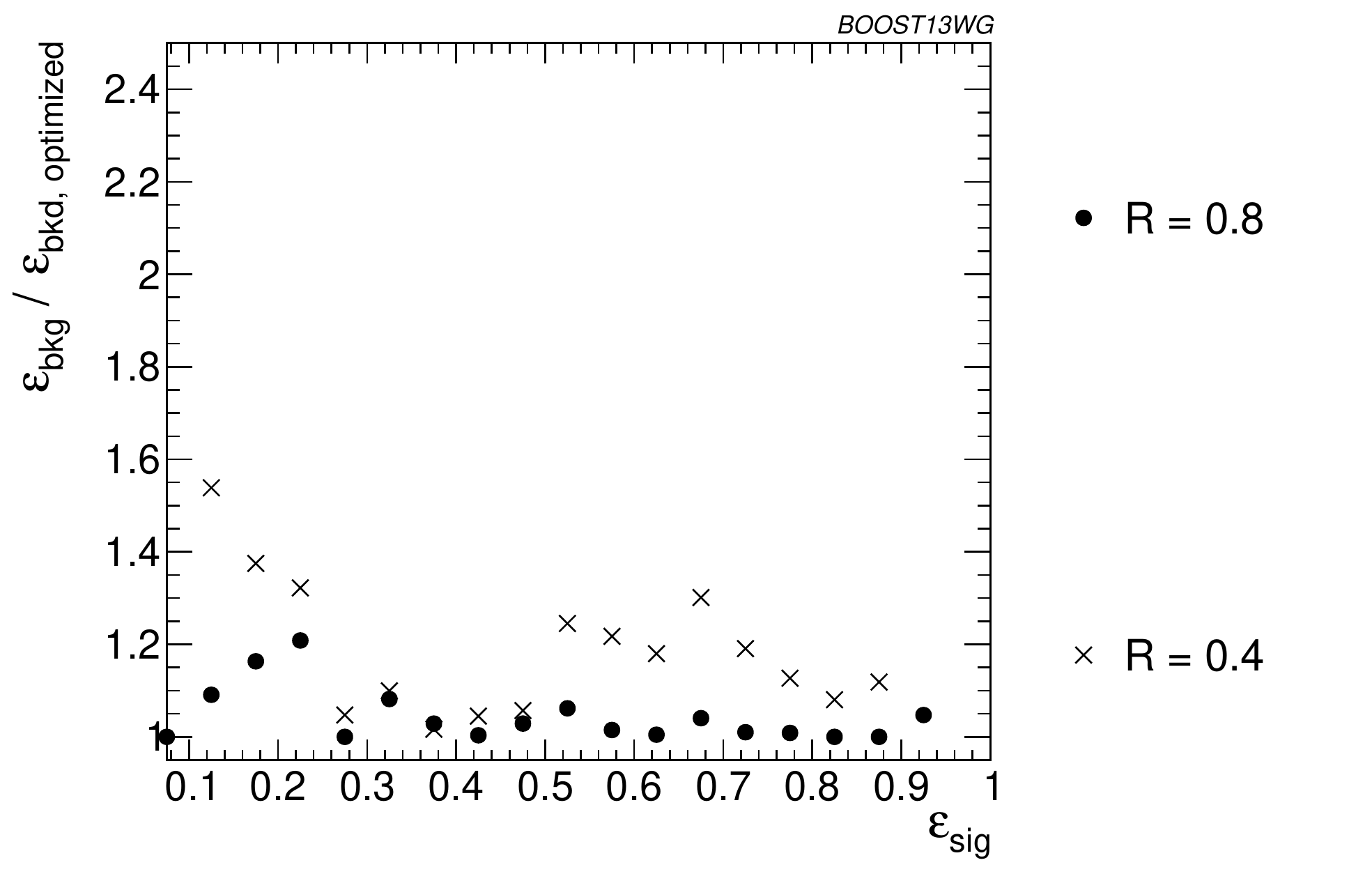}}
\caption{Comparison of tagger performance at different $R$ in \pt = 1.5-1.6 \TeV bin. For each tagger/groomer, all output variables are combined in a BDT, and the tagger inputs are set to the optimum value for $R=1.2$, and the performance is normalized to the performance using the optimized tagger inputs at each $R$.}
\label{fig:Rcomparison_top_optOnce}
\end{figure*}

The same holds true for the BDT combinations of the full tagger outputs, shown in Figure~\ref{fig:Rcomparison_top_optOnce}. The performance for the sub-optimal taggers is still within an $O(1)$ factor of the optimized performance, and the HEPTopTagger performs better with the combination of all of its outputs relative to the performance with just \topmass. The same behaviour holds for the BDT combinations of tagger outputs and shape variables. \\

\noindent {\bf Optimizing at a single efficiency:}
The strongest assumption we have made so far is that the taggers can be re-optimized for each signal efficiency point. This is useful for making a direct comparison of the power of different top-tagging algorithms, but is not particularly practical for LHC analyses. We now consider the scenario in which the tagger inputs are optimized once, in the $\varepsilon_{\rm sig}=0.3$-$0.35$ bin, and then used for all signal efficiencies. We do this in the  \pt = 1.0-1.1 \TeV bin and with $R=0.8$.

\begin{figure*}
\centering
\includegraphics[width=0.49\textwidth]{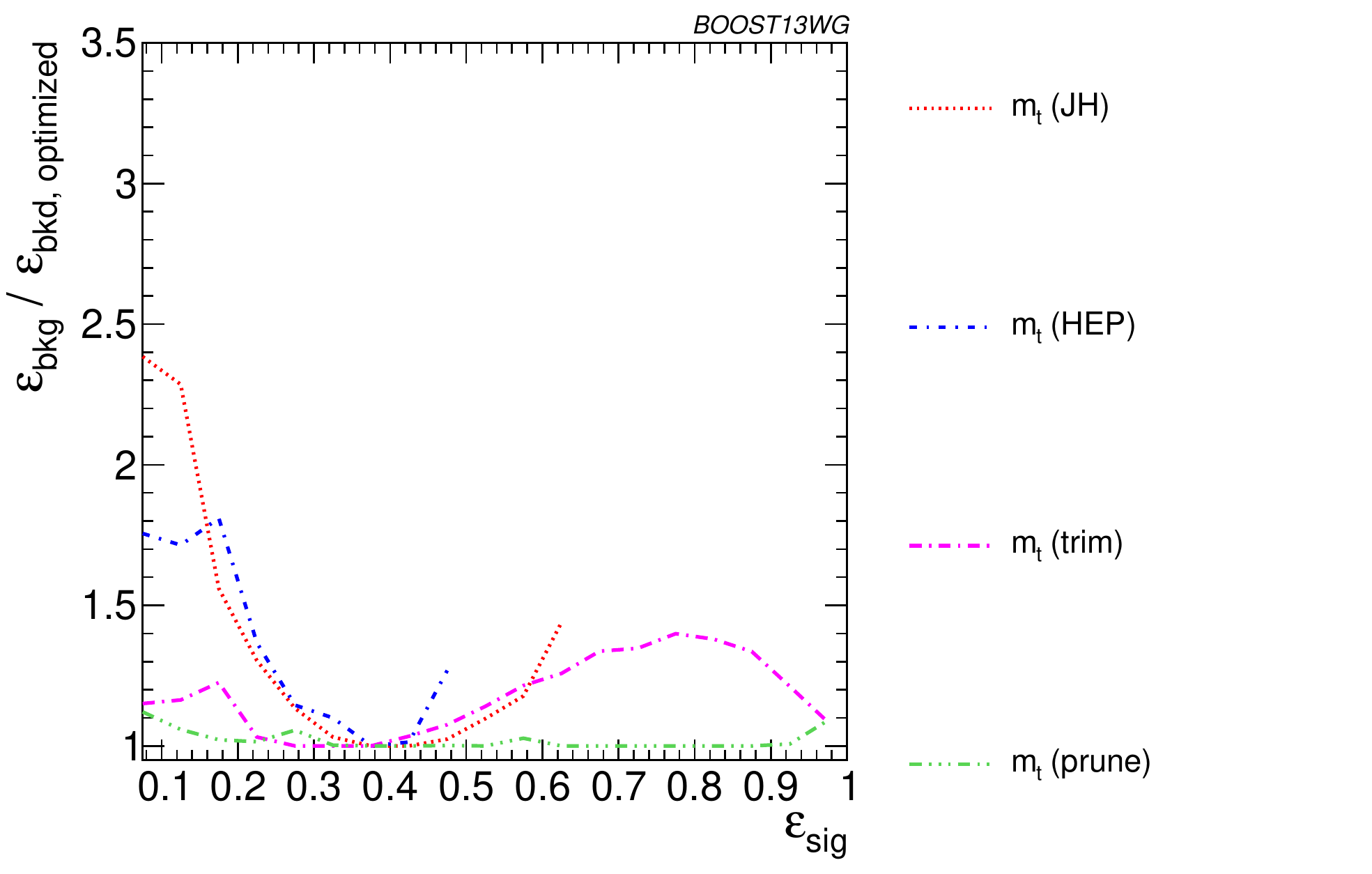}
\caption{Comparison of  top-tagging performance with $\topmass$ in the $\pt= 1-1.1$ GeV bin using the anti-\kT, $R=0.8$ algorithm. The inputs for each tagger are optimized for the $\varepsilon_{\rm sig}=0.3-0.35$ bin, and the performance is normalized to the performance using the optimized tagger inputs at each $\varepsilon_{\rm sig}$.}
\label{fig:single_variable_ROC_eps0_35}
\end{figure*}

\begin{figure*}
\centering
\subfigure[Tagger-Groomer comparison]{\includegraphics[width=0.48\textwidth]{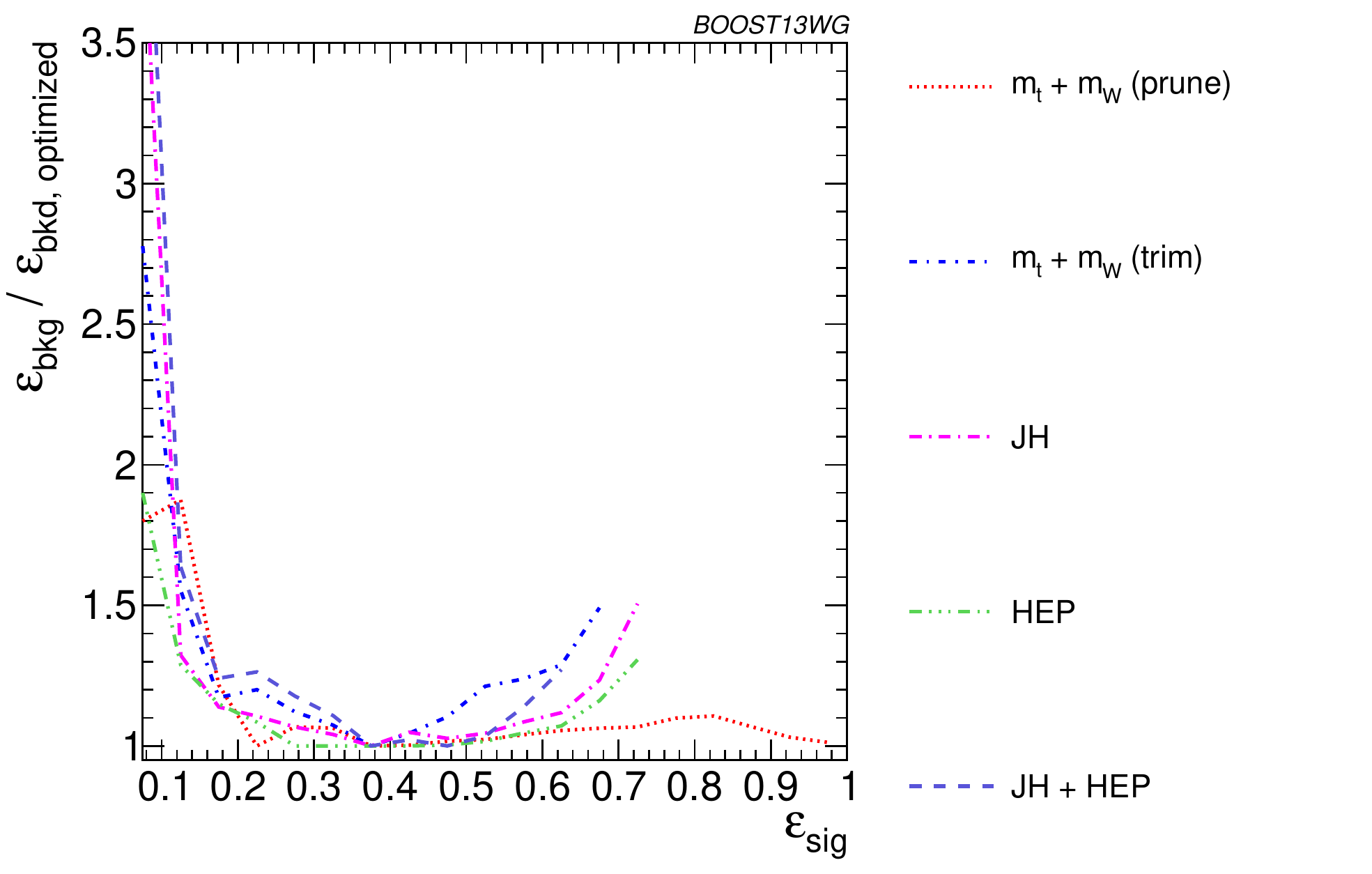}}
\subfigure[HEPTopTagger + Shape]{\includegraphics[width=0.48\textwidth]{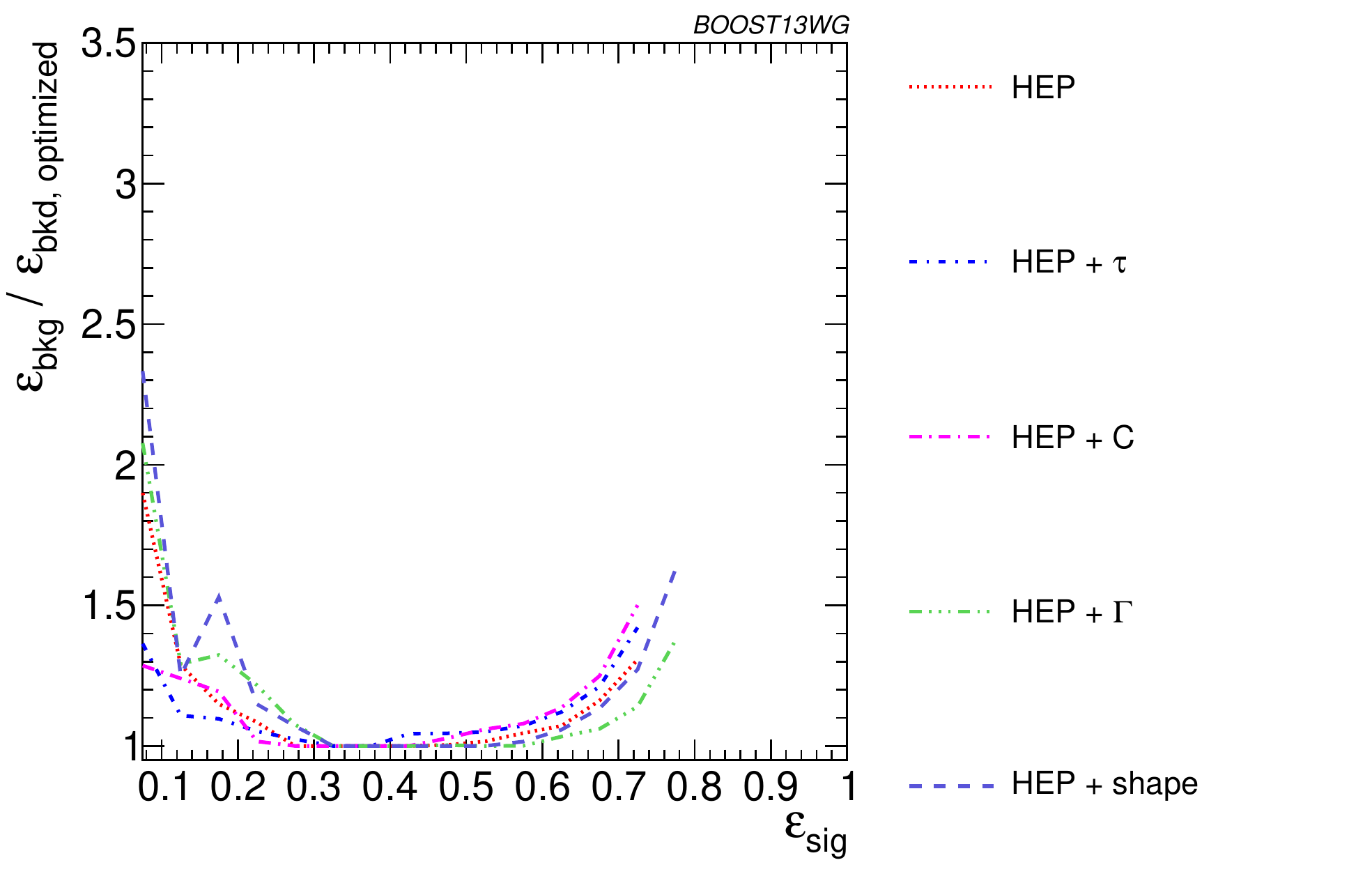}}
\subfigure[Johns Hopkins Tagger + shape]{\includegraphics[width=0.48\textwidth]{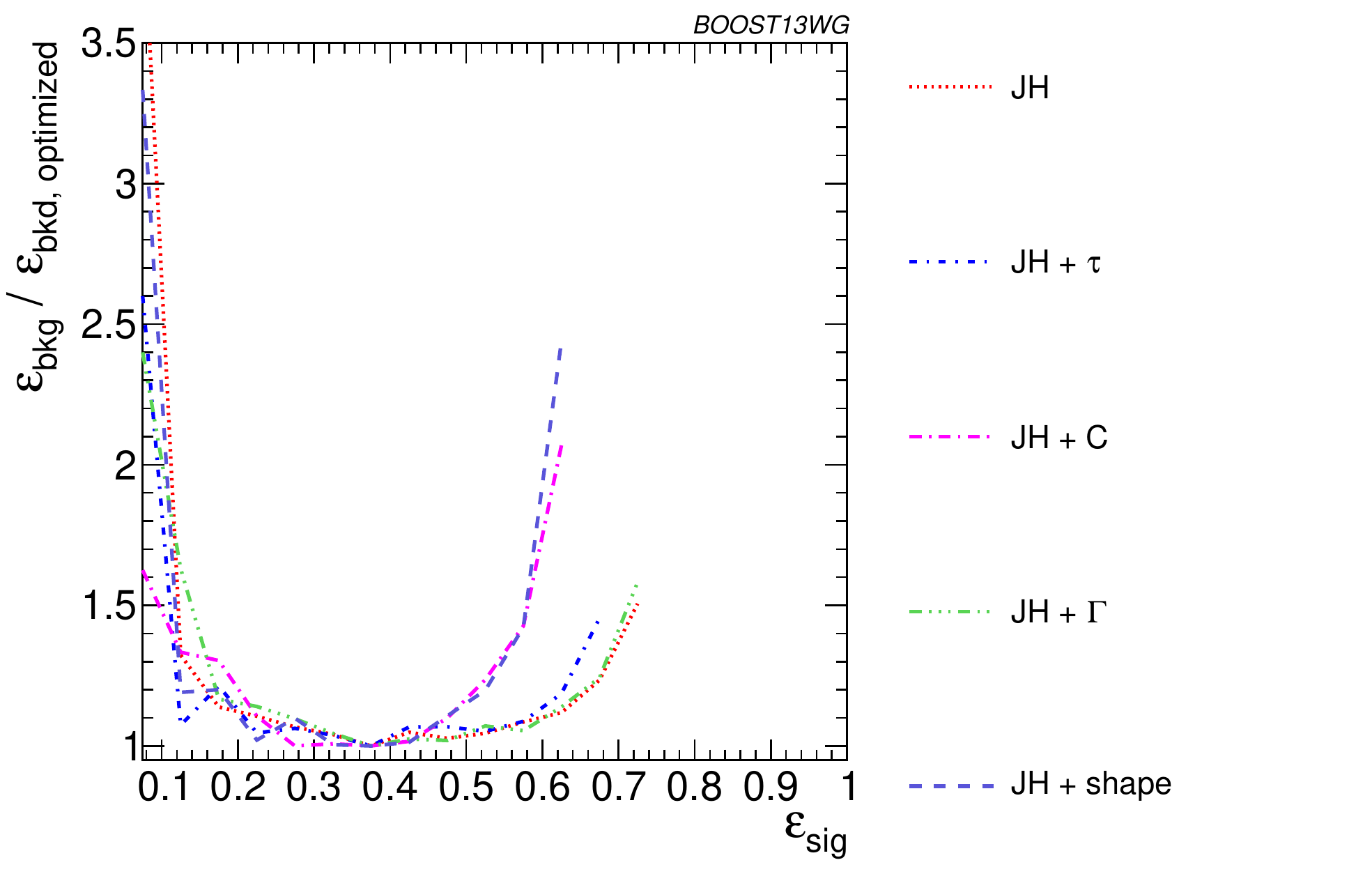}}
\subfigure[HEP vs.~JH comparison (incl. shape)]{\includegraphics[width=0.48\textwidth]{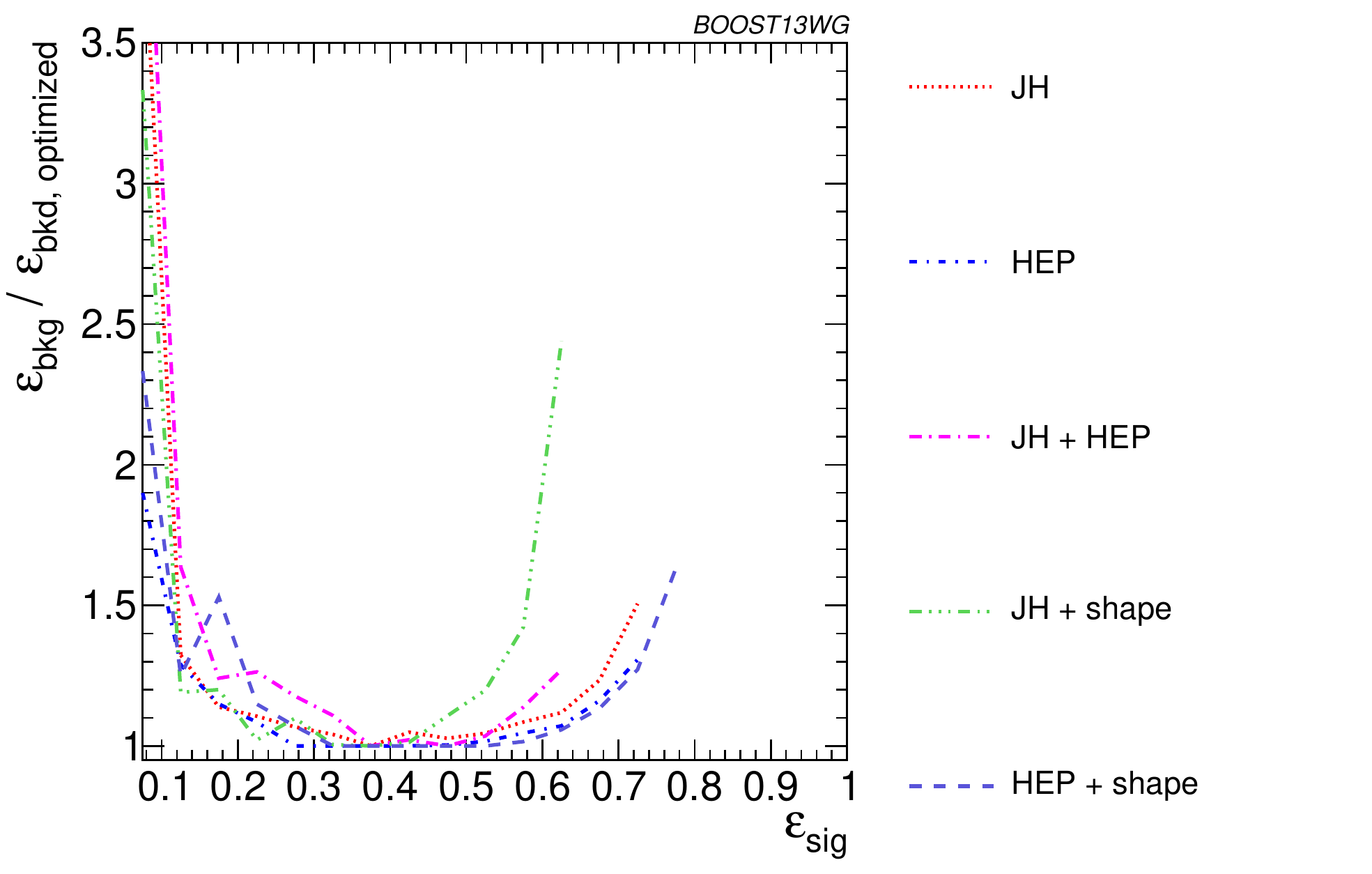}}
\subfigure[Pruning + Shape]{\includegraphics[width=0.48\textwidth]{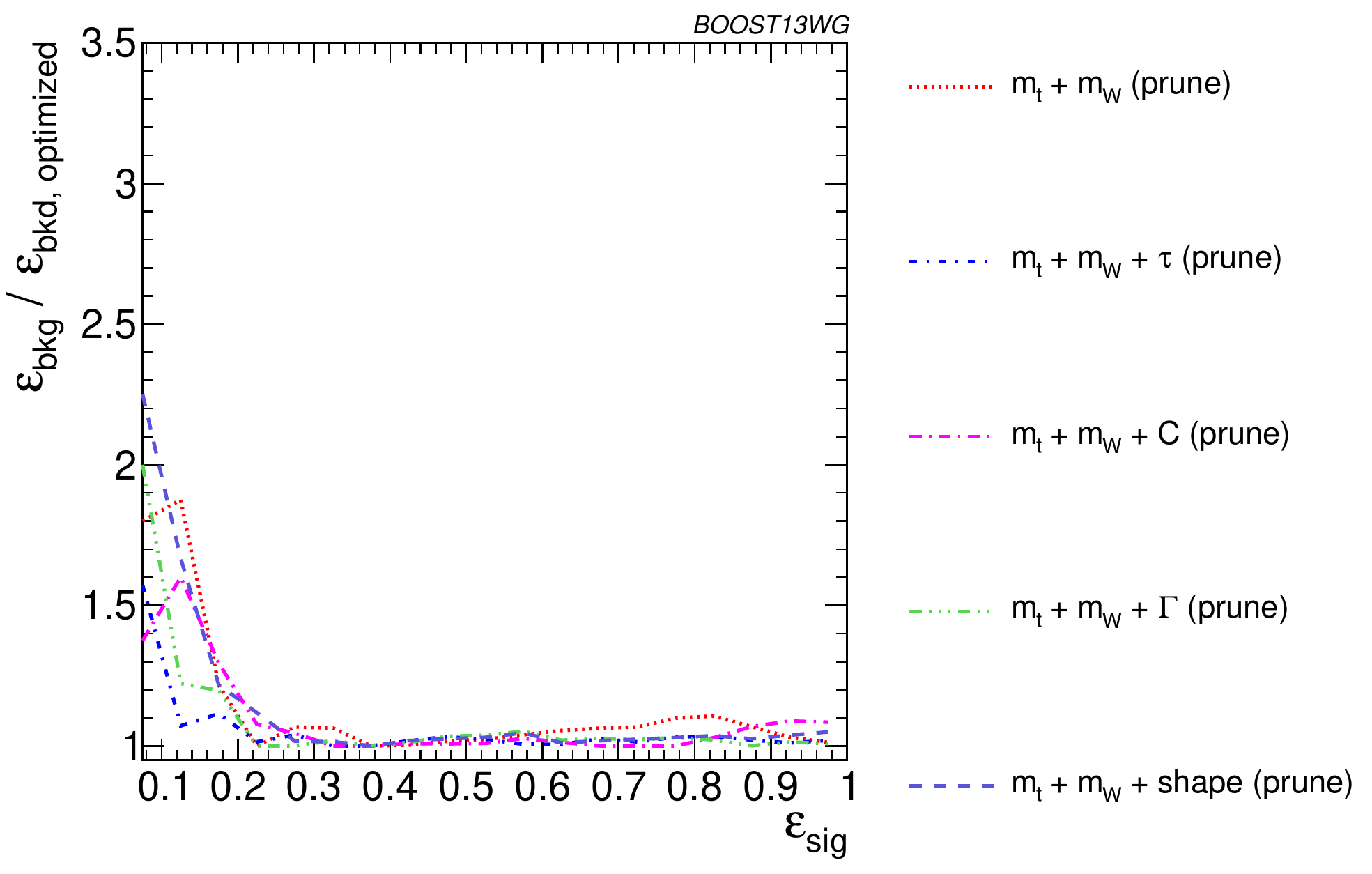}}
\subfigure[Trimming + Shape]{\includegraphics[width=0.48\textwidth]{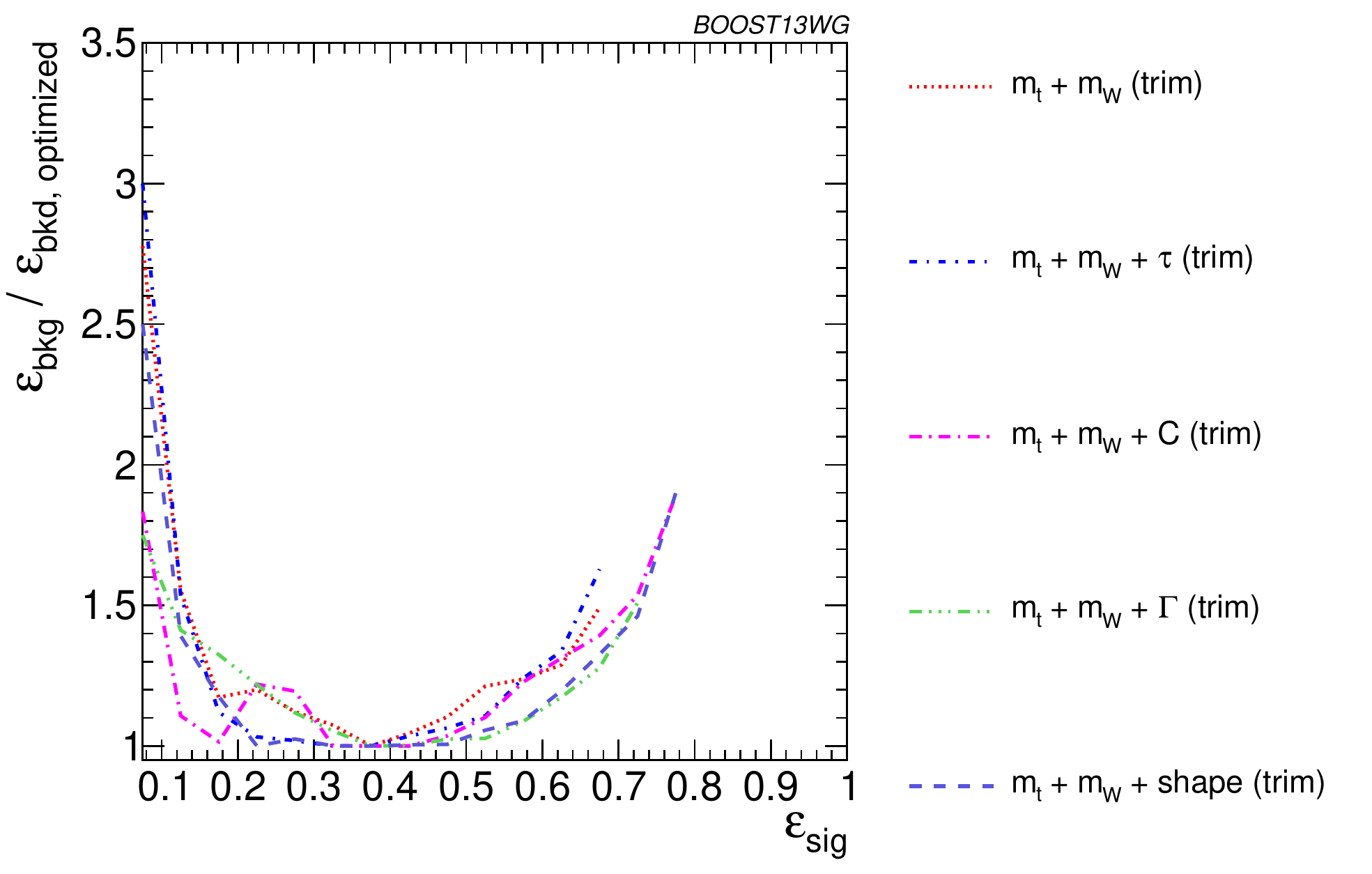}}
\subfigure[Trim vs.~Prune comparison (incl. shape)]{\includegraphics[width=0.48\textwidth]{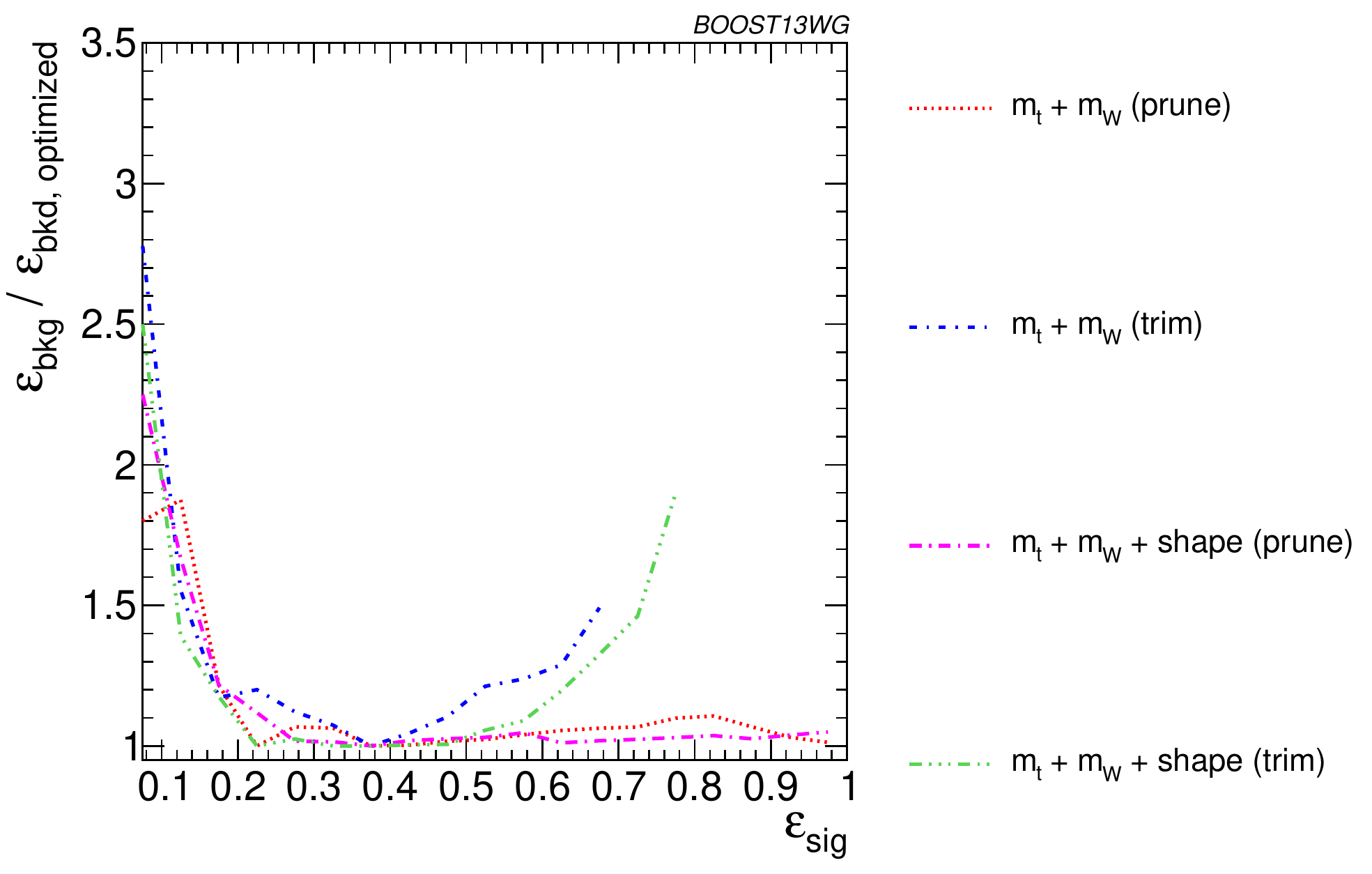}}
\subfigure[Comparison of all Tagger+Shape]{\includegraphics[width=0.48\textwidth]{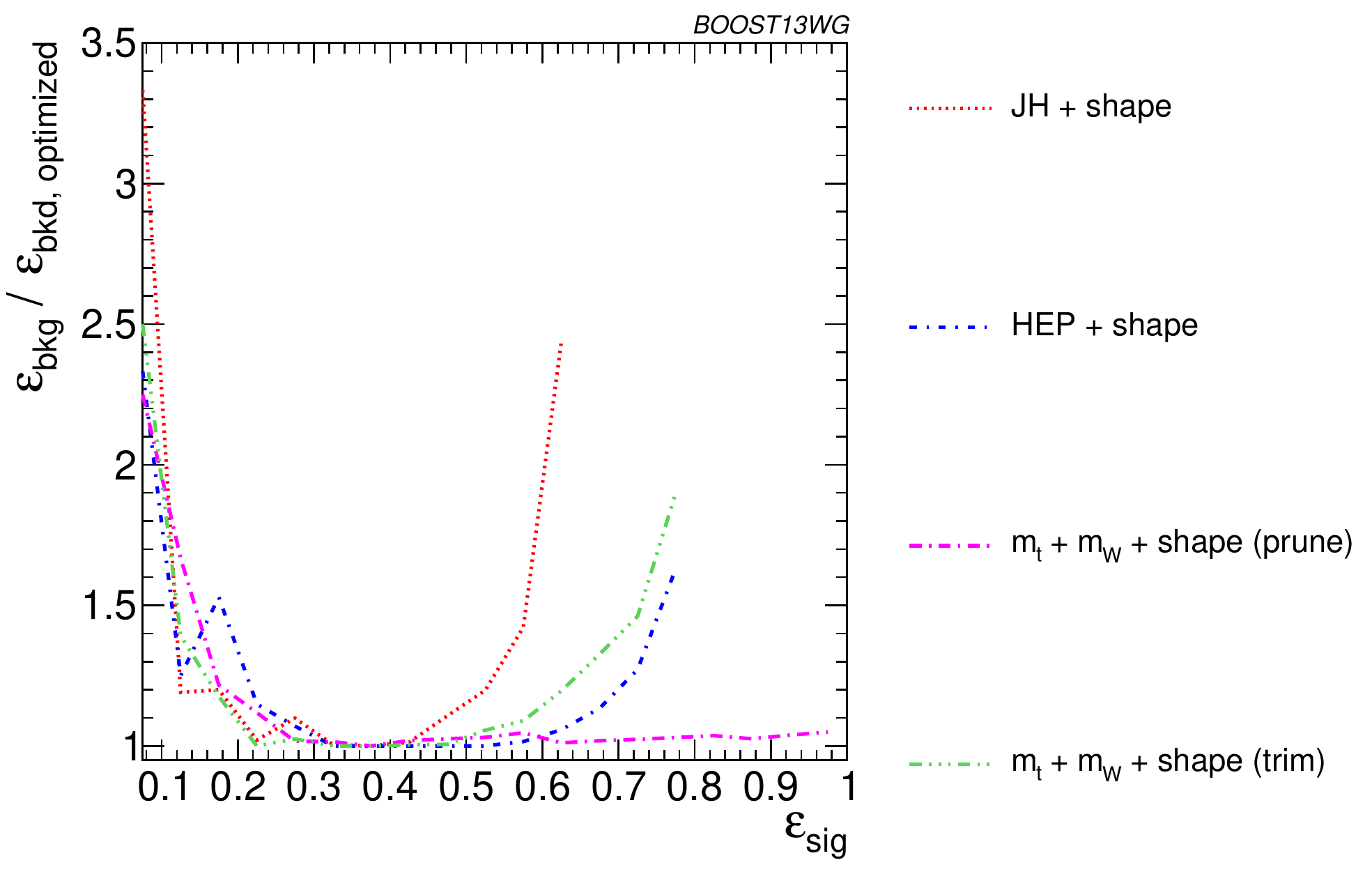}}
\caption{The BDT combinations in the $\pt = 1-1.1$ TeV bin using the anti-\kT $R=0.8$ algorithm. Taggers are combined with the following shape variables: $\tau_{21}^{\beta=1}+\tau_{32}^{\beta=1}$, $C_{2}^{\beta=1}+C_{3}^{\beta=1}$, $\Gamma_{\rm Qjet}$, and all of the above (denoted ``shape''). The inputs for each tagger are optimized for the $\varepsilon_{\rm sig}=0.3-0.35$ bin, and the performance is normalized to the performance using the optimized tagger inputs at each $\varepsilon_{\rm sig}$.}
\label{fig:pt1000_allcompare_AKt_R08_eps0_35}
\end{figure*}

The performance of each tagger, normalized to its performance optimized in each signal efficiency bin, is shown in Figure~\ref{fig:single_variable_ROC_eps0_35} for cuts on the top mass and $W$ mass, and in Figure~\ref{fig:pt1000_allcompare_AKt_R08_eps0_35} for BDT combinations of tagger outputs and shape variables. In both plots, it is apparent that optimizing the taggers in the $\varepsilon_{\rm sig}=0.3$-$0.35$ efficiency bin gives comparable performance over efficiencies ranging from 0.2-0.5, although performance degrades at substantially different signal efficiencies. Pruning appears to give especially robust signal-background discrimination without re-optimization, most likely due to the fact that there are no absolute distance or \pt scales that appear in the algorithm. Figures~\ref{fig:single_variable_ROC_eps0_35} and~\ref{fig:pt1000_allcompare_AKt_R08_eps0_35} suggest that, while optimization at all signal efficiencies is a useful tool for comparing different algorithms, it is not crucial to achieve good top-tagging performance in experiments.

\subsection{Conclusions}

We have studied the performance of various jet substructure variables, groomed masses, and top taggers to study the performance of top tagging with different \pt and jet radius parameters. At each \pt, $R$, and signal efficiency working point, we optimize the parameters for those variables with tuneable inputs. Overall, we have found that these techniques, individually and in combination, continue to perform well at high $\pt$, at least at the particle-level, which is important for future LHC running. In general, the John Hopkins tagger performs best, while jet grooming algorithms under-perform relative to the best top taggers due to the lack of an optimized $W$-identification step. 
Tagger performance can be improved by a further factor of 2-4 through combination with jet substructure variables such as $\tau_{32}$, $C_3$, and $\Gamma_{\rm Qjet}$. When combined with jet substructure variables, the performance of various groomers and taggers becomes very comparable, suggesting that, taken together, the variables studied are sensitive to nearly all of the physical differences between top and QCD jets at particle-level. A small improvement is also found by combining the Johns Hopkins and HEPTopTaggers, indicating that different taggers are not fully correlated. The degree to which these findings continue to hold under more realistic pile-up and detector configurations is, however, not addressed in this analysis and left to future study.

Comparing results at different \pt and $R$, top-tagging performance is generally better at smaller $R$ due to less contamination from uncorrelated radiation. Similarly, most variables perform better at larger $\pt$ due to the higher degree of collimation of radiation. Some variables fare worse at higher $\pt$, such as the $N$-subjettiness ratio $\tau_{32}$ and the Qjet mass volatility $\Gamma_{\rm Qjet}$, as higher-$\pt$ QCD jets have more and harder emissions that fake the top-jet substructure. The standard HEPTopTagger algorithm is also worse at high $\pt$ due to the tendency of the tagger to shape backgrounds around the top mass. This is unsurprising, given that the HepTopTagger was specifically designed for a lower \pT range than that considered here; recently proposed updates may improve performance at high \pt and $R$  \cite{Anders:2013oga,Kasieczka:2015jma}. The \pt- and $R$-dependence of the multivariable combinations is dominated by the \pt- and $R$-dependence of the top mass reconstruction component of the tagger/groomer.

Finally, we consider the performance of various tagger and jet substructure variable combinations under the more realistic assumption that the input parameters are only optimized at a single $\pt$, $R$, or signal efficiency, and then the same inputs are used at other working points. Remarkably, the performance of all variables is typically within a factor of 2 of the fully optimized inputs, suggesting that while optimization can lead to substantial gains in performance, the general behavior found in the fully optimized analyses extends to more general applications of each variable. In particular, the performance of pruning typically varies the least when comparing sub-optimal working points to the fully optimized tagger due to the scale-invariant nature of the pruning algorithm.

%% file: conclusions.tex
Furthering our understanding of jet substructure is crucial to enhancing the prospects for the discovery of new physical processes at Run II of the LHC. In this report we have studied the performance of jet substructure techniques over a wide range of kinematic regimes that will be encountered in Run II of the LHC. The performance of observables and their correlations have been studied by combining the variables into Boosted Decision Tree (BDT) discriminants, and comparing the background rejection power of this discriminant to the rejection power achieved by the individual variables. The performance of ``all variables'' BDT discriminants has also been investigated, to understand the potential of the ``ultimate'' tagger where ``all'' available particle-level information (at least, all of that provided by the variables considered) is used.

We focused on the discrimination of quark jets from gluon jets, and the discrimination of boosted $W$ bosons and top quarks from the QCD backgrounds. For each, we have identified the best-performing jet substructure observables at particle level, both individually and in combination with other observables. In doing so, we have also provided a physical picture of why certain sets of observables are (un)correlated. Additionally, we have investigated how the performance of jet substructure observables varies with $R$ and \pt, identifying observables that are particularly robust against or susceptible to these changes. In the case of $q/g$ tagging, it seems that the ideal performance can be nearly achieved by combining the most powerful discriminant, the number of constituents of a jet, with just one other variable, $C_1^{\beta =1}$ (or $\tau_1^{\beta=1}$). Many of the other variables considered are highly correlated and provide little additional discrimination. For both top and $W$ tagging, the groomed mass is a very important discriminating variable, but one that can be substantially improved in combination with other variables. There is clearly a rich and complex relationship between the variables considered for $W$ and top tagging, and the performance and correlations between these variables can change considerably with changing jet \pt and $R$. In the case of $W$ tagging, even after combining groomed mass with two other substructure observables, we are still some way short of the ultimate tagger performance, indicating the complexity of the information available, and the complementarity between the observables considered. In the case of top tagging, we have shown that the performance of both the John Hopkins and HEPTopTagger can be improved when their outputs are combined with substructure observables such as $\tau_{32}$ and $C_{3}$, and that the performance of a discriminant built from groomed mass information plus substructure observables is very comparable to the performance of the taggers.  We have optimized the top taggers for particular values of \pt, $R$, and signal efficiency, and studied their performance at other working points. We have found that the performance of observables remains within at most a factor of two of the optimized value, suggesting that the performance of jet substructure observables is not significantly degraded when tagger parameters are only optimized for a few select benchmark points.

In all of $q/g$, $W$ and top tagging, we have observed that the tagging performance improves with increasing \pt. However, whereas for $q/g$ and top tagging the performance improves with decreasing $R$ (for the range of $R$ considered here), the dependence on $R$ for $W$ tagging is more complex, with a peak performance at $R=0.8$ for each \pt bin considered. 

Our analyses were performed with ideal detector and pile-up conditions in order to most clearly elucidate the underlying physical scaling with \pt and $R$. At higher boosts, detector resolution effects will become more important, and with the higher pile-up expected at Run II of the LHC, pile-up mitigation will be crucial for future jet substructure studies. Future studies will be needed to determine which of the observables we have studied are most robust against pile-up and detector effects, and our analyses suggest particularly useful combinations of observables to consider in such studies. 

At the new energy frontier of Run II of the LHC, boosted jet substructure techniques will be more central to our searches for new physics than ever before. By achieving a deeper understanding of the underlying structure of quark, gluon, $W$ and top-initiated jets, as well as the relations between observables sensitive to their respective structures, it is hoped that more sophisticated analyses can be performed that will maximally extend the reach for new physics.\\

{\bf Acknowledgments:}~We thank the Department of Physics at the University of Arizona for hosting 
and providing support for the BOOST 2013 workshop, and the US Department of Energy for their support of the workshop.  We especially thank Vivian Knight (University of Arizona)
for her help with the organization of the of the workshop.  
We also thank Prof. J. Boelts of the University of Arizona School of Art VisCom program and his Fall 2012 ART 465 class for organizing the design competition for the workshop 
poster. In particular, we thank the winner of the competition, Ms. Hallie Bolonkin, for creating the final design.